\documentclass[11pt,a4paper,twoside]{report}

\usepackage[latin1]{inputenc} 
\usepackage[dvips]{graphicx} 
\usepackage{a4wide}
\usepackage{latexsym}
\usepackage{amsmath}
\usepackage{amssymb}

\title{\vspace{-5.5cm}
\begin{flushleft}
\includegraphics[width=2.5cm]{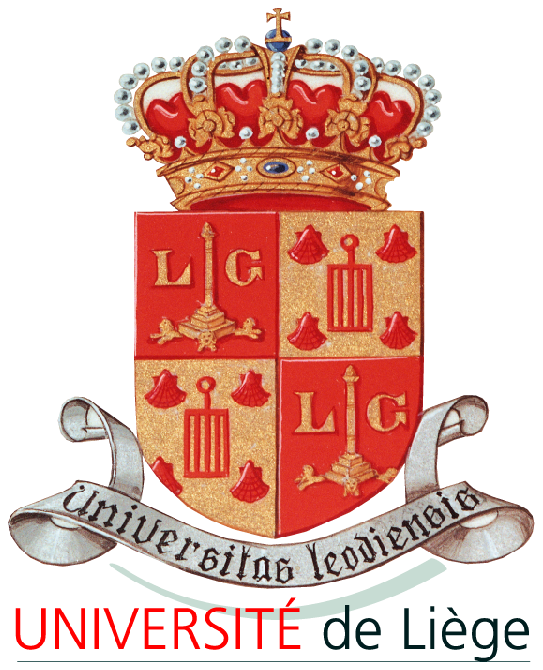}\\
\small{University of Liège\\
Faculty of Sciences\\
AGO Department\\
IFPA}
\end{flushleft} \vspace{-4cm}
\begin{flushright}
Academic year 2007-2008
\end{flushright}
\vspace{8cm}{\huge{\textbf{Exotic and Non-Exotic Baryon Properties\\
on the Light Cone}}}}

\author{\huge{Cédric Lorcé}\\\\
\emph{E-mail: C.Lorce@ulg.ac.be}\\\\
\large{Supervisor: Maxim Polyakov}}
\date{
\vspace{6.5cm}
\large{Thesis presented in fulfillment of\\
      the requirements for the Degree of\\
      Doctor of Science}}

\newcommand{\ud}{\mathrm{d}}

\newcommand{\uM}{\mathcal{M}}
\newcommand{\umo}{m_\mathcal{O}}
\newcommand{\umd}{M_\mathcal{D}}
\newcommand{\uZ}{\mathcal{Z}}
\newcommand{\uQcal}{\mathcal{Q}}
\newcommand{\uN}{\mathcal{N}}

\newcommand{\pslash}{p\!\!\!/}

\newcommand{\qslash}{q\!\!\!/}

\newcommand{\utau}{\boldsymbol{\tau}}
\newcommand{\usigma}{\boldsymbol{\sigma}}

\newcommand{\upi}{\boldsymbol{\pi}}
\newcommand{\uPi}{\boldsymbol{\Pi}}

\newcommand{\ux}{\mathbf{x}}

\newcommand{\ur}{\mathbf{r}}
\newcommand{\un}{\mathbf{n}}
\newcommand{\up}{\mathbf{p}}
\newcommand{\uP}{\mathbf{P}}
\newcommand{\uq}{\mathbf{q}}
\newcommand{\uQ}{\mathbf{Q}}
\newcommand{\uk}{\mathbf{k}}

\newcommand{\mup}{G_M^{(p)u}}
\newcommand{\mdp}{G_M^{(p)d}}
\newcommand{\msp}{G_M^{(p)s}}

\begin{document}
\maketitle  \pagestyle{empty} \cleardoublepage \rule{2cm}{0pt}
\begin{minipage}[]{15cm}
\rule{0pt}{1cm}
\subsection*{\huge{Acknowledgment}}

\rule{0pt}{0.5cm}

First I would like to warmly thank my supervisor M. Polyakov for
accepting me as a Ph.D. student, for his faith in me and especially
for his kindness. It was a genuine pleasure to work and discuss with
him. Thanks to him I have learned a lot on physics and the
scientific world and met many very interesting people. Next I would
like to express my gratitude towards J. Cugnon for his help, advice
and support in all the necessary steps linked to the present thesis.
I am especially grateful to K. Goeke and M. Hacke for the two years
passed in Ruhr Universität Bochum (Germany) where the biggest part
of the present thesis has been achieved. I will forever remember the
welcoming people of Theoretische Physik II group. I have
particularly appreciated everyday K. Goeke's good humour. He was
always whistling, laughing or in raptures about other physicists'
research. Living in Germany was a unique and extraordinary
experience for me. That is the reason why I address this sentence to
all people met there: Ich danke euch für alles. I would also like to
thank D. Diakonov for his help, patience and disponibility. Beside
being a great physicist he is also a man full of humanity. The other
part of the present thesis has been achieved in Liège University. I
am thankful to members of the IFPA group. I am in fact also indebted
to all my teachers and professors for the share of their knowledge
and passion.
\newline

A thesis work does not only consist of passing all the days in front
of a computer, papers or drafts. Discussions about physics,
pataphysics and metaphysics are also essential to proceed. So I am
thankful for all the entertaining, pleasant, interesting, serious
and less serious moments passed with friends and other people met. I
cannot of course cite all of them: Alice D., Aline D., Christophe
B., Daniel H., Danielle R., Delphine D., Denis F., Frédéric K.,
Geoffrey M., Ghil-Soek Y., Grégory A., Jacqueline M., Jean-Paul M.,
Jérémie G., Kirill S., Lionel H., Luc L., Michel W., Nina G.,
Pauline M., Quentin J., Renaud V., Sophie P., Stéphane T., Thibaut
M., Tim L., Virginie C., Virgine D., Xavier C. and many others.
\newline

Finally I would like to express my gratitude to my family: my
mother, my sister and my father for their love, presence, help and
for giving me the possibility to reach my dreams. I have also
special thanks for Pierrot Di Marco and his two sons Christophe and
Sébastien. Thanks for all what they did for us. I am really happy
and proud to consider them as genuine part of our family.
\end{minipage}
\cleardoublepage \pagestyle{headings} \tableofcontents
\newpage\thispagestyle{empty}\cleardoublepage
\chapter{Introduction}

\section{Hadron structure and QCD}

One of the main objectives in Physics is to understand the structure
and properties of matter. The dream would be to find the ultimate
constituents with which one can build the whole universe. These
ultimate constituents have, by definition, no internal structure and
are called fundamental particles. The properties of any matter
should in principle be deduced and/or explained starting from the
fundamental particles and their dynamics.

The question of hadron structure is a rather old one. Fermi-Yang
\cite{FermiYang} and Sakata models \cite{Sakata} might be considered
as first studies of this question. Both models considered the proton
as a fundamental particle. Later, thanks to Deep Inelastic
Scattering (DIS) experiments \cite{DIS} at SLAC, it was realized
that the nucleon was not a fundamental particle. Indeed, the
observation in DIS of scaling phenomenon\footnote{Electron and muon
are ideal probes to study the internal nucleon structure. The
virtual photon emitted by the lepton interacts with the target
nucleon. The cross section of the process is related to two
unpolarized $F_1$ and $F_2$ and two longitudinally polarized
structure functions $g_1$ and $g_2$. They depend in general on two
kinematical variables $Q^2=-q^2$ and $x=Q^2/2p.q$ where $q$ is the
virtual photon momentum and $p$ is the nucleon momentum. These
structure functions provide important clues to internal nucleon
structure \cite{Nuclstruct}. Bj{\o}rken scaling phenomenon refers to
the fact that these structure functions are almost independent of
$Q^2$, \emph{i.e.} independent of the resolution. This indicates
that the photon scatters on structureless objects inside the proton.
The cross section is calculated by the lepton scattering on
individual quarks with incoherent impulse approximation which is
supposed to be valid at large $Q^2$ in the sense that virtual-photon
interaction time with a quark is fairly small compared with the
interaction time among quarks.} predicted by Bj{\o}rken
\cite{Bjorkenpred} was the first direct evidence for the existence
of point-like constituents in the nucleon. These point-like
constituents were found to be charged spin-1/2 particles and were
called \emph{partons}. A simple and intuitive picture for explaining
the scaling behavior is the parton model proposed by Feynman
\cite{Feynman} in which the electron-nucleon DIS is described as an
incoherent sum of elastic electron-parton scattering. The nature of
these partons was however not specified. The uncalculable nucleon
structure functions can then be expressed in terms of Parton
Distribution Functions (PDF).

On the spectroscopic side, in view of the huge number of observed
hadrons, Gell-mann \cite{GellMann} and Zweig \cite{Zweig} proposed
in 1964 the quark model of hadrons. In this model hadrons can be
grouped together in multiplets of a flavor $SU(3)$ symmetry. The
(non-trivial) fundamental representation has dimension 3 and its
elements are called quarks. Baryons then appear as systems of three
strongly bound quarks and mesons as systems of a quark strongly
bound to an antiquark.

Partons observed in DIS were initially identified with quarks. They
have however a qualitative different behavior. While quarks are
strongly bound in hadrons, partons appeared in DIS as almost free
particles. This difference did not hurt much at that time. On the
contrary, the phenomenological success of $SU(6)$ Naive Quark Model
(NQM) in explaining hadron properties and the evidence for the
existence of partons inside hadrons motivated the development of a
new theory of strong interaction in 1973, namely the Quantum
ChromoDynamics (QCD) \cite{QCD}. The weakly interacting partons
revealed in DIS and the fact that no free quark has been discovered
in all the experiments performed are explained in QCD thanks to the
asymptotic freedom and confinement properties respectively. The weak
interacting high energy processes can be calculated and tested
thanks to the asymptotic freedom property and gave a strong support
to establish QCD as the correct theory of strong interaction.

By solving QCD one could thus in principle understand the structure
as well as low-energy interactions and properties of all hadrons in
terms of quarks and gluons. Unfortunately they cannot be easily
calculated since the confinement property of QCD forbids an obvious
and standard perturbative approach. The proposed way out is to study
QCD numerically on a lattice\footnote{QCD is in fact studied in its
Euclidean version on a lattice, obtained after a Wick rotation of
space-time.}. Many results have been obtained but are not quite
reliable because of many numerical uncertainties due to lattice size
and spacing, unprecise extrapolations to physical masses, \ldots One
of the major problems is of course the required computation time.

\section{Models and degrees of freedom}

Due to the huge difficulty encountered in solving QCD in the
non-perturbative regime many models have been developed to
understand and predict as far as possible hadron properties. For
example, perturbative QCD can predict only the $Q^2$ dependence of
PDF whereas it can say nothing about the PDF at a prescribed energy
scale. These PDF are thus expected to be given by a low-energy model
of QCD. The present models are more or less inspired from QCD and
differ by the effective degrees of freedom they emphasize
\cite{Models,Witten}. While QCD plays with quarks and gluons as
fundamental degrees of freedom, they could be inappropriate for a
low-energy description.

\subsection{Constituent quark models}

The Naive Quark Model (NQM) is among the most successful models in
explaining hadron properties \cite{hadronprop} and hadron
interactions \cite{hadronint}. This is also the most popular and
intuitive picture of hadron internal structure. The most striking
feature of NQM is that it gives a very simple but quite successful
explanation of the static baryon properties, \emph{e.g.} baryon
spectroscopy and magnetic moments, by means of effective
\emph{constituent} quarks and nothing else. These constituent quarks
are needed in hadron spectroscopy but have mass much larger than
\emph{current} quarks revealed in DIS experiments. The relation
between constituent and current quarks can be considered as the holy
grail of hadron physics. In NQM constituent quarks are
non-relativistic (they are all considered to be in the $s$ state)
and the baryon spin-flavor structure is given by $SU(6)$ symmetry.
Many variations of NQM exist and are collectively called Constituent
Quark Models (CQM). All these models, based on the effective degrees
of freedom of valence constituent quarks and on $SU(6)$ spin-flavor
symmetry, also contain a long-range linear confining potential and a
$SU(6)$-breaking term like One-Gluon-Exchange (OGE),
Goldstone-Boson-Exchange (GBE) or even Instanton-Induced (II)
interaction.

While they are able to give good results for the static properties
of the hadrons (spectrum, magnetic moments), they all fail to
reproduce the dynamic ones, like electromagnetic transition form
factors at low $Q^2$. A systematic lack of strength is observed at
low $Q^2$. This seems to be a problem of degrees of freedom. Indeed,
the region of low $Q^2$ corresponds to high distance, in which the
creation of quark-antiquark pair degrees of freedom has a higher
probability.

\subsection{Quark-antiquark pairs and the nucleon sea}

DIS experiments have shown a large enhancement of the cross sections
at small Bj{\o}rken $x$, the fraction of nucleon momentum carried by
the partons. This is related to the fact that the structure function
$F_2(x)$ approaches a constant value as $x\to 0$ \cite{Limit}. If
the proton consists of only three valence quarks or any
\emph{finite} number of quarks, $F_2(x)$ is expected to vanish as
$x\to 0$. It was then realized that valence quarks alone are not
sufficient. Bj{\o}rken and Pascho \cite{BP} therefore assumed that
the nucleon consists of three quarks in a background of an infinite
number of quark-antiquark pairs. Kuti and Weisskopf \cite{Kuti}
further included gluons among the constituents of nucleons in order
to account for the missing momentum not carried by the quarks and
antiquarks.

Quark-antiquark pairs are very important in the nucleon. This is in
sharp contrast with the atomic system where particle-antiparticle
pairs play a relatively minor role. In strong interaction,
quark-antiquark pairs are readily produced as a result of the
relatively large magnitude of the strong coupling constant
$\alpha_S$. In CQM they are however not considered as degrees of
freedom. Constituent quarks can be viewed as non-perturbative
objects, current quarks dressed by a cloud of quark-antiquark pairs
and gluons. This picture is however not realistic since CQM, which
are supposed to model QCD at low energy, completely forget a very
important approximate symmetry of QCD, namely \emph{chiral}
symmetry.

\subsection{Chiral symmetry of QCD}

The six observed quark flavors can be separated into light ($u,d,s$)
and heavy flavors ($c,b,t$). As the masses of heavy and light quarks
are separated by the same scale ($\simeq 1$ GeV) as the perturbative
and non-perturbative regime, one may expect different physics
associated with those two kinds of quarks. It appeared that physics
of light quarks is governed by chiral symmetry. Since we are
interested in this thesis only in light baryons, we will completely
forget about the heavy flavors. Light baryons being composed of
light quarks, chiral symmetry is expected to be crucial in the study
of (light) baryon properties.

If the masses of light quarks are put to zero, then QCD Lagrangian
becomes invariant under $SU(3)_R\times SU(3)_L$, the chiral flavor
group. This symmetry implies that left- and right-handed quarks
independently undergo a chiral rotation under the action of the
group. According to Noether's theorem \cite{Noether} every
continuous symmetry of the Lagrangian is associated to a
four-current whose four-divergence vanishes. This in turn implies a
conserved charge as a constant of motion. There are consequently
sixteen conserved charges: eight vector and eight (pseudoscalar)
chiral charges $Q_5^a$. One has
\begin{equation}
\left[Q_5^a,H_\textrm{QCD}\right]=0
\end{equation}
meaning that the chiral charges are conserved and that QCD
Hamiltonian $H_\textrm{QCD}$ is chirally invariant. Under parity
transformation axial charges change sign $Q_5^a\to -Q_5^a$. One
expects thus (nearly) degenerate parity doublets in nature which do
not exist empirically. The splitting in mass between particles of
opposite parities is too large to be explained by the small current
quark masses which break \emph{explicitly} chiral symmetry
($m_u\simeq 4$ MeV, $m_d\simeq 7$ MeV and $m_s\simeq 150$ MeV).

The only explanation is that chiral symmetry is \emph{spontaneously}
broken. This means that the QCD Hamiltonian is invariant under
chiral transformations whereas QCD ground state (\emph{i.e.} the
vacuum $|\Omega\rangle$) is \emph{not} chirally invariant
$Q_5^a|\Omega\rangle\neq 0$. For this reason, there must exist a
non-vanishing vacuum expectation value (VEV), the chiral or quark
condensate
\begin{equation}
\langle\bar\psi\psi\rangle=\langle\bar\psi_R\psi_L+\bar\psi_L\psi_R\rangle\simeq-(250\textrm{
MeV})^3
\end{equation}
at the scale of a few hundred MeV. This condensate is not chirally
invariant since it mixes left (L) and right (R) components and
therefore serves as an order parameter of the symmetry breaking.

Goldstone theorem \cite{Goldstone} states that to any spontaneously
broken symmetry generator is associated a massless boson with the
quantum numbers of this generator. Since we have eight spontaneously
broken chiral generators, we can expect that in massless QCD there
should exist an octet of massless pseudoscalar mesons. In real QCD
current quarks have masses and the pseudoscalar mesons are expected
to be also massive but relatively light. These Goldstone bosons are
identified to the lightest meson octet ($\pi^0,\pi^\pm,K^0,\bar
K^0,K^\pm,\eta$).

Spontaneous Chiral Symmetry Breaking (SCSB) implies thus that QCD
vacuum is non-trivial: it must contain quark-antiquark pairs with
spins and momenta aligned in a way consistent with vacuum quantum
numbers. It also implies that a massless quark develops a non-zero
dynamical mass in this non-trivial vacuum. This mass depends in
general on the momentum $p$. At small momentum it can be estimated
to one half of the $\rho$ meson mass or one third of the nucleon
mass $M(0)\approx 350$ MeV. Constituent quarks can then be seen as
current quarks dressed by the mechanism of SCSB explaining the
origin of 93\% of light baryon masses. Let us also emphasize another
important consequence of SCSB, the fact that quarks get a strong
coupling with pions $g_{\pi qq}(0)= M(0)/F_\pi\simeq 4$ which is
roughly one third of the pion-nucleon coupling constant $g_{\pi
NN}\simeq 13.3$.

Let us stress that chiral symmetry has nothing to say about the
mechanism of confinement which is presumably a totally different
story. This is reflected in the instanton model of QCD vacuum
\cite{Instantonvac} which explains many facts of low-energy hadronic
physics but is known not to yield confinement. It is therefore
possible that confinement is not particularly relevant for the
understanding of hadron structure.

Application of QCD sum rules \cite{QCDSR} to nucleons pioneered by
B.L. Ioffe \cite{Ioffe2} provided several important lessons. One is
that the physics of nucleons is heavily dominated by effects of the
SCSB. This can be seen by the fact that all Ioffe's formulae for
nucleon observables, including nucleon mass itself, are expressed
through the SCSB order parameter $\langle\bar\psi\psi\rangle$. It is
therefore hopeless to build a realistic theory of the nucleon
without taking into due account the SCSB.

\subsection{Importance of pions in models}

As we have just seen, pions\footnote{We will often use the term
``pions'' to refer in fact to the whole lightest pseudoscalar meson
octet.} or quark-antiquark pairs are required both experimentally
and theoretically. A more realistic picture of the hadron would be a
system of three valence quarks surrounded by a pion cloud. This pion
cloud is in fact also needed from a phenomenological point of view.
Here is a short list of the phenomenological hints supporting the
pion cloud:
\begin{enumerate}
\item The nucleon strong interactions, particularly the long-range
part of the nucleon-nucleon interaction, have been described by
means of meson exchange. The development of a low energy
nucleon-nucleon potential has gone for many years \cite{longrange}
with the long-range part in particular requiring a dominant role for
the pion exchange. There have been attempts to generate this
interaction from QCD-inspired models \cite{QCDinspired} but without
quantitative success \cite{echec}. Meson exchanges are thus needed
to account for medium- and long-range parts of the nucleon-nucleon
interaction.

\item The requirement that the nucleon axial-vector current to be partially
conserved (PCAC) requires the pion to be an active participant in
the nucleon. Employing PCAC one can easily derive the
Goldberger-Treiman relation \cite{GTR}
\begin{equation}
g_A^{(3)}=\frac{F_\pi\,g_{\pi NN}}{M_p}
\end{equation}
where $F_\pi$ is the pion decay constant $F_\pi=92.42\pm 0.26$ MeV,
$g_{\pi NN}$ is the pion-nucleon coupling constant and $M_p$ the
proton mass. This yields to a value for $g_A^{(3)}$ that is $(3.8\pm
2.5)\%$ too high, not inconsistent with what is expected from the
explicit breaking of chiral symmetry. The value of the induced
pseudoscalar from factor $g_p$ is also directly dependent on the
pionic field of the nucleon. The PCAC gives \cite{PCAC}
$g_p^{PCAC}=8.44\pm 0.23$ consistent with the measured value
$g_p^{exp}=8.7\pm 2.9$.

\item Many properties of light hadrons and especially of nucleon
seem to be correctly described only when the pion cloud is taken
into account. Since pions are light they are expected to dominate at
long range, \emph{i.e.} at low $Q^2$. Among these properties, let us
mention the reduction of quark contribution to baryon spin due to a
redistribution of the angular momentum in favor of non-valence
degrees of freedom, the increased value of the magnetic dipole
moment and the non-zero electric quadrupole moment in the $\gamma
N\to \Delta$ transition. These properties and the effects of the
pion cloud will be further emphasized when discussing the results
obtained in the present thesis.
\end{enumerate}
For an overview of the importance of pions in hadrons, see
\emph{e.g.} \cite{Bernstein}. In conclusion, pions or
quark-antiquark pairs are genuine participants in the baryon
structure and properties. We are however left with the problem of
how this pion cloud should be implemented in a model.

\section{Baryon properties and experimental surprises}

After the question concerning the nature of the baryon constituents
and relevant degrees of freedom at a given scale comes the question
of their distribution in the baryon and their individual
contributions to the baryon properties. Without exhausting the set
of questions let us mention the following interesting ones:
\begin{itemize}
\item How many
quarks and antiquarks of a given flavor $f$ do we have in a given
baryon?
\item How is the total baryon spin distributed among its
constituents?
\item Is there any hidden flavor contribution to observables?
\item How large are the relativistic effects?
\item What is the intrinsic shape of a given baryon?
\item Is there any exotic baryon, \emph{i.e.} that cannot be made up of
three quarks only?
\item \ldots
\end{itemize}
NQM has simple answers to these questions. However it turned out
that all these NQM answers were in contradiction with the
experimental observations.

A large part of these questions amounts to study PDF which give the
probability to find a parton, say a quark, inside the baryon with a
given fraction $x$ of the total longitudinal momentum, a given
flavor $f$ and in a given spin/helicity state. PDF are defined in
QCD by the light-cone Fourier transform of field-operator products
\cite{Jaffe}. At the leading twist, \emph{i.e.} leading order in
$Q^{-1}$ or $\mathcal{O} (P^+)$ in the IMF language (representing
the asymptotic freedom domain), only three light-cone quark
correlation functions are required $f_1,g_1,h_1$ for a complete
quark-parton model of the baryon spin structure. $f_1$ is a
spin-average distribution which measures the probability to find a
quark in a baryon independent of its spin orientation, $g_1$ is
chiral-even spin distribution which measures the polarization
asymmetry in a longitudinally polarized baryon and $h_1$ is
chiral-odd spin distribution which measures the polarization
asymmetry in a transversely polarized baryon. First moments of these
distributions correspond to vector, axial and tensor charges
respectively. They encode information on quark distribution, quark
polarization and relativistic effects due to quark motion. These
charges are easily obtained by computing forward baryon matrix
element of the corresponding quark current. Part of the present
thesis has been devoted to compute these charges for all the
lightest baryon multiplets within a fairly realistic and successful
model presented in Chapter 3.

Most of the present unsolved questions concerning baryons in the
low-energy regime can be related to one of the following four
topics: proton spin crisis, strangeness in nucleon and Dirac sea,
shape of baryons and exotic baryons.

\subsection{Proton spin crisis}

High-energy experiments are best suited to answer the question of
spin repartition inside the nucleon because quarks and gluons behave
as (almost) free particles at energy/momentum-scales
$Q\gg\Lambda_\textrm{QCD}$. The predominant role in the development
of understanding the spin structure of nucleons is played by the
deep inelastic leptoproduction processes ($lN\to l'X$ where $X$ is
undetected) because of their unique simplicity. Their significance
has been anticipated by Bj{\o}rken \cite{Bjorken} and others
\cite{Others}.

The nucleon spin can be decomposed as follows
\cite{Protondecomposition}
\begin{equation}
J=\frac{1}{2}\,\Delta\Sigma+L_q+\Delta G+L_G
\end{equation}
where we have on the lhs the spin $J=+1/2$ of a polarized nucleon
state and on the rhs the decomposition in terms of the quark spin
contribution $\Delta\Sigma$, gluon spin contribution $\Delta G$ and
quark and gluon orbital angular momentum contribution $L_q+L_G$. The
quark spin contribution $\Delta\Sigma$ can be further decomposed
into the contributions from the various quark species $\Delta
u+\Delta d+\Delta s+\Delta\bar u+\Delta\bar d+\Delta\bar s$.
Unfortunately the decomposition cannot directly be measured in
experiments. Instead various combinations of these terms appear in
experimental observables. In the NQM which uses only one-body
axial-vector currents one obtains a clear answer, namely
$J=\Delta\Sigma/2=1/2$, \emph{i.e.} the nucleon spin is just the sum
of the three constituent quark spins and nothing else. This has to
be contrasted with the Skyrme model. This model describes a nucleon
as a soliton of the pion field in the limit of a large number of
colors $N_C\to\infty$ and concludes that the nucleon spin is due to
orbital momentum $\Delta\Sigma=0$ \cite{Skyrmeresult}.

The EMC experiment \cite{EMC} challenged NQM since it showed that
only one third of the proton spin is due to the quark spins. One may
wonder why this is a problem, given that the nucleon mass is not
carried by the quark masses, why should the nucleon spin be carried
by the quark spins? The answer \cite{QCDspin} is in fact that what
is surprising is the violation of the OZI rule\footnote{Okubo, Zweig
and Iizuka \cite{Zweig,OZI} independently suggested in 1960´s that
strong interaction processes where the final states can only be
reached through quark-antiquark annihilation are suppressed in order
to explain the observation that $\phi$ meson ($\bar ss$) decayed
(strongly) into kaons more often than expected. }: $g^{(0)}_A\ll
\sqrt{3}g^{(8)}_A$.

Explanations of this phenomenon fall in two broad classes: either
the singlet $g_A^{(0)}$ is special because it can couple to gluons
or the octet $g_A^{(8)}$ is special because strangeness in the
nucleon is much larger than one might expect. Missing spin of the
proton is then understood as due either to the large strangeness of
the sea or to a large gluon contribution. The latter point of view
is adopted for example by the valon model \cite{Hwa} where the sea
contribution is small $\Delta q_\textrm{sea}\approx 0$ and the gluon
contribution is large $\approx 60\%$.

The present-day data claim that the first moment of the polarized
gluon is likely to be positive though the gluon spin is nowhere near
as large as would be required to explain the spin crisis. The most
recent measurements of inclusive $\pi^0$ jets at RHIC are best fit
with $\Delta G=0$ \cite{RHIC} and Bianchi reported $\Delta G/G\sim
0.08$ \cite{Bianchi}. On the contrary the total strangeness
contribution to nucleon spin is likely to be negative and quite
large. For an experimental status, see the short experimental review
\cite{Forte}. It is now well accepted that the neglected sea
contribution is very important to understand the suppression of the
quark spin contribution and that there is a sizeable amount of
strange quarks with polarization antiparallel to the proton
polarization. For a review on nucleon spin structure, see
\cite{Spincrisis}.

\subsection{Strangeness in nucleon and Dirac sea}

Quark-antiquark pairs are usually thought to be mainly produced in
the perturbative process of gluon splitting. Since there is no
explicit strangeness in the nucleon the study of nucleon strangeness
is considered as a unique approach to study the nucleon sea.
Experiments have indicated that strange quarks play a fundamental
role in understanding properties of the nucleon \cite{Sexp}. For
example, by combining parity-violating $\vec{e}p$ forward-scattering
elastic asymmetry data with the $\nu p$ elastic cross section data
one can extract the strangeness contribution to vector and axial
nucleon form factors. Traditionally the investigation on the role of
strange quarks played in ``non-strange'' baryons have taken place in
the context of DIS where we have seen that a sizeable amount of
strange quarks contribute to the nucleon spin.

There have also been strong efforts to measure the strange quark
contribution to the elastic form factors of the proton, in
particular the vector (electric and magnetic) form factors. These
experiments \cite{SAMPLEM,PVA4M,HAPPEXM,G0M} exploit an interference
between the $\gamma$- and $Z$-exchange amplitudes in order to
measure weak elastic form factors $G_E^{Z,p}$ and $G_M^{Z,p}$ which
are the weak-interaction analogs of the more traditional
electromagnetic elastic form factors $G_E^{\gamma,p}$ and
$G_M^{\gamma,p}$. The interference term is observable as a
parity-violating asymmetry in elastic $\vec{e}p$ scattering, with
the electron longitudinally polarized. By combining all these form
factors one may separate the $u$, $d$ and $s$ quark contributions.
However, in elastic $\vec{e}p$ scattering, the axial form factor
does not appear as a pure weak-interaction process. There are
significant radiative corrections which carry non-trivial
theoretical uncertainties. The result is that, while the measurement
of parity-violating asymmetries in $\vec{e}p$ elastic scattering is
well suited to a measurement of $G_E^s$ and $G_M^s$ these
experiments cannot cleanly extract $G^s_A(Q^2=0)=\Delta s$. Most of
QCD-inspired models seem to favor a negative value of the strange
magnetic moment in the range $-0.6\leq G^{(p)s}_M\leq 0.0\,\mu_N$
\cite{ModelsGMS}. The first experimental results from the SAMPLE
\cite{SAMPLEM}, PVA4 \cite{PVA4M}, HAPPEX \cite{HAPPEXM} and G0
\cite{G0M} collaborations have shown evidence for a non-vanishing
strange quark contribution to the structure of the nucleon. In
particular, the strangeness content of the proton magnetic moment
was found to be positive \cite{HAPPEXM}, suggesting that strange
quarks reduce the proton magnetic moment.

The growing interest in Semi-Inclusive Deep Inelastic Scattering
(SIDIS) with longitudinally polarized beams and target is due to the
fact that they provide an additional information on the spin
structure of the nucleon compared to inclusive DIS measurements.
They allow one to separate valence and sea contributions to the
nucleon spin. The present experimental results
\cite{SMC,HERMES,COMPASS} favor an asymmetric structure of the light
nucleon sea $\Delta\bar u\simeq-\Delta\bar d$. This is in
contradiction with the earliest parton models which assumed that the
nucleon sea was flavor symmetric even though valence quark
distributions are clearly flavor asymmetric. This assumption implies
that the sea is independent of the valence quark composition and
thus that the proton sea is the same as the neutron sea. This
assumption was however not based on any known physics and remained
to be tested by experiments. From experimental data for the
muon-nucleon DIS, Drell-Yan process (DY) and SIDIS we also know that
$\bar u(x,Q^2)<\bar d(x,Q^2)$, for reviews see \cite{ExpGott}. The
analysis of the muon-nucleon DIS data performed by the NMC
collaboration \cite{NMCGott} gives $I_G=0.235\pm 0.026$ at $Q^2=4$
GeV$^2$ which is violation of the Gottfried Sum Rule (GSR)
\cite{GSR} $I_G=1/3$ at the $4\sigma$ level.

Another different experimental indication of the presence of hidden
strangeness in the nucleon comes from the pion-nucleon sigma term
$\sigma_{\pi N}$ \cite{Koch} which measures the nucleon mass due to
current quarks and thus the explicit breaking of chiral symmetry.
Recent data \cite{sigmaterm} suggest that its value is $\sigma_{\pi
N}\simeq(60$-$80)$ MeV. Such a large value implies a surprisingly
large strangeness content of the nucleon in contrast to what one
would expect on the basis of the OZI rule. Let us also mention a QCD
fit to the CCFR and NuTeV dimuon data which indicates an asymmetry
in the strange quark distributions $s(x)\neq \bar s(x)$
\cite{QCDFIT}.

In short, independent experiments point out the existence of a
significant strangeness in the nucleon. In order to describe
correctly nucleon properties, strange quarks have to be taken into
account properly in models. The amount of these strange quarks
cannot be understood by purely perturbative processes. There is a
sizeable non-perturbative amount which has still to be explained.
For a lecture on the topic of strange spin, see \cite{Lecture}.

\subsection{Shape of baryons}

The question of hadron shape is a natural one. Hadrons are composite
particle and nothing prevents them to deviate from spherical shape.
The attention is then focused on the existence of quadrupolar
deformation. The nucleon being a spin-$1/2$ particle, no intrinsic
quadrupole moment can be directly measured because angular momentum
conservation forbids a non-zero element of a ($L=2$) quadrupole
operator between spin-$1/2$ states. On the contrary, $\Delta$ is a
spin-$3/2$ particle where such a quadrupole can be in principle
measured. That is the reason why the octet-to-decuplet transition
magnetic moments have especially focused attention since 1979.

It is now well confirmed experimentally \cite{Exp} that
non-spherical amplitudes do exist in hadrons and this has motivated
intense experimental and theoretical studies (for reviews see
\cite{RevND1}). The electromagnetic transition $\gamma N\to\Delta$
allows one to access to quadrupole moments of both proton and
$\Delta$. Only three multipole contributions to the transition are
not forbidden by spin and parity conservation: magnetic dipole
($M1$), electric quadrupole ($E2$) and Coulomb quadrupole ($C2$).

In NQM where $SU(6)$ spin-flavor symmetry is unbroken, one predicts
$E2/M1=0$ \cite{UnbrokenSU6} and the dominant multipole $M1$ is
$\approx 30\%$ below experimental values
\cite{RevND2,Quarkpredictions}. Non-spherical amplitudes in nucleon
and $\Delta$ are caused by non-central, tensor interaction between
quarks \cite{Quarkinterpretation}. If one adds a $d$-wave component
in nucleon and/or $\Delta$ wave function $E2$ and $C2$ are now
non-vanishing \cite{Quarkpredictions} but are at least one order of
magnitude too small. Moreover the $M1$ prediction is worse than in
the $SU(6)$ symmetry limit \cite{Discrepancy}. It is likely due to
the fact that quark models do not respect chiral symmetry, whose
spontaneous breaking leads to strong emission of virtual pions
\cite{RevND3}. The latter couple to nucleon as $\usigma\cdot\up$
where $\usigma$ is the nucleon spin and $\up$ is the pion momentum.
The coupling is strong in $p$-waves and mixes in non-zero angular
momentum components. As the pion is the lightest hadron, one indeed
expects it to dominate the long distance behavior of hadron wave
functions and to yield characteristic signatures in the low-momentum
transfer hadronic form factors. Since $\Delta(1232)$ resonance
nearly entirely decays into $\pi N$, one has another indication that
pions appear to be of particular relevance in the electromagnetic
$\gamma N\to\Delta$ transition.

Experimental ratios $E2/M1$ and $C2/M1$ are small and negative,
$|E2/M1|$ smaller than 5\%. With broken $SU(6)$ values range from 0
to $-2\%$ \cite{BrokenSU6}. Models such as Skyrme and large $N_C$
limit of QCD also find a small and negative ratio \cite{SkyrmeNc}.
Since $\Delta$ decays almost entirely into a nucleon and a pion, it
is not surprising that chiral bag models tend to agree well with
experimental data \cite{ChiralBag}. In recent years chiral effective
field theories were quite popular and gave precise results
\cite{ChEFT}. Lattice calculations predict a ratio to be around
$-3\%$ \cite{LattCalc}. For a recent review summarizing the various
theoretical approaches, see \cite{RevTh}.

\subsection{Exotic baryons}

The simple and unrealistic though quite successful in baryon
spectroscopy NQM describes all light baryons as made of three quarks
only\footnote{For the sake of simplicity we will use in the present
thesis the shorter expression $nQ$ for $n$ quarks (particle and
antiparticle). A $5Q$ state indicates thus that we have four quarks
and one antiquark.}. Group theory then tells us that light baryons
belong to singlet, octet and decuplet representations of the flavor
$SU(3)$ group
$\mathbf{3}\otimes\mathbf{3}\otimes\mathbf{3}=\mathbf{1}\oplus\mathbf{8}\oplus\mathbf{8}\oplus\mathbf{10}$.
Phenomenological observation tells us that the lightest baryon
multiplets are the octet with spin 1/2
$\left(\mathbf{8},\frac{1}{2}^+\right)$ and the decuplet with spin
3/2 $\left(\mathbf{10},\frac{3}{2}^+\right)$ both with positive
parity.

\begin{figure}[h]\begin{center}\includegraphics[width=11cm]{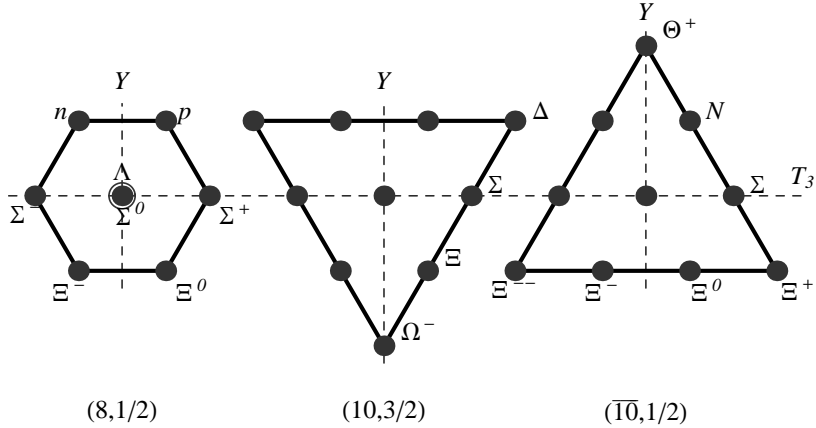}\caption{\small{The lightest baryon multiplets: octet $\mathbf{8}$, decuplet $\mathbf{10}$ and hypothetic antidecuplet $\overline{\mathbf{10}}$.}}\label{Multiplets}
\end{center}
\end{figure}

Let us stress however that QCD does not forbid states made of more
than $3Q$ as long as they are colorless. The next simplest colorless
quark structure is $QQQQ\bar Q$. States described by such a
structure are called \emph{pentaquarks}. It was first expected that
pentaquarks have wide widths \cite{Wide width} and thus difficult to
observe experimentally. Later, some theorists have suggested that
particular quark structures might exist with a narrow width
\cite{Small width,DPP}. The experimental status on the existence of
the exotic $\Theta^+$ pentaquark is still unclear. Even though most
of the latest experiments suggest that it does not exist, no
definitive answer can be given \cite{explanation}. There are many
experiments in favor (mostly low energy and low statistics) and
against (mostly high energy and high statistics). For reviews on the
experimental status of pentaquarks, see \cite{Experiment1}.
Concerning the experiments in favor, they all agree that the
$\Theta^+$ width is small but give only upper values. It turns out
that if it exists, the exotic $\Theta^+$ has a width of the order of
a few MeV or maybe even less than 1 MeV, a really curious property
since usual resonance widths are of the order of 100 MeV. In the
paper \cite{DPP} that actually motivated experimentalists to look
for a pentaquark, Diakonov, Petrov and Polyakov have estimated the
$\Theta^+$ width to be less than 15 MeV and claimed that pentaquarks
belong to an antidecuplet with spin 1/2
$\left(\overline{\mathbf{10}},\frac{1}{2}^+\right)$, see Fig.
\ref{Multiplets}.

More recently, Diakonov and Petrov with a technique based on
light-cone baryon wave functions used in the present thesis have
estimated more accurately the width and have found that it turns out
to be $\sim 4$ MeV \cite{DiaPet}. However, many approximations have
been used such as non-relativistic limit and omission of some $5Q$
contributions (exchange diagrams). The authors expected that these
have high probability to reduce further the width.

Exotic members of the antidecuplet can easily be recognized because
their quantum numbers cannot be obtained from $3Q$ only. The problem
is the identification of a nucleon resonance to a non-exotic or
crypto-exotic member of this antidecuplet. It is then interesting to
study the electromagnetic transition between octet and antidecuplet.
From simple flavor $SU(3)$ symmetry considerations, the existence of
antidecuplet would imply a sizeable breaking of isospin symmetry in
the excitation of an octet nucleon into an antidecuplet nucleon. The
magnetic transition between octet proton and crypto-exotic proton
should be suppressed compared to the neutron case \cite{SU3pred}.

Candidates for the nucleon-like members of the antidecuplet have
recently been discussed in the literature. The Partial Wave Analysis
(PWA) of pion-nucleon scattering presented two candidates for
$N_\mathbf{\overline{10}}$ with masses 1680 MeV and 1730 MeV
\cite{Candidates}. Experimental evidence for a new nucleon resonance
with mass near 1670 MeV has recently been obtained in the $\eta$
photoproduction on nucleon by the GRAAL collaboration \cite{Graal}.
A resonance peak is seen in the $\gamma n\to n\eta$ and is absent in
the $\gamma p\to p\eta$ process. This resonance structure has a
narrow width $\Gamma_{N^*\to\eta N}\simeq 40$ MeV. When the
Fermi-motion corrections are taken into account the width may become
even narrower $\Gamma_{N^*\to\eta N}\simeq 10$ MeV
\cite{Fermimotion}. Such a narrow width naturally reminds pentaquark
baryons. Even more recently the Tokohu LNS \cite{LNS} and
CB/TAPS@ELSA \cite{CBELSA} reported $\eta$ photoproduction from the
deuteron target and concluded on the same asymmetry.

The question of pentaquark is a very intriguing and confusing one.
The predicted pentaquarks have very special properties such as
unusual small width and large isospin breaking of nucleon
photoexcitation. On the experimental side the situation is far from
being clear and simple. While part of the original positive
sightings have been refuted by further more accurate experiments,
some striking positive signals persist and cannot be \emph{a priori}
understood as statistical fluctuations. Further experiments are
therefore needed. Finally, let us emphasize that even if the
existence of pentaquark is not confirmed we will have learned much
on the problem of experimental resolution, techniques allowing one
to detect a narrow resonance, validity of many theoretical
assumptions, \ldots On a more theoretical side, the absence of the
predicted pentaquark will probably and definitely invalidate the
rigid rotator quantization scheme for exotic states. Pentaquarks
with narrow width may simply not exist. There can be however
pentaquarks with very large width or with masses in a completely
different range. There could also be no $5Q$ state at all but this
would need some restriction due to QCD not known hitherto.

\section{Motivations and Plan of the thesis}

As we have seen understanding the baryon structure is still an open
and challenging problem. The correct low-energy QCD model should in
principle at the same time explain experimental data on baryon
structure and properties, predict the unmeasured ones in a reliable
manner, incorporate all relevant degrees of freedom, relate cleanly
constituent and current quarks, be in some sense directly derived
from QCD, \ldots No present model fulfills all these requirements.
That much is not in fact expected from models. We hope at least that
they deal with the relevant degrees of freedom, reproduce the
correct dynamics leading to the observed baryon structure and
properties and of course give reliable predictions.

Many questions both on the experimental and theoretical sides have
to be answered. Part of them have been shortly discussed in the
present introduction because they are related to the results of our
studies in the context of this thesis. Later they will be discussed
a bit further but without pretending to be complete and exhaust the
topics. For the interested reader many references to papers, reviews
and lectures are given throughout the text.

The Chiral Quark-Soliton Model ($\chi$QSM) is among the most
successful models in describing low-energy QCD. Recently it has been
formulated on the light cone \cite{DiaPet,PetPol} where the concept
of wave function is well defined. The basic formula have been
derived and the general technique developed. Then the axial decay
constant of the nucleon and the pentaquark width have been
investigated in the non-relativistic limit up to the $5Q$ Fock
component.

The aim of the present thesis was to further explore this new
approach to the model. One part of the work has been devoted to the
estimation of corrections coming from previously neglected diagrams,
relativity (quark angular momentum) and higher Fock components. The
second part has been devoted to study in details light baryon
properties and structure, extract the individual contributions due
to each quark flavor and separate the valence contributions from the
sea contributions. Let us stress that in this thesis we have
performed only \emph{ab initio} calculations, no fit to experimental
data has been made.

This work is very interesting for many reasons. First of all, as
mentioned earlier, this is a detailed study of baryon structure and
properties in terms of valence, sea and flavor contributions. The
values obtained are compared with the present experimental knowledge
and many predictions for the unmeasured baryon properties are given.
Due to the approximations specific to the approach and the model all
the predictions should not be considered as \emph{quantitatively}
reliable but at least give some \emph{qualitative} information. This
work is also interesting since we have estimated the impact of many
effects on the observables: quark angular momentum, quark-antiquark
pairs, \dots This allows one to emphasize the importance and role of
each degree of freedom.
\newline

The approach to $\chi$QSM we used is based on light-cone techniques.
In Chapter 2 we give a short introduction to the light-cone
approach. We remind why the light cone is appealing when describing
baryons and how they are studied usually in light-cone models.

Then in Chapter 3 we give a short introduction to $\chi$QSM. The
general baryon wave function is presented and all quantities needed
in this thesis are defined and explicit expressions are given. The
general technique for extracting baryon observables is also
presented.

Our whole work has been done in the flavor $SU(3)$ limit. Before
presenting the results obtained we discuss the implication of this
symmetry on observables, introduce the parametrization used in the
results and compare with the non-relativistic $SU(6)$ symmetry of
the usual CQM in Chapter 4.

In Chapters 5, 6, 7 and 8 we collect our results for normalizations,
vector, axial and tensor charges, and magnetic moments of all
lightest baryon multiplets (octet, decuplet and antidecuplet). They
are discussed and compared with the present experimental data. Part
of these results have already been published \cite{Moi} or submitted
on the web \cite{Moi3,moi4} waiting for publication. The remaining
results (especially concerning magnetic moments) are collected in
other papers in preparation \cite{moi2}.

We conclude this work in Chapter 9. We remind the important points
and results of the thesis and give tracks for further studies.

We join to this work two appendices. The first one contains all the
group integrals needed and explains how they can be obtained. The
second one gives general tools for simplifying the problem of
contracting the creation-annihilation operators leading to the
identification and weight of the diagrams involved in a given Fock
sector.
\newpage\thispagestyle{empty}\cleardoublepage
\chapter{Light-cone approach}

\section{Forms of dynamics}

Particle physics needs a synthesis of special relativity and quantum
mechanics. A quantum treatment is obvious since particle physics
plays at scales several order of magnitude smaller than in atomic
physics. These scales also require a relativistic formulation. Let
us consider for example a typical hadronic scale of 1 fm which
corresponds to momenta of the order $p\sim\hbar c/1$ fm $\simeq 200$
MeV. For particles with masses $M\lesssim 1$ GeV this implies
sizable velocities $v\simeq p/M\gtrsim 0.2$ c and thus
non-negligible relativistic effects.

A relativistic quantum mechanics requires the state vectors of a
system to transform according to a unitary representation of the
Poincaré group. The subgroup of continuous transformations, called
the proper group, has ten generators satisfying a set of commutation
relations called the proper Poincaré algebra.

A state vector $|t\rangle$ describes the system at a given ``time''
$t$. The evolution in ``time'' of is driven by the Hamiltonian $H$
operator of the system.  As defined by Dirac \cite{Dirac2}, the
Hamiltonian $H$ is that operator whose action on the state vector
$|t\rangle$ of a physical system has the same effect as taking the
partial derivative with respect to time $t$
\begin{equation}
H|t\rangle=i\frac{\partial}{\partial t}|t\rangle.
\end{equation}
Its expectation value $\langle t|H|t\rangle$ is a constant of motion
and is called ``energy'' of the system.

Time and space are however not separate issues. In a covariant
theory they are only different aspects of the four-dimensional
space-time. These concepts of space and time can be generalized in
an operational sense. One can define ``space'' as that hypersurface
in four-space on which one chooses the initial field configurations
in accord with microcausality, \emph{i.e.} a light emitted from any
point on the hypersurface must not cross the hypersurface. The
remaining fourth coordinate can be thought as being normal to that
hypersurface and understood as ``time''. There are many possible
parametrizations\footnote{The only condition is the existence of
inverse $x(\tilde{x})$.} or \emph{foliation} of space-time. A change
in parametrization $\tilde{x}(x)$ implies a change in metric in
order to conserve the arc length $\ud s^2$. This means that the
covariant $x_\mu$ and contravariant components $x^\mu$ can be quite
different and can have rather different interpretations.

We have then a certain freedom in describing the dynamics of a
system. One should however exclude all parametrizations accessible
by a Lorentz transformation. This limits considerably the freedom.
Following Dirac \cite{Dirac} there are basically three different
parametrizations or ``forms'' of dynamics: instant, front and point
forms. They cannot be mapped on each other by a Lorentz
transformation. They differ by the hypersurface $\Sigma$ in
Minkowski four-space on which the initial conditions of the fields
are given. To characterize the state of the system unambiguously,
$\Sigma$ must intersect every world-line once and only once. One has
then correspondingly different ``times''. The instant form is the
most familiar one with its hypersurface $\Sigma$ given at instant
time $x^0=t=0$. In the front form the hypersurface $\Sigma$ is a
tangent plane to the light cone defined at the light-cone time
$x^+=(t+z)/\sqrt{2}=0$. There seems here to be problems with
microcausality. Note however that a signal carrying information
moves with the group velocity always smaller than phase velocity
$c=1$. Thus if no information is carried by the signal, points on
the light cone cannot communicate. In the point form the time-like
coordinate is identified with the eigentime of the physical system
and the hypersurface has a hyperboloid shape. In principle all these
three forms yield the same physical results since physics should not
depend on how we parametrize space-time\footnote{In actual model
calculations differences arise because of approximations. Only a
complete and exact treatment would lead to the same physical results
in any parametrization.}. The choice of the form depends on the
amount of work needed to solve the physical problem. Let us note
that in the non-relativistic limit $c\to \infty$ only one foliation
is possible, the instant form and the absolute time is Galilean.
This is due to the fact that particles can have any velocity and
thus any slope of the hypersurface can be obtained by Lorentz boost.

Among the ten generators of the Poincaré algebra, there are some
that map $\Sigma$ into itself, not affecting the time evolution.
They form the so-called \emph{stability} subgroup and are referred
to as \emph{kinematical} generators. The others drive the evolution
of the system and contain the entire dynamics. They are called
\emph{dynamical} generators or Hamiltonians.

The generic four-vector $A^\mu$ is written in Cartesian
contravariant components as
\begin{equation}
A^\mu=(A^0,A^1,A^2,A^3)=(A^0,\mathbf{A}).
\end{equation}
Using Kogut and Soper convention, the light-cone components are
defined as
\begin{equation}
A^\mu=(A^+,\mathbf{A}_\perp,A^-),\qquad\textrm{where }
A^\pm=\frac{1}{\sqrt{2}}\,(A^0\pm A^3).
\end{equation}
The norm of this four-vector is then given by
\begin{equation}
A^2=(A^0)^2-\mathbf{A}^2=2A^+A^--\mathbf{A}_\perp^2
\end{equation}
and the scalar product of two four-vectors $A^\mu$ and $B^\mu$ by
\begin{equation}
A\cdot
B=A^0B^0-\mathbf{A}\cdot\mathbf{B}=A^+B^-+A^-B^+-\mathbf{A}_\perp\cdot\mathbf{B}_\perp.
\end{equation}
In the usual instant form the Hamiltonian operator $P_0$ is a
constant of motion which acts as the displacement operator in
instant time $x^0\equiv t$. In the light-cone approach or front form
the Hamiltonian operator $P_+$ is a constant of motion which acts as
the displacement operator in light-cone time $x^+\equiv
(t+z)/\sqrt{2}$. Let emphasize that $\partial_+=\partial^-$ is a
time-like derivative $\partial/\partial x^+=\partial/\partial x_-$
while $\partial_-=\partial^+$ is a space-like derivative
$\partial/\partial x^-=\partial/\partial x_+$. Correspondingly
$P_+=P^-$ is the Hamiltonian while $P_-=P^+$ is the longitudinal
space-like momentum.

\section{Advantages of the light-cone approach}

Representations of the Poincaré group are labeled by eigenvalues of
two Casimir operator $P^2$ and $W^2$. $P^\mu$ is the energy-momentum
operator, $W^\mu$ is the Pauli-Lubanski operator \cite{PLO}
constructed from $P^\mu$ and the angular-momentum operator
$M^{\mu\nu}$
\begin{equation}
W^\mu=-\frac{1}{2}\,\epsilon^{\mu\nu\rho\sigma}M_{\nu\rho}P_\sigma.
\end{equation}
Their eigenvalues are respectively $m^2$ and $-m^2s(s+1)$ with $m$
the mass and $s$ the spin the particle. The states of a Dirac
particle $s=1/2$ are eigenvectors of $P^\mu$ and polarization
operator $\Pi\equiv-W\cdot s/m$
\begin{eqnarray}
P^\mu|p,s\rangle&=&p^\mu|p,s\rangle,\\
-\frac{W\cdot s}{m}\,|p,s\rangle&=&\pm\frac{1}{2}\,|p,s\rangle
\end{eqnarray}
where $s^\mu$ is the spin (or polarization) vector of the particle
with properties
\begin{equation}
s^2=-1,\qquad s\cdot p=0.
\end{equation}
It can be written in general as
\begin{equation}
s^\mu=\left(\frac{\up\cdot\un}{m},\un+\frac{(\up\cdot\un)\up}{m(m+p^0)}\right)
\end{equation}
where $\un$ is a unit vector identifying a generic space direction.

Since the Lagrangian of a system is frame-independent there must be
ten conserved current corresponding to the ten Poincaré generators.
Integrating these currents over a three-dimensional hypersurface of
a hypersphere, embedded in the four-dimensional space-time,
generates conserved charges. The proper Poincaré group has then ten
conserved charges or constants of motion: the four components of the
energy momentum tensor $P^\mu$ and the six components of the
boost-angular momentum tensor $M^{\mu\nu}$. These ten constants of
motion are observables and are thus hermitian operators with real
eigenvalues. It is therefore advantageous to construct
representations\footnote{The problem of constructing Poincaré
representations is equivalent to the problem of looking for the
different forms of dynamics.} in which these constants of motion are
diagonal. Unfortunately one cannot diagonalize all the ten
simultaneously because they do not commute.

In the usual instant form dynamics the initial conditions are set at
some instant of time and the hypersurfaces $\Sigma$ are flat
three-dimensional surfaces only containing directions that lie
outside the light cone. The generators of rotations and space
translations leave the instant invariant, \emph{i.e.} do not affect
the dynamics. There are then six generators constituting the
kinematical subgroup in the instant form: three momentum $P_i$ and
three angular momentum generators $J_i=\frac{1}{2}\,\epsilon_{ijk}
M_{jk}$. The remaining four generators are dynamical and therefore
involve interaction: three boost $K_i=M_{i0}$ and one
time-translation generators $P_0$.

In the front form dynamics one considers instead three-dimensional
surfaces in space-time formed by a plane-wave front advancing at the
velocity of light, \emph{e.g.} $x^+=0$. In this case seven
generators are kinematical
$P_1,P_2,P_-,M_{12},M_{+-},M_{1-},M_{2-}$. The three remaining ones
$P_+,M_{1+},M_{2+}$ are then dynamical. This corresponds in fact to
best one can do \cite{Dirac}. One cannot diagonalize simultaneously
more than seven Poincaré generators. Components of the
energy-momentum operator are easily interpreted as generators of
space $P_1,P_2,P_-$ and time translations $P_+$. Kogut and Soper
\cite{Kogut} have written the components of the angular momentum
operator in terms of boosts and angular momenta. They introduced the
transversal vector $\mathbf{B}_\perp$
\begin{equation}
B_1=M_{1-}=\frac{1}{\sqrt{2}}\,(K_1+J_2),\qquad
B_2=M_{2-}=\frac{1}{\sqrt{2}}\,(K_2-J_1).
\end{equation}
They are kinematical and boost the system in the $x$ and $y$
direction respectively. The other kinematical operators $M_{12}=J_3$
and $M_{+-}=K_3$ rotate the system in the $x$-$y$ plane and boost it
in the longitudinal direction respectively. The remaining dynamical
operators are combined in a transversal angular-momentum vector
$\mathbf{S}_\perp$
\begin{equation}
S_1=M_{1+}=\frac{1}{\sqrt{2}}\,(K_1-J_2),\qquad
S_2=M_{2+}=\frac{1}{\sqrt{2}}\,(K_2+J_1).
\end{equation}

Light-cone calculations for relativistic CQM are convenient as they
allow to boost quark wave functions independently of the details of
the interaction. Unlike the traditional instant form Hamiltonian
formalism where the internal and center-of-mass motion of
relativistic interacting particles cannot be separated in principle,
the light-cone Hamiltonian formalism can be formulated without
reference to a specific Lorentz frame. The drawback is however that
the construction of states with good total angular momentum becomes
interaction dependent. Except for the free theory, it is very hard
to write down states with good angular momentum as diagonalizing
$\mathbf{L}^2$ is as difficult as solving the Schrödinger equation.
This is the notorious problem of angular momentum of the light-cone
approach\footnote{A way to formulate covariantly the plane is by
defining a light-like four-vector $\omega$ and the plane equation by
$\omega\cdot x=0$ which is invariant under any Lorentz
transformation of both $\omega$ and $x$. Exact on-shell physical
amplitudes should not depend on the orientation of the light-front
plane. However, in practice, this dependence survives due to
approximations. Results are spoiled by unphysical form factors.
Poincaré invariance is destroyed as soon as truncation of the Fock
space or regularizations of Fock sectors are implemented
\cite{Poincaredes}.} \cite{Fuda}.

The useful concept of wave function borrowed from non-relativistic
quantum mechanics is not well defined in instant form since the
particle number of a state is neither bounded nor fixed.
Quark-antiquark pairs are constantly popping in and out the vacuum.
This means that even the ground state is complicated. One of biggest
advantages of the front form is that the vacuum structure is much
simpler. In many cases the vacuum state of the free Hamiltonian is
also an eigenstate of the full light-cone Hamiltonian. Contrary to
$P_z$ the operator $P^+$ is positive, having only positive
eigenvalues. Each Fock state is eigenstate of the operators $P^+$
and $\uP_\perp$. The eigenvalues are
\begin{equation}
\uP_\perp=\sum_{i=1}^n\up_{\perp i},\qquad P^+=\sum_{i=1}^n p^+_i
\end{equation}
with $p^+_i>0$ for massive quanta, $n$ being the number of particles
in the Fock state. The vacuum has eigenvalue 0, \emph{i.e.}
$\uP_\perp|0\rangle=\mathbf{0}$ and $P^+|0\rangle=0$. The
restriction $p^+_i>0$ for massive quanta is the key difference
between light-cone and ordinary equal-time quantization. In the
latter the state of a parton is specified by its ordinary
three-momentum $\up$. Since each component of the momentum can be
either positive or negative there exists an infinite number of Fock
states with zero total momentum. The physical vacuum
$|\Omega\rangle$ is thus complicated. In the former particles have
non-zero longitudinal momentum and the vacuum is
identified\footnote{This simplification works only for massive
particles. The restriction $p^+_i>0$ cannot be applied to massless
particles. This leads to the zero-mode problem of the light-cone
vacuum.} to the zero-particle state $|\Omega\rangle=|0\rangle$.

The Fock expansion constructed on this vacuum provides thus a
complete relativistic many-particle basis for the baryon states.
This means that all constituents are directly related to the baryon
state and not do disconnected vacuum fluctuations. The concept of
wave function is then well defined on the light cone. The light-cone
wave functions are frame independent and can be expressed by means
of relative coordinates only because the boosts are kinematical. For
example, Lorentz boost in the third direction is diagonal.
Light-cone time and space do not get mixed but are just rescaled.
Since $p^+_i>0$ and $P^+>0$ one can define boost-invariant
longitudinal momentum fractions $z_i=p^+_i/P^+$ with $0<z_i<1$. In
the intrinsic frame $\uP_\perp=\mathbf{0}$ we have the constraints
\begin{equation}
\sum_{i=1}^n z_i=1,\qquad \sum_{i=1}^n\up_{\perp i}=\mathbf{0}.
\end{equation}

These light-cone wave functions are very important and useful
objects as they encode hadronic properties. In the context of QCD
their relevance relies on the concept of factorization. Processes
with hadrons at sufficiently high energy/momentum transfer can be
divided into two parts: a hard part which can be calculated
according to perturbative QCD and a soft part usually encoded in
soft functions, parton distributions, fragmentation functions,
\ldots This soft part can in principle be expressed in terms of
light-cone wave functions. For example PDF are forward matrix of
non-local operator and can be obtained by squaring the wave function
and integrating over some transverse momenta. With electromagnetic
probes one has
\begin{equation}
f(x)\propto \int\ud\lambda\,e^{i\lambda x}\langle
P|\bar\psi(0)\gamma_\mu\psi(\lambda)|P\rangle\sim\int\ud\uk_\perp\,\psi^\dag(x,\uk_\perp)\,\psi(x,\uk_\perp).
\end{equation}
Form factors (FF) are off-forward matrix elements of local operator
and can be obtained from an overlap of light-cone wave functions
\begin{equation}
F(Q^2)\propto\langle
P'|\bar\psi(0)\gamma_\mu\psi(0)|P\rangle\sim\int_{-1}^1\ud
x\int\ud\uk_\perp\,\psi^\dag(x,\uk_\perp+\mathbf{Q}_\perp/2)\,\psi(x,\uk_\perp-\mathbf{Q}_\perp/2).
\end{equation}
Generalized Parton Distributions (GPD) provide a natural
interpolation between PDF and FF and are relevant in processes like
Deeply Virtual Compton Scattering (DVCS) and hard meson production
\cite{GPDs}. They are off-forward matrix elements of non-local
operator and can also be easily presented in terms of light-cone
wave functions \cite{GPD}
\begin{equation}
GPD(x,\xi,Q^2)\propto\int\ud\lambda\,e^{i\lambda x}\langle
P'|\bar\psi(-\lambda/2)\gamma_\mu\psi(\lambda/2)|P\rangle\sim\int\ud\uk_\perp\,\psi^\dag(x+\xi,\uk_\perp+\mathbf{Q}_\perp/2)\,\psi(x-\xi,\uk_\perp-\mathbf{Q}_\perp/2).
\end{equation}

The light-cone calculation of nucleon form factors has been
pioneered by Berestetsky and Terentev \cite{Berestetsky} and more
recently developed by Chung and Coester \cite{CC}. Form factors are
generally constructed from hadronic matrix elements of the current
$\langle P+q|j^\mu(0)|P\rangle$. In the interaction picture one can
identify the fully interacting Heisenberg current $J^\mu$ with the
free current $j^\mu$ at the space-time point $x^\mu=0$. The
computation of these hadronic matrix elements is greatly simplified
in the so-called Drell-Yan-West (DYW) frame \cite{DYW}, \emph{i.e.}
in the limit $q^+=0$ where $q$ is the light-cone longitudinal
transfer momentum. Matrix elements of the $+$ component of the
current are diagonal in particle number $n'=n$, \emph{i.e.} the
transitions between Fock states with different particle numbers are
vanishing. The current can neither create nor annihilate
quark-antiquark pairs. Such a simplification can be seen using
projectors on ``good'' and ``bad'' components of a Dirac
four-spinor. The operator $\mathcal{P}_+=\gamma^-\gamma^+/2$
projects the four-component Dirac spinor $\psi$ onto the
two-dimensional subspace of ``good'' light-cone components which are
canonically independent fields \cite{Kogut}. Likewise
$\mathcal{P}^-=\gamma^+\gamma^-/2$ projects on the two-dimensional
subspace of ``bad'' light-cone components which are interaction
dependent fields and should not enter at leading twist.

Finally, instant form has also a practical disadvantage. For
example, consider the wave function of an atom with $n$ electrons.
An experiment which specifies the initial wave function would
require simultaneous measurement of the position of all the bounded
electrons. In contrast, the initial wave function at fixed
light-cone time only requires an experiment which scatters one
plane-wave laser beam since the signal reaches each of the $n$
electrons at the same light-cone time.

\section{Light cone \emph{v.s.} Infinite Momentum Frame}

Dirac's legacy has been forgotten and re-invented many times with
other names. The Infinite Momentum Frame (IMF) first appeared in the
work of Fubini and Furlan \cite{Fubini} in connection with current
algebra as the limit of a reference frame moving with almost the
speed of light. Weinberg \cite{Weinberg} considered the
infinite-momentum limit of old-fashioned perturbation diagrams for
scalar meson theories and showed that the vacuum structure of these
theories simplified in this limit. Later, Susskind \cite{Susskind}
showed that the infinities which occur among the generators of the
Poincaré group when they are boosted in the IMF can be scaled or
substracted out consistently. The result is essentially a change in
variables. With these new variables he drew the attention to the
(two-dimensional) Galilean subgroup of the Poincaré group. Bardakci
and Halpern \cite{Bardakci} further analyzed the structure of
theories in IMF. They viewed the infinite-momentum limit as a change
of variables from the laboratory time $t$ and space coordinate $z$
to a new ``time'' $\tau=(t+z)/\sqrt{2}$ and a new ``space''
$\zeta=(t-z)/\sqrt{2}$. Kogut and Soper \cite{Kogut} have examined
the formal foundations of Quantum ElectroDynamics (QED) in the IMF.
Finally Drell and others \cite{DYW,Drell+} have recognized that the
formalism could serve as kind of natural tool for formulating the
quark-parton model.

Let us consider two particles with three-momenta $\up_1$ and $\up_2$
and use the variables $\uP=(\up_1+\up_2)/2$ and $\uq=\up_2-\up_1$.
The IMF prescription is to take the limit $|\uP|\to\infty$ and
impose the condition $\uP\cdot\uq=0$, \emph{i.e.} momentum transfer
has to be orthogonal to the (very large) mean momentum which
guarantees that the momentum transfer has no time component
\cite{currents}. This prescription introduces from the outside an
infinite factor in the covariant normalization for the physical
states
\begin{equation}
\langle
\up_2|\up_1\rangle=(2\pi)^32E\,\delta^{(3)}(\up_1-\up_2)\longrightarrow(2\pi)^32|\uP|\,\delta^{(3)}(\up_1-\up_2).
\end{equation}
Thus the ``natural'' $|\uP|$ power in an expansion is actually
reduced by one unit. Any vector $\mathbf{v}$ can be decomposed into
a longitudinal component $\mathbf{v}_L$ which is along the direction
of $\uP$ and a transverse component $\mathbf{v}_\perp$ which is
orthogonal to $\uP$. Let us consider in the following that $\uP$
defines the $z$ direction.

Currents can be decomposed into ``good'' and ``bad'' components
referring to their behavior in the limit $P_z\to\infty$. The
``good'' components behave like $P_z$ while ``bad'' components are
of order $\mathcal{O}(1)$. The scalar $S$, pseudoscalar $P$, vector
$V_\mu$, axial vector $A_\mu$ and tensor $T_{\mu\nu}$ operators have
the most immediate relevance in elementary particle physics.
``Good'' components correspond to free quarks. Creation-annihilation
of quark-antiquark pairs are suppressed. On the contrary, ``bad''
components correspond to interacting quarks. Creation-annihilation
of quark-antiquark pairs are important. In the IMF the ``good''
operators appeared to be $V_0,V_3,A_0,A_3,T_{0\perp},T_{3\perp}$ and
the ``bad'' ones to be
$S,P,V_\perp,A_\perp,T_{00},T_{03},T_{33},T_{\perp\perp'}$. This
means that it is simple to compute the zeroth and third components
of the vector and axial vector current in the IMF. Moreover these
zeroth and third components coincide in the leading order in $P_z$.
On the contrary scalar and pseudoscalar currents as well as
transverse components of the vector and axial-vector currents are
difficult because the interaction is involved.

These features naturally remind the light-cone approach in the DYW
frame. The light-cone and IMF approaches are indeed identified in
the literature. For example one defines the light-cone wave function
as the instant-form wave function boosted to the IMF
\cite{BoostIMF}. However unboosting the wave function from IMF is
generally impossible. For a qualitative picture, all the physical
processes in the IMF become as slow as possible because of time
dilatation in this system of reference. The investigation of the
wave function is equivalent to make a snapshot of as system not
spoiled by vacuum fluctuations. Note  also that in the IMF, there is
no distinction between the quark helicity and its spin projection
$S_z$. That is why both these two terms will be used without
distinction.

\section{Standard model approach based on Melosh rotation}

As we have just seen, light-cone wave functions are obtained by
boosting the rest-frame wave function. The usual approach is to use
a $3Q$ rest-frame wave function ideally fitted to the baryon
spectrum. The spin $S$ of a particle is not Lorentz invariant. Only
the total angular momentum $J=L+S$ is the meaningful quantity. Its
decomposition into spin $S$ and orbital angular momentum $L$ depends
on the reference frame. This means that boosting a particle induces
a change in its spin orientation.

The conventional spin three-vector $\mathbf{s}$ of a moving particle
with finite mass $m$ and four-momentum $p_\mu$ can be defined by
transforming its Pauli-Lubanski four-vector $W_\mu$ to its rest
frame \emph{via} a rotationless Lorentz boost $L(p)$ which satisfies
$L(p)p=(m,\mathbf{0})$. One has \cite{Rest}
\begin{equation}
(0,\mathbf{s})=L(p)W/m.
\end{equation}
Under an arbitrary Lorentz transformation $\Lambda$ a particle of
spin $\mathbf{s}$ and four-momentum $p_\mu$ will be mapped onto the
state of spin $\mathbf{s}'$ and four-momentum $p'_\mu$ given by
\begin{equation}
\mathbf{s}'=\mathcal{R}_W(\Lambda,p)\mathbf{s},\qquad p'=\Lambda p
\end{equation}
where $\mathcal{R}_W(\Lambda,p)=L(\Lambda p)\Lambda L^{-1}(p)$ is a
pure rotation known as Wigner rotation.

So when a baryon is boosted \emph{via} a rotationless Lorentz
transformation along its spin direction from the rest frame to a
frame where it is moving, each quark will undergo a Wigner rotation.
Specified to the spin-1/2 case the Wigner rotation reduces to the
Melosh rotation \cite{Melosh}
\begin{equation}
\psi^i_{LC,\lambda}=\frac{[(m+z_i\uM)\bold
1+i\un\cdot(\usigma\times\up_i)]^{\lambda'}_\lambda}{\sqrt{(m+z_i\uM)^2+\up^2_{i\perp}}}\,\psi^i_{\lambda'}
\end{equation}
where $\un=(0,0,1)$. This transformation assures that the baryon is
an eigenfunction of $J$ and $J_z$ in its rest frame \cite{Rest}.
This rotation transforms rest-frame quark states $\psi^i_{\lambda'}$
into light-cone quark states $\psi^i_{LC,\lambda}$, with $i=1,2,3$.
Here is the explicit expression for the Melosh rotated states
\begin{eqnarray}
\psi^i_{LC,+}&=&\frac{(m_q+z_i\uM)\psi^i_\uparrow+p_i^R\psi^i_\downarrow}{\sqrt{(m_q+z_i\uM)^2+\up_{i\perp}^2}},\\
\psi^i_{LC,-}&=&\frac{-p_i^L\psi^i_\uparrow+(m_q+z_i\uM)\psi^i_\downarrow}{\sqrt{(m_q+z_i\uM)^2+\up_{i\perp}^2}}
\end{eqnarray}
where $p_i^{R,L}=p_i^x\pm ip_i^y$ and $\uM$ is the invariant mass
$\uM^2=\sum_{i=1}^3(\up_i^2+m_q^2)/z_i$ with the constraints
$\sum_{i=1}^3 z_i=0$ and $\sum_{i=1}^3\up_{i\perp}=0$. The internal
transverse-momentum dependence of the light-cone wave function also
affects its helicity structure \cite{Brod}. The zero-binding limit
$z_i\uM\to p^+_i$ is not a justified approximation for QCD bound
states. This rotation mixes the helicity states due to a nonzero
transverse momentum $\up_{i\perp}$. The light-cone spinor with
helicity $+$ corresponds to \emph{total} angular momentum projection
$J_z=1/2$ and is thus constructed from a spin $\uparrow$ state with
orbital angular momentum $L_z=0$ and a spin $\downarrow$ state with
orbital angular momentum $L_z=+1$ expressed by the factor $p^R$.
Similarly the light-cone spinor with helicity $-$ corresponds to
\emph{total} angular momentum projection $J_z=-1/2$ and is thus
constructed from a spin $\uparrow$ state with orbital angular
momentum $L_z=-1$ expressed by the factor $p^L$ and a spin
$\downarrow$ state with orbital angular momentum $L_z=0$. Note
however that the general form of a light-cone wave function
\cite{LFWF} must contain two functions
\begin{equation}\label{LFWF}
\psi^\sigma_{\sigma_1}=\chi^\dagger_{\sigma_1}\left(f_1+i\,\un\cdot\frac{(\usigma\times\up)}{|\up|}f_2\right)\chi^\sigma.
\end{equation}
The additional $f_2$ term represents a separate dynamical
contribution to be contrasted with the purely kinematical
contribution of angular momentum from Melosh rotations.

For a review on the light-cone topic, see \cite{Brodsky}.
\newpage\thispagestyle{empty}\cleardoublepage
\chapter{The Chiral Quark-Soliton Model}
\section{Introduction}

As mentioned in the thesis introduction we know that a realistic
description of the nucleon should incorporate the Spontaneous Chiral
Symmetry Breaking (SCSB). This idea is one of the basics of the
Chiral Quark-Soliton Model ($\chi$QSM) and plays a dominant role in
the dynamics of the nucleon bound state.

As in the Skyrme model, $\chi$QSM is essentially based on a $1/N_C$
expansion where $N_C$ is the number of colors in QCD. It is a
general QCD theorem that at large $N_C$ the nucleon is heavy and can
be viewed as a classical soliton \cite{Witten}. While the dynamical
realization given by the Skyrme model \cite{Skyrme} is based on
unrealistic effective chiral Lagrangian, a far more realistic one
has been proposed later \cite{Action}. This NJL-type Lagrangian has
been derived from the instanton model of the QCD vacuum which
provides a natural mechanism of chiral symmetry breaking. Based on
this Lagrangian, the $\chi$QSM model \cite{Profile
function,Approximation} has been proposed and describes baryon
properties better than the Skyrme model. For a recent status of this
model see the reviews \cite{Baryon properties,cQSMrev}. Let us also
mention that Generalized Parton Distributions (GPD) \cite{cQSMGPD}
have also recently been computed in the model at a low normalization
point.

A distinguishable feature of $\chi$QSM as compared with many other
effective models of baryons, like NQM or MIT bag model, is that it
is field theoretical model which takes into account not only three
valence quarks but also the whole Dirac sea as degrees of freedom.
It is also almost the only effective model that can give reliable
predictions for the quark and antiquark distribution functions of
the nucleon satisfying the fundamental field theoretical
restrictions like positivity of the antiquark distribution
\cite{Theoryconstraints,Cutoff}. $\chi$QSM is often seen as the
interpolation between two drastically different pictures of the
nucleon, namely the NQM where we have only valence quarks and Skyrme
model where we have only the pion field.

\begin{figure}[h]\begin{center}\includegraphics[width=7cm]{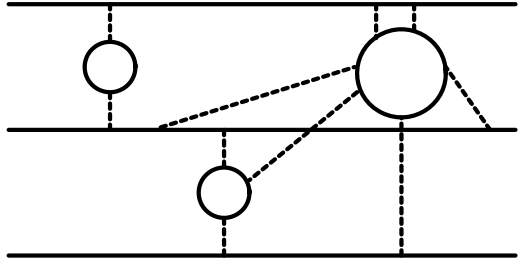}
\caption{\small{Picture of the nucleon arising from models based on
non-linear chiral Lagrangians. Quarks (solid lines) interact via
pion fields (dashed lines) which propagate through quark loops
inducing many-quark interactions.}}\label{Nucleon}\end{center}
\end{figure}

An important difference between $\chi$QSM and the Chiral Quark Model
($\chi$QM) is that in the former a non-trivial topology is
introduced which is crucial for stabilizing the soliton whereas in
the latter the $\chi$QM fields are treated as a perturbation.
$\chi$QSM differs also from the linear $\sigma$-model \cite{Level
wave function} in that no kinetic energy at tree level is associated
to the chiral fields. Pions propagate only through quark loops.
Furthermore quark loops induce many-quark interactions, see Fig.
\ref{Nucleon}. Consequently the emerging picture is rather far from
a simple one-pion exchange between the constituent quarks:
non-linear effects in the pion field are not at all suppressed.
Note also that the chiral fields are effective degrees of freedom,
totally equivalent to the quark-antiquark excitations of the Dirac
sea (no problem of double counting) \cite{PetPol}.

\subsection{The effective action of $\chi$QSM}

$\chi$QSM is assumed to mimic low-energy QCD thanks to an effective
action describing constituent quarks with a momentum dependent
dynamical mass $M(p)$ interacting with the scalar $\Sigma$ and
pseudoscalar $\uPi$ fields. The chiral circle condition
$\Sigma^2+\uPi^2=1$ is invoked. Due to its momentum dependence
$M(p)$ serves as a form factor for the constituent quarks and
provides also the effective theory with the UV cutoff. At the same
time, it makes the theory non-local as one can see in the action
\begin{equation}\label{Effective lagrangian}
S_\textrm{eff}=\int\frac{\ud^4p\,\ud^4p'}{(2\pi)^8}\,\bar\psi(p)\left[\pslash\,(2\pi)^4\delta^{(4)}(p-p')-\sqrt{M(p)}(\Sigma(p-p')+i\Pi(p-p')\gamma_5)\sqrt{M(p')}\right]\psi(p')
\end{equation}
where $\psi$ and $\bar\psi$ are quark fields. This action has been
originally derived in the instanton model of the QCD vacuum
\cite{Action}. After reproducing masses and decay constants in the
mesonic sector, the only free parameter left to be fixed in the
baryonic sector is the constituent quark mass. The number of gluons
is suppressed in the instanton vacuum by the parameter $(M\rho)^2\ll
1$ where $\rho$ is the instanton size, so gluons in this model do
not participate in the formation of the nucleon wave function. Note
that oppositely to the naive bag picture, this action
(\ref{Effective lagrangian}) is fully relativistic and supports all
general principles and sum rules for conserved quantities.

The form factors $\sqrt{M(p)}$ cut off momenta at some
characteristic scale which corresponds in the instanton picture to
the inverse average size of instantons $1/\bar\rho\approx 600$ MeV.
One can then consider the scale of this model to be $Q^2_0\approx
0.36$ GeV$^2$. This means that in the range of quark momenta $p\ll
1/\bar\rho$ one can neglect the non-locality. We use the standard
approach: the constituent quark mass is replaced by a constant
$M=M(0)$ and we mimic the decreasing function $M(p)$ by the UV
Pauli-Villars cutoff \cite{Cutoff}
\begin{equation}
S_\textrm{eff}=\int\frac{\ud^4p}{(2\pi)^4}\,\bar\psi(p)(\pslash-MU^{\gamma_5})\psi(p)
\end{equation}
with $U^{\gamma_5}$ a $SU(3)$ matrix
\begin{equation}
U^{\gamma_5}=\left(\begin{array}{cc}
                     U_0 & 0 \\
                     0 & 1
                   \end{array}
\right),\qquad U_0=e^{i\upi\cdot\utau\gamma_5}=e^{i\pi\gamma_5}
\end{equation}
and $\tau^a$ the usual $SU(2)$ Pauli matrices.

In the following we expose the general technique from \cite{DiaPet}
allowing one to derive the (light-cone) baryon wave functions.

\section{Explicit baryon wave function}\label{Section trois}

In $\chi$QSM it is easy to define the baryon wave function in the
rest frame. Indeed this model represents quarks in the Hartree
approximation in the self-consistent pion field. The baryon is then
described as $N_C$ valence quarks + Dirac sea in that
self-consistent external field. It has been shown \cite{PetPol} that
the wave function of the Dirac sea is the coherent exponential of
the quark-antiquark pairs
\begin{equation}\label{Coherent exponential}
|\Omega\rangle=\exp\left(\int(\ud\up)(\ud\up')\,a^\dag(\up)W(\up,\up')b^\dag(\up')\right)|\Omega_0\rangle
\end{equation}
where $|\Omega_0\rangle$ is the vacuum of quarks and antiquarks
$a,b\,|\Omega_0\rangle=0$, $\langle\Omega_0|\,a^\dag,b^\dag=0$,
$(\ud\up)=\ud^3\up/(2\pi)^3$ and $W(\up,\up')$ is the quark Green
function at equal times in the background $\Sigma, \uPi$ fields
\cite{PetPol,Green function} (its explicit expression is given in
Subsection 3.4).

\begin{figure}[h]\begin{center}\begin{minipage}[c]{8cm}\begin{center}\includegraphics[width=8cm]{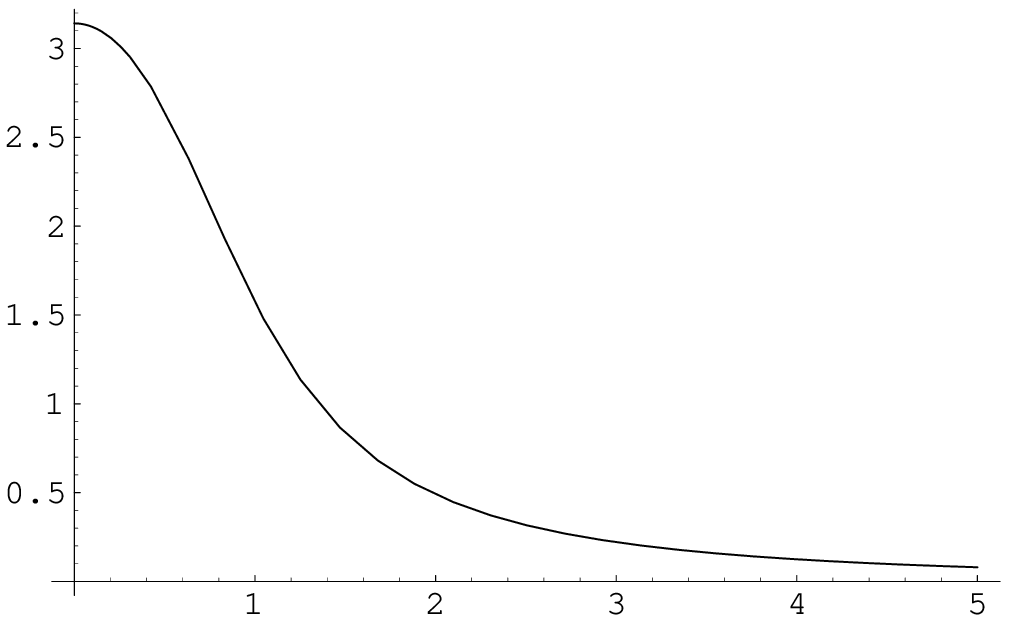}\caption{\small{Profile of the self-consistent
chiral field $P(r)$ in light baryons. The horizontal axis unit is
$r_0=0.8/M=0.46$ fm.\newline\newline\newline}}\label{Profile}
\end{center}\end{minipage}\hspace{0.5cm}
\begin{minipage}[c]{8cm}\begin{center}\includegraphics[width=8cm]{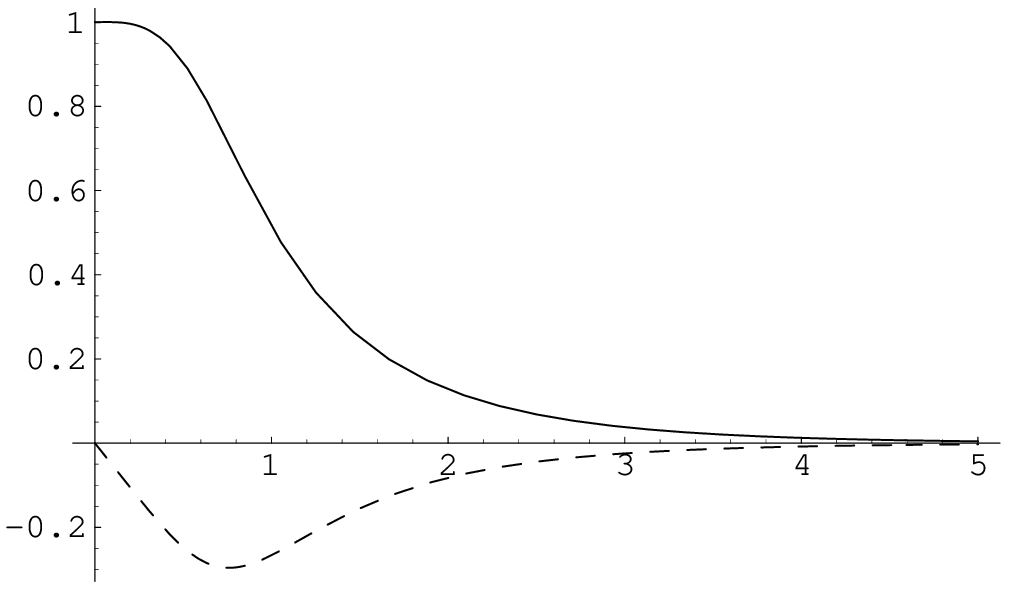}\caption{\small{Upper $s$-wave component $h(r)$ (solid) and lower $p$-wave component
$j(r)$ (dashed) of the bound-state quark level in light baryons.
Each of the three valence quarks has energy $E_\textrm{lev}=200$
MeV. Horizontal axis has units of $1/M=0.57$ fm.}}\label{Level}
\end{center}\end{minipage}\end{center}
\end{figure}

The saddle-point or mean-field approximation is invoked to obtain
the stationary pion field corresponding to the nucleon at rest. A
mean field approach is usually justified by the large number of
participants. For example, the Thomas-Fermi model of atoms is
justified at large $Z$ \cite{Thomas-Fermi}. For baryons, the number
of colors $N_C$ has been used as such parameter \cite{Witten}. Since
$N_C=3$ in the real world, one can wonder how accurate is the
mean-field approach. The chiral field experiences fluctuations about
its mean-field value of the order of $1/N_C$. These are loop
corrections which are further suppressed by factors of $1/2\pi$
yielding to corrections typically of the order of 10\% which are
simply ignored. In the mean-field approximation the chiral field is
replaced by the following spherically-symmetric self-consistent
field
\begin{equation}\label{Self-consistent field}
\pi(\ur)=\un\cdot\utau\,P(r),\qquad\un=\ur/r.
\end{equation}
We then have on the chiral circle $\Pi(\ur)=\un\cdot\utau\sin P(r)$,
$\Sigma(\ur)=\Sigma(r)=\cos P(r)$ with $P(r)$ being the profile
function of the self-consistent field. The latter is fairly
approximated by \cite{Profile function, Approximation} (see Fig.
\ref{Profile})
\begin{equation}\label{Profile function}
P(r)=2\arctan\left(\frac{r_0^2}{r^2}\right),\qquad
r_0\approx\frac{0.8}{M},\qquad M\approx 345\textrm{ MeV}.
\end{equation}
Consequently, in this approach, most of low-energy properties of
light baryons follow from the shape of the mean chiral field in the
classical baryon.

Such a chiral field creates a bound-state level for quarks, whose
wave function $\psi_\textrm{lev}$ satisfies the static Dirac
equation with eigenenergy $E_\textrm{lev}$ in the $K^p=0^+$ sector
with $K=T+J$ \cite{Profile function,Level wave function,Valence
level}
\begin{equation}
\psi_\textrm{lev}(\ur)=\left(\begin{array}{c}\epsilon^{ji}h(r)\\-i\epsilon^{jk}(\un\cdot\sigma)^i_k\,
j(r)\end{array}\right),\qquad\left\{\begin{array}{c}h'+h\,M\sin
P-j(M\cos P+E_\textrm{lev})=0\\j'+2j/r-j\,M\sin P-h(M\cos
P-E_\textrm{lev})=0\end{array}\right.
\end{equation}
where $i=1,2=\uparrow,\downarrow$ and $j=1,2$ are respectively spin
and isospin indices. Solving those equations with the
self-consistent field (\ref{Self-consistent field}) one finds that
``valence'' quarks are tightly bound ($E_\textrm{lev}=200$ MeV)
along with a lower component $j(r)$ smaller than the upper one
$h(r)$ (see Fig. \ref{Level}).

For the valence quark part of the baryon wave function it is
sufficient to write the product of $N_C$ quark creation operators
that fill in the discrete level \cite{PetPol}
\begin{equation}\label{Valence part}
\prod_{\textrm{color}=1}^{N_C}\int(\ud\up)F(\up)a^\dag(\up)
\end{equation}
where $F(\up)$ is obtained by expanding and commuting
$\psi_\textrm{lev}(\up)$ with the coherent exponential
(\ref{Coherent exponential})
\begin{equation}\label{Discrete level}
F(\up)=\int(\ud\up')\sqrt{\frac{M}{\epsilon's}}[\bar
u(\up)\gamma_0\psi_\textrm{lev}(\up)(2\pi)^3\delta^{(3)}(\up-\up')-W(\up,\up')\bar
v(\up')\gamma_0\psi_\textrm{lev}(-\up')].
\end{equation}
One can see from the second term that the distorted Dirac sea
contributes to the one-quark wave function. For the plane-wave Dirac
bispinor $u_\sigma(\up)$ and $v_\sigma(\up)$ we used the standard
basis
\begin{equation}
u_\sigma(\up)=\left(\begin{array}{c}\sqrt{\frac{\epsilon+M}{2M}}s_\sigma\\
\sqrt{\frac{\epsilon-M}{2M}}\frac{\up\cdot\sigma}{|\up|}s_\sigma\end{array}\right),
\qquad
v_\sigma(\up)=\left(\begin{array}{c}\sqrt{\frac{\epsilon-M}{2M}}\frac{\up\cdot\sigma}{|\up|}s_\sigma\\
\sqrt{\frac{\epsilon+M}{2M}}s_\sigma\end{array}\right),\qquad \bar
u^{\sigma'} u_\sigma=\delta^{\sigma'}_\sigma=-\bar
v^{\sigma'}v_\sigma
\end{equation}
where $\epsilon=+\sqrt{\up^2+M^2}$ and $s_\sigma$ are two
2-component spinors normalized to unity
\begin{equation}
s_1=\left(\begin{array}{c}1\\0\end{array}\right),\qquad
s_2=\left(\begin{array}{c}0\\1\end{array}\right).
\end{equation}
The complete baryon wave function is then given by the product of
the valence part (\ref{Valence part}) and the coherent exponential
(\ref{Coherent exponential})
\begin{equation}
|\Psi_B\rangle=\prod_{\textrm{color}=1}^{N_C}\int(\ud\up)F(\up)a^\dag(\up)\exp\left(\int(\ud\up)(\ud\up')\,a^\dag(\up)W(\up,\up')b^\dag(\up')\right)|\Omega_0\rangle.
\end{equation}

We remind that the saddle-point of the self-consistent pion field is
degenerate in global translations and global $SU(3)$ flavor
rotations (the $SU(3)$-breaking strange mass can be treated
perturbatively later). These zero modes must be handled with care.
The result is that integrating over translations leads to momentum
conservation which means that the sum of all quarks and antiquarks
momenta have to be equal to the baryon momentum. As first pointed
out by Witten \cite{Witten} and then derived using different
techniques by a number of authors \cite{Quantization}, the
quantization rule for the rotations of the mean chiral field in the
ordinary and flavor spaces is such that the lowest baryon multiplets
are the octet with spin 1/2 and the decuplet with spin 3/2 followed
by the exotic antidecuplet with spin 1/2. All of those multiplets
have same parity. The lowest baryons appear just as rotational
excitations of the same mean chiral field (soliton). They are
distinguished by their specific rotational wave functions given
explicitly in Section 3.3. Let us note that in $\chi$QSM $\Theta^+$
pentaquark is light because it is not the sum of constituent quark
masses but rather a collective excitation of the mean chiral field
inside baryons.

Since rotations of the chiral field are not slow we integrate
\emph{exactly} over $SU(3)$ rotations $R$ in this thesis. This has
to be contrasted with the usual slowly-rotating approach used in
former studies of $\chi$QSM in the instant form. This leads to the
projection of the flavor state of all quarks and antiquarks onto the
spin-flavor state $B(R)$ specific to any particular baryon from the
$\left({\bf 8},\frac{1}{2}^+\right)$, $\left({\bf
10},\frac{3}{2}^+\right)$ and $\left({\bf
\overline{10}},\frac{1}{2}^+\right)$ multiplets.

If we restore color ($\alpha=1,2,3$), flavor ($f=1,2,3$), isospin
($j=1,2$) and spin ($\sigma=1,2$) indices, we obtain the following
quark wave function of a particular baryon $B$ with spin projection
$k$ \cite{PetPol, Green function}
\begin{eqnarray}
|\Psi_k(B)\rangle&=&\int\ud
R\,B_k^*(R)\epsilon^{\alpha_1\alpha_2\alpha_3}\prod_{n=1}^3\int(\ud\up_n)R_{j_n}^{f_n}F^{j_n\sigma_n}(\up_n)a^\dag_{\alpha_n f_n \sigma_n}(\up_n)\nonumber \\
&\times&\exp\left(\int(\ud\up)(\ud\up')a^\dag_{\alpha f\sigma}(\up)
R^f_j W^{j\sigma}_{j'\sigma'}(\up,\up') R^{\dag
j'}_{f'}b^{\dag\alpha f'\sigma'}(\up')
\right)|\Omega_0\rangle.\label{Full glory}
\end{eqnarray}
The three $a^\dag$ create three valence quarks with the same wave
function $F$ while the rest of $a^\dag$'s, $b^\dag$'s create
\emph{any} number of additional quark-antiquark pairs whose wave
function is $W$. One can notice that the valence quarks are
antisymmetric in color whereas additional quark-antiquark pairs are
color singlets. One can obtain the spin-flavor structure of a
particular baryon by projecting a general $QQQ+n\,Q\bar Q$ state
onto the quantum numbers of the baryon under consideration. This
projection is an integration over all spin-flavor rotations $R$ with
the rotational wave function $B^*_k(R)$ unique for a given baryon.

Expanding the coherent exponential allows one to get the $3Q$, $5Q$,
$7Q$, \ldots wave functions of a particular baryon. Explicit
expressions for the baryon rotational wave functions $B(R)$, the
$Q\bar Q$ pair wave function in a baryon
$W^{j\sigma}_{j'\sigma'}(\up,\up')$ and the valence wave function
$F^{j\sigma}(\up)$ are given in the next sections.

\section{Baryon rotational wave functions}\label{Rotational section}

Baryon rotational wave functions are in general given by the $SU(3)$
Wigner finite-rotation matrices \cite{Wigner matrices} and any
particular projection can be obtained by a $SU(3)$ Clebsch-Gordan
technique. In order to see the symmetries of the quark wave
functions explicitly, we keep the expressions for $B(R)$ and
integrate over the Haar measure $\int\ud R$ in eq. (\ref{Full
glory}).

The rotational $D$-functions for the $\left({\bf
8},\frac{1}{2}^+\right)$, $\left({\bf 10},\frac{3}{2}^+\right)$ and
$\left({\bf \overline{10}},\frac{1}{2}^+\right)$ multiplets are
listed below in terms of the product of the $R$ matrices. Since the
projection onto a particular baryon in eq. (\ref{Full glory})
involves the conjugated rotational wave function, we list the latter
one only. The unconjugated ones are easily obtained by hermitian
conjugation.

\subsection{The octet $\left({\bf
8},\frac{1}{2}^+\right)$}

From the $SU(3)$ group point of view, the octet transforms as
$(p,q)=(1,1)$, \emph{i.e.} the rotational wave function can be
composed of a quark (transforming as $R$) and an antiquark
(transforming as $R^\dag$). Then the (conjugated) rotational wave
function of an octet baryon having spin index $k=1,2$ is
\begin{equation}
\left[D^{(8,\frac{1}{2})*}(R)\right]^g_{f,k}\sim\epsilon_{kl}R^{\dag
l}_f R^{g}_3.
\end{equation}
The flavor part of this octet tensor $P^g_f$ represents the
particles as follows
\begin{eqnarray}
&P^3_1=N^+_8,\qquad P^3_2=N^0_8,\qquad P^2_1=\Sigma^+_8,\qquad
P^1_2=\Sigma^-_8,&\nonumber\\
&P^1_1=\frac{1}{\sqrt{2}}\,\Sigma^0_8+\frac{1}{\sqrt{6}}\,\Lambda^0_8,\qquad
P^2_2=-\frac{1}{\sqrt{2}}\,\Sigma^0_8+\frac{1}{\sqrt{6}}\,\Lambda^0_8,&\\
&P^3_3=-\sqrt{\frac{2}{3}}\,\Lambda^0_8,\qquad P^2_3=\Xi^0_8,\qquad
P^1_3=-\Xi^-_8.&\nonumber\label{Octet}
\end{eqnarray}
For example, the proton ($f=1,g=3$) and neutron ($f=2,g=3$)
rotational wave functions are
\begin{equation}
p^{+*}_k(R)=\sqrt{8}\,\epsilon_{kl}R^{\dag l}_1 R^3_3,\qquad
n^{0*}_k(R)=\sqrt{8}\,\epsilon_{kl}R^{\dag l}_2 R^3_3.
\end{equation}

\subsection{The decuplet $\left({\bf
10},\frac{3}{2}^+\right)$}

The decuplet transforms as $(p,q)=(3,0)$, \emph{i.e.} the rotational
wave function can be composed of three quarks. The rotational wave
functions are then labeled by a triple flavor index $\{f_1f_2f_3\}$
symmetrized in flavor and by a triple spin index $\{k_1k_2k_3\}$
symmetrized in spin
\begin{equation}
\left[D^{(10,\frac{3}{2})*}(R)\right]_{\{f_1f_2f_3\}\{k_1k_2k_3\}}\sim\epsilon_{k'_1k_1}\epsilon_{k'_2k_2}\epsilon_{k'_3k_3}R^{\dag
k'_1}_{f_1}R^{\dag k'_2}_{f_2}R^{\dag k'_3}_{f_3}\Big|_{\textrm{sym
in }\{f_1f_2f_3\}}.
\end{equation}
The flavor part of this decuplet tensor $D_{f_1f_2f_3}$ represents
the particles as follows
\begin{equation}
\begin{array}{llll}
D_{111}=\sqrt{6}\,\Delta^{++}_{10},&
D_{112}=\sqrt{2}\,\Delta^+_{10},& D_{122}=\sqrt{2}\,\Delta^0_{10},&
D_{222}=\sqrt{6}\,\Delta^-_{10},\\
D_{113}=\sqrt{2}\,\Sigma^+_{10},& D_{123}=-\Sigma^0_{10},&
D_{223}=-\sqrt{2}\,\Sigma^-_{10},&
D_{133}=\sqrt{2}\,\Xi^0_{10},\\
&D_{233}=\sqrt{2}\,\Xi^-_{10},& D_{333}=-\sqrt{6}\,\Omega^-_{10}.&
\end{array}\label{Decuplet}
\end{equation}
For example, the $\Delta^{++}$ with spin projection 3/2
($f_1=1,f_2=1,f_3=1$) and $\Delta^0$ with spin projection 1/2
($f_1=1,f_2=2,f_3=2$) rotational wave functions are
\begin{equation}
\Delta^{++*}_{\uparrow\uparrow\uparrow}(R)=\sqrt{10}\,R^{\dag
2}_1R^{\dag 2}_1R^{\dag 2}_1,\qquad
\Delta^{0*}_\uparrow(R)=\sqrt{10}\,R^{\dag 2}_2(2R^{\dag 2}_1R^{\dag
1}_2+R^{\dag 2}_2R^{\dag 1}_1).
\end{equation}

\subsection{The antidecuplet $\left({\bf\overline{
10}},\frac{1}{2}^+\right)$}

The antidecuplet transforms as $(p,q)=(0,3)$, \emph{i.e.} the
rotational wave function can be composed of three antiquarks. The
rotational wave functions are then labeled by a triple flavor index
$\{f_1f_2f_3\}$ symmetrized in flavor
\begin{equation}
\left[D^{(\overline{10},\frac{1}{2})*}(R)\right]^{\{f_1f_2f_3\}}_k\sim
R^{f_1}_3R^{f_2}_3R^{f_3}_k\Big|_{\textrm{sym in }\{f_1f_2f_3\}}.
\end{equation}
The flavor part of this antidecuplet tensor $T^{f_1f_2f_3}$
represents the particles as follows
\begin{equation}
\begin{array}{llll}
T^{111}=\sqrt{6}\,\Xi^{--}_{\overline{10}},&
T^{112}=-\sqrt{2}\,\Xi^-_{\overline{10}},&
T^{122}=\sqrt{2}\,\Xi^0_{\overline{10}},&
T^{222}=-\sqrt{6}\,\Xi^+_{\overline{10}},\\
T^{113}=\sqrt{2}\,\Sigma^-_{\overline{10}},&
T^{123}=-\Sigma^0_{\overline{10}},&
T^{223}=-\sqrt{2}\,\Sigma^+_{\overline{10}},&
T^{133}=\sqrt{2}\,N^0_{\overline{10}},\\
&T^{233}=-\sqrt{2}\,N^+_{\overline{10}},&
T^{333}=\sqrt{6}\,\Theta^+_{\overline{10}}.&\label{Antidecuplet}
\end{array}
\end{equation}
For example, the $\Theta^+$ ($f_1=3,f_2=3,f_3=3$) and crypto-exotic
neutron ($f_1=1,f_2=3,f_3=3$) rotational wave functions are
\begin{equation}
\Theta^{+*}_k(R)=\sqrt{30}\,R^3_3R^3_3R^3_k,\qquad
n^{0*}_{\overline{10},k}(R)=\sqrt{10}\,R^3_3(2R^1_3R^3_k+R^3_3R^1_k).
\end{equation}

All examples of rotational wave functions above have been normalized
in such a way that for any (but the same) spin projection we have
\begin{equation}
\int\ud R\,B^*_\textrm{spin}(R)B^\textrm{spin}(R)=1,
\end{equation}
the integral being zero for different spin projections. Note that
rotational wave functions belonging to different baryons are also
orthogonal. This can be easily checked using the group integrals in
Appendix A. The particle representations (\ref{Octet}),
(\ref{Decuplet}) and (\ref{Antidecuplet}) have been found in
\cite{Tensor structure}.

\section{Formulation in the Infinite Momentum Frame}

As explained earlier the formulation in the IMF or equivalently on
the light cone is very appealing. Thanks to the particularly simple
structure of the vacuum the concept of wave function (borrowed from
quantum mechanics) is well defined. By definition \cite{BoostIMF} a
light-cone wave function is the wave function in the Infinite
Momentum Frame, \emph{i.e.} in the frame where the particle is
travelling with almost the speed of light. Usually one cannot start
with the instant form wave function and boost it to the IMF because
boosts involve interaction. However as the effective chiral
Lagrangian is relativistically invariant, we are guaranteed that
there are infinitely many solutions of saddle-point equations of
motion which describe the nucleon moving in some direction with
speed $V$. The IMF is obtained when $V\to 1$. The corresponding pion
field becomes time-dependent and can be obtained from the stationary
field by a Lorentz transformation \cite{PetPol}.

\subsection{$Q\bar Q$ pair wave function}\label{Section cinq}

In \cite{PetPol, Green function} it is explained that the pair wave
function $W^{j\sigma}_{j'\sigma'}(\up,\up')$ is expressed by means
of the finite-time quark Green function at equal times in the
external static chiral field (\ref{Self-consistent field}). The
Fourier transforms of this chiral field will be needed
\begin{equation}\label{Fourier tranform of mean field}
\Pi(\uq)^j_{j'}=\int\ud^3\ux\,
e^{-i\uq\cdot\ux}(\un\cdot\utau)^j_{j'}\sin
P(r),\qquad\Sigma(\uq)^j_{j'}=\int\ud^3\ux\, e^{-i\uq\cdot\ux}(\cos
P(r)-1)\delta^j_{j'}
\end{equation}
where $\Pi(\uq)$ is purely imaginary and odd and $\Sigma(\uq)$ is
real and even. They can be rewritten as follows
\begin{eqnarray}
&&\Pi(\uq)^j_{j'}=i\,\frac{(\uq\cdot\utau)^j_{j'}}{|\uq|}\,\Pi(q),\qquad
\Pi(q)=\frac{4\pi}{q^2}\int_0^\infty\ud r\,\sin P(r)\left(qr\cos
qr-\sin qr\right)\\
&&\Sigma(\uq)^j_{j'}=\delta^j_{j'}\,\Sigma(q),\qquad\qquad\quad
\Sigma(q)=\frac{4\pi}{q}\int_0^\infty\ud r\,r\left(\cos
P(r)-1\right)\sin qr
\end{eqnarray}
where the radial functions are depicted in Fig. \ref{Self-consistent
field plot}.
\begin{figure}[h]\begin{center}\includegraphics[width=7cm]{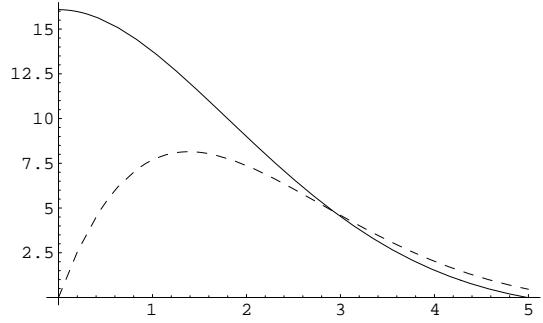}
\caption{\small{The self-consistent pseudoscalar $-|\uq|\Pi(\uq)$
(solid) and scalar $-|\uq|\Sigma(\uq)$ (dashed) fields in baryons.
The horizontal axis unit is $M$.}}\label{Self-consistent field plot}
\end{center}
\end{figure}

A simplified interpolating approximation for the pair wave function
$W$ has also been derived and becomes exact in three limiting cases:
\begin{enumerate}
\item small pion field $P(r)$, \item slowly varying $P(r)$ and \item
fast varying $P(r)$.
\end{enumerate}
Since the model is relativistically invariant, this wave function
can be translated to the infinite momentum frame (IMF). In this
particular frame, the result is a function of the fractions of the
baryon longitudinal momentum carried by the quark $z$ and antiquark
$z'$ of the pair and their transverse momenta $\up_\perp$,
$\up'_\perp$
\begin{equation}\label{Pair wave function}
W^{j\sigma}_{j'\sigma'}(z,\up_\perp;z',\up'_\perp)=\frac{M\uM}{2\pi
Z}\left\{\Sigma^j_{j'}(\uq)[M(z'-z)\tau_3+\uQ_\perp\cdot\utau_\perp]^\sigma_{\sigma'}-i\Pi^j_{j'}(\uq)[-M(z'+z)\bold
1+i\uQ_\perp\times\utau_\perp]^\sigma_{\sigma'}\right\}
\end{equation}
where $\uq=((\up+\up')_\perp,(z+z')\uM)$ is the three-momentum of
the pair as a whole transferred from the background fields
$\Sigma(\uq)$ and $\Pi(\uq)$, $\tau_{1,2,3}$ are Pauli matrices,
$\uM$ is the baryon mass and $M$ is the constituent quark mass. In
order to simplify the notations we used
\begin{equation}\label{Notation}
Z=\uM^2zz'(z+z')+z(p'^2_\perp+M^2)+z'(p^2_\perp+M^2),\qquad
\uQ_\perp=z\up'_\perp-z'\up_\perp.
\end{equation}

This pair wave function $W$ is normalized in such a way that the
creation-annihilation operators satisfy the following
anticommutation relations
\begin{equation}\label{Anticommutation}
\{a^{\alpha_1f_1\sigma_1}(z_1,\up_{1\perp}),a^\dag_{\alpha_2f_2\sigma_2}(z_2,\up_{2\perp})\}=\delta^{\alpha_1}_{\alpha_2}\delta^{f_1}_{f_2}\delta^{\sigma_1}_{\sigma_2}
\delta(z_1-z_2)(2\pi)^2\delta^{(2)}(\up_{1\perp}-\up_{2\perp})
\end{equation}
and similarly for $b$, $b^\dag$, the integrals over momenta being
understood as $\int\ud z\int\ud^2\up_\perp/(2\pi)^2$.

A more compact form for this wave function can be obtained by means
of the following two variables
\begin{equation}
y=\frac{z'}{z+z'},\qquad\uQcal_\perp=\frac{z\up'_\perp-z'\up_\perp}{z+z'}.
\end{equation}
The pair wave function then takes the form
\begin{equation}\label{Pair wavefunction}
W^{j,\sigma}_{j'\sigma'}(y,\uq,\uQcal_\perp)=\frac{M\uM}{2\pi
}\frac{\Sigma^j_{j'}(\uq)[M(2y-1)\tau_3+\uQcal_\perp\cdot\utau_\perp]^\sigma_{\sigma'}-i\Pi^j_{j'}(\uq)[-M\mathbf{1
}+i\uQcal_\perp\times\utau_\perp]^\sigma_{\sigma'}}{\uQcal^2_\perp+M^2+y(1-y)\uq^2}.
\end{equation}

\subsection{Discrete-level wave function}\label{Section six}

\begin{figure}[h]\begin{center}\includegraphics[width=7cm]{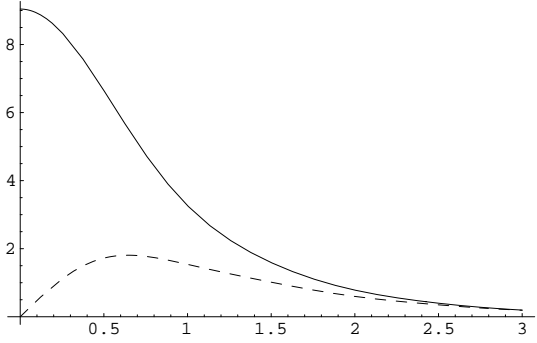}\caption{\small{Fourier transforms of the upper $s$-wave component $h(p)$ (solid) and lower $p$-wave component
$j(p)$ (dashed) of the bound-state quark level in light baryons.
Horizontal axis has units of $M$.}}\label{LevelFT}
\end{center}
\end{figure}

We see from eq. (\ref{Discrete level}) that the discrete-level wave
function
$F^{j\sigma}(\up)=F^{j\sigma}_\textrm{lev}(\up)+F^{j\sigma}_\textrm{sea}(\up)$
is the sum of two parts: the one is directly the wave function of
the valence level and the other is related to the change of the
number of quarks at the discrete level due to the presence of the
Dirac sea. I t is a relativistic effect and can be ignored in the
non-relativistic limit ($E_\textrm{lev}\approx M$) together with the
small $L=1$ lower component $j(r)$. Indeed, in the baryon rest frame
$F^{j\sigma}_\textrm{lev}$ gives
\begin{equation}
F^{j\sigma}_\textrm{lev}=\epsilon^{j\sigma}\left(\sqrt{\frac{E_\textrm{lev}+M}{2E_\textrm{lev}}}h(p)+\sqrt{\frac{E_\textrm{lev}-M}{2E_\textrm{lev}}}j(p)\right)
\end{equation}
where $h(p)$ and $j(p)$ are the Fourier transforms of the valence
wave function, see Fig. \ref{LevelFT}
\begin{eqnarray}
h(p)&=&\int\ud^3\ux\, e^{-i\up\cdot\ux}h(r)=4\pi\int_0^\infty\ud
r\,r^2\frac{\sin pr}{pr}h(r),\\
j^a(p)&=&\int\ud^3\ux\,
e^{-i\up\cdot\ux}(-in^a)j(r)=\frac{p^a}{|\up|}\,j(p),\qquad
j(p)=\frac{4\pi}{p^2}\int_0^\infty\ud r\,(pr\cos pr-\sin pr)j(r).
\end{eqnarray}

In the non-relativistic limit the second term is double-suppressed:
first due to the kinematical factor and second due to the smallness
of the $L=1$ wave $j(r)$ compared to the $L=0$ wave $h(r)$.

Switching to the IMF one obtains \cite{PetPol, Green function}
\begin{equation}\label{Discrete level IMF}
F^{j\sigma}_\textrm{lev}(z,\up_\perp)=\sqrt{\frac{\uM}{2\pi}}\left[\epsilon^{j\sigma}h(p)+(p_z\bold
1+i\up_\perp\times\utau_\perp)^\sigma_{\sigma'}\epsilon^{j\sigma'}\frac{j(p)}{|\up|}\right]_{p_z=z\uM-E_\textrm{lev}}.
\end{equation}
As expected for a covariant light-cone wave function two
distributions are involved $h$ and $j$, see eq. (\ref{LFWF})

The ``sea'' part of the discrete-level wave function gives in the
IMF
\begin{equation}\label{Discrete level IMF 2}
F^{j\sigma}_\textrm{sea}(z,\up_\perp)=-\sqrt{\frac{\uM}{2\pi}}\int\ud
z'\frac{\ud^2\up'_\perp}{(2\pi)^2}\,W^{j\sigma}_{j'\sigma'}(z,\up_\perp;z',\up'_\perp)\,\epsilon^{j'\sigma''}\left[(\tau_3)^{\sigma'}_{\sigma''}h(p')-(\up'\cdot\tau)^{\sigma'}_{\sigma''}\frac{j(p')}{|\up'|}\right]_{p'_z=z'\uM-E_\textrm{lev}}.
\end{equation}
This sea part will be ignored in the present thesis. It is difficult
to estimate its impact without an explicit computation.

In the work made by Diakonov and Petrov \cite{DiaPet}, the
relativistic effects in the discrete-level wave function were
neglected. One can then use only the first term in (\ref{Discrete
level IMF})
\begin{equation}\label{Approximation}
F^{j\sigma}(z,\up_\perp)\approx\sqrt{\frac{\uM}{2\pi}}\,\epsilon^{j\sigma}h(p)\big|_{p_z=z\uM-E_\textrm{lev}}.
\end{equation}
In the following the function $h(p)$ and $j(p)$ are understood with
the condition $p_z=z\uM-E_\textrm{lev}$.

\section{Baryon Fock components}

In a realistic picture baryons cannot be made of three (valence)
quarks only. It has been soon realized that pions or quark-antiquark
pairs are also present but the naive idea was that they can be in
some sense integrated out and their effects encoded in valence quark
non-trivial form factors. This idea was supported by the fact that
axial decay constants and especially magnetic moments seem well
described with three quarks only. However more recent experiments
revealed the presence of hidden flavor in nucleons. Even though the
number of strange quarks and antiquarks is the same, the strangeness
contribution to nucleon spin and magnetic moment is non-zero. This
indicates that the sea of quark-antiquarks or the pion cloud has to
be somehow implemented explicitly in models.

In the present approach light baryons are explicitly described as an
infinite tower of Fock states thanks to the coherent exponential
(\ref{Coherent exponential}). The latter can then be expanded to
obtain any baryon $nQ$ Fock component. In this thesis we have
expanded the wave function up to the $7Q$ component. We will see
that the higher is the Fock state the smaller is its contribution to
observables.

\subsection{$3Q$ component of baryons}\label{Section sept}

We will show in this section how to derive systematically the $3Q$
component of the octet and decuplet baryons (antidecuplet baryons
have no such component). On the top of that we will also show that
they become in the non-relativistic limit similar to the well-known
$SU(6)$ wave functions of the constituent quark model.

\begin{figure}[h]\begin{center}\begin{minipage}[c]{8cm}\begin{center}\includegraphics[width=2.5cm]{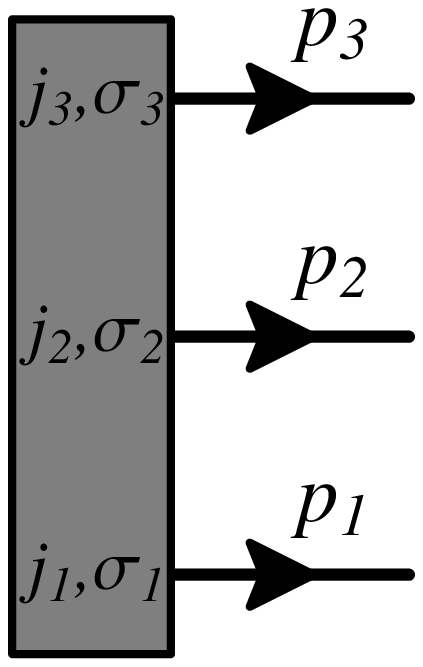}
\caption{\small{Schematic representation of the $3Q$ component of
baryon wave functions. The dark gray rectangle stands for the three
discrete-level wave functions
$F^{j_i\sigma_i}(\up_i)$.}}\label{Threequarks}
\end{center}\end{minipage}\hspace{0.5cm}
\begin{minipage}[c]{8cm}\begin{center}\includegraphics[width=2cm]{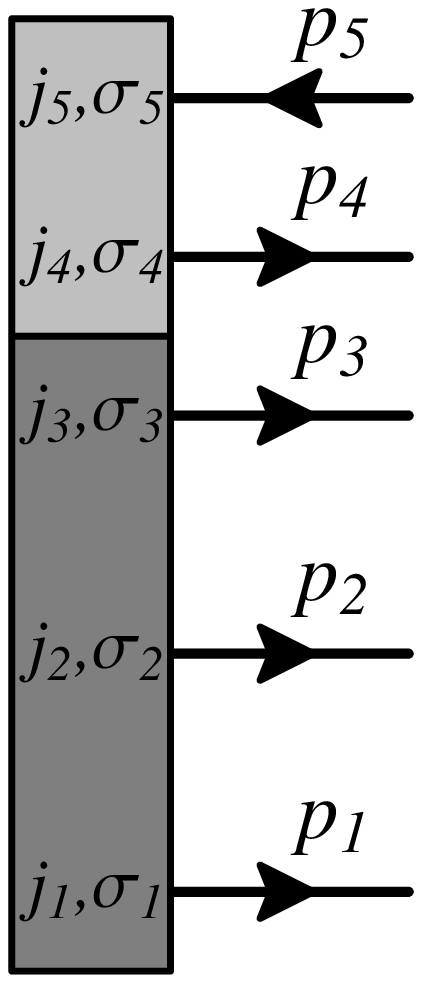}
\caption{\small{Schematic representation of the $5Q$ component of
baryon wave functions. The light gray rectangle stands for the pair
wave function $W^{j_i\sigma_i}_{j_k\sigma_k}(\up_i,\up_k)$ where the
reversed arrow represents the antiquark.}}\label{Fivequarks}
\end{center}\end{minipage}\end{center}
\end{figure}

An expansion of the coherent exponential gives access to all Fock
components of the baryon wave function. Since we are interested in
the present case only in the $3Q$ component, this coherent
exponential is just ignored (since it has to be expanded to the
zeroth order, \emph{i.e.} $e^{\int \ud\up\,\ud\up'W}\sim 1$). One
can see from eq. (\ref{Full glory}) that the three valence quarks
are rotated by the $SU(3)$ matrices $R^f_j$ where $f=1,2,3\equiv
u,d,s$ is the flavor and $j=1,2$ is the isospin index. The
projection of the $3Q$ state onto the quantum numbers of a specific
baryon leads to the following group integral
\begin{equation}\label{T tensor}
T(B)^{f_1f_2f_3}_{j_1j_2j_3,k}\equiv\int\ud
R\,B^*_k(R)R^{f_1}_{j_1}R^{f_2}_{j_2}R^{f_3}_{j_3}.
\end{equation}
The group integrals can be found in Appendix A. This tensor $T$ must
be contracted with the three discrete-level wave functions to obtain
the $3Q$ baryon wave function
\begin{equation}
F^{j_1\sigma_1}(p_1)F^{j_2\sigma_2}(p_2)F^{j_3\sigma_3}(p_3).
\end{equation}
The wave function is schematically represented on Fig.
\ref{Threequarks}.

Let us consider, for example, the non-relativistic $3Q$ wave
function of the neutron in the coordinate space
\begin{eqnarray}
(|n\rangle_k)^{f_1f_2f_3,\sigma_1\sigma_2\sigma_3}(\ur_1,\ur_2,\ur_3)&=&\frac{\sqrt{8}}{24}\,\epsilon^{f_1f_2}\epsilon^{\sigma_1\sigma_2}\delta^{f_3}_2\delta^{\sigma_3}_kh(r_1)h(r_2)h(r_3)\nonumber\\
&+&\textrm{
cyclic permutations of (1,2,3)}
\end{eqnarray}
times the antisymmetric tensor $\epsilon^{\alpha_1\alpha_2\alpha_3}$
in color. This equation means that in the non-relativistic $3Q$
picture the whole neutron spin $k$ is carried by a $d$ quark
$\delta^{f_3}_2\delta^{\sigma_3}_k$ while the $ud$ pair is in the
spin- and isospin-zero combination
$\epsilon^{f_1f_2}\epsilon^{\sigma_1\sigma_2}$. This is similar to
the better known non-relativistic $SU(6)$ wave function of the
neutron
\begin{eqnarray}
|n\uparrow\rangle&=&2 |d\uparrow(r_1)\rangle |d\uparrow(r_2)\rangle
|u\downarrow(r_3)\rangle-|d\uparrow(r_1)\rangle
|u\uparrow(r_2)\rangle |d\downarrow(r_3)\rangle-
|u\uparrow(r_1)\rangle |d\uparrow(r_2)\rangle
|d\downarrow(r_3)\rangle\nonumber\\&+&\textrm{ cyclic permutations
of (1,2,3)}.
\end{eqnarray}
There are, of course, many relativistic corrections arising from the
exact discrete-level wave function (\ref{Discrete level
IMF})+(\ref{Discrete level IMF 2}) and the additional
quark-antiquark pairs, both effects being potentially not small.

\subsection{$5Q$ component of baryons}\label{section huit}

The $5Q$ component of the baryon wave functions is obtained by
expanding the coherent exponential (\ref{Coherent exponential}) to
the first order in the $Q\bar Q$ pair. The projection involves now
along with the three $R$'s from the discrete level two additional
matrices $R\,R^\dag$ that rotate the quark-antiquark pair in the
$SU(3)$ space
\begin{equation}
T(B)^{f_1f_2f_3f_4,j_5}_{j_1j_2j_3j_4,f_5,k}\equiv\int\ud
R\,B^*_k(R)R^{f_1}_{j_1}R^{f_2}_{j_2}R^{f_3}_{j_3}\left(R^{f_4}_{j_4}R^{\dag
j_5}_{f_5}\right).
\end{equation}
Components $i=1,2,3$ refer then to the valence part, $i=4$ to the
quark of the sea and $i=5$ to the antiquark. One obtains the
following $5Q$ component of the neutron wave function in the
momentum space
\begin{eqnarray}
(|n\rangle_k)^{f_1f_2f_3f_4,\sigma_1\sigma_2\sigma_3\sigma_4}_{f_5,\sigma_5}(p_1\ldots p_5)&=&\frac{\sqrt{8}}{360}\,F^{j_1\sigma_1}(p_1)F^{j_2\sigma_2}(p_2)F^{j_3\sigma_3}(p_3)W^{j_4\sigma_4}_{j_5\sigma_5}(p_4,p_5)\nonumber\\
&\times&\epsilon_{k'k}\left\{\epsilon^{f_1f_2}\epsilon_{j_1j_2}\left[\delta^{f_3}_2\delta^{f_4}_{f_5}\left(4\delta^{j_5}_{j_4}\delta^{k'}_{j_3}-\delta^{j_5}_{j_3}\delta^{k'}_{j_4}\right)+\delta^{f_4}_2\delta^{f_3}_{f_5}\left(4\delta^{j_5}_{j_3}\delta^{k'}_{j_4}-\delta^{j_5}_{j_4}\delta^{k'}_{j_3}\right)\right]\right.\nonumber\\
&+&\epsilon^{f_1f_4}\epsilon_{j_1j_4}\left[\delta^{f_2}_2\delta^{f_3}_{f_5}\left(4\delta^{j_5}_{j_3}\delta^{k'}_{j_2}-\delta^{j_5}_{j_2}\delta^{k'}_{j_3}\right)+\delta^{f_3}_2\delta^{f_2}_{f_5}\left(4\delta^{j_5}_{j_2}\delta^{k'}_{j_3}-\delta^{j_5}_{j_3}\delta^{k'}_{j_2}\right)\right]\nonumber\\
&+&\textrm{cyclic permutations of (1,2,3)}\Big\}.
\end{eqnarray}
The color degrees of freedom are not explicitly written but the
three valence quarks (1,2,3) are still antisymmetric in color while
the quark-antiquark pair (4,5) is a color singlet. The wave function
is schematically represented on Fig. \ref{Fivequarks}. Let us
concentrate on the flavor part of this wave function. The
quark-antiquark pair introduces explicitly the hidden strange flavor
thanks to the terms like $\delta^{f_4}_{f_5}$ for the particular
component $f_4=f_5=3\equiv s$. Moreover one can notice that
strangeness is also allowed to access to the valence level thanks to
terms like $\delta^{f_i}_{f_5}$ with $i=1,2,3$. The flavor structure
of the neutron at the $5Q$ level is then
\begin{equation}\label{hidden}
|n\rangle=A|udd(u\bar u)\rangle+B|udd(d\bar d)\rangle+C|udd(s\bar
s)\rangle+D|uud(d\bar u)\rangle+E|uds(d\bar s)\rangle+F|ddd(u\bar d)\rangle+G|dds(u\bar s)\rangle
\end{equation}
where the three first flavors belong to the valence sector and the
last two to the quark-antiquark pair.

Exotic baryons from the
$\left({\bf\overline{10}},\frac{1}{2}^+\right)$ multiplet, despite
the absence of a $3Q$ component, have such a $5Q$ component in their
wave function. Here is for example the $5Q$ wave function for the
$\Theta^+$ pentaquark
\begin{eqnarray}
(|\Theta^+\rangle_k)^{f_1f_2f_3f_4,\sigma_1\sigma_2\sigma_3\sigma_4}_{f_5,\sigma_5}(p_1\ldots p_5)&=&\frac{\sqrt{30}}{180}\,F^{j_1\sigma_1}(p_1)F^{j_2\sigma_2}(p_2)F^{j_3\sigma_3}(p_3)W^{j_4\sigma_4}_{j_5\sigma_5}(p_4,p_5)\nonumber\\
&\times&\{\epsilon^{f_1f_2}\epsilon^{f_3f_4}\epsilon_{j_1j_2}\epsilon_{j_3j_4}\delta^3_{f_5}\delta^{j_5}_k\nonumber\\
&+&\textrm{ cyclic permutations of (1,2,3)}\}.
\end{eqnarray}
The color structure is here very simple:
$\epsilon^{\alpha_1\alpha_2\alpha_3}\delta^{\alpha_4}_{\alpha_5}$.
Like in the nucleon we have a $ud$ pair in the spin- and
isospin-zero combination $\epsilon^{f_1f_2}\epsilon_{j_1j_2}$.

\subsection{$7Q$ component of baryons}

The $7Q$ component of the baryon wave functions is obtained by
expanding the coherent exponential (\ref{Coherent exponential}) to
the second order in the $Q\bar Q$ pair. The projection involves now
along with the three $R$'s from the discrete level four additional
matrices $\left(R\,R^\dag\right)\left(R\,R^\dag\right)$ that rotate
the two quark-antiquark pairs in the $SU(3)$ space
\begin{equation}
T(B)^{f_1f_2f_3f_4f_6,j_5j_7}_{j_1j_2j_3j_4j_6,f_5f_7,k}\equiv\int\ud
R\,B^*_k(R)R^{f_1}_{j_1}R^{f_2}_{j_2}R^{f_3}_{j_3}\left(R^{f_4}_{j_4}R^{\dag
j_5}_{f_5}\right)\left(R^{f_6}_{j_6}R^{\dag j_7}_{f_7}\right).
\end{equation}
Components $i=1,2,3$ refer then to the valence part, $i=4,6$ to the
quarks of the sea and $i=5,7$ to the antiquarks. The $7Q$ component
of the neutron wave function in the momentum space is quite
complicated but the three valence quarks (1,2,3) are still
antisymmetric in color while the quark-antiquark pairs (4,5) and
(6,7) are color singlets.

By analogy with the $5Q$ component of ordinary baryons, the $7Q$
component of pentaquark modifies the flavor structure in the valence
sector. Let us consider for example $\Theta^+$ whose valence
structure in the $5Q$ sector is $uud$ and $udd$. The $7Q$ component
introduces four new possibilities $uuu$, $ddd$, $uus$ and $dds$ and
thus even though $\Theta^+$ has strangeness $S=+1$ it can contain a
valence strange quark. The flavor structure of the valence sector is
especially interesting for the tensor charges. Since the tensor
operator is chiral odd only valence quarks can contribute and thus a
non-zero strange contribution to nucleon tensor charge would
indicate the presence of strange quarks in the valence sector, which
is forbidden in the $3Q$ picture. This will be discussed further in
the chapter dedicated to tensor charges.

\subsection{$nQ$ component of baryons}

It is easy to generalize to the case $nQ$ with $n\geq 3$ and odd.
The $nQ$ component of the baryon wave functions is obtained by
expanding the coherent exponential (\ref{Coherent exponential}) to
the $(n-3)/2$th order in the $Q\bar Q$ pair. The projection involves
now along with the three $R$'s from the discrete level $(n-3)/2$
additional pairs of matrices $R\,R^\dag$ that rotate the $(n-3)/2$
quark-antiquark pairs in the $SU(3)$ space
\begin{equation}
T(B)^{f_1f_2f_3f_4f_6\dots f_{n-1},j_5j_7\dots
j_n}_{j_1j_2j_3j_4j_6\dots j_{n-1},f_5f_7\dots f_n,k}\equiv\int\ud
R\,B^*_k(R)R^{f_1}_{j_1}R^{f_2}_{j_2}R^{f_3}_{j_3}\left(R^{f_4}_{j_4}R^{\dag
j_5}_{f_5}\right)\left(R^{f_6}_{j_6}R^{\dag
j_7}_{f_7}\right)\dots\left(R^{f_{n-1}}_{j_{n-1}}R^{\dag
j_n}_{f_n}\right).
\end{equation}
Components $i=1,2,3$ refer then to the valence part,
$i=4,6,\ldots,n-1$ to the quarks of the sea and $i=5,7,\ldots,n$ to
the antiquarks.

\section{Matrix elements, normalization and charges}

The normalization of the $nQ$ Fock component of a specific baryon
$B$ wave function is obtained by
\begin{equation}\label{Normalization}
\uN^{(n)}(B)=\langle\Psi^{(n)k}(B)|\Psi^{(n)}_k(B)\rangle\qquad\textrm{no
summation on $k$!}
\end{equation}
where $k$ is the spin projection on the $z$ direction. One has to
drag all annihilation operators in $\Psi^{(n)\dag k}(B)$ to the
right and the creation operators in $\Psi^{(n)}_k(B)$ to the left so
that the vacuum state $|\Omega_0\rangle$ is nullified. One then gets
a non-zero result due to the anticommutation relations
(\ref{Anticommutation}) or equivalently to the ``contractions'' of
the operators.

Nucleon properties are characterized by its parton distributions in
hard processes. At the leading twist level there have been
considerable efforts both theoretically and experimentally to
determine the unpolarized $f_1(x)$ and longitudinally polarized (or
helicity) $g_1(x)$ quark-spin distributions. In fact a third
structure function exists and is called the transversity
distribution $h_1(x)$ \cite{Ralston}. The functions $f_1,g_1,h_1$
are respectively spin-average, chiral-even and chiral-odd spin
distributions. Only $f_1$ and $g_1$ contribute to deep-inelastic
scattering (DIS) when small quark-mass effects are ignored. The
function $h_1$ can be measured in certain physical processes such as
polarized Drell-Yan processes \cite{Ralston} and other exclusive
hard reactions \cite{Jaffe,Hard,Collins}. Let us stress however that
$h_1(x)$ does not represent the quark transverse spin distribution.
The transverse spin operator does not commute with the free-particle
Hamiltonian. In the light-cone formalism the transverse spin
operator is a bad operator and depends on the dynamics. This would
explain why the interest in transversity distributions is rather
recent. The interested reader can find a review of the subject in
\cite{Review}.

Vector, axial and tensor charges, which are first moment of the
leading twist distributions, are examples of typical physical
observables that can be obtained by means of the matrix element of
some operator (preferably written in terms of quark
annihilation-creation operators $a$, $b$, $a^\dag$, $b^\dag$)
sandwiched between the initial and final baryon wave functions. In
the present thesis we consider four types of charges: vector, axial,
tensor and magnetic. A chapter is dedicated to each charge where it
is treated explicitly and discussed.

The big advantage of the IMF is that the number of $Q\bar Q$ pairs
is not changed by the current. Hence there will only be diagonal
transitions in the Fock space, \emph{i.e.} the charges can be
decomposed into the sum of the contributions from all Fock
components
\begin{equation}
Q=\sum_nQ^{(n)}.
\end{equation}

Note that the matrix elements have to be properly normalized as in
the following example
\begin{equation}
Q(B_1\to B_2)=\frac{Q^{(3)}(B_1\to B_2)+Q^{(5)}(B_1\to
B_2)+\ldots}{\sqrt{\uN^{(3)}(B_1)+\uN^{(5)}(B_1)+\dots}\sqrt{\uN^{(3)}(B_2)+\uN^{(5)}(B_2)+\ldots}}
\end{equation}
in order to get the physical values. The current used may change the
nature of the particle and so the initial and final baryons are not
necessarily the same.

\subsection{$3Q$ contribution}

\begin{figure}[h]\begin{center}\includegraphics[width=17cm]{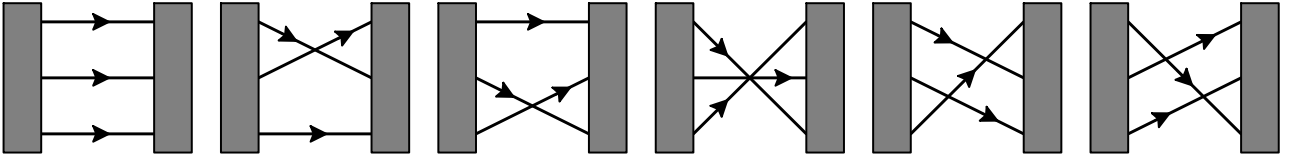}
\caption{\small{Schematic representation of the $3Q$ contribution to
matrix elements. Each quark line stands for the color, flavor and
spin contractions
$\delta^{\alpha_i}_{\alpha'_i}\delta^{f_i}_{f'_i}\delta^{\sigma_i}_{\sigma'_i}
\int\ud
z'_i\,\ud^2\up'_{i\perp}\delta(z_i-z'_i)\delta^{(2)}(\up_{i\perp}-\up'_{i\perp})$
with the primed variables referring to $\Psi^{(3)\dag}(B)$ (right
rectangle) and the unprimed ones to $\Psi^{(3)}(B)$ (left
rectangle). A crossing of these quark lines corresponds to an
anticommutation of two creation or annihilation operators and thus
introduces a \emph{minus} sign. This sign is however compensated by
the one coming from the contraction of $\epsilon$ tensors with color
indices. Since only valence quarks are involved all those diagrams
are equivalent to the first one.}}\label{Three-q normalization}
\end{center}
\end{figure}

To get the normalization of the $3Q$ sector one has to contract the
three creation operators $a_1^\dag a_2^\dag a_3^\dag$ of
$\Psi^{(3)}(B)$ with the three annihilation operators $a_1a_2a_3$ of
$\Psi^{(3)\dag}(B)$. Since the three valence quarks are equivalent
the $3!$ possible contractions give identical contributions. So only
one diagram is needed to represent the $3Q$ component and we chose
the simplest one represented in Fig. \ref{Three-q normalization}.
The contraction in color gives an additional factor of
$3!=\epsilon^{\alpha_1\alpha_2\alpha_3}\epsilon_{\alpha_1\alpha_2\alpha_3}$.

From eqs. (\ref{T tensor}) and (\ref{Normalization}) one can express
the normalization of the $3Q$ component of baryon wave functions as
\begin{eqnarray}
\uN^{(3)}(B)&=&36\,T(B)^{f_1f_2f_3}_{j_1j_2j_3,k}T(B)_{f_1f_2f_3}^{l_1l_2l_3,k}\int\ud
z_{1,2,3}\,\frac{\ud^2\up_{1,2,3\perp}}{(2\pi)^6}\,\delta(z_1+z_2+z_3-1)(2\pi)^2\delta^{(2)}(\up_{1\perp}+\up_{2\perp}+\up_{3\perp})\nonumber\\
&\times&F^{j_1\sigma_1}(p_1)F^{j_2\sigma_2}(p_2)F^{j_3\sigma_3}(p_3)F^\dag_{l_1\sigma_1}(p_1)F^\dag_{l_2\sigma_2}(p_2)F^\dag_{l_3\sigma_3}(p_3)\qquad\textrm{no
summation on $k$!}\label{Normalization 3q}
\end{eqnarray}
where $F^{j\sigma}(p)\equiv F^{j\sigma}(z,\up_\perp)$ is the
discrete-level wave function (\ref{Discrete level
IMF})+(\ref{Discrete level IMF 2}).

All charges considered in this thesis are obtained by means of
one-quark operators. These charges are computed by inserting the
corresponding operator in each quark line. In the $3Q$ sector there
is no antiquark which means that the $b^\dag b$ part of the operator
does not play. As in the $3Q$ normalization one gets the factor 36
from all contractions. Since valence quarks are equivalent the
insertion of the operator in all three quark lines gives three times
the same result. Let the third quark line be the one where the
operator is inserted, see Fig. \ref{Three-q charge}. If we denote by
$\int(\ud p_{1-3})$ the integrals over momenta with the
$\delta$-functions as in eq. (\ref{Normalization 3q}) one obtains
the following expression for matrix element of the charge $Q$
\begin{eqnarray}
Q^{(3)}(1\to 2)&=&36\,T(1)^{f_1f_2f_3}_{j_1j_2j_3,k}T(2)_{f_1f_2g_3}^{l_1l_2l_3,l}\int(\ud p_{1-3})\nonumber\\
&\times&\Big[F^{j_1\sigma_1}(p_1)F^{j_2\sigma_2}(p_2)F^{j_3\sigma_3}(p_3)\Big]\left[F^\dag_{l_1\sigma_1}(p_1)F^\dag_{l_2\sigma_2}(p_2)F^\dag_{l_3\tau_3}(p_3)\right]\left[3M^{\tau_3}_{\sigma_3}J^{g_3}_{f_3}\right].\label{Vector
charge 3q}
\end{eqnarray}
where $J^{g_3}_{f_3}$ is the flavor content of the operator and
$M^{\tau_3}_{\sigma_3}$ is the action of the operator on the quark
spin. For example the vector operator is blind concerning the quark
spin and thus $M^{\tau_3}_{\sigma_3}=\delta^{\tau_3}_{\sigma_3}$.
The axial operator gives different signs to quarks with spin up and
spin down and thus
$M^{\tau_3}_{\sigma_3}=(\sigma_3)^{\tau_3}_{\sigma_3}$. We consider
here for simplicity only matrix elements in the limit of zero
momentum transfer.

\begin{figure}[h]\begin{center}\begin{minipage}[c]{10cm}\begin{center}\includegraphics[width=4cm]{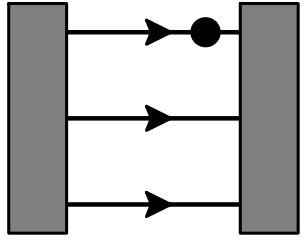}
\caption{\small{Schematic representation of the $3Q$ contribution to
a charge. The black dot stands for the one-quark operator. Since all
three quark lines are equivalent one has three times this specific
contribution.}}\label{Three-q
charge}\end{center}\end{minipage}\end{center}
\end{figure}

\subsection{$5Q$ contributions}\label{fivequarks diagrams}

\begin{figure}[h]\begin{center}\begin{minipage}[c]{10cm}\begin{center}\includegraphics[width=3cm]{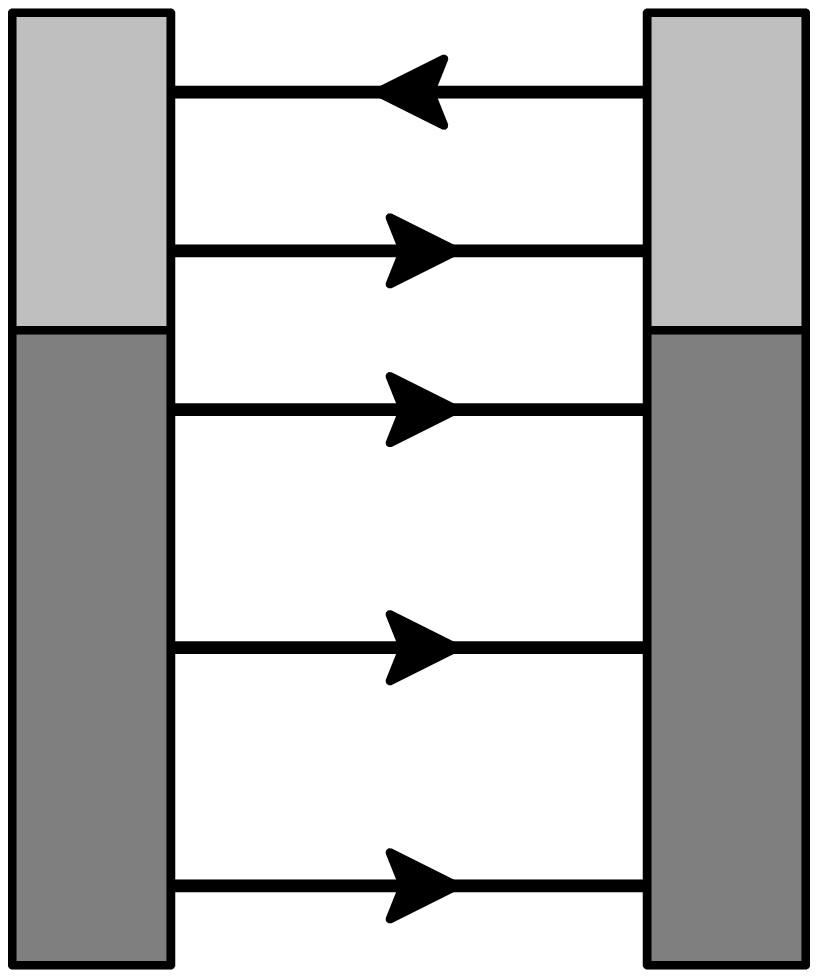}\hspace{1cm}\includegraphics[width=3cm]{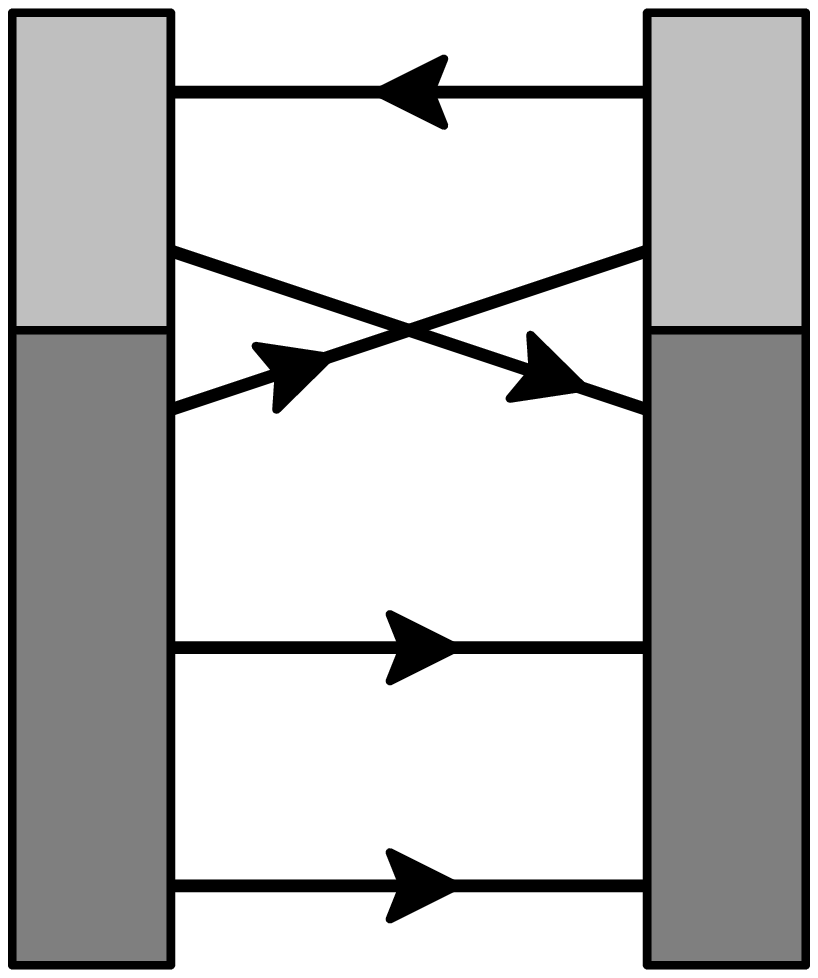}
\caption{\small{Schematic representation of the $5Q$ direct (left)
and exchange (right) contributions to matrix
elements.}}\label{Five-q normalization}
\end{center}\end{minipage}\end{center}\end{figure}

In the $5Q$ sector due to the presence of a quark-antiquark pair
more diagrams are possible. Using the fact that valence quarks are
equivalent only two types of diagrams survive: the direct and the
exchange ones (see Fig. \ref{Five-q normalization}).
In the former one contracts the $a^\dag$ from the pair wave function
with the $a$ in the conjugate pair and all the valence operators are
contracted with each other. As in the $3Q$ case there are 6
equivalent possibilities but the contractions in color give now a
factor of $6\cdot
3=\epsilon^{\alpha_1\alpha_2\alpha_3}\epsilon_{\alpha_1\alpha_2\alpha_3}\delta^{\alpha}_{\alpha}$
because of the sum over color in the pair, giving then a total
factor of 108. In the exchange contribution one contracts the
$a^\dag$ from the pair with one of the three $a$'s from the
conjugate discrete level. \emph{Vice versa}, the $a$ from the
conjugate pair is contracted with one of the three $a^\dag$'s from
the discrete level. There are at all 18 equivalent possibilities but
the contractions in color give only a factor of
$6=\epsilon^{\alpha_1\alpha_2\alpha}\epsilon_{\alpha_1\alpha_2\alpha_3}\delta^{\alpha_3}_{\alpha}$
and so one gets also a global factor of 108 for the exchange
contribution but with an additional minus sign because one has to
anticommute fermion operators to obtain exchange terms. We thus
obtain the following expression for the $5Q$ normalization
\begin{eqnarray}
\uN^{(5)}(B)&=&108\,T(B)^{f_1f_2f_3f_4,j_5}_{j_1j_2j_3j_4,f_5,k}T(B)_{f_1f_2g_3g_4,l_5}^{l_1l_2l_3l_4,f_5,k}\int(\ud p_{1-5})\nonumber\\
&\times&F^{j_1\sigma_1}(p_1)F^{j_2\sigma_2}(p_2)F^{j_3\sigma_3}(p_3)W^{j_4\sigma_4}_{j_5\sigma_5}(p_4,p_5)F^\dag_{l_1\sigma_1}(p_1)F^\dag_{l_2\sigma_2}(p_2)\nonumber\\
&\times&\left[F^\dag_{l_3\sigma_3}(p_3)W_{c\,l_4\sigma_4}^{l_5\sigma_5}(p_4,p_5)\delta^{g_3}_{f_3}\delta^{g_4}_{f_4}-F^\dag_{l_3\sigma_4}(p_4)W_{c\,l_4\sigma_3}^{l_5\sigma_5}(p_3,p_5)\delta^{g_3}_{f_4}\delta^{g_4}_{f_3}\right]\nonumber\\
&&\textrm{no summation on $k$!}\label{Normalization 5q}
\end{eqnarray}
where we have denoted
\begin{equation}
\int(\ud p_{1-5})=\int\ud
z_{1-5}\,\delta(z_1+\ldots+z_5-1)\int\frac{\ud^2\up_{1-5\perp}}{(2\pi)^{10}}\,(2\pi)^2\delta^{(2)}(\up_{1\perp}+\ldots+\up_{5\perp}).
\end{equation}

Concerning the charges we have three types of direct contributions
(antiquark, sea quark and valence quarks) and four types of exchange
contributions (antiquark, exchange of the sea quark with a valence
quark and other valence quarks).
\begin{figure}[h]\begin{center}\begin{minipage}[c]{6.86cm}\begin{center}\includegraphics[width=6.86cm]{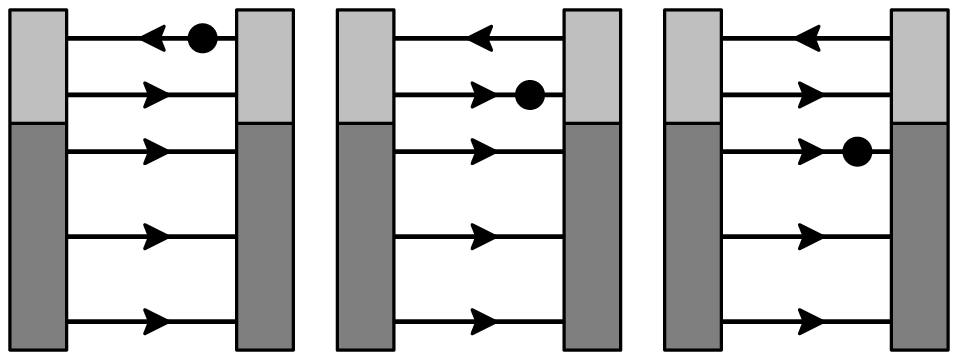}
\caption{\small{Schematic representation of the three types of $5Q$
direct contributions to the charges.}}\label{Direct charges}
\end{center}\end{minipage}\hspace{0.5cm}
\begin{minipage}[c]{9.14cm}\begin{center}\includegraphics[width=9.14cm]{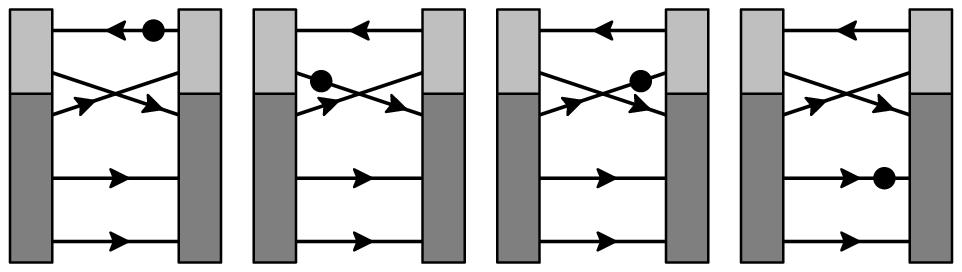}
\caption{\small{Schematic representation of the four types of $5Q$
exchange contributions to the charges.\newline}}\label{Exchange
charges}
\end{center}\end{minipage}\end{center}
\end{figure}
From the schematic representations of these contributions  (see
Figs. \ref{Direct charges} and \ref{Exchange charges}) it is easy to
write the transitions.\newline\newline Direct diagram:
\begin{eqnarray}
Q^{(5)\textrm{direct}}(1\to 2)&=&108\,T(1)^{f_1f_2f_3f_4,j_5}_{j_1j_2j_3j_4,f_5,k}T(2)_{f_1f_2g_3g_4,l_5}^{l_1l_2l_3l_4,g_5,l}\int(\ud p_{1-5})\nonumber\\
&\times&F^{j_1\sigma_1}(p_1)F^{j_2\sigma_2}(p_2)F^{j_3\sigma_3}(p_3)W^{j_4\sigma_4}_{j_5\sigma_5}(p_4,p_5)F^\dag_{l_1\sigma_1}(p_1)F^\dag_{l_2\sigma_2}(p_2)F^\dag_{l_3\tau_3}(p_3)W_{c\,l_4\tau_4}^{l_5\tau_5}(p_4,p_5)\nonumber\\
&\times&\left[-\delta^{g_3}_{f_3}\delta^{g_4}_{f_4}
J^{f_5}_{g_5}\delta^{\tau_3}_{\sigma_3}\delta^{\tau_4}_{\sigma_4}M_{\tau_5}^{\sigma_5}+\delta^{g_3}_{f_3}
J^{g_4}_{f_4}\delta^{f_5}_{g_5}\delta^{\tau_3}_{\sigma_3}M^{\tau_4}_{\sigma_4}\delta_{\tau_5}^{\sigma_5}+3
J^{g_3}_{f_3}\delta^{g_4}_{f_4}\delta^{f_5}_{g_5}M^{\tau_3}_{\sigma_3}\delta^{\tau_4}_{\sigma_4}\delta_{\tau_5}^{\sigma_5}\right].\label{Direct}
\end{eqnarray}
Exchange diagram:
\begin{eqnarray}
Q^{(5)\textrm{exchange}}(1\to 2)&=&-108\,T(1)^{f_1f_2f_3f_4,j_5}_{j_1j_2j_3j_4,f_5,k}T(2)_{f_1g_2g_3g_4,l_5}^{l_1l_2l_3l_4,g_5,l}\int(\ud p_{1-5})\nonumber\\
&\times&F^{j_1\sigma_1}(p_1)F^{j_2\sigma_2}(p_2)F^{j_3\sigma_3}(p_3)W^{j_4\sigma_4}_{j_5\sigma_5}(p_4,p_5)F^\dag_{l_1\sigma_1}(p_1)F^\dag_{l_2\tau_2}(p_2)F^\dag_{l_3\tau_3}(p_3)W_{c\,l_4\tau_4}^{l_5\tau_5}(p_3,p_5)\nonumber\\
&\times&\left[-\delta^{g_2}_{f_2}\delta^{g_4}_{f_3}\delta^{g_3}_{f_4}
J^{f_5}_{g_5}\delta^{\tau_2}_{\sigma_2}\delta^{\tau_4}_{\sigma_3}\delta^{\tau_3}_{\sigma_4}M_{\tau_5}^{\sigma_5}+\delta^{g_2}_{f_2}\delta^{g_4}_{f_3}
J^{g_3}_{f_4}\delta^{f_5}_{g_5}\delta^{\tau_2}_{\sigma_2}\delta^{\tau_4}_{\sigma_3}M^{\tau_3}_{\sigma_4}\delta_{\tau_5}^{\sigma_5}\right.\nonumber\\
&&+\left.\delta^{g_2}_{f_2}
J^{g_4}_{f_3}\delta^{g_3}_{f_4}\delta^{f_5}_{g_5}\delta^{\tau_2}_{\sigma_2}M^{\tau_4}_{\sigma_3}\delta^{\tau_3}_{\sigma_4}\delta_{\tau_5}^{\sigma_5}
+2
J^{g_2}_{f_2}\delta^{g_4}_{f_3}\delta^{g_3}_{f_4}\delta^{f_5}_{g_5}M^{\tau_2}_{\sigma_2}\delta^{\tau_4}_{\sigma_3}\delta^{\tau_3}_{\sigma_4}\delta_{\tau_5}^{\sigma_5}\right].\label{Exchange}
\end{eqnarray}

\subsection{$7Q$ contributions}\label{fivequarks diagrams}

\begin{figure}[h]\begin{center}\begin{minipage}[c]{7cm}\begin{center}\includegraphics[width=7cm]{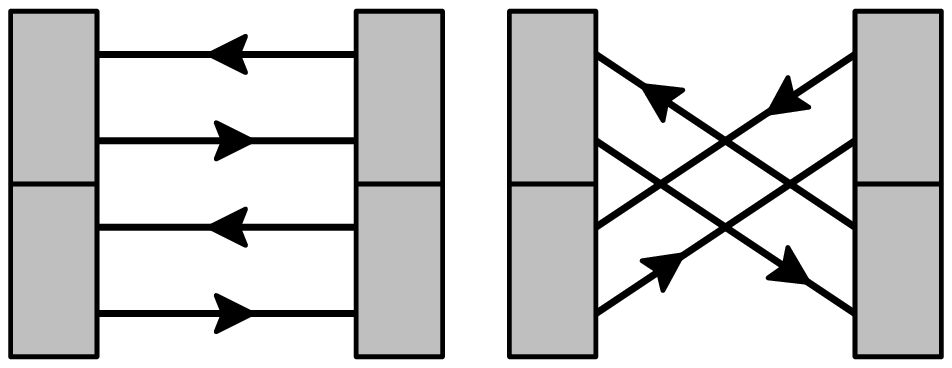}
\caption{\small{Contractions of two quark-antiquark pairs leaving
them
unchanged.}}\label{unchanged}\end{center}\end{minipage}\hspace{1cm}\begin{minipage}[c]{7cm}\begin{center}\includegraphics[width=7cm]{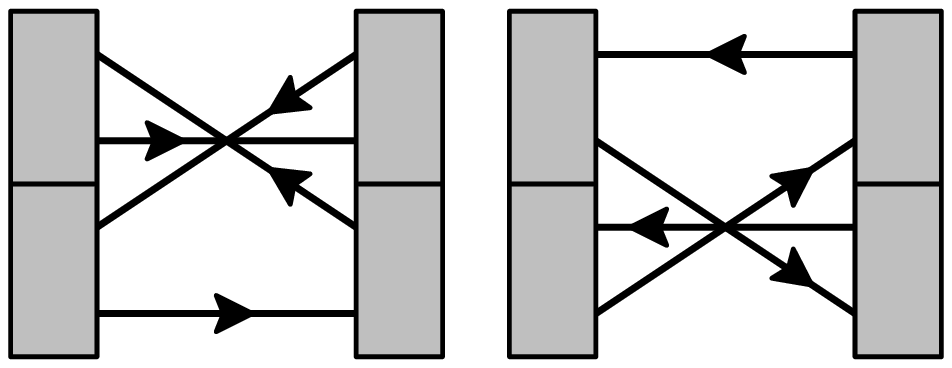}
\caption{\small{Contractions of two quark-antiquark pairs where one
of the constituents is
exchanged.}}\label{exchanged}\end{center}\end{minipage}\end{center}
\end{figure}

In the $7Q$ sector where two quark-antiquark pairs are involved the
number of possible diagrams grows. These two pairs may remain
unchanged (see Fig. \ref{unchanged}) or exchange one of their
constituents (see Fig. \ref{exchanged}).
The pairs are in fact also equivalent and can be exchanged without
any change in the result. This means that diagrams in Fig.
\ref{unchanged} are equivalent, the same for those in Fig.
\ref{exchanged}. Finally after all contractions only five
non-equivalent diagrams survive, see Fig. \ref{Seven-q
normalization}.
\begin{figure}[h]\begin{center}\includegraphics[width=15cm]{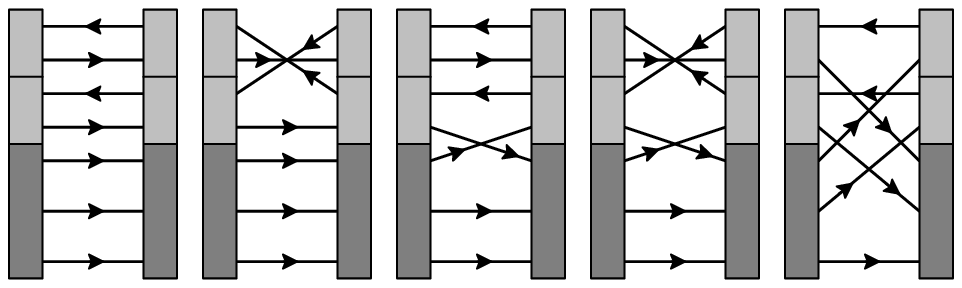}
\caption{\small{Schematic representation of the $7Q$ contributions
to matrix elements.}}\label{Seven-q normalization}\end{center}
\end{figure}
The factor associated to each of these diagrams can easily be
obtained since it is just a game of combinatorics. We give in
Appendix B some general tools to find all these diagrams and their
respective factor in any $nQ$ Fock sector. The $7Q$ factors are
treated as explicit examples. With these tools one can obtain
systematically all the contributions with the factor and the sign by
means of diagrams only. One avoids thus the tedious work of
contracting all creation-annihilation operators.

The $7Q$ normalization has the following form
\begin{eqnarray}
\uN^{(7)}(B)&=&216\,T(B)^{f_1f_2f_3f_4f_6,j_5j_7}_{j_1j_2j_3j_4j_6,f_5f_7,k}T(B)_{f_1g_2g_3g_4g_6,l_5l_7}^{l_1l_2l_3l_4l_6,g_5g_7,k}\int(\ud p_{1-7})\nonumber\\
&\times&F^{j_1\sigma_1}(p_1)F^{j_2\sigma_2}(p_2)F^{j_3\sigma_3}(p_3)W^{j_4\sigma_4}_{j_5\sigma_5}(p_4,p_5)W^{j_6\sigma_6}_{j_7\sigma_7}(p_6,p_7)F^\dag_{l_1\sigma_1}(p_1)\nonumber\\
&\times&\Big[3F^\dag_{l_2\sigma_2}(p_2)F^\dag_{l_3\sigma_3}(p_3)W_{c\,l_4\sigma_4}^{l_5\sigma_5}(p_4,p_5)W_{c\,l_6\sigma_6}^{l_7\sigma_7}(p_6,p_7)\delta^{g_2}_{f_2}\delta^{g_3}_{f_3}\delta^{g_4}_{f_4}\delta^{f_5}_{g_5}\delta^{g_6}_{f_6}\delta^{f_7}_{g_7}\nonumber\\
&-&F^\dag_{l_2\sigma_2}(p_2)F^\dag_{l_3\sigma_3}(p_3)W_{c\,l_4\sigma_4}^{l_5\sigma_7}(p_4,p_7)W_{c\,l_6\sigma_6}^{l_7\sigma_5}(p_6,p_5)\delta^{g_2}_{f_2}\delta^{g_3}_{f_3}\delta^{g_4}_{f_4}\delta^{f_7}_{g_5}\delta^{g_6}_{f_6}\delta^{f_5}_{g_7}\nonumber\\
&-&6F^\dag_{l_2\sigma_2}(p_2)F^\dag_{l_3\sigma_4}(p_4)W_{c\,l_4\sigma_3}^{l_5\sigma_5}(p_3,p_5)W_{c\,l_6\sigma_6}^{l_7\sigma_7}(p_6,p_7)\delta^{g_2}_{f_2}\delta^{g_3}_{f_4}\delta^{g_4}_{f_3}\delta^{f_5}_{g_5}\delta^{g_6}_{f_6}\delta^{f_7}_{g_7}\nonumber\\
&+&2F^\dag_{l_2\sigma_2}(p_2)F^\dag_{l_3\sigma_4}(p_4)W_{c\,l_4\sigma_3}^{l_5\sigma_7}(p_3,p_7)W_{c\,l_6\sigma_6}^{l_7\sigma_5}(p_6,p_5)\delta^{g_2}_{f_2}\delta^{g_3}_{f_4}\delta^{g_4}_{f_3}\delta^{f_7}_{g_5}\delta^{g_6}_{f_6}\delta^{f_5}_{g_7}\nonumber\\
&+&2F^\dag_{l_2\sigma_4}(p_4)F^\dag_{l_3\sigma_6}(p_6)W_{c\,l_4\sigma_2}^{l_5\sigma_5}(p_2,p_5)W_{c\,l_6\sigma_3}^{l_7\sigma_7}(p_3,p_7)\delta^{g_2}_{f_4}\delta^{g_3}_{f_6}\delta^{g_4}_{f_2}\delta^{f_5}_{g_5}\delta^{g_6}_{f_3}\delta^{f_7}_{g_7}\Big]\nonumber\\
&&\textrm{no summation on $k$!}\label{Normalization 7q}
\end{eqnarray}

Here is the explicit expression for the contribution to charges
represented by the first diagram in Fig. \ref{Seven-q normalization}
\begin{eqnarray}
Q^{(7)\textrm{direct}}(1\to 2)&=&648\,T(1)^{f_1f_2f_3f_4f_6,j_5j_7}_{j_1j_2j_3j_4j_6,f_5f_7,k}T(2)_{f_1f_2g_3f_4g_6,l_5l_7}^{l_1l_2l_3l_4l_6,f_5g_7,l}\int(\ud p_{1-7})\nonumber\\
&\times&F^{j_1\sigma_1}(p_1)F^{j_2\sigma_2}(p_2)F^{j_3\sigma_3}(p_3)W^{j_4\sigma_4}_{j_5\sigma_5}(p_4,p_5)F^\dag_{l_1\sigma_1}(p_1)F^\dag_{l_2\sigma_2}(p_2)F^\dag_{l_3\tau_3}(p_3)W_{c\,l_4\tau_4}^{l_5\tau_5}(p_4,p_5)\nonumber\\
&\times&\left[-2\delta^{g_3}_{f_3}\delta^{g_6}_{f_6}
J^{f_7}_{g_7}\delta^{\tau_3}_{\sigma_3}\delta^{\tau_6}_{\sigma_6}M_{\tau_7}^{\sigma_7}+2\delta^{g_3}_{f_3}
J^{g_6}_{f_6}\delta^{f_7}_{g_7}\delta^{\tau_3}_{\sigma_3}M^{\tau_6}_{\sigma_6}\delta_{\tau_7}^{\sigma_7o}+3
J^{g_3}_{f_3}\delta^{g_6}_{f_6}\delta^{f_7}_{g_7}M^{\tau_3}_{\sigma_3}\delta^{\tau_6}_{\sigma_6}\delta_{\tau_7}^{\sigma_7}\right].\label{Direct}
\end{eqnarray}
The factor 2 in front of the first and second terms reflects the
fact that the action of the operator on both quark-antiquark pairs
is the same. The contributions of the other diagrams can easily be
obtained but have been neglected in the present thesis. We will
discuss this point later.

\section{Scalar overlap integrals}

The computation of matrix elements has been done in two steps. The
first step is the contraction over all flavor $(f,g)$, isospin
$(j,l)$ and spin $(\sigma,\tau)$ indices. All charges are then
reduced to linear combinations of a finite set of scalar integrals
over longitudinal $z$ and transverse $\up_\perp$ momenta. These
integrals are just overlaps of valence and pair wave functions. The
integrals over relative transverse momenta in the quark-antiquark
pair are generally UV divergent. This divergence should be cut by
the momentum-dependent dynamical quark mass $M(p)$, see eq.
(\ref{Effective lagrangian}). Following the authors of \cite{Cutoff}
we mimic the fall-off of $M(p)$ by the Pauli-Villars cutoff at
$M_\textrm{PV}=556.8$ MeV (this value being chosen from the
requirement that the pion decay constant $F_\pi=93$ MeV is
reproduced from $M(0)=345$ MeV).

The complexity of these integrals is directly related to the
complexity of the diagram. The simplest ones are the ``direct''
diagrams where no exchange of quarks is involved. In this case
valence quarks and sea pairs keep their identity. This is reflected
by the fact that the integrals can be performed in many steps and
that valence and sea pairs variables almost decouple.

\subsection{$3Q$ scalar integrals}

For convenience we introduce the probability distribution
$\Phi^I(z,\uq_\perp)$ that three valence quarks leave longitudinal
fraction $z=q_z/\uM$ and transverse momentum $\uq_\perp$ to the
quark-antiquark pair(s) with $I=V,A,T,M$ referring to the vector,
axial, tensor or magnetic case
\begin{equation}\label{Probability3q}
\Phi^I(z,\uq_\perp)=\int\ud
z_{1,2,3}\frac{\ud^2\up_{1,2,3\perp}}{(2\pi)^6}\,\delta(z+z_1+z_2+z_3-1)(2\pi)^2\delta^{(2)}(\uq_\perp+\up_{1\perp}+\up_{2\perp}+\up_{3\perp})D^I(p_1,p_2,p_3).
\end{equation}
The function $D^I(p_1,p_2,p_3)$ is given in terms of the upper and
lower valence wave functions $h(p)$ and $j(p)$ and is constructed
from the product of all valence wave functions $F$ and the current
operator. Its explicit form in the vector, axial, tensor or magnetic
case can be found in the corresponding chapters.

In the $3Q$ sector there is no quark-antiquark pair. This means that
the whole baryon momentum is carried by the valence quarks. The $3Q$
scalar integrals are thus simply $\Phi^I(0,0)$.

\subsection{$5Q$ direct scalar integrals}

In the $5Q$ sector there is one quark-antiquark pair. Thanks to the
simplicity of the direct diagram the corresponding scalar overlap
integrals can be written in two parts: purely valence $\Phi^I$ and
sea $G_J$
\begin{equation}
K^I_J=\frac{M^2}{2\pi}\int \frac{\ud^3
\uq}{(2\pi)^3}\,\Phi^I\left(\frac{q_z}{\uM},\uq_\perp\right)\theta(q_z)\,q_z\,G_J(q_z,\uq_\perp)
\end{equation}
where $G_J$ is a quark-antiquark probability distribution and
$J=\pi\pi,33,332,\sigma\sigma,3\sigma$. These distributions are
obtained by contracting two quark-antiquark wave functions $W$, see
eq. (\ref{Pair wavefunction}) and regularized by means of
Pauli-Villars procedure
\begin{eqnarray}
G_{\pi\pi}(q_z,\uq_\perp)&=&\Pi^2(\uq)\int^1_0\ud
y\int\frac{\ud^2\uQcal_\perp}{(2\pi)^2}\frac{\uQcal^2_\perp+M^2}{(\uQcal^2_\perp+M^2+y(1-y)\uq^2)^2}-(M\to
M_\textrm{PV}),\label{Direct1}
\\
G_{33}(q_z,\uq_\perp)&=&\frac{q_z^2}{\uq^2}\,G_{\pi\pi}(q_z,\uq_\perp),\label{Direct2}
\\
G_{332}(q_z,\uq_\perp)&=&\frac{q_z}{\uq^2}\,G_{\pi\pi}(q_z,\uq_\perp),\label{Direct2}
\\
G_{\sigma\sigma}(q_z,\uq_\perp)&=&\Sigma^2(\uq)\int^1_0\ud
y\int\frac{\ud^2\uQcal_\perp}{(2\pi)^2}\frac{\uQcal^2_\perp+M^2(2y-1)^2}{(\uQcal^2_\perp+M^2+y(1-y)\uq^2)^2}-(M\to
M_\textrm{PV}),\label{Direct3}
\\
G_{3\sigma}(q_z,\uq_\perp)&=&\frac{q_z}{|\uq|}\,\Pi(\uq)\Sigma(\uq)\int^1_0\ud
y\int\frac{\ud^2\uQcal_\perp}{(2\pi)^2}\frac{\uQcal^2_\perp+M^2(2y-1)}{(\uQcal^2_\perp+M^2+y(1-y)\uq^2)^2}-(M\to
M_\textrm{PV})\label{Direct4}
\end{eqnarray}
where $q_z=z\uM=(z_4+z_5)\uM$ and
$\uq_\perp=\up_{4\perp}+\up_{5\perp}$. In summary, the upper index
$I$ refers to the valence part and the lower index $J$ to the sea
part. Axial integrals may thus have the vector index $V$ since it
refers only the valence structure. Index $J$ refers to the
transition experienced by the quark-antiquark pair.

\begin{center}
\begin{tabular}{c|r@{$\leftrightarrow$\,}l}
Index $J$&\multicolumn{2}{c}{Transition}\\ \hline
$\pi\pi,\pi 3,33$&pseudoscalar&pseudoscalar\\
$\sigma\sigma$&scalar&scalar\\
$\pi\sigma,3\sigma$&pseudoscalar&scalar
\end{tabular}\end{center}

\subsection{$5Q$ exchange scalar integrals}

The exchange diagram mixes valence and non-valence quarks. Therefore
the integrals cannot be decomposed into a purely valence part and a
sea part. However since valence quarks are equivalent a
decomposition is still possible. One part is the distribution of two
valence quarks $\phi$ and the other part is the rest, \emph{i.e.}
the third valence quark entangled with the pair. This diagram has
been studied in the non-relativistic limit $j=0$ to keep things as
simple as possible. The integrals have then the following structure
\begin{equation}
K_J=\frac{M^2}{2\pi}\int(\ud p_{3,4,5})
\,\phi\left(\uZ,\uP_\perp\right)\frac{\uM^2}{2\pi
Z'Z}\,I_J(z_{3,4,5},\up_{3,4,5\perp})h(p_3)h(p_4),
\end{equation}
where $\uZ=z_3+z_4+z_5$, $\uP_\perp=(\up_3+\up_4+\up_5)_\perp$, $Z$
is given by eq. (\ref{Notation}) with $z=z_4$,
$\up_\perp=\up_{4\perp}$ and $z'=z_5$, $\up'_\perp=\up_{5\perp}$
while $Z'$ is the same but with the replacement $4\to 3$. The
function $I_J(z_{3,4,5},\up_{3,4,5\perp})$ stands for the thirteen
integrands
\begin{eqnarray}
I_1&=&\Sigma(\uq')\Sigma(\uq)\left(\uQ'_\perp\cdot
\uQ_\perp+M^2(z_5-z_3)(z_5-z_4)\right),\label{Exchange begin}\\
I_2&=&\Pi(\uq')\Pi(\uq)\,\frac{q'\cdot
q}{|\uq'||\uq|}\left(\uQ'_\perp\cdot
\uQ_\perp+M^2(z_5+z_3)(z_5+z_4)\right),\\
I_3&=&\Pi(\uq')\Pi(\uq)\,\frac{\uq'_\perp\times
\uq_\perp}{|\uq'||\uq|}\,(\uQ'_\perp\times
\uQ_\perp),\\
I_4&=&\Pi(\uq')\Pi(\uq)\,\frac{M(\uq'_\perp q_z-\uq_\perp
q'_z)}{|\uq'||\uq|}\cdot\left(\uQ_\perp-
\uQ'_\perp\right),\\
I_5&=&\Pi(\uq')\Pi(\uq)\,\frac{q'_zq_z}{|\uq'||\uq|}\left(\uQ'_\perp\cdot
\uQ_\perp+M^2(z_5+z_3)(z_5+z_4)\right),\\
I_6&=&\Sigma(\uq')\Pi(\uq)\,\frac{q_z}{|\uq|}\left(\uQ'_\perp\cdot
\uQ_\perp+M^2(z_5-z_3)(z_5+z_4)\right),
\end{eqnarray}
\begin{eqnarray}
I_7&=&\Sigma(\uq')\Pi(\uq)\,\frac{M\uq_\perp}{|\uq|}\cdot\left(\uQ'_\perp
(z_5+z_4)-\uQ_\perp(z_5-z_3)\right),\\
I_8&=&\Sigma(\uq')\Sigma(\uq)\left(\uQ'_\perp\cdot
\uQ_\perp-M^2(z_5-z_3)(z_5-z_4)\right),\\
I_9&=&\Pi(\uq')\Pi(\uq)\,\frac{q'\cdot
q}{|\uq'||\uq|}\left(\uQ'_\perp\cdot
\uQ_\perp-M^2(z_5+z_3)(z_5+z_4)\right),\\
I_{10}&=&\Pi(\uq')\Pi(\uq)\,\frac{M(\uq'_\perp q_z+\uq_\perp
q'_z)}{|\uq'||\uq|}\cdot\left(\uQ_\perp+ \uQ'_\perp\right),\\
I_{11}&=&\Pi(\uq')\Pi(\uq)\,\frac{q'_zq_z}{|\uq'||\uq|}\left(\uQ'_\perp\cdot
\uQ_\perp-M^2(z_5+z_3)(z_5+z_4)\right),\\
I_{12}&=&\Sigma(\uq')\Pi(\uq)\,\frac{q_z}{|\uq|}\left(\uQ'_\perp\cdot
\uQ_\perp-M^2(z_5-z_3)(z_5+z_4)\right),\\
I_{13}&=&\Sigma(\uq')\Pi(\uq)\,\frac{M\uq_\perp}{|\uq|}\cdot\left(\uQ'_\perp
(z_5+z_4)+\uQ_\perp(z_5-z_3)\right)\label{Exchange end}
\end{eqnarray}
where $\uq=((\up_4+\up_5)_\perp,(z_4+z_5)\uM)$ and
$\uQ_\perp=z_4\up_{5\perp}-z_5\up_{4\perp}$. The primed variables
stand for the same as the unprimed ones but with the replacement
$4\to 3$. The regularization of those integrals is done exactly in
the same way as for the direct contributions.

The function $\phi(\uZ,\uP_\perp)$ stands for the probability that
two valence quarks ``leave'' the longitudinal fraction
$\uZ=z_3+z_4+z_5$ and the transverse momentum
$\uP_\perp=\up_{3\perp}+\up_{4\perp}+\up_{5\perp}$ to the rest of
the partons. In the non-relativistic limit we have
\begin{equation}\label{probability2q}
\phi(\uZ,\uP_\perp)=\int\ud
z_{1,2}\frac{\ud^2\up_{1,2\perp}}{(2\pi)^4}\,\delta(\uZ+z_1+z_2-1)(2\pi)^2\delta^{(2)}(\uP_\perp+\up_{1\perp}+\up_{2\perp})h^2(p_1)h^2(p_2).
\end{equation}

We have kept of course the same non-relativistic normalization of
the discrete-level wave function $h(p)$ as in the direct
contributions, \emph{i.e.} such that $\Phi_{NR}(0,0)=\int(\ud
p)\,\phi(z,\up_\perp)h^2(p)=1$.

\subsection{$7Q$ scalar integrals}

In the $7Q$ sector there are two quark-antiquark pairs. Thanks to
the simplicity of the direct diagram the corresponding scalar
overlap integrals can be written in two parts: purely valence
$\Phi^I$ and sea $G_J$
\begin{equation}
K^I_J=\frac{M^4}{(2\pi)^2}\int \frac{\ud^3 \uq}{(2\pi)^3}\frac{\ud^3
\uq'}{(2\pi)^3}\,\Phi^I\left(\frac{(q_z+q'_z)}{\uM},\uq_\perp+\uq'_\perp\right)\theta(q_z)\,\theta(q'_z)\,q_z\,q'_z\,G_J(q_z,q'_z,\uq_\perp,\uq'_\perp).
\end{equation}
where $J=\pi\pi\pi\pi,\pi\pi\pi\pi 2,\pi\pi 33,3333,\pi 3\pi
3,\sigma\sigma\pi\pi,\sigma\sigma 33,\sigma\sigma\sigma\sigma,\pi\pi
3\sigma,333\sigma,\pi 3\pi\sigma,\sigma\sigma 3\sigma$. These
distributions are obtained by contracting four quark-antiquark wave
functions $W$, see eq. (\ref{Pair wavefunction}) and regularized by
means of Pauli-Villars procedure. They can in fact be expressed in
terms of $G_J(q_z,\uq_\perp)$ since in direct diagrams $Q\bar Q$
pairs keep their identity. Here are then the distributions in the
$7Q$ sector
\begin{eqnarray}
G_{\pi\pi\pi\pi}(q_z,q'_z,\uq_\perp,\uq'_\perp)&=&G_{\pi\pi}(q_z,\uq_\perp)\,G_{\pi\pi}(q'_z,\uq'_\perp),\label{int1}
\\
G_{\pi\pi\pi\pi
2}(q_z,q'_z,\uq_\perp,\uq'_\perp)&=&\frac{(\uq\cdot\uq')^2}{\uq^2\uq'^2}\,G_{\pi\pi}(q_z,\uq_\perp)\,G_{\pi\pi}(q'_z,\uq'_\perp),\label{int2}
\\
G_{\pi\pi
33}(q_z,q'_z,\uq_\perp,\uq'_\perp)&=&G_{\pi\pi}(q_z,\uq_\perp)\,G_{33}(q'_z,\uq'_\perp),\label{int3}
\\
G_{3333}(q_z,q'_z,\uq_\perp,\uq'_\perp)&=&G_{33}(q_z,\uq_\perp)\,G_{33}(q'_z,\uq'_\perp),\label{int4}
\\
G_{\pi 3\pi
3}(q_z,q'_z,\uq_\perp,\uq'_\perp)&=&\frac{q_zq'_z(\uq\cdot\uq')}{\uq^2\uq'^2}\,G_{\pi\pi}(q_z,\uq_\perp)\,G_{\pi\pi}(q'_z,\uq'_\perp),\label{int5}
\\
G_{\sigma\sigma\pi\pi}(q_z,q'_z,\uq_\perp,\uq'_\perp)&=&G_{\sigma\sigma}(q_z,\uq_\perp)\,G_{\pi\pi}(q'_z,\uq'_\perp),\label{int6}
\end{eqnarray}
\begin{eqnarray}
G_{\sigma\sigma
33}(q_z,q'_z,\uq_\perp,\uq'_\perp)&=&G_{\sigma\sigma}(q_z,\uq_\perp)\,G_{33}(q'_z,\uq'_\perp),\label{int7}
\\
G_{\sigma\sigma\sigma\sigma}(q_z,q'_z,\uq_\perp,\uq'_\perp)&=&G_{\sigma\sigma}(q_z,\uq_\perp)\,G_{\sigma\sigma}(q'_z,\uq'_\perp),\label{int8}
\\
G_{\pi\pi
3\sigma}(q_z,q'_z,\uq_\perp,\uq'_\perp)&=&G_{\pi\pi}(q_z,\uq_\perp)\,G_{3\sigma}(q'_z,\uq'_\perp),\label{int9}
\\
G_{333\sigma}(q_z,q'_z,\uq_\perp,\uq'_\perp)&=&G_{33}(q_z,\uq_\perp)\,G_{3\sigma}(q'_z,\uq'_\perp),\label{int10}
\\
G_{\pi
3\pi\sigma}(q_z,q'_z,\uq_\perp,\uq'_\perp)&=&\frac{q_z(\uq\cdot\uq')}{q'_z\uq^2}\,G_{\pi\pi}(q_z,\uq_\perp)\,G_{3\sigma}(q'_z,\uq'_\perp),\label{int11}
\\
G_{\sigma\sigma
3\sigma}(q_z,q'_z,\uq_\perp,\uq'_\perp)&=&G_{\sigma\sigma}(q_z,\uq_\perp)\,G_{3\sigma}(q'_z,\uq'_\perp)\label{int12}
\end{eqnarray}
where $q_z=z\uM=(z_4+z_5)\uM$, $q'_z=z\uM=(z_6+z_7)\uM$,
$\uq_\perp=\up_{4\perp}+\up_{5\perp}$ and
$\uq'_\perp=\up_{6\perp}+\up_{7\perp}$.

Scalar integrals arising from the four exchange diagrams in the $7Q$
sector have not been computed. As we will show later, exchange
diagrams give only negligible contributions and can therefore be
ignored. Naively one could indeed expect exchange contributions to
be smaller than direct ones. Direct diagrams correspond to the
simple case where nothing really happens, all partons keep their
role. On the contrary exchange diagrams describe modifications in
the roles played by the partons and implies thus some correlations
among quarks.
\newpage\thispagestyle{empty}\cleardoublepage
\chapter{Symmetry relations and parametrization}\label{Section cinq}

In this work we have studied baryons properties in flavor $SU(3)$
symmetry. Even though this symmetry is broken in nature, it gives
quite a good estimation. With such an assumption all particles in a
given representation are on the same footing and are related through
pure flavor $SU(3)$ transformations. In other words, we need to
concentrate only on, say, proton properties. Properties of the other
octet members can be obtained from the proton one.

The naive non-relativistic quark model is based on a larger group
$SU(6)$ that imbeds $SU(3)_F\times SU(2)_S$. In this approach, octet
and decuplet baryons belong to the same supermultiplet. This yields
relations \emph{between} different $SU(3)$ multiplets and new ones
within multiplets.

As one can see symmetry is very useful and convenient. In the next
sections we give the explicit relations among baryons properties.

\section{General flavor $SU(3)$ symmetry relations}

Let us consider a charge $Q$, \emph{e.g.} the vector, axial or
tensor charge or even the electric charge and magnetic form factor.
If the contribution of each flavor is known for a member in a given
multiplet, then flavor $SU(3)$ symmetry allows one to find those
contributions for all other members of the same multiplet. One could
then use these flavor contributions as parameters for the given
multiplet. We chose however to use another parametrization in order
to emphasize some properties. The number of parameters depends on
the multiplet under consideration.

\begin{table}[h!]\begin{center}\caption{\small{$SU(3)$ octet
relations.}}
\begin{tabular}{c|ccc} \hline\hline
$B$&$Q^{(B)u}$&$Q^{(B)d}$&$Q^{(B)s}$\rule{0pt}{3ex}\\\hline
\rule{0pt}{3ex}
$p_\mathbf{8}^+$&$\alpha+\gamma$&$\beta+\gamma$&$\gamma$\\\rule{0pt}{3ex}
$n_\mathbf{8}^0$&$\beta+\gamma$&$\alpha+\gamma$&$\gamma$\\\rule{0pt}{3ex}
$\Lambda_\mathbf{8}^0$&$\frac{1}{6}(\alpha+4\beta)+\gamma$&$\frac{1}{6}(\alpha+4\beta)+\gamma$&$\frac{1}{3}(2\alpha-\beta)+\gamma$\\\rule{0pt}{3ex}
$\Sigma_\mathbf{8}^+$&$\alpha+\gamma$&$\gamma$&$\beta+\gamma$\\\rule{0pt}{3ex}
$\Sigma_\mathbf{8}^0$&$\frac{1}{2}\alpha+\gamma$&$\frac{1}{2}\alpha+\gamma$&$\beta+\gamma$\\\rule{0pt}{3ex}
$\Sigma_\mathbf{8}^-$&$\gamma$&$\alpha+\gamma$&$\beta+\gamma$\\\rule{0pt}{3ex}
$\Xi_\mathbf{8}^0$&$\beta+\gamma$&$\gamma$&$\alpha+\gamma$\\\rule{0pt}{3ex}
$\Xi_\mathbf{8}^-$&$\gamma$&$\beta+\gamma$&$\alpha+\gamma$\rule[-2ex]{0pt}{5ex}\\\hline\hline
\end{tabular}\label{Octetcontent}\end{center}
\end{table}
For baryon octet one needs to use three parameters, \emph{e.g.}
$\alpha$, $\beta$ and $\gamma$ while for baryon decuplet and
antidecuplet only two are needed, \emph{e.g.} $\alpha'$, $\beta'$
and $\alpha''$, $\beta''$. Tables \ref{Octetcontent},
\ref{Decupletcontent} and \ref{Antidecupletcontent} give for each
flavor the parametrization of the contribution to the baryon
charges.
\begin{table}[h!]\begin{center}
\begin{minipage}[c]{6cm}\caption{\small{$SU(3)$ decuplet relations.}}
\begin{tabular}{c|ccc}
\hline\hline
$B$&$Q^{(B)u}$&$Q^{(B)d}$&$Q^{(B)s}$\rule{0pt}{3ex}\\\hline
\rule{0pt}{3ex}
$\Delta_\mathbf{10}^{++}$&$3\alpha'+\beta'$&$\beta'$&$\beta'$\\\rule{0pt}{3ex}
$\Delta_\mathbf{10}^+$&$2\alpha'+\beta'$&$\alpha'+\beta'$&$\beta'$\\\rule{0pt}{3ex}
$\Delta_\mathbf{10}^0$&$\alpha'+\beta'$&$2\alpha'+\beta'$&$\beta'$\\\rule{0pt}{3ex}
$\Delta_\mathbf{10}^-$&$\beta'$&$3\alpha'+\beta'$&$\beta'$\\\rule{0pt}{3ex}
$\Sigma_\mathbf{10}^+$&$2\alpha'+\beta'$&$\beta'$&$\alpha'+\beta'$\\\rule{0pt}{3ex}
$\Sigma_\mathbf{10}^0$&$\alpha'+\beta'$&$\alpha'+\beta'$&$\alpha'+\beta'$\\\rule{0pt}{3ex}
$\Sigma_\mathbf{10}^-$&$\beta'$&$2\alpha'+\beta'$&$\alpha'+\beta'$\\\rule{0pt}{3ex}
$\Xi_\mathbf{10}^0$&$\alpha'+\beta'$&$\beta'$&$2\alpha'+\beta'$\\\rule{0pt}{3ex}
$\Xi_\mathbf{10}^-$&$\beta'$&$\alpha'+\beta'$&$2\alpha'+\beta'$\\\rule{0pt}{3ex}
$\Omega_\mathbf{10}^-$&$\beta'$&$\beta'$&$3\alpha'+\beta'$\rule[-2ex]{0pt}{5ex}\\
\hline\hline
\end{tabular}\label{Decupletcontent}\end{minipage}\hspace{2cm}\begin{minipage}[c]{7cm}\caption{\small{$SU(3)$ antidecuplet relations.}}
\begin{tabular}{c|ccc}
\hline\hline
$B$&$Q^{(B)u}$&$Q^{(B)d}$&$Q^{(B)s}$\rule{0pt}{3ex}\\\hline
\rule{0pt}{3ex}
$\Theta_\mathbf{\overline{10}}^+$&$2\alpha''+\beta''$&$2\alpha''+\beta''$&$-\alpha''+\beta''$\\\rule{0pt}{3ex}
$p_\mathbf{\overline{10}}^+$&$2\alpha''+\beta''$&$\alpha''+\beta''$&$\beta''$\\\rule{0pt}{3ex}
$n_\mathbf{\overline{10}}^0$&$\alpha''+\beta''$&$2\alpha''+\beta''$&$\beta''$\\\rule{0pt}{3ex}
$\Sigma_\mathbf{\overline{10}}^+$&$2\alpha''+\beta''$&$\beta''$&$\alpha''+\beta''$\\\rule{0pt}{3ex}
$\Sigma_\mathbf{\overline{10}}^0$&$\alpha''+\beta''$&$\alpha''+\beta''$&$\alpha''+\beta''$\\\rule{0pt}{3ex}
$\Sigma_\mathbf{\overline{10}}^-$&$\beta''$&$2\alpha''+\beta''$&$\alpha''+\beta''$\\\rule{0pt}{3ex}
$\Xi_\mathbf{\overline{10}}^+$&$2\alpha''+\beta''$&$-\alpha''+\beta''$&$2\alpha''+\beta''$\\\rule{0pt}{3ex}
$\Xi_\mathbf{\overline{10}}^0$&$\alpha''+\beta''$&$\beta''$&$2\alpha''+\beta''$\\\rule{0pt}{3ex}
$\Xi_\mathbf{\overline{10}}^-$&$\beta''$&$\alpha''+\beta''$&$2\alpha''+\beta''$\\\rule{0pt}{3ex}
$\Xi_\mathbf{\overline{10}}^{--}$&$-\alpha''+\beta''$&$2\alpha''+\beta''$&$2\alpha''+\beta''$\rule[-2ex]{0pt}{5ex}\\
\hline\hline
\end{tabular}\label{Antidecupletcontent}\end{minipage}\end{center}
\end{table}
\begin{table}[h!]\begin{center}\caption{\small{$SU(3)$ octet vector and axial transition
relations.}}
\begin{tabular}{r@{$\to$\,}l|c||r@{$\to$\,}l|c} \hline\hline
\multicolumn{2}{c|}{Transitions}&$g_{V,A}$&\multicolumn{2}{c|}{Transitions}&$g_{V,A}$\rule{0pt}{3ex}\\\hline
\rule{0pt}{3ex}
$n_\mathbf{8}^0$&$p_\mathbf{8}^+$&$\alpha-\beta$&$\textrm{
}\Sigma_\mathbf{8}^-$&$n_\mathbf{8}^0$&$-\beta$\\\rule{0pt}{3ex}
$\Sigma_\mathbf{8}^-$&$\Sigma_\mathbf{8}^0$&$\alpha/\sqrt{2}$&$\Xi_\mathbf{8}^-$&$\Sigma_\mathbf{8}^0$&$(\beta-\alpha)/\sqrt{2}$\\\rule{0pt}{3ex}
$\Sigma_\mathbf{8}^-$&$\Lambda_\mathbf{8}^0$&$(\alpha-2\beta)/\sqrt{6}$&$\Xi_\mathbf{8}^-$&$\Lambda_\mathbf{8}^0$&$-(\alpha+\beta)/\sqrt{6}$\\\rule{0pt}{3ex}
$\Sigma_\mathbf{8}^0$&$\Sigma_\mathbf{8}^+$&$-\alpha/\sqrt{2}$&$\Sigma_\mathbf{8}^0$&$p_\mathbf{8}^+$&$-\beta/\sqrt{2}$\\\rule{0pt}{3ex}
$\Lambda_\mathbf{8}^0$&$\Sigma_\mathbf{8}^+$&$(\alpha-2\beta)/\sqrt{6}$&$\Lambda_\mathbf{8}^0$&$p_\mathbf{8}^+$&$(\beta-2\alpha)/\sqrt{6}$\\\rule{0pt}{3ex}
$\Xi_\mathbf{8}^-$&$\Xi_\mathbf{8}^0$&$\beta$&$\Xi_\mathbf{8}^0$&$\Sigma_\mathbf{8}^+$&$\alpha-\beta$\rule[-2ex]{0pt}{5ex}\\\hline\hline
\end{tabular}\label{axialdecay}
\end{center}
\end{table}
This specific parametrization shows explicitly that one cannot
access to all electric and magnetic parameters from total electric
charges and magnetic moments alone\footnote{Indeed one has
$Q^{tot}=(2Q^u-Q^d-Q^s)/3$ and thus $(2x-x-x)/3=0$ with
$x=\gamma,\beta',\beta''$.} or to all axial content from the
isovector or octet charges alone\footnote{Indeed one has
$Q^{(3)}=Q^u-Q^d$, $\sqrt{3}Q^{(8)}=Q^u+Q^d-2Q^s$ and thus $x-x=0$,
$x+x-2x=0$ with $x=\gamma,\beta',\beta''$.}. It is also clear that
the singlet vector, axial and tensor charges are the same within a
multiplet\footnote{Indeed one has $Q^{(0)}=Q^u+Q^d+Q^s$ and thus
$\alpha+\beta+3\gamma$ for all octet members, $3(\alpha'+\beta')$
for all decuplet members and $3(\alpha''+\beta'')$ for all
antidecuplet members.}.

Some parameters can be interpreted if one considers the $3Q$ sector
only. It implies that $\gamma,\beta',\beta''=0$ and thus
$\alpha,\beta$ are the contribution of (valence) $u,d$ quarks in the
proton and $\alpha'$ is the contribution of (valence) $u$ quarks in
the $\Delta^{++}$. From the structure in Tables \ref{Octetcontent},
\ref{Decupletcontent} and \ref{Antidecupletcontent} one could
naively think that $\gamma,\beta',\beta''$ represent the
contribution of the $SU(3)$ symmetric sea. In fact these relations
hold separately for valence quarks, sea quarks and antiquarks.

A few octet baryon decay constants are known from experiment. It is
then interesting to express them in terms of our parameters $\alpha$
and $\beta$ ($\gamma$ disappears), see Table \ref{axialdecay}. In
the literature they are usually expressed in terms of the Cabibbo
parameters $F\& D$ \cite{Cabibbo}. Here is the link between both
parametrization
\begin{equation}
\alpha=2F,\qquad \beta=F-D.
\end{equation}

\section{Flavor $SU(3)$ symmetry and magnetic moments}

Let us discuss a little bit further flavor $SU(3)$ symmetry in
relation with magnetic moments and transition magnetic moments. We
have seen that magnetic form factors within a multiplet are related
by the flavor symmetry and so are the magnetic moments. The total
magnetic moments are obtained by the formula
\begin{equation}
\mu_B=e_u\,G_M^{(B)u}+e_d\,G_M^{(B)d}+e_s\,G_M^{(B)s}
\end{equation}
where $e_u$, $e_d$ and $e_s$ are quark electric charges, \emph{i.e.}
2/3, -1/3 and -1/3 respectively. If one considers magnetic
transitions between multiplets, flavor symmetry will also impose
relations and will even forbid some transitions. The $SU(3)$
prediction for magnetic moments and transition magnetic moments is
greatly simplified by the use of the concept of $U$-spin. Let us
briefly recall this concept.

\subsection{Charge and $U$-spin}

\begin{figure}[h]\begin{center}\includegraphics[width=10cm]{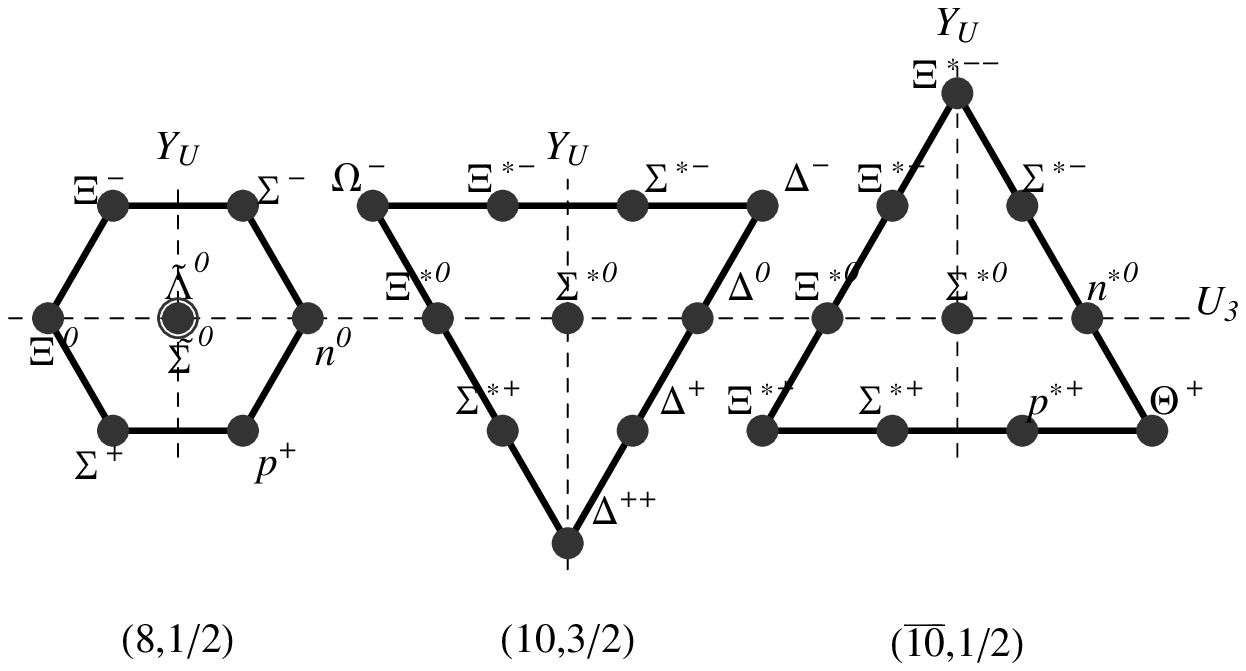}
\caption{\small{$SU(3)$ multiplets in the $(U_3,Y_U)$-basis.
}}\label{Uspinfig}\end{center}
\end{figure}

Flavor $SU(3)$ multiplets are usually represented in the
$(I_3,Y)$-basis, \emph{i.e.} to each member are associated two
numbers: the third component of its isospin $I_3$ and its
hypercharge $Y$. Unfortunately the electromagnetic current
\begin{equation}
J_\mu=\frac{2}{3}\bar u\gamma_\mu u-\frac{1}{3}\bar d\gamma_\mu
d-\frac{1}{3}\bar s\gamma_\mu s
\end{equation}
contains $I=0$ and $I=1$ components and then transforms in a
complicated way under isospin rotations. One can instead choose to
work in the $(U_3,Y_U)$-basis \cite{Uspin}
\begin{equation}
U_3=-\frac{1}{2}I_3+\frac{3}{4}Y, \qquad Y_U=-Q.
\end{equation}
The multiplets in this basis are represented in Fig. \ref{Uspinfig}.
In this figure, $\tilde\Sigma^0=(-\Sigma^0+\sqrt{3}\Lambda^0)/2$ and
$\tilde\Lambda^0=-(\sqrt{3}\Sigma^0+\Lambda^0)/2$. Now $\bar
u\gamma_\mu u$ is the $U$-spin singlet and the sum $\bar d\gamma_\mu
d+\bar s\gamma_\mu s$ is invariant under $U$-spin rotations.
From the assumption of $U$-spin conservation\footnote{The assumption
of $U$-spin conservation is weaker than flavor SU(3) symmetry
($U$-spin is embedded in flavor $SU(3)$ symmetry). That's the reason
why all flavor $SU(3)$ relations are not obtained.}, magnetic
moments and electric charges of all members of the same $U$-spin
multiplet are equal. One can also predicts that transition magnetic
moments between different $U$-spin multiplets are forbidden. From
Fig. \ref{Uspinfig} one can see for example that the magnetic
transition of negatively charged particles between octet and
decuplet is forbidden while it is allowed between octet and
antidecuplet. This simple and useful rule is important to understand
the observed isospin asymmetry in the eta photoproduction on
nucleon. This point will be discussed later.

\subsection{More about $SU(3)$ octet magnetic moments}

In flavor $SU(3)$ symmetry limit the nine octet quantities (eight
baryon magnetic moments and one transition magnetic moment) are
related \cite{ColGla,GuzPol} by $U$-spin conservation and an
auxiliary isospin relation
$\mu_{\Sigma^0_\mathbf{8}}=(\mu_{\Sigma^+_\mathbf{8}}+\mu_{\Sigma^-_\mathbf{8}})/2$.
One then obtains the seven Coleman-Glashow relations
\begin{eqnarray}
&\mu_{\Sigma^+_\mathbf{8}}=\mu_{p^+_\mathbf{8}},&\\
&2\,\mu_{\Lambda^0_\mathbf{8}}=-2\,\mu_{\Sigma^0_\mathbf{8}}=-\frac{2}{\sqrt{3}}\,\mu_{\Sigma^0_\mathbf{8}\Lambda^0_\mathbf{8}}=\mu_{\Xi^0_\mathbf{8}}=\mu_{n^0_\mathbf{8}},&\\
&\mu_{\Sigma^-_\mathbf{8}}=\mu_{\Xi^-_\mathbf{8}}=-(\mu_{p^+_\mathbf{8}}+\mu_{n^0_\mathbf{8}}).&
\end{eqnarray}

Baryon octet magnetic moments then depend on two parameters only,
say, $\mu_{p^+_\mathbf{8}}$ and $\mu_{n^0_\mathbf{8}}$. One can find
in the literature (see \emph{e.g.} in \cite{group}) the $F\&D$
parametrization where all magnetic moments are expressed in terms of
$\mu_F$ and $\mu_D$. The conversion into this set of parameters is
obtained by means of the relations
\begin{equation}
\mu_{p^+_\mathbf{8}}=\mu_F+\frac{1}{3}\,\mu_D,
\qquad\mu_{n^0_\mathbf{8}}=-\frac{2}{3}\,\mu_D.
\end{equation}

In large $N_C$ one uses the $a\&b$ parametrization, $a$ being of
order $N_C^0$ and $b$ of order $N_C^{-1}$ (see \emph{e.g.}
\cite{Flores}). It is related to the $F\&D$ parametrization as
follows
\begin{equation}
\mu_F=\frac{2}{3}\,a+b, \qquad \mu_D=a.
\end{equation}

Due to flavor $SU(3)$ symmetry one can relate all octet magnetic
moments to the proton one only, provided that each flavor
contribution is known. Since there are three light flavors, only two
linear combinations of $\mup$, $\mdp$ and $\msp$ can be extracted
from experimental octet magnetic moments. One then needs to combine
this with the nucleon response to the weak neutral vector current in
order to extract the individual flavor contributions \cite{Kaplan}.
Recent experiments used parity-violating elastic electron-proton
scattering to probe the contribution of the $s$ quark
\cite{SAMPLEM}.

Table \ref{Param} gives explicitly the expression for octet magnetic
moments in the four parametrizations mentioned above.
\begin{table}[h!]\begin{center}\caption{\small{Parametrizations of octet magnetic moments in the flavor $SU(3)$ symmetry limit.}}
\begin{tabular}{l|cccc}
\hline\hline&$\mu_p$, $\mu_n$&$\mu_F$, $\mu_D$&$a$, $b$&$\mup$,
$\mdp$, $\msp$\rule{0pt}{3ex}\\\hline \rule{0pt}{3ex}
$\mu_{p^+_\mathbf{8}}$&$\mu_p$&$\mu_F+\frac{1}{3}\,\mu_D$&$a+b$&$\frac{2}{3}\,\mup-\frac{1}{3}\,\mdp-\frac{1}{3}\,\msp$\\\rule{0pt}{3ex}
$\mu_{n^0_\mathbf{8}}$&$\mu_n$&$-\frac{2}{3}\,\mu_D$&$-\frac{2}{3}\,a$&$-\frac{1}{3}\,\mup+\frac{2}{3}\,\mdp-\frac{1}{3}\,\msp$\\\rule{0pt}{3ex}
$\mu_{\Lambda^0_\mathbf{8}}$&$\frac{1}{2}\,\mu_n$&$-\frac{1}{3}\,\mu_D$&$-\frac{1}{3}\,a$&$-\frac{1}{6}\,\mup+\frac{1}{3}\,\mdp-\frac{1}{6}\,\msp$\\\rule{0pt}{3ex}
$\mu_{\Sigma^+_\mathbf{8}}$&$\mu_p$&$\mu_F+\frac{1}{3}\,\mu_D$&$a+b$&$\frac{2}{3}\,\mup-\frac{1}{3}\,\mdp-\frac{1}{3}\,\msp$\\\rule{0pt}{3ex}
$\mu_{\Sigma^0_\mathbf{8}}$&$-\frac{1}{2}\,\mu_n$&$\frac{1}{3}\,\mu_D$&$\frac{1}{3}\,a$&$\frac{1}{6}\,\mup-\frac{1}{3}\,\mdp+\frac{1}{6}\,\msp$\\\rule{0pt}{3ex}
$\mu_{\Sigma^-_\mathbf{8}}$&$-\mu_p-\mu_n$&$-\mu_F+\frac{1}{3}\,\mu_D$&$-\frac{1}{3}\,a-b$&$-\frac{1}{3}\,\mup-\frac{1}{3}\,\mdp+\frac{2}{3}\,\msp$\\\rule{0pt}{3ex}
$\mu_{\Xi^0_\mathbf{8}}$&$\mu_n$&$-\frac{2}{3}\,\mu_D$&$-\frac{2}{3}\,a$&$-\frac{1}{3}\,\mup+\frac{2}{3}\,\mdp-\frac{1}{3}\,\msp$\\\rule{0pt}{3ex}
$\mu_{\Xi^-_\mathbf{8}}$&$-\mu_p-\mu_n$&$-\mu_F+\frac{1}{3}\,\mu_D$&$-\frac{1}{3}\,a-b$&$-\frac{1}{3}\,\mup-\frac{1}{3}\,\mdp+\frac{2}{3}\,\msp$\\\rule{0pt}{3ex}
$\mu_{\Sigma^0_\mathbf{8}\Lambda^0_\mathbf{8}}$&$-\frac{\sqrt{3}}{2}\,\mu_n$&$\frac{1}{\sqrt{3}}\,\mu_D$&$\frac{1}{\sqrt{3}}\,a$&$\frac{1}{2\sqrt{3}}\,\mup-\frac{1}{\sqrt{3}}\,\mdp+\frac{1}{2\sqrt{3}}\,\msp$
\rule[-2ex]{0pt}{5ex}\\ \hline\hline
\end{tabular}\label{Param}\end{center}
\end{table}
We remind that magnetic form factors $\mup$, $\mdp$ and $\msp$ are
related to our parameters $\alpha_M$, $\beta_M$ and $\gamma_M$ as
indicated in Table \ref{Octetcontent}.

\subsection{More about $SU(3)$ decuplet and antidecuplet magnetic moments}

$U$-spin symmetry tells us that all particles in the same $U$-spin
multiplet have the same magnetic moment. Flavor $SU(3)$ symmetry
imposes a stronger condition. As mentioned earlier, on the one hand
decuplet and antidecuplet total magnetic moments are proportional to
a unique parameter $\alpha'_M$ and $\alpha''_M$ respectively. On the
other hand decuplet and antidecuplet electric charges are
proportional (with the same proportionality factors as in the
magnetic case) to another unique parameter $\alpha'_E$ and
$\alpha''_E$ respectively. Flavor $SU(3)$ symmetry then tells us
that within the decuplet and antidecuplet, magnetic moments are
proportional to the electric charge of the baryon
\begin{equation}
\mu_\mathbf{10}\propto Q_\mathbf{10},\qquad
\mu_\mathbf{\overline{10}}\propto Q_\mathbf{\overline{10}}.
\end{equation}
Since particles in the same $U$-spin multiplet have the same charge,
the $SU(3)$ relation includes the $U$-spin relation as it should be.

\subsection{More about $SU(3)$ transition magnetic moments}

In flavor $SU(3)$ symmetry limit two of the eight octet-to-decuplet
transition magnetic moments are identically zero because of $U$-spin
conservation. The six others are all related to each other as
follows
\begin{eqnarray}
&\mu_{p^+_\mathbf{8}\Delta^+_\mathbf{10}}=-\mu_{\Sigma^+_\mathbf{8}\Sigma^+_\mathbf{10}}=\mu_{n^0_\mathbf{8}\Delta^0_\mathbf{10}}=-2\,\mu_{\Sigma^0_\mathbf{8}\Sigma^0_\mathbf{10}}=\frac{2}{\sqrt{3}}\,\mu_{\Lambda^0_\mathbf{8}\Sigma^0_\mathbf{10}}=-\mu_{\Xi^0_\mathbf{8}\Xi^0_\mathbf{10}},&\\
&\mu_{\Sigma^-_\mathbf{8}\Sigma^-_\mathbf{10}}=\mu_{\Xi^-_\mathbf{8}\Xi^-_\mathbf{10}}=0.&
\end{eqnarray}
This means that only one parameter is sufficient to describe all
transition magnetic moments, \emph{e.g.} the proton-to-Delta
transition magnetic moment
$\mu_{p^+_\mathbf{8}\Delta^+_\mathbf{10}}$. Since decuplet baryons
are spin-$3/2$ particles besides the magnetic dipole transition, an
electric quadrupole transition may be allowed. The relations are
exactly the same as for the magnetic dipole. Table
\ref{OctetDecupletcontent} gives the contribution of each flavor to
the total moments for all octet-to-decuplet transitions expressed in
terms of the parameters $\alpha_{M,E}^{8\to 10}$.
\begin{table}[h!]\begin{center}\caption{\small{$SU(3)$
octet-to-decuplet relations.}}
\begin{tabular}{r@{$\to$\,}l|ccc}
\hline\hline
\multicolumn{2}{c|}{Transition}&$G^{*u}_{M,E}$&$G^{*d}_{M,E}$&$G^{*s}_{M,E}$\rule{0pt}{3ex}\\\hline
\rule{0pt}{3ex}
$p^+_\mathbf{8}$&$\Delta^+_\mathbf{10}$&$\alpha_{M,E}^{8\to
10}$&$-\alpha_{M,E}^{8\to 10}$&$0$\\\rule{0pt}{3ex}
$n^0_\mathbf{8}$&$\Delta^0_\mathbf{10}$& $\alpha_{M,E}^{8\to
10}$&$-\alpha_{M,E}^{8\to 10}$&$0$\\\rule{0pt}{3ex}
$\Sigma^+_\mathbf{8}$&$\Sigma^+_\mathbf{10}$&$-\alpha_{M,E}^{8\to
10}$&$0$&$\alpha_{M,E}^{8\to 10}$\\\rule{0pt}{3ex}
$\Lambda^0_\mathbf{8}$&$\Sigma^0_\mathbf{10}$&$\frac{\sqrt{3}}{2}\,\alpha_{M,E}^{8\to
10}$&$\frac{-\sqrt{3}}{2}\,\alpha_{M,E}^{8\to
10}$&$0$\\\rule{0pt}{3ex}
$\Sigma^0_\mathbf{8}$&$\Sigma^0_\mathbf{10}$&$\frac{-1}{2}\,\alpha_{M,E}^{8\to
10}$&$\frac{-1}{2}\,\alpha_{M,E}^{8\to 10}$&$\alpha_{M,E}^{8\to
10}$\\\rule{0pt}{3ex}
$\Sigma^-_\mathbf{8}$&$\Sigma^-_\mathbf{10}$&$0$&$-\alpha_{M,E}^{8\to
10}$&$\alpha_{M,E}^{8\to 10}$\\\rule{0pt}{3ex}
$\Xi^0_\mathbf{8}$&$\Xi^0_\mathbf{10}$&$-\alpha_{M,E}^{8\to
10}$&$0$&$\alpha_{M,E}^{8\to 10}$\\\rule{0pt}{3ex}
$\Xi^-_\mathbf{8}$&$\Xi^-_\mathbf{10}$&$0$&$-\alpha_{M,E}^{8\to 10}$&$\alpha_{M,E}^{8\to 10}$\rule[-2ex]{0pt}{5ex}\\
\hline\hline
\end{tabular}\label{OctetDecupletcontent}\end{center}
\end{table}

Concerning the octet-to-antidecuplet transitions the situation is
similar. In this case the magnetic transitions $p^+_\mathbf{8}\to
p^+_\mathbf{\overline{10}}$ and $\Sigma^+_\mathbf{8}\to
\Sigma^+_\mathbf{\overline{10}}$ are forbidden. This is once more a
consequence of $U$-spin symmetry. Flavor $SU(3)$ symmetry imposes
that all six other magnetic transitions are proportional to each
other
\begin{eqnarray}
&\mu_{n^0_\mathbf{8}n^0_\mathbf{\overline{10}}}=2\,\mu_{\Sigma^0_\mathbf{8}\Sigma^0_\mathbf{\overline{10}}}=-\frac{2}{\sqrt{3}}\,\mu_{\Lambda^0_\mathbf{8}\Sigma^0_\mathbf{\overline{10}}}=-\mu_{\Xi^0_\mathbf{8}\Xi^0_\mathbf{\overline{10}}}=\mu_{\Sigma^-_\mathbf{8}\Sigma^-_\mathbf{\overline{10}}}=-\mu_{\Xi^-_\mathbf{8}\Xi^-_\mathbf{\overline{10}}},&\\
&\mu_{p^+_\mathbf{8}p^+_\mathbf{\overline{10}}}=\mu_{\Sigma^+_\mathbf{8}\Sigma^+_\mathbf{\overline{10}}}=0.&
\end{eqnarray}
Table \ref{OctetAntiDecupletcontent} gives the contribution of each
flavor to the total magnetic moments for all octet-to-antidecuplet
transitions expressed in terms of the parameter $\alpha_M^{8\to
\overline{10}}$.
\begin{table}[h!]\begin{center}\caption{\small{$SU(3)$
octet-to-antidecuplet relations.}}
\begin{tabular}{r@{$\,\to\,$}l|ccc|c}
\hline\hline
\multicolumn{2}{c|}{Transition}&$G_M^{u}$&$G_M^{d}$&$G_M^{s}$&$\mu_{8\to
\overline{10}}^{tot}$\rule[-1.3ex]{0pt}{4.2ex}\\\hline\rule{0pt}{3ex}
$p^+_\mathbf{8}$&$p^+_\mathbf{\overline{10}}$&$0$&$\alpha_M^{8\to
\overline{10}}$&$-\alpha_M^{8\to
\overline{10}}$&$0$\\
$n^0_\mathbf{8}$&$n^0_\mathbf{\overline{10}}$&$\alpha_M^{8\to
\overline{10}}$&$0$&$-\alpha_M^{8\to \overline{10}}$&$\alpha_M^{8\to
\overline{10}}$\rule{0pt}{3ex}\\
$\Sigma^+_\mathbf{8}$&$\Sigma^+_\mathbf{\overline{10}}$&$0$&$\alpha_M^{8\to
\overline{10}}$&$-\alpha_M^{8\to
\overline{10}}$&$0$\rule{0pt}{3ex}\\
$\Sigma^0_\mathbf{8}$&$\Sigma^0_\mathbf{\overline{10}}$&$\frac{1}{2}\,\alpha_M^{8\to
\overline{10}}$&$\frac{1}{2}\,\alpha_M^{8\to
\overline{10}}$&$-\alpha_M^{8\to
\overline{10}}$&$\frac{1}{2}\,\alpha_M^{8\to
\overline{10}}$\rule{0pt}{3ex}\\
$\Lambda^0_\mathbf{8}$&$\Sigma^0_\mathbf{\overline{10}}$&$-\frac{\sqrt
3}{2}\,\alpha_M^{8\to \overline{10}}$&$\frac{\sqrt
3}{2}\,\alpha_M^{8\to \overline{10}}$&$0$&$-\frac{\sqrt
3}{2}\,\alpha_M^{8\to
\overline{10}}$\rule{0pt}{3ex}\\
$\Sigma^-_\mathbf{8}$&$\Sigma^-_\mathbf{\overline{10}}$&$\alpha_M^{8\to
\overline{10}}$&$0$&$-\alpha_M^{8\to \overline{10}}$&$\alpha_M^{8\to
\overline{10}}$\rule{0pt}{3ex}\\
$\Xi^0_\mathbf{8}$&$\Xi^0_\mathbf{\overline{10}}$&$-\alpha_M^{8\to
\overline{10}}$&$\alpha_M^{8\to \overline{10}}$&$0$&$-\alpha_M^{8\to
\overline{10}}$\rule{0pt}{3ex}\\
$\Xi^-_\mathbf{8}$&$\Xi^-_\mathbf{\overline{10}}$&$-\alpha_M^{8\to
\overline{10}}$&$\alpha_M^{8\to \overline{10}}$&$0$&$-\alpha_M^{8\to
\overline{10}}$\rule[-2ex]{0pt}{5ex}\\
\hline\hline
\end{tabular}\label{OctetAntiDecupletcontent}\end{center}
\end{table}

The last magnetic transitions to be discussed are the
decuplet-to-antidecuplet ones. $U$-spin conservation authorizes
transitions between electrically neutral particles only. Flavor
$SU(3)$ symmetry in fact forbids any magnetic transition between
decuplet and antidecuplet. This is reflected by the fact that each
flavor contribution to the transition magnetic moments is
identically zero while it was not the case for transition from
octet, even in the case of vanishing total transition magnetic
moments like $\mu_{\Sigma^-_\mathbf{8}\Sigma^-_\mathbf{10}}$,
$\mu_{\Xi^-_\mathbf{8}\Xi^-_\mathbf{10}}$,
$\mu_{p^+_\mathbf{8}p^+_\mathbf{\overline{10}}}$ and
$\mu_{\Sigma^+_\mathbf{8}\Sigma^+_\mathbf{\overline{10}}}$. We have
then
\begin{equation}
\mu_{\Delta^+_\mathbf{10}p^+_\mathbf{\overline{10}}}=\mu_{\Sigma^+_\mathbf{10}\Sigma^+_\mathbf{\overline{10}}}=\mu_{\Delta^0_\mathbf{10}n^0_\mathbf{\overline{10}}}=\mu_{\Sigma^0_\mathbf{10}\Sigma^0_\mathbf{\overline{10}}}=\mu_{\Xi^0_\mathbf{10}\Xi^0_\mathbf{\overline{10}}}=\mu_{\Sigma^-_\mathbf{10}\Sigma^-_\mathbf{\overline{10}}}=\mu_{\Xi^-_\mathbf{10}\Xi^-_\mathbf{\overline{10}}}=0.
\end{equation}

\section{Specific $SU(6)$ symmetry relations}

The imbedding of the flavor $SU(3)$ symmetry into a larger group
$SU(6)$ implies stronger symmetry relations. The naive quark model
(NQM) assumes that baryons are made of three non-relativistic
valence quarks, their spin-flavor wave functions being given by
$SU(6)\supset SU(3)_F\times SU(2)_S$ symmetry. This simple picture
explains rather well masses and the ratio between proton and neutron
magnetic moments.

Within the assumption of $SU(6)$ symmetry octet and decuplet baryons
belong to the same supermultiplet. This means that such a symmetry
relates octet properties to decuplet ones and thus reduces the
number of parameters needed compared to flavor $SU(3)$ symmetry
only. In this section we remind these $SU(6)$ relations for octet
and decuplet \footnote{Pentaquarks may also be described in a
$SU(6)$ scheme but there is no obvious choice concerning the
supermultiplet \cite{SU6pentaquark}}. Later we will show that they
are satisfied by the $3Q$ Fock sector but explicitly broken by the
higher ones.

The naive $SU(6)$ quark model describes octet and decuplet baryons
as a system of three valence quarks only. This means that all
parameters with indices $q_\textrm{s}$ (contribution from quarks of
the sea) and $\bar q$ (contribution from antiquarks) vanish. It also
imposes that only explicit flavors, \emph{i.e.} flavors that are not
hidden, contribute leading to
\begin{equation}
\gamma_{I,q_\textrm{val}}=\beta'_{I,q_\textrm{val}}=0,\qquad
I=V,A,T,M.
\end{equation}

In NQM vector charges just count the number of valence quark of each
flavor and is blind concerning their spins. This means that we have
\begin{equation}
\alpha_{V,q_\textrm{val}}=2\qquad\beta_{V,q_\textrm{val}}=1
\end{equation} since the proton is seen as a system of two $u$ and one
$d$ quarks. For the decuplet we have
\begin{equation}
\alpha'_{V,q_\textrm{val}}=1
\end{equation}
since $\Delta^{++}$ is seen as a system of three $u$ quarks. The
other charges, \emph{i.e.} axial charges, tensor charges and
magnetic moments, depend on the quark spins through a difference in
orientation. $SU(6)$ symmetry relates the three parameters
$\alpha_{q_\textrm{val}},\beta_{q_\textrm{val}},\alpha'_{q_\textrm{val}}$
in the same manner
\begin{equation}
\alpha_{I,q_\textrm{val}}=-4\beta_{I,q_\textrm{val}}=\frac{4}{3}\,\alpha'_{I,q_\textrm{val}},\qquad
I=A,T,M.
\end{equation}
As a result, decuplet magnetic moments are proportional to the
proton magnetic moment.

NQM is a non-relativistic model where valence quark are in a purely
$s$ state. Rotational invariance implies axial and tensor charges to
be equal $\Delta q=\delta q$
\begin{equation}
\alpha_{A,q_\textrm{val}}=\alpha_{T,q_\textrm{val}}=\frac{4}{3}.
\end{equation}
On the top of that there cannot be any electric quadrupole
transition between octet and decuplet
\begin{equation}
\alpha^{8\to 10}_{E,q_\textrm{val}}=0.
\end{equation}

Let us now concentrate on octet and decuplet magnetic and transition
magnetic moments. We have collected in Table \ref{SU6} the $SU(3)$
relations and added the new ones imposed by $SU(6)$.
\begin{table}[h!]\begin{center}\caption{\small{$SU(3)$ and $SU(6)$ relations for magnetic and transition magnetic moments \cite{Beg}. Only the new relations compared to flavor $SU(3)$ are listed for $SU(6)$.}}
\begin{tabular}{c|c|c}
\hline\hline Multiplets&$SU(3)$&$SU(6)$\rule{0pt}{3ex}\\\hline
\rule{0pt}{3ex}
&Coleman-Glashow relations:&\\
&$\mu_{\Sigma^+_\mathbf{8}}=\mu_{p^+_\mathbf{8}}$&\\\rule{0pt}{3ex}
$\mathbf{8}$&$2\mu_{\Lambda^0_\mathbf{8}}=-2\,\mu_{\Sigma^0_\mathbf{8}}=-\frac{2}{\sqrt{3}}\,\mu_{\Sigma^0_\mathbf{8}\Lambda^0_\mathbf{8}}=\mu_{\Xi^0_\mathbf{8}}=\mu_{n^0_\mathbf{8}}$&$\frac{\mu_{p^+_\mathbf{8}}}{\mu_{n^0_\mathbf{8}}}=-\frac{3}{2}$\\\rule{0pt}{3ex}
&$\mu_{\Sigma^-_\mathbf{8}}=\mu_{\Xi^-_\mathbf{8}}=-(\mu_{p^+_\mathbf{8}}+\mu_{n_\mathbf{8}})$&\\\rule{0pt}{3ex}
&&\\\rule{0pt}{3ex} $\mathbf{10}$&$\mu_{\bf 10}\propto
Q_B$&$\mu_{\bf 10}=Q_B\,\mu_{p^+_\mathbf{8}}$\\\rule{0pt}{3ex}
&&\\\rule{0pt}{3ex}
$\mathbf{8}\to\mathbf{10}$&$\mu_{p^+_\mathbf{8}\Delta^+_\mathbf{10}}=-\mu_{\Sigma^+_\mathbf{8}\Sigma^+_\mathbf{10}}=\mu_{n^0_\mathbf{8}\Delta^0_\mathbf{10}}=-2\,\mu_{\Sigma^0_\mathbf{8}\Sigma^0_\mathbf{10}}=\frac{2}{\sqrt{3}}\,\mu_{\Lambda^0_\mathbf{8}\Sigma^0_\mathbf{10}}=-\mu_{\Xi^0_\mathbf{8}\Xi^0_\mathbf{10}}$&$\mu_{p^+_\mathbf{8}\Delta^+_\mathbf{10}}=\frac{2\sqrt{2}}{3}\,\mu_{p^+_\mathbf{8}}$\\\rule{0pt}{3ex}
&$\mu_{\Sigma^-_\mathbf{8}\Sigma^-_\mathbf{10}}=\mu_{\Xi^-_\mathbf{8}\Xi^-_\mathbf{10}}=0$&\rule[-2ex]{0pt}{5ex}\\
\hline\hline
\end{tabular}\label{SU6}\end{center}
\end{table}
One can see that only one magnetic moment is needed in $SU(6)$, say
$\mu_{p^+_\mathbf{8}}$ while in the $SU(3)$ case four are needed.

In conclusion NQM is a very simple model for octet and decuplet
baryons and is very predictive since only a few parameters are left
undetermined. One of the best successes of the $SU(6)$ symmetry is
the prediction of the proton-to-neutron magnetic moments ratio.
However, as the time passed by, more and more experiments gave
results in contradiction with the NQM predictions. Among the
discrepancies let us mention the overestimation of the nucleon axial
charges, the underestimation of the nucleon-to-Delta transition
magnetic moment, the absence of strangeness in the nucleon and of
electric quadrupole transition between nucleon and Delta. This
indicates that something is missing in NQM. Nowadays it is clear
that relativity and quark motion have to be taken into account to
understand correctly axial and tensor charges. On the top of that
the picture of baryons made of three quarks only is too simple. A
full description would involve an indefinite number of
quark-antiquark pairs. These pairs should in principle be
implemented somehow in a realistic model. This is done either
explicitly by incorporating quark-antiquark pairs as a new degree of
freedom ($\sim$ pion cloud) or implicitly by considering that the
effect of quark-antiquark pairs can be described in an effective way
by means of constituent quark form factors. A combination of these
two approaches is of course also possible.

\chapter{Vector charges and normalization}
\section{Introduction}

The vector charges of a baryon are defined as forward matrix
elements of the vector current
\begin{equation}
\langle
B(p)|\bar\psi\gamma_\mu\lambda^a\psi|B(p)\rangle=g^{(a)}_V\bar
u(p)\gamma_\mu u(p)
\end{equation}
where $a=0,3,8$ and $\lambda^3,\lambda^8$ are Gell-Mann matrices,
$\lambda^0$ is just in this context the $3\times3$ unit matrix.
These vector charges are related to the first moment of the
unpolarized quark distributions
\begin{equation}\label{Nucleon vector charges}
g^{(3)}_V=u-d,\qquad g^{(8)}_V=\frac{1}{\sqrt{3}}(u+d-2s),\qquad
g^{(0)}_V=u+d+s
\end{equation}
where $q\equiv\int_0^1\ud z\left[q_+(z)+q_-(z)-\bar q_+(z)-\bar
q_-(z)\right]$ with $q=u,d,s$ and $\pm$ referring to the helicity
state. We split the vector charges into valence quark, sea quark and
antiquark contributions
\begin{equation}
q=q_\textrm{val}+q_\textrm{sea}, \qquad
q_\textrm{sea}=q_\textrm{s}-\bar q
\end{equation}
where index s refers to the quarks in the sea pairs.

The vector charges can be understood as follows. They count the
total number of quarks $q_{tot}=q_\textrm{val}+q_\textrm{s}$ with
$q_{\textrm{val},s+}+q_{\textrm{val},s-}$ \emph{minus} the total
number of antiquarks $\bar q=\bar q_++\bar q_-$, irrespective of
their spin. The vector charges $\bar q\gamma_\mu q$ then give the
\emph{net} number of quarks of flavor $q=u,d,s$ in the baryon.

Since there could be an infinite number of quark-antiquark pairs in
the nucleon the meaningful quantity is this difference between the
number of quarks and antiquarks which is restricted by the baryon
number and charge. In the literature the net number of quarks is
identified with the number of valence quarks $q_v=q_{tot}-\bar q=q$.
This identification is due to the NQM picture of the baryon and the
fact that quark-antiquark pairs are commonly thought to be mainly
produced in the perturbative process of gluon splitting leading to
$q_\textrm{s}=\bar q$, \emph{i.e.} a vanishing contribution of the
sea $q_\textrm{sea}=0$. We stress that this definition of valence
quarks does not coincide with our definition where valence quarks
are quarks filling the discrete level (\ref{Discrete level IMF}).
For a given quark flavor $f$ the restriction $q_\textrm{s}=\bar q$
does not hold and thus $q_\textrm{sea}\neq 0$ leading to
$q_\textrm{val}\neq q_v$. This is due to the fact that starting from
the $5Q$ Fock sector there are components with a different valence
composition than the $3Q$ sector, \emph{e.g.} $|udd(u\bar d)\rangle$
in the proton. Perturbative gluon splitting is flavor symmetric
while non-perturbative processes such as pion emission are not.

Since the quark mass differences are fairly small compared with a
typical energy scale in DIS, the gluon splitting processes are
expected to occur almost equally for the three flavor and thus
generate a flavor symmetric sea. The Gottfried Sum Rule (GSR) of
charged lepton-nucleon DIS \cite{GSR}
\begin{equation}
I_G(Q^2)=\int_0^1\frac{\ud
x}{x}\left[F_2^{lp}(x,Q^2)-F_2^{ln}(x,Q^2)\right]
\end{equation}
can provide important information on the possible existence of a
light antiquark flavor asymmetry in the nucleon sea. The sum rule
can be expressed in terms of PDF as follows
\begin{equation}
I_G(Q^2)=\int_0^1\ud
x\left[\frac{1}{3}\left(u_v(x,Q^2)-d_v(x,Q^2)\right)+\frac{2}{3}\left(\bar
u(x,Q^2)-\bar d(x,Q^2)\right)\right].
\end{equation}
A symmetric sea scenario implies thus that $\bar u(x,Q^2)=\bar
d(x,Q^2)$ leading to $I_G=1/3$. The analysis of muon-nucleon DIS
data performed by NMC \cite{NMCGott} gives $I_G(Q^2=4$
GeV$^2)=0.235\pm 0.026$ which is significantly smaller than
symmetric scenario prediction. This deviation indicates the
existence of a non-zero integrated light quark flavor asymmetry
\begin{equation}
\int_0^1\ud x\left[\bar d(x,Q^2)-\bar u(x,Q^2)\right]=0.147\pm
0.039.
\end{equation}
Perturbative QCD corrections to GSR are really small and cannot be
responsible for the violation of the flavor-symmetric prediction
\cite{QCDcorr}.

The analysis of the Drell-Yan production in proton-proton and
proton-deuteron scattering by the E866 collaboration \cite{E866Coll}
concluded to a similar value at $Q^2=56$ GeV$^2$ as well as the
HERMES experiment using Semi-Inclusive Deep-Inelastic Scattering
(SIDIS) measurement of charged pions from hydrogen and deuterium
targets \cite{HERMESGott} at $Q^2=2.3$ GeV$^2$. It has then been
concluded that the integrated value of the light quark flavor
asymmetry is almost independent of $Q^2$ over a wide range of the
momentum transfer. This demonstrates in part its non-perturbative
origin.

From the simple perturbative gluon splitting process one also
expects to have a strangeness contribution to the nucleon sea.
According to the usual definition of valence quarks nucleon
strangeness is restricted to the sea only. It is then claimed that
studying nucleon strangeness would give important lessons on the
nucleon sea. Neutrino-dimuon events of neutrino-nucleon scattering
revealed that the strange sea is roughly half of the $u$ and $d$ sea
\cite{Ssea}. The perturbative gluon splitting also implies a
vanishing asymmetry of the strange distribution $s(x)-\bar s(x)=0$.
The possible asymmetry of $s(x)$ and $\bar s(x)$ has however been
discussed by Signal and Thomas \cite{ST} and further explored by
others \cite{STplus}. While the analysis of related experimental
data seems not conclusive \cite{expdata}, a refreshed interest is
due to the ``NuTeV anomaly'' \cite{NuTeVanom}, a $3\sigma$ deviation
of the NuTeV measured value of $\sin^2\theta_W=0.2277\pm 0.0013\pm
0.0009$ \cite{ST,SAexp} from the world average of other measurements
$\sin^2\theta_W=0.2227\pm 0.0004$ with $\theta_W$ the Weinberg angle
of the Standard Model.

\section{Vector charges on the light cone}

On the light-cone vector charges are obtained from the \emph{plus}
component of the vector current operator $\bar\psi\gamma^+\psi$
\begin{equation}
q=\frac{1}{2P^+}\langle
P,\frac{1}{2}|\bar\psi_{LC}\gamma^+\psi_{LC}|P,\frac{1}{2}\rangle.
\end{equation}
Using the Melosh rotation of the standard approach one can see that
$q_{LC}$ and $q_{NR}$ are related as follows
\begin{equation}
q_{LC}=\langle M_V\rangle q_{NR}
\end{equation}
where
\begin{equation}\label{Defq}
M_V=1.
\end{equation}
The relativistic effect introduced by the Melosh rotation affects
the quark spins. Since the vector charges count quarks irrespective
of their spin these charges are not affected by the rotation.

In the IMF language one has to use the ``good'' components $\mu=0,3$
of the vector current operator. This operator does not flip the spin
and counts quarks irrespective of their spin. We have then
$M^{\tau}_{\sigma}=\delta^{\tau}_{\sigma}$.

\section{Scalar overlap integrals and quark distributions}

From the expression (\ref{Discrete level IMF}) and if we concentrate
on the spin part the contraction of two valence wave functions $F$
gives
\begin{equation}\label{VectOQL}
F^\dag F\propto h^2(p)+2h(p)\frac{p_z}{|\up|}j(p)+j^2(p).
\end{equation}
The physical interpretation is straightforward. The first term just
describes a valence quark staying in a $s$ state. The second term
describes the transition of a valence quark from a $s$ state into a
$p$ state and \emph{vice versa}. Angular momentum conservation
forces the $p$ state to have $J_z=0$ which is expressed by the
factor $p_z$. Note that it is still implicitly understood that
$p_z=z\uM-E_\textrm{lev}$. The last term describes a valence quark
staying in the same $p$ state. The vector operator is blind
concerning the spin. This means that the spin structure of a valence
quark line with or without vector operator acting on it is
(\ref{VectOQL}).

\begin{figure}[h]\begin{center}\begin{minipage}[c]{6cm}\begin{center}\includegraphics[width=5.3cm]{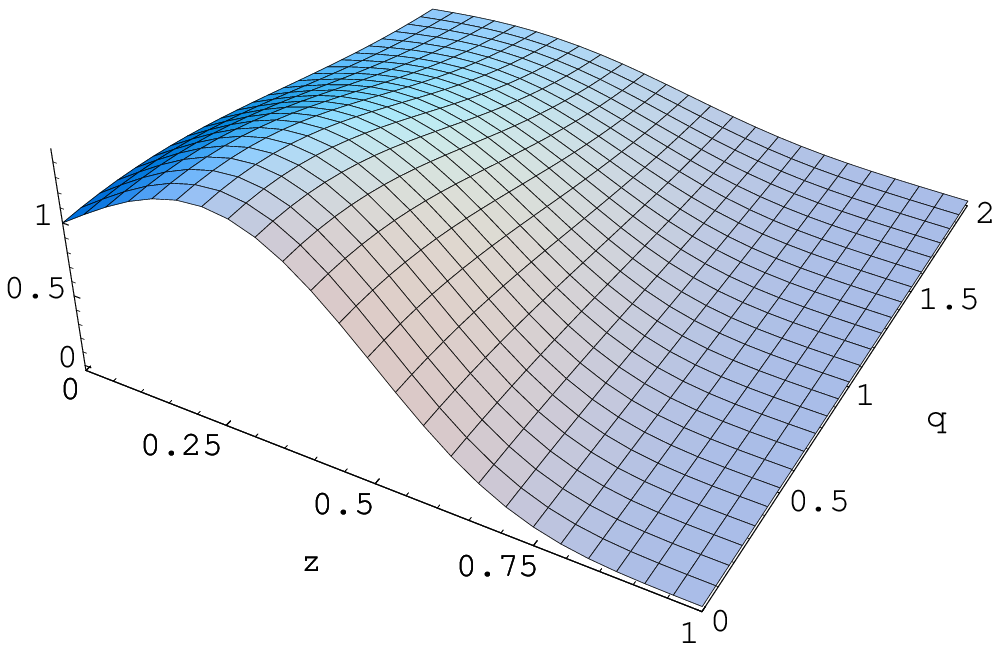}
\end{center}\end{minipage}\hspace{1cm}\begin{minipage}[c]{6cm}\begin{center}\includegraphics[width=5.3cm]{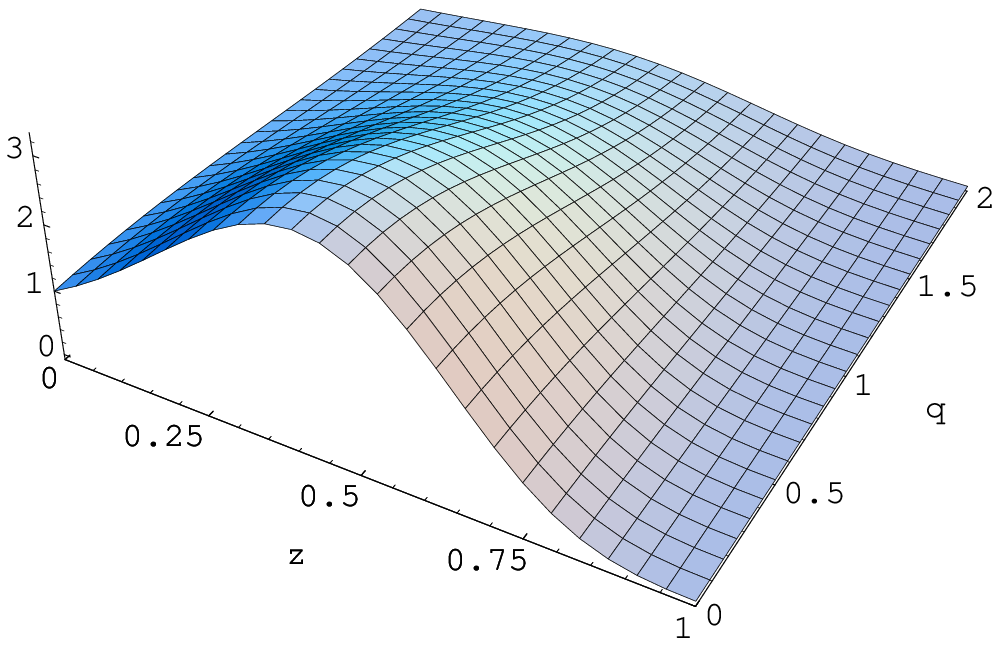}
\end{center}\end{minipage}\caption{\small{Vector probability distribution $\Phi^V(z,\uq_\perp)$ that three valence quarks leave the fraction $z$ of the baryon momentum and
transverse momentum $\uq_\perp$ to the quark-antiquark pair(s) in
the relativistic (left) and non-relativistic (right) cases plotted
in units of $M$ and normalized to unity for $z=\uq_\perp=0$.
}}\label{Phiplot}\end{center}
\end{figure}

The vector valence quark distribution is obtained by the
multiplication of three factor with this structure where the momenta
are respectively $p_1$, $p_2$ and $p_3$. The expansion gives the
following function $D$
\begin{eqnarray}
&D^V(p_1,p_2,p_3)=h^2(p_1)h^2(p_2)h^2(p_3)+6h^2(p_1)h^2(p_2)\left[h(p_3)\frac{p_{3z}}{|\up_3|}j(p_3)\right]+3h^2(p_1)h^2(p_2)j^2(p_3)&\nonumber\\
&+12h^2(p_1)\left[h(p_2)\frac{p_{2z}}{|\up_2|}j(p_2)\right]\left[h(p_3)\frac{p_{3z}}{|\up_3|}j(p_3)\right]+12h^2(p_1)\left[h(p_2)\frac{p_{2z}}{|\up_2|}j(p_2)\right]j^2(p_3)&\nonumber\\
&+8\left[h(p_1)\frac{p_{1z}}{|\up_1|}j(p_1)\right]\left[h(p_2)\frac{p_{2z}}{|\up_2|}j(p_2)\right]\left[h(p_3)\frac{p_{3z}}{|\up_3|}j(p_3)\right]+3h^2(p_1)j^2(p_2)j^2(p_3)&\\
&+12\left[h(p_1)\frac{p_{1z}}{|\up_1|}j(p_1)\right]\left[h(p_2)\frac{p_{2z}}{|\up_2|}j(p_2)\right]j^2(p_3)+6\left[h(p_1)\frac{p_{1z}}{|\up_1|}j(p_1)\right]j^2(p_2)j^2(p_3)&\nonumber\\
&+j^2(p_1)j^2(p_2)j^2(p_3)&\nonumber\label{Phi}
\end{eqnarray}
that is needed in the expression of the valence quark distribution
(\ref{Probability3q}). In the non-relativistic limit $j=0$ this
function $D$ is reduced to
\begin{equation}
D^V_{NR}(p_1,p_2,p_3)=h^2(p_1)h^2(p_2)h^2(p_3).
\end{equation}
The vector valence probability distribution $\Phi^V(z,\uq_\perp)$ is
then obtained by integration over the valence quark momenta, see eq.
(\ref{Probability3q}) and are depicted in Fig. \ref{Phiplot} in the
relativistic and non-relativistic cases.
Relativistic corrections (quark angular momentum) clearly shift the
bump in the probability distribution to smaller values of $z$
meaning that it leaves less longitudinal momentum fraction to the
quark-antiquark pair(s). This can be easily understood from the
shape of the discrete-level wave function $h(p)$ and $j(p)$, see
Fig. \ref{LevelFT}. While the $s$-wave $h(p)$ is maximum at $p=0$,
the $p$-wave $j(p)$ has a node. The maximum value of the latter is
then obtained for non-vanishing $p$. Relativistic valence quarks
then need more momentum than non-relativistic ones explaining the
fact that less momentum is left for the quark-antiquark pair(s) in
the relativistic case compared to the non-relativistic one.

In the following we give the integrals appearing in each Fock sector
and the numerical values obtained for them. In the evaluation of the
scalar overlap integrals we have used the constituent quark mass
$M=345$ MeV, the Pauli-Villars mass $M_\textrm{PV}=556.8$ MeV for
the regularization of (\ref{Direct1})-(\ref{Direct4}),
(\ref{Exchange begin})-(\ref{Exchange end}) and of
(\ref{int1})-(\ref{int12}) and the baryon mass $\uM=1207$ MeV as it
follows for the ``classical'' mass in the mean field approximation
\cite{Approximation}.

\subsection{$3Q$ scalar integral}

In the $3Q$ sector there is no quark-antiquark pair and thus only
one integral is involved. It corresponds to the valence quark
distribution without momentum left to the sea $\Phi^V(0,0)$. Notice
that we have the freedom to choose the normalization of the
discrete-level wave functions $h$ and $j$. In other words we have
the freedom to choose in particular
\begin{equation}
\Phi^V(0,0)=1.
\end{equation}
Since the relativistic and non-relativistic vector valence quark
distributions have different expressions the normalization
$\Phi^V(0,0)=1$ implies different normalizations for $h$ and
$h_{NR}$. Diakonov and Petrov commented in \cite{DiaPet} that the
lower component $j(p)$ is ``substantially'' smaller than the upper
one $h(p)$. More quantitatively it turned out that the $j(p)$
contribution to the normalization of the discrete-level wave
function $\psi_\textrm{lev}(\up)$ is still 20\% (result in
accordance with \cite{Lower component}). This with the combinatoric
factors in eq. (\ref{Phi}) shows that taking the lower component $j$
into account can have a non-negligible impact on the estimations.
The nucleon is thus definitely a relativistic system.

\subsection{$5Q$ scalar integrals}

In the $5Q$ sector there is one quark-antiquark pair. Contractions
given by the direct diagram lead to three different integrals
$J=\pi\pi,33,\sigma\sigma$. Only the scalar-to-scalar
$\Sigma^\dag\Sigma$ and pseudoscalar-to-pseudoscalar $\Pi^\dag\Pi$
transitions are allowed.

In the non-relativistic case we have obtained
\begin{equation}
K^V_{\pi\pi,NR}=0.06237,\qquad K^V_{33,NR}=0.02842,\qquad
K^V_{\sigma\sigma,NR}=0.03731
\end{equation}
while in the relativistic case we have obtained
\begin{equation}
K^V_{\pi\pi}=0.03652,\qquad K^V_{33}=0.01975,\qquad
K^V_{\sigma\sigma}=0.01401.
\end{equation}
As one can expect from the comparison of both probability
distributions in Fig. \ref{Phiplot} relativistic corrections reduce
strongly (about one half) the values of the $5Q$ scalar overlap
integrals.

Contractions given by the exchange diagram lead to seven different
integrals $J=1$-$7$ in the non-relativistic limit
\begin{eqnarray}
&K_1=0.00560,\qquad K_2=0.00968,\qquad K_3=-0.00077,\qquad K_4=0.00470,&\nonumber\\
&K_5=0.00857,\qquad K_6=0.00423,\qquad K_7=0.00288.&
\end{eqnarray}
Anticipating on the final results, these exchange contributions have
only a small impact on the observables and can be reasonably
neglected. That's the reason why they haven't been computed with
relativistic corrections.

\subsection{$7Q$ scalar integrals}

In the $7Q$ sector there are two quark-antiquark pairs. Contractions
given by the direct diagram give eight different integrals
$J=\pi\pi\pi\pi,\pi\pi\pi\pi 2,\pi\pi 33,3333,\pi 3\pi
3,\sigma\sigma\pi\pi,\sigma\sigma 33,\sigma\sigma\sigma\sigma$.
Since the $5Q$ sector taught us that relativistic effects are
important these integrals have been evaluated with the relativistic
valence probability distribution only
\begin{eqnarray}
&K^V_{\pi\pi\pi\pi}=0.00082,\qquad K^V_{\pi\pi\pi\pi
2}=0.00026,\qquad
K^V_{\pi\pi 33}=0.00039,\qquad K^V_{3333}=0.00019,&\nonumber\\
&K^V_{\pi 3\pi 3}=0.00017,\qquad
K^V_{\sigma\sigma\pi\pi}=0.00027,\qquad K^V_{\sigma\sigma
33}=0.00012,\qquad K^V_{\sigma\sigma\sigma\sigma}=0.00009.&
\end{eqnarray}
Even though these values are smaller than the $5Q$ exchange scalar
integrals the contribution of the $7Q$ component is larger due to
large combinatoric factors, see next section.

By analogy with the $5Q$ sector, exchange diagrams contributions are
neglected and thus have not been computed.

\section{Combinatoric results}

Normalizations and vector matrix elements are linear combinations of
the vector scalar overlap integrals. These specific combinations are
obtained by contracting the baryon rotational wave functions with
the vector operator. In the following we give for each multiplet the
combinations obtained.

\subsection{Octet baryons}

Here are the expressions for the octet baryons normalization. They
are obtained by contracting the octet baryon wave functions without
any charge acting on the quark lines. The upper indices $3,5,7$
refer to the $3Q$, $5Q$ and $7Q$ Fock sectors.\newline The
contributions to the octet normalization are
\begin{eqnarray}
\uN^{(3)}(B_{\bf 8})&=&9\,\Phi^V(0,0),\\
\uN^{(5)}(B_{\bf
8})&=&\frac{18}{5}\left(11K^V_{\pi\pi}+23K^V_{\sigma\sigma}\right),\\
\uN^{(5)\textrm{exch}}(B_{\bf
8})&=&\frac{-12}{5}\left(9K_1+4K_3+4K_4-17K_6-17K_7\right),\\
\uN^{(7)}(B_{\bf
8})&=&\frac{144}{5}\left(15K^V_{\pi\pi\pi\pi}+5K^V_{\pi\pi\pi\pi
2}+52K^V_{\sigma\sigma\pi\pi}+54K^V_{\sigma\sigma\sigma\sigma}\right).
\end{eqnarray}
In the $3Q$ sector there is no quark-antiquark pair and thus only
valence quarks contribute to the vector charges
\begin{equation}
\alpha_{V,q_\textrm{val}}^{(3)}=18\,\Phi^V(0,0),\qquad
\beta_{V,q_\textrm{val}}^{(3)}=9\,\Phi^V(0,0),\qquad
\gamma_{V,q_\textrm{val}}^{(3)}=0.
\end{equation}
In the $5Q$ sector one has for the direct diagram
\begin{eqnarray}
&\alpha_{V,q_\textrm{val}}^{(5)}=\frac{18}{5}\left(15K^V_{\pi\pi}+43K^V_{\sigma\sigma}\right),\quad
\alpha_{V,q_\textrm{s}}^{(5)}=\frac{132}{5}\left(K^V_{\pi\pi}+K^V_{\sigma\sigma}\right),\quad
\alpha_{V,\bar
q}^{(5)}=\frac{6}{5}\left(K^V_{\pi\pi}+13K^V_{\sigma\sigma}\right),&\\
&\beta_{V,q_\textrm{val}}^{(5)}=\frac{72}{25}\left(12K^V_{\pi\pi}+25K^V_{\sigma\sigma}\right),\quad
\beta_{V,q_\textrm{s}}^{(5)}=\frac{24}{25}\left(13K^V_{\pi\pi}+22K^V_{\sigma\sigma}\right),\quad
\beta_{V,\bar
q}^{(5)}=\frac{6}{25}\left(31K^V_{\pi\pi}+43K^V_{\sigma\sigma}\right),&\\
&\gamma_{V,q_\textrm{val}}^{(5)}=\frac{36}{25}\left(7K^V_{\pi\pi}+5K^V_{\sigma\sigma}\right),\quad
\gamma_{V,q_\textrm{s}}^{(5)}=\frac{6}{25}\left(K^V_{\pi\pi}+49K^V_{\sigma\sigma}\right),\quad
\gamma_{V,\bar
q}^{(5)}=\frac{6}{25}\left(43K^V_{\pi\pi}+79K^V_{\sigma\sigma}\right).&
\end{eqnarray}
Concerning the exchange diagram, since one cannot disentangle
valence quarks from sea quarks we can decompose the parameters into
quark $q_\textrm{val+s}$ and antiquark $\bar q$ contributions only
\begin{eqnarray}
\alpha_{V,q_\textrm{val+s}}^{(5)\textrm{exch}}&=&\frac{-24}{25}\left(57K_1+22K_3+22K_4-101K_6-101K_7\right),\\
\alpha_{V,\bar q}^{(5)\textrm{exch}}&=&\frac{-48}{25}\left(6K_1+K_3+K_4-8K_6-8K_7\right),\\
\beta_{V,q_\textrm{val+s}}^{(5)\textrm{exch}}&=&\frac{-4}{25}\left(171K_1+91K_3+91K_4-353K_6-353K_7\right),\\
\beta_{V,\bar q}^{(5)\textrm{exch}}&=&\frac{-4}{25}\left(36K_1+31K_3+31K_4-98K_6-98K_7\right),\\
\gamma_{V,q_\textrm{val+s}}^{(5)\textrm{exch}}&=&\gamma_{V,\bar
q}^{(5)\textrm{exch}}=\frac{-4}{25}\left(27K_1+17K_3+17K_4-61K_6-61K_7\right).
\end{eqnarray}
In the $7Q$ sector the combinations are
\begin{eqnarray}
\alpha_{V,q_\textrm{val}}^{(7)}&=&\frac{48}{5}\left(49K^V_{\pi\pi\pi\pi}+38K^V_{\pi\pi\pi\pi
2}+200K^V_{\sigma\sigma\pi\pi}+285K^V_{\sigma\sigma\sigma\sigma}\right),\\
\alpha_{V,q_\textrm{s}}^{(7)}&=&\frac{48}{5}\left(47K^V_{\pi\pi\pi\pi}+2K^V_{\pi\pi\pi\pi
2}+144K^V_{\sigma\sigma\pi\pi}+99K^V_{\sigma\sigma\sigma\sigma}\right),\\
\alpha_{V,\bar
q}^{(7)}&=&\frac{96}{5}\left(3K^V_{\pi\pi\pi\pi}+5K^V_{\pi\pi\pi\pi
2}+16K^V_{\sigma\sigma\pi\pi}+30K^V_{\sigma\sigma\sigma\sigma}\right),\\
\beta_{V,q_\textrm{val}}^{(7)}&=&\frac{48}{25}\left(181K^V_{\pi\pi\pi\pi}+41K^V_{\pi\pi\pi\pi
2}+626K^V_{\sigma\sigma\pi\pi}+618K^V_{\sigma\sigma\sigma\sigma}\right),\\
\beta_{V,q_\textrm{s}}^{(7)}&=&\frac{96}{25}\left(61K^V_{\pi\pi\pi\pi}+22K^V_{\pi\pi\pi\pi
2}+201K^V_{\sigma\sigma\pi\pi}+198K^V_{\sigma\sigma\sigma\sigma}\right),\\
\beta_{V,\bar
q}^{(7)}&=&\frac{96}{25}\left(39K^V_{\pi\pi\pi\pi}+5K^V_{\pi\pi\pi\pi
2}+124K^V_{\sigma\sigma\pi\pi}+102K^V_{\sigma\sigma\sigma\sigma}\right),\\
\gamma_{V,q_\textrm{val}}^{(7)}&=&\frac{48}{25}\left(83K^V_{\pi\pi\pi\pi}-2K^V_{\pi\pi\pi\pi
2}+238K^V_{\sigma\sigma\pi\pi}+129K^V_{\sigma\sigma\sigma\sigma}\right),\\
\gamma_{V,q_\textrm{s}}^{(7)}&=&\frac{48}{25}\left(31K^V_{\pi\pi\pi\pi}+32K^V_{\pi\pi\pi\pi
2}+146K^V_{\sigma\sigma\pi\pi}+243K^V_{\sigma\sigma\sigma\sigma}\right),\\
\gamma_{V,\bar
q}^{(7)}&=&\frac{288}{25}\left(19K^V_{\pi\pi\pi\pi}+5K^V_{\pi\pi\pi\pi
2}+64K^V_{\sigma\sigma\pi\pi}+62K^V_{\sigma\sigma\sigma\sigma}\right).
\end{eqnarray}

One can easily check that the obvious sum rules for the proton
\begin{equation}
\int\ud z\,[u(z)-\bar u(z)]=2, \qquad \int\ud z\,[d(z)-\bar d(z)]=1,
\qquad \int\ud z\,[s(z)-\bar s(z)]=0
\end{equation}
are satisfied separately in each sector. They are translated in our
parametrization as follows
\begin{equation}
\alpha_{V,q_\textrm{val}}^{(i)}+\alpha_{V,q_\textrm{s}}^{(i)}-\alpha_{V,\bar
q}^{(i)}=2\uN^{(i)}(B_{\bf 8}),\quad
\beta_{V,q_\textrm{val}}^{(i)}+\beta_{V,q_\textrm{s}}^{(i)}-\beta_{V,\bar
q}^{(i)}=\uN^{(i)}(B_{\bf 8}),\quad
\gamma_{V,q_\textrm{val}}^{(i)}+\gamma_{V,q_\textrm{s}}^{(i)}-\gamma_{V,\bar
q}^{(i)}=0 \quad \forall i.
\end{equation}

\subsection{Decuplet baryons}

Here are the expressions for the decuplet baryons normalization.
They are obtained by contracting the decuplet baryon wave functions
without any charge acting on the quark lines. The upper indices
$i=3,5$ refer to the $3Q$ and $5Q$ Fock sectors while the lower ones
$3/2,1/2$ refer to the $z$-component of the decuplet baryon spin.
Notice that in the $5Q$ sector only direct contributions are given.
In our study of exchange diagrams only octet and antidecuplet were
considered. Since the conclusion is that exchange contributions are
negligible we did not compute them when we studied the
decuplet.\newline The contributions to the decuplet normalization
are
\begin{eqnarray}
\uN^{(3)}_{3/2}(B_{\bf 10})&=&\uN^{(3)}_{1/2}(B_{\bf 10})=\frac{18}{5}\,\Phi^V(0,0),\\
\uN^{(5)}_{3/2}(B_{\bf
10})&=&\frac{9}{5}\left(15K^V_{\pi\pi}-6K^V_{33}+17K^V_{\sigma\sigma}\right),\\
\uN^{(5)}_{1/2}(B_{\bf
10})&=&\frac{9}{5}\left(11K^V_{\pi\pi}+6K^V_{33}+17K^V_{\sigma\sigma}\right).
\end{eqnarray}
In the $3Q$ sector there is no quark-antiquark pair and thus only
valence quarks contribute to the vector charges
\begin{equation}
\alpha'^{(3)}_{V,q_\textrm{val},3/2}=\alpha'^{(3)}_{V,q_\textrm{val},1/2}=\frac{18}{5}\,\Phi^V(0,0),\qquad
\beta'^{(3)}_{V,q_\textrm{val},3/2}=\beta'^{(3)}_{V,q_\textrm{val},1/2}=0.
\end{equation}
In the $5Q$ sector one has
\begin{eqnarray}
&\alpha'^{(5)}_{V,q_\textrm{val},3/2}=\frac{9}{20}\left(33K^V_{\pi\pi}-6K^V_{33}+67K^V_{\sigma\sigma}\right),\quad
\alpha'^{(5)}_{V,q_\textrm{s},3/2}=\frac{3}{20}\left(57K^V_{\pi\pi}-30K^V_{33}+19K^V_{\sigma\sigma}\right),&\nonumber\\
&\alpha'^{(5)}_{V,\bar
q,3/2}=\frac{-6}{5}\left(3K^V_{\pi\pi}-3K^V_{33}-2K^V_{\sigma\sigma}\right),&\\
&\alpha'^{(5)}_{V,q_\textrm{val},1/2}=\frac{9}{20}\left(29K^V_{\pi\pi}+6K^V_{33}+67K^V_{\sigma\sigma}\right),\quad
\alpha'^{(5)}_{V,q_\textrm{s},1/2}=\frac{3}{20}\left(37K^V_{\pi\pi}+30K^V_{33}+19K^V_{\sigma\sigma}\right),&\nonumber\\
&\alpha'^{(5)}_{V,\bar
q,1/2}=\frac{-6}{5}\left(K^V_{\pi\pi}+3K^V_{33}-2K^V_{\sigma\sigma}\right),&\\
&\beta'^{(5)}_{V,q_\textrm{val},3/2}=\frac{9}{20}\left(27K^V_{\pi\pi}-18K^V_{33}+K^V_{\sigma\sigma}\right),\quad
\beta'^{(5)}_{V,q_\textrm{s},3/2}=\frac{3}{20}\left(3K^V_{\pi\pi}+6K^V_{33}+49K^V_{\sigma\sigma}\right),&\nonumber\\
&\beta'^{(5)}_{V,\bar
q,3/2}=\frac{3}{5}\left(21K^V_{\pi\pi}-12K^V_{33}+13K^V_{\sigma\sigma}\right),&\\
&\beta'^{(5)}_{V,q_\textrm{val},1/2}=\frac{9}{20}\left(15K^V_{\pi\pi}+18K^V_{33}+K^V_{\sigma\sigma}\right),\quad
\beta'^{(5)}_{V,q_\textrm{s},1/2}=\frac{3}{20}\left(7K^V_{\pi\pi}-6K^V_{33}+49K^V_{\sigma\sigma}\right),&\nonumber\\
&\beta'^{(5)}_{V,\bar
q,1/2}=\frac{3}{5}\left(13K^V_{\pi\pi}+12K^V_{33}+13K^V_{\sigma\sigma}\right).&
\end{eqnarray}
The $7Q$ sector of the decuplet has not been computed due to its far
bigger complexity.

One can easily check that the obvious sum rules for $\Delta^{++}$
\begin{equation}
\int\ud z\,[u(z)-\bar u(z)]=3, \qquad \int\ud z\,[d(z)-\bar d(z)]=0,
\qquad \int\ud z\,[s(z)-\bar s(z)]=0
\end{equation}
are satisfied separately in each sector. They are translated in our
parametrization as follows
\begin{equation}
\alpha'^{(i)}_{V,q_\textrm{val},J}+\alpha'^{(i)}_{V,q_\textrm{s},J}-\alpha'^{(i)}_{V,\bar
q,J}=\uN^{(i)}_J(B_{\bf 10}),\quad
\beta'^{(i)}_{V,q_\textrm{val},J}+\beta'^{(i)}_{V,q_\textrm{s},J}-\beta'^{(i)}_{V,\bar
q,J}=0 \quad \forall i \textrm{ and } J=3/2,1/2.
\end{equation}

Let us emphasize an interesting observation. If the decuplet was
made of three quarks only then one would have the following relation
between spin-3/2 and spin-1/2 vector contributions
\begin{equation}\label{sphericityV}
V_{3/2}=V_{1/2}.
\end{equation}
This picture presents the $\Delta$ as a spherical particle. Things
change in the $5Q$ sector. One notices directly that the relations
are broken by a unique structure $(K^V_{\pi\pi}-3K^V_{33})$ in the
vector case and normalizations. Going back to the definition of
those integrals this amounts in fact to a structure like
$(q^2-3q_z^2)$ coming from the coupling to pions in $p$ waves. This
naturally reminds the expression of a quadrupole
\begin{equation}
Q_{ij}=\int\ud^3r\,\rho(\ur)\,(3r_ir_j-r^2\delta_{ij})
\end{equation}
specified to the component $i=j=z$. Remarkably the present approach
shows \emph{explicitly} that the pion field is responsible for the
deviation of the $\Delta$ from spherical symmetry. This discussion
will be resumed in the chapter dedicated to magnetic moments,
especially concerning the $\gamma N\Delta$ transition.

\subsection{Antidecuplet baryons}

Here are the expressions for the antidecuplet baryons normalization.
They are obtained by contracting the antidecuplet baryon wave
functions without any charge acting on the quark lines. The upper
indices $5,7$ refer to the $5Q$ and $7Q$ Fock sectors\footnote{We
remind that there is no $3Q$ component in pentaquarks.}.\newline The
contributions to the antidecuplet normalization are
\begin{eqnarray}
\uN^{(5)}(B_{\bf\overline{10}})&=&\frac{36}{5}\left(K^V_{\pi\pi}+K^V_{\sigma\sigma}\right),\\
\uN^{(5)\textrm{exch}}(B_{\bf\overline{10}})&=&\frac{-12}{5}\left(K_3+K_4-2K_6-2K_7\right),\\
\uN^{(7)}(B_{\bf\overline{10}})&=&\frac{72}{5}\left(9K^V_{\pi\pi\pi\pi}+K^V_{\pi\pi\pi\pi
2}+26K^V_{\sigma\sigma\pi\pi}+18K^V_{\sigma\sigma\sigma\sigma}\right).
\end{eqnarray}
In the $5Q$ sector one has for the direct diagram
\begin{eqnarray}
&\alpha''^{(5)}_{V,q_\textrm{val}}=\frac{18}{5}\left(K^V_{\pi\pi}+K^V_{\sigma\sigma}\right),\quad
\alpha''^{(5)}_{V,q_\textrm{s}}=\frac{6}{5}\left(K^V_{\pi\pi}+K^V_{\sigma\sigma}\right),\quad
\alpha''^{(5)}_{V,\bar
q}=\frac{-12}{5}\left(K^V_{\pi\pi}+K^V_{\sigma\sigma}\right),&\\
&\beta''^{(5)}_{V,q_\textrm{val}}=\frac{18}{5}\left(K^V_{\pi\pi}+K^V_{\sigma\sigma}\right),\quad
\beta''^{(5)}_{V,q_\textrm{s}}=\frac{6}{5}\left(K^V_{\pi\pi}+K^V_{\sigma\sigma}\right),\quad
\beta''^{(5)}_{V,\bar
q}=\frac{24}{5}\left(K^V_{\pi\pi}+K^V_{\sigma\sigma}\right).&
\end{eqnarray}
The $5Q$ exchange diagram gives
\begin{eqnarray}
\alpha''^{(5)\textrm{exch}}_{V,q_\textrm{val+s}}&=&\frac{-8}{5}\left(K_3+K_4-2K_6-2K_7\right),\\
\alpha''^{(5)\textrm{exch}}_{V,\bar q}&=&\frac{4}{5}\left(K_3+K_4-2K_6-2K_7\right),\\
\beta''^{(5)\textrm{exch}}_{V,q_\textrm{val+s}}&=&\beta''^{(5)\textrm{exch}}_{V,\bar
q}=\frac{-8}{5}\left(K_3+K_4-2K_6-2K_7\right).
\end{eqnarray}
Compared to octet and decuplet baryons, antidecuplet baryons have a
rather simple $5Q$ component. This is of course related to the fact
that there is no $3Q$ component. The minimal content of a baryon is
simple while higher Fock states introduce more complicated
structures.

In the $7Q$ sector one has
\begin{eqnarray}
\alpha''^{(7)}_{V,q_\textrm{val}}&=&\frac{12}{5}\left(22K^V_{\pi\pi\pi\pi}+5K^V_{\pi\pi\pi\pi
2}+68K^V_{\sigma\sigma\pi\pi}+51K^V_{\sigma\sigma\sigma\sigma}\right),\\
\alpha''^{(7)}_{V,q_\textrm{s}}&=&\frac{12}{5}\left(17K^V_{\pi\pi\pi\pi}+2K^V_{\pi\pi\pi\pi
2}+42K^V_{\sigma\sigma\pi\pi}+27K^V_{\sigma\sigma\sigma\sigma}\right),\\
\alpha''^{(7)}_{V,\bar
q}&=&\frac{-12}{5}\left(15K^V_{\pi\pi\pi\pi}-K^V_{\pi\pi\pi\pi
2}+46K^V_{\sigma\sigma\pi\pi}+30K^V_{\sigma\sigma\sigma\sigma}\right),\\
\beta''^{(7)}_{V,q_\textrm{val}}&=&\frac{12}{5}\left(32K^V_{\pi\pi\pi\pi}+K^V_{\pi\pi\pi\pi
2}+88K^V_{\sigma\sigma\pi\pi}+57K^V_{\sigma\sigma\sigma\sigma}\right),\\
\beta''^{(7)}_{V,q_\textrm{s}}&=&\frac{12}{5}\left(19K^V_{\pi\pi\pi\pi}+2K^V_{\pi\pi\pi\pi
2}+62K^V_{\sigma\sigma\pi\pi}+45K^V_{\sigma\sigma\sigma\sigma}\right),\\
\beta''^{(7)}_{V,\bar
q}&=&\frac{36}{5}\left(17K^V_{\pi\pi\pi\pi}+K^V_{\pi\pi\pi\pi
2}+50K^V_{\sigma\sigma\pi\pi}+34K^V_{\sigma\sigma\sigma\sigma}\right).
\end{eqnarray}

One can easily check that the obvious sum rules for $\Theta^+$
\begin{equation}
\int\ud z\,[u(z)-\bar u(z)]=2, \qquad \int\ud z\,[d(z)-\bar d(z)]=2,
\qquad \int\ud z\,[s(z)-\bar s(z)]=-1
\end{equation}
are satisfied separately in each sector. They are translated in our
parametrization as follows
\begin{equation}
\alpha''^{(i)}_{V,q_\textrm{val}}+\alpha''^{(i)}_{V,q_\textrm{s}}-\alpha''^{(i)}_{V,\bar
q}=\uN^{(i)}(B_{\bf\overline{10}}),\qquad
\beta''^{(i)}_{V,q_\textrm{val}}+\beta''^{(i)}_{V,q_\textrm{s}}-\beta''^{(i)}_{V,\bar
q}=0 \qquad \forall i.
\end{equation}

\section{Numerical results and discussion}

Let us start the discussion of our results with the normalizations.
They allow us to estimate which fraction of proton is actually made
of $3Q$, $5Q$ and $7Q$.

In the non-relativistic limit $j(p)=0$ we have computed up to $5Q$
sector the octet composition, see Table \ref{pourcent1}. In this
limit the proton consist of $2/3$ state with three quarks and $1/3$
state with five quarks. The exchange diagram does not change
significantly these ratios and contribute only up to $1\%$. Already
at this stage we can expect reasonably that neglecting the exchange
diagram would not alter noticeably the results.
\begin{table}[h!]\begin{center}\caption{\small{Non-relativistic octet baryons fractions with and without exchange diagram contribution.\newline}}
\begin{tabular}{c|cc} \hline\hline
dir&$3Q\equiv
\frac{\uN^{(3)}(B)}{\uN^{(3)}(B)+\uN^{(5)}(B)}$&$5Q\equiv\frac{\uN^{(5)}(B)}{\uN^{(3)}(B)+\uN^{(5)}(B)}$\rule[-2ex]{0pt}{5.5ex}\\\hline
\rule{0pt}{3ex} $B_{\bf
8}$&$65\%$&$35\%$\rule[-2ex]{0pt}{5ex}\\\hline\hline
dir+exch&$3Q\equiv
\frac{\uN^{(3)}(B)}{\uN^{(3)}(B)+\uN^{(5)}(B)+\uN^{(5)\textrm{exch}}(B)}$&$5Q\equiv\frac{\uN^{(5)}(B)+\uN^{(5)\textrm{exch}}(B)}{\uN^{(3)}(B)+\uN^{(5)}(B)+\uN^{(5)\textrm{exch}}(B)}$\rule[-2ex]{0pt}{5.5ex}\\\hline
\rule{0pt}{3ex}$B_{\bf
8}$&$64\%$&$36\%$\rule[-2ex]{0pt}{5ex}\\\hline\hline
\end{tabular}\label{pourcent1}\end{center}
\end{table}

In Table \ref{pourcent2} we give the composition of octet and
decuplet baryons still up to the $5Q$ sector but with the
relativistic correction $j(p)\neq 0$. Let us first compare the
relativistic result for the octet with the non-relativistic one.
Quark angular momentum clearly reduces the impact of the $5Q$
component. The $5Q$ scalar overlap integrals being smaller in the
relativistic case than in the non-relativistic one the conclusion
drawn is not surprising. This effect is not negligible since the
$5Q$ weight drops from 1/3 to less than 1/4. Let us compare now the
octet and decuplet fractions. They appear to be quite similar but
notice that the $J_z=1/2$ component of the decuplet has a slightly
larger $5Q$ component than the $J_z=3/2$ one.

\begin{table}[h!]\begin{center}\caption{\small{Comparison of octet and decuplet baryons fractions up to the $5Q$ sector.}}
\begin{tabular}{c|cc} \hline\hline
&$3Q\equiv
\frac{\uN^{(3)}(B)}{\uN^{(3)}(B)+\uN^{(5)}(B)}$&$5Q\equiv\frac{\uN^{(5)}(B)}{\uN^{(3)}(B)+\uN^{(5)}(B)}$\rule[-2ex]{0pt}{5.5ex}\\\hline
\rule{0pt}{3ex} $B_{\bf 8}$&$77.5\%$&$22.5\%$\\\rule{0pt}{3ex}
$B_{\mathbf{10},3/2}$&$75\%$&$25\%$\\\rule{0pt}{3ex}
$B_{\mathbf{10},1/2}$&$72.5\%$&$27.5\%$\rule[-2ex]{0pt}{5ex}\\\hline\hline
\end{tabular}\label{pourcent2}\end{center}
\end{table}

A more precise description of baryons would involve the $7Q$ sector.
By analogy with the $5Q$ sector of ordinary baryons, pentaquarks are
expected to have a non-negligible $7Q$ component. In Table
\ref{pourcent3} one observes that the dominant component in
pentaquarks is smaller ($\sim 60\%$) than the dominant one in
ordinary baryons ($\sim 75\%$). This would indicate that when
considering a pentaquark one should care more about higher Fock
contributions than in ordinary baryons. Concerning these ordinary
baryons, it is interesting to notice that the $7Q$ component is not
that negligible since $7.5\%$ of the proton is a system with seven
quarks. It is however not surprising that the adjunction of the $7Q$
sector reduces the weight of the other ones.
\begin{table}[h!]\begin{center}
\caption{\small{Comparison of octet and antidecuplet baryons
fractions up to the $7Q$ sector.}}
\begin{tabular}{c|ccc}
\hline\hline &$3Q\equiv
\frac{\uN^{(3)}(B)}{\uN^{(3)}(B)+\uN^{(5)}(B)+\uN^{(7)}(B)}$&$5Q\equiv\frac{\uN^{(5)}(B)}{\uN^{(3)}(B)+\uN^{(5)}(B)+\uN^{(7)}(B)}$&$7Q\equiv\frac{\uN^{(7)}(B)}{\uN^{(3)}(B)+\uN^{(5)}(B)+\uN^{(7)}(B)}$\rule[-2ex]{0pt}{5.5ex}\\\hline
\rule{0pt}{3ex}
$B_{\mathbf{8}}$&$71.7\%$&$20.8\%$&$7.5\%$\\\rule{0pt}{3ex}
$B_{\mathbf{\overline{10}}}$&$0\%$&$60.6\%$&$39.4\%$\rule[-2ex]{0pt}{5ex}\\
\hline\hline
\end{tabular}\label{pourcent3}\end{center}
\end{table}
We now proceed with our results concerning baryon vector content.

\subsection{Octet content}

The first Table \ref{OctetresultV1} contains the non-relativistic
contributions to proton vector charges
\begin{table}[h!]\begin{center}\caption{\small{Our non-relativistic vector content of the proton.}}
\begin{tabular}{c|cc|cc|cc}
\hline\hline
Vector&\multicolumn{2}{c|}{$u$}&\multicolumn{2}{c|}{$d$}&\multicolumn{2}{c}{$s$}\rule{0pt}{3ex}\\
&$\bar q$&$q_s+q_{val}$&$\bar q$&$q_s+q_{val}$&$\bar
q$&$q_s+q_{val}$\rule{0pt}{3ex}\\\hline \rule{0pt}{3ex}
$3Q$&0&2&0&1&0&0\\\rule{0pt}{3ex} $3Q+5Q$
(dir)&0.123&2.123&0.140&1.140&0.086&0.086\\\rule{0pt}{3ex}
$3Q+5Q$ (dir+exch)&0.125&2.125&0.143&1.143&0.087&0.087\rule[-2ex]{0pt}{5ex}\\
\hline\hline
\end{tabular}\label{OctetresultV1}\end{center}
\end{table}
while relativistic results can be found in Table
\ref{OctetresultV2}. Since the $5Q$ component is larger in the
non-relativistic it is not surprising to find more antiquarks in
this limit. Exchange contributions are of order 1-2\% and are thus
clearly negligible. A coincidence makes that the non-relativistic
proton up to the $5Q$ picture seems equivalent to the relativistic
proton up to the $7Q$ picture concerning the vector properties.
\begin{table}[h!]\begin{center}\caption{\small{Our vector content of the proton compared with NQM.}}
\begin{tabular}{c|ccc|ccc|ccc}
\hline\hline
Vector&\multicolumn{3}{c|}{$u$}&\multicolumn{3}{c|}{$d$}&\multicolumn{3}{c}{$s$}\rule{0pt}{3ex}\\
&$\bar q$&$q_s$&$q_{val}$&$\bar q$&$q_s$&$q_{val}$&$\bar
q$&$q_s$&$q_{val}$\rule{0pt}{3ex}\\\hline \rule{0pt}{3ex}
NQM&0&0&2&0&0&1&0&0&0\\\rule{0pt}{3ex}
$3Q$&0&0&2&0&0&1&0&0&0\\\rule{0pt}{3ex}
$3Q+5Q$&0.078&0.130&1.948&0.091&0.080&1.012&0.055&0.015&0.040\\\rule{0pt}{3ex}
$3Q+5Q+7Q$&0.125&0.202&1.924&0.145&0.128&1.017&0.088&0.028&0.060\rule[-2ex]{0pt}{5ex}\\
\hline\hline
\end{tabular}\label{OctetresultV2}\end{center}
\end{table}

Considering Fock states beyond the $3Q$ sector naturally generates
antiquarks with the three flavors. As discussed previously the sea
is not $SU(3)$ symmetric as sometimes assumed in models. $SU(3)$
symmetry does not force neither $\bar q$ and $q_\textrm{s}$ to be
equal nor that the $u$, $d$ and $s$ sea to have the same magnitude.
The population in the sea is affected by the population in the
valence sector. They cannot be treated independently. Another
interesting comment is that as also discussed previously, hidden
flavor(s) can access to the valence level. Our computations show the
existence of valence strange quarks in the proton. This simple
observation can be used to understand the fact that even if the
effective number of strange quarks $s-\bar s$ is zero the strange
quark and antiquark distributions are not necessarily equal
$s(z)-\bar s(z)\neq 0$ as revealed by experiments \cite{SAexp}.

The corrections due to the $7Q$ component go in the same direction
as the ones due to the $5Q$ component. The former are of course (and
fortunately) small but not that negligible. Results are
\emph{quantitatively} but not \emph{qualitatively} changed. While
the $5Q$ component is essential in order to produce a sea
contribution, exploratory studies do not need absolutely this $7Q$
component. Only a fine quantitative estimation would have to take it
into account. The problem is that it is difficult to estimate the
theoretical errors of the actual approach hitherto. Further work is
thus needed.

Violation of Gottfried sum rule \cite{GSR} allows one to study also
the vector content of the sea. Experiments suggest that $\bar d$ is
dominant over $\bar u$. This can physically be understood by
considering some simple Pauli-blocking effect. Since there are
already two valence $u$ quarks and only one valence $d$ quark in the
proton, the presence of $\bar d d$ pair will be favored compared to
$\bar u u$. The E866 collaboration \cite{E866Coll} gives $\bar
d-\bar u=0.118\pm 0.012$ while we have obtained $\bar d-\bar
u=0.019$. We indeed confirm an excess of $\bar d$ over $\bar u$ but
the magnitude is one order of magnitude too small.

\subsection{Decuplet content}

In Tables \ref{DecupletresultV3/2} and \ref{DecupletresultV1/2} one
can find the $\Delta^{++}$ vector content with respectively
$J_z=3/2,1/2$.
\begin{table}[h!]\begin{center}\caption{\small{Our vector content of the $\Delta^{++}$ with spin projection $J_z=3/2$ compared with NQM.}}
\begin{tabular}{c|ccc|ccc|ccc}
\hline\hline
Vector&\multicolumn{3}{c|}{$u$}&\multicolumn{3}{c|}{$d$}&\multicolumn{3}{c}{$s$}\rule{0pt}{3ex}\\
$J_z=3/2$&$\bar q$&$q_s$&$q_{val}$&$\bar q$&$q_s$&$q_{val}$&$\bar
q$&$q_s$&$q_{val}$\rule{0pt}{3ex}\\\hline \rule{0pt}{3ex}
NQM&0&0&3&0&0&0&0&0&0\\\rule{0pt}{3ex}
$3Q$&0&0&3&0&0&0&0&0&0\\\rule{0pt}{3ex}
$3Q+5Q$&0.072&0.193&2.879&0.089&0.029&0.060&0.089&0.029&0.060\rule[-2ex]{0pt}{5ex}\\
\hline\hline
\end{tabular}\label{DecupletresultV3/2}\end{center}
\end{table}
\begin{table}[h!]\begin{center}\caption{\small{Our vector content of the $\Delta^{++}$ with spin projection $J_z=1/2$ compared with NQM.}}
\begin{tabular}{c|ccc|ccc|ccc}
\hline\hline
Vector&\multicolumn{3}{c|}{$u$}&\multicolumn{3}{c|}{$d$}&\multicolumn{3}{c}{$s$}\rule{0pt}{3ex}\\
$J_z=1/2$&$\bar q$&$q_s$&$q_{val}$&$\bar q$&$q_s$&$q_{val}$&$\bar
q$&$q_s$&$q_{val}$\rule{0pt}{3ex}\\\hline \rule{0pt}{3ex}
NQM&0&0&3&0&0&0&0&0&0\\\rule{0pt}{3ex}
$3Q$&0&0&3&0&0&0&0&0&0\\\rule{0pt}{3ex}
$3Q+5Q$&0.059&0.225&2.834&0.108&0.025&0.083&0.108&0.025&0.083\rule[-2ex]{0pt}{5ex}\\
\hline\hline
\end{tabular}\label{DecupletresultV1/2}\end{center}
\end{table}

To the best of our knowledge there is no experimental results
concerning the vector content of decuplet baryons. Our results can
then be considered as just theoretical predictions, at least
qualitatively. As discussed in the previous section, it is clear
that eq. (\ref{sphericityV}) is not satisfied indicating a deviation
from spherical shape .

Note however that the $3Q$ sector reproduces all the octet and
decuplet vector content predicted by NQM. Higher Fock sectors change
these results by breaking explicitly $SU(6)$ symmetry.

\subsection{Antidecuplet content}

The study of the $7Q$ sector has mainly been motivated by the
pentaquark. In previous results we have seen that the $5Q$ component
of usual baryons has non-negligible and interesting effects on the
vector quantities. In the same spirit, since there is no $3Q$
component in pentaquarks, it would be interesting to see what
happens when considering the $7Q$ component. In Table
\ref{AntidecupletresultV} one can find the $\Theta^+$ vector
content.
\begin{table}[h!]\begin{center}\caption{\small{Our vector content of the $\Theta^+$.}}
\begin{tabular}{c|ccc|ccc|ccc}
\hline\hline
Vector&\multicolumn{3}{c|}{$u$}&\multicolumn{3}{c|}{$d$}&\multicolumn{3}{c}{$s$}\rule{0pt}{3ex}\\
&$\bar q$&$q_s$&$q_{val}$&$\bar q$&$q_s$&$q_{val}$&$\bar
q$&$q_s$&$q_{val}$\rule{0pt}{3ex}\\\hline \rule{0pt}{3ex}
$5Q$&0&1/2&3/2&0&1/2&3/2&1&0&0\\\rule{0pt}{3ex}
$5Q+7Q$&0.153&0.680&1.474&0.153&0.680&1.474&1.088&0.035&0.053\rule[-2ex]{0pt}{5ex}\\
\hline\hline
\end{tabular}\label{AntidecupletresultV}\end{center}
\end{table}

We did not provide a table with direct and exchange $5Q$
contributions. The minimal pentaquark content lies in the $5Q$
sector. This means that the structure is simple and is the same for
both type of diagrams, emph{i.e.} the exchange diagram does not
change the minimal vector content of pentaquarks. The sole
restriction is that exchange diagram forbids a clear distinction
between valence quarks and quarks from the sea.

The $7Q$ component introduces $\bar u$, $\bar d$ and $s$ in
$\Theta^+$. In accordance with the normalizations this $7Q$
contribution has a stronger impact on pentaquarks than the $5Q$
contribution on ordinary baryons. A precise study of pentaquarks
needs thus to take $7Q$ contributions into account.

\chapter{Axial charges}
\section{Introduction}

The axial charges of a baryon are defined as forward matrix elements
of the axial vector current
\begin{equation}
\langle
B(p)|\bar\psi\gamma_\mu\gamma_5\lambda^a\psi|B(p)\rangle=g^{(a)}_A\bar
u(p)\gamma_\mu\gamma_5 u(p)
\end{equation}
where $a=0,3,8$ and $\lambda^3,\lambda^8$ are Gell-Mann matrices,
$\lambda^0$ is just in this context the $3\times3$ unit matrix. In
principle we could add in the definition of the axial-vector
operator $\bar\psi\gamma_\mu\gamma_5\lambda^a\psi$ a factor $g_{Aq}$
which is the quark axial-vector current coupling constant. As
commonly assumed we use $g_{Aq}=1$, \emph{i.e.} the same as for the
structureless QCD quarks. These axial charges are related to the
first moment of the longitudinally polarized quark distributions
\begin{equation}\label{Nucleon axial charges}
g^{(3)}_A=\Delta u-\Delta d,\qquad
g^{(8)}_A=\frac{1}{\sqrt{3}}(\Delta u+\Delta d-2\Delta s),\qquad
g^{(0)}_A=\Delta u+\Delta d+\Delta s
\end{equation}
where $\Delta q\equiv\int_0^1\ud z\left[q_+(z)-q_-(z)+\bar
q_+(z)-\bar q_-(z)\right]$ with $q=u,d,s$. Isovector $g_A^{(3)}$ and
octet $g_A^{(8)}$ axial charges are independent of the
renormalization point. On the contrary, the flavor singlet axial
charge $g_A^{(0)}$ depends on the renormalization scale at which it
is measured\footnote{The gluon spin contributions $\Delta g$ are
admixed to the quark spin contributions in leading order
perturbation theory because of the axial gluon anomaly of QCD.
Therefore the DIS experiments actually measure $\Delta q(Q^2)=\Delta
q-\alpha_S(Q^2)\,\Delta g(Q^2)$, where $\alpha_S$ is the running QCD
coupling constant. This $Q^2$ dependence is canceled in the
combinations $g_A^{(3)}$ and $g_A^{(8)}$ but not in $g_A^{(0)}$. The
$Q^2$ dependence is very soft in the perturbative regime but its
evolution down to the confinement scale is not known}. Because of
isospin symmetry we expect $g_A^{(3)}$ in proton to be equal to the
axial charge of the transition
$p^+_\mathbf{8}\to\pi^+n^0_\mathbf{8}$. We split the axial charges
into valence quark, sea quark and antiquark contributions
\begin{equation}
\Delta q=\Delta q_\textrm{val}+\Delta q_\textrm{sea}, \qquad \Delta
q_\textrm{sea}=\Delta q_\textrm{s}+\Delta\bar q
\end{equation}
where index s refers to the quarks in the sea pairs.

The axial charges can be understood as follows. They count the total
number of quarks with spin aligned $q_+$ \emph{minus} the total
number of quarks with spin antialigned $q_-$ with the baryon spin,
irrespective of their quark $q_\textrm{val,s}$ or antiquark $\bar q$
nature. The axial charges $\bar q\gamma_\mu\gamma_5 q$ then give the
contribution of quarks spin with flavor $q=u,d,s$ to the total
baryon spin.

The proton polarized structure function $g_1^p(x)$ has been measured
by EMC in 1987 \cite{EMC}. The value obtained by the collaboration
implies that only a small part of the nucleon spin is carried by
quarks demonstrating that the Ellis-Jaffe Sum Rule (EJSR)
\cite{Others} based on $\Delta s=\Delta\bar s=0$ did not hold true.
This result was not anticipated in conventional quark models and is
often referred to as the \emph{proton spin crisis} in the
literature, see \emph{e.g.} the review \cite{Spincrisis}. Subsequent
measurements at CERN and SLAC supported the initial EMC measurements
and a global analysis \cite{Globalanalysis} of these data suggested
$\Delta s\approx -0.15$. It carries however with it an unknown
theoretical uncertainty because DIS must be extrapolated to $x=0$
and an assumption of flavor $SU(3)$ symmetry must be invoked.

In DIS one is probing the baryon in IMF where the relativistic
many-body problem is suitably described. There is consequently a
significant change in the vector sum of quark spins, arising from
relativistic effects due to internal quark motions. So $\Delta q$
measured in DIS has to be interpreted as the net spin polarization
of quarks in the IMF which is different from the net spin vector sum
of quark spins in the rest frame. The reason for this reduction of
spin contribution can be ascribed to a negative spin contribution
from the lower component of the Dirac spinor when the quark
transversal motions are considered. A quantitative estimation of
this effect can be obtained using the light-cone CQM. The results is
that the correction is significative but not sufficient. The missing
spin has thus to be carried by non-valence degrees of freedom,
\emph{i.e.} quark-antiquark pairs and gluons.

It is very important to study axial charges since a lot of physics
is involved. Many ingredients have to be incorporated in a realistic
model: relativistic quark description with orbital motion,
non-valence degrees of freedom, strangeness, \ldots We draw the
attention to the fact that the values of $\Delta u$ and $\Delta d$
may have astrophysical knock-on effects \cite{Astro}. A precise
determination of their value is thus highly desired. Let us also
note that axial charges are affected by the first-order flavor
$SU(3)$ symmetry breaking while vector charges are safe as stated by
the Ademollo-Gatto theorem \cite{Ademollo}. Longitudinally polarized
SIDIS \cite{HERMES} are subject to a growing interest as they
provide an additional information on the spin structure of the
nucleon compared to inclusive DIS measurements. They allow one to
separate valence and sea contributions to the nucleon spin. Recent
data suggest an asymmetry between $\Delta\bar u(x)$ and $\Delta\bar
d(x)$ \cite{HERMES,SMC,COMPASS}. Flavor structure and spin structure
of the nucleon sea are closely related \cite{Peng}. All these points
explain the interest in these quantities in both theoretical and
experimental sides.

\section{Axial charges on the light cone}

Axial charges are obtained from the \emph{plus} component of the
axial-vector current operator $\bar\psi\gamma^+\gamma_5\psi$
\begin{equation}
\Delta q=\frac{1}{2P^+}\langle
P,\frac{1}{2}|\bar\psi_{LC}\gamma^+\gamma_5\psi_{LC}|P,\frac{1}{2}\rangle.
\end{equation}
Using the Melosh rotation one can see that $\Delta q_{LC}$ and
$\Delta q_{NR}$ are related as follows \cite{facteuraxial}
\begin{equation}
\Delta q_{LC}=\langle M_A\rangle\Delta q_{NR}
\end{equation}
where
\begin{equation}\label{DefMq}
M_A=\frac{(m_q+z_3\uM)^2-\up_{3\perp}^2}{(m_q+z_3\uM)^2+\up_{3\perp}^2}
\end{equation}
and $\langle M_A\rangle$ is its expectation value
\begin{equation}
\langle M\rangle=\int\ud^3p\,M|\Psi(p)|^2
\end{equation}
with $\Psi(p)$ a simple normalized momentum wave function. The
calculation with two different wave functions (harmonic oscillator
and power-law fall off) gave $\langle M_A\rangle=0.75$ \cite{Mq}.
Relativity implies that quarks may have non-zero orbital angular
momentum. The total baryon spin is thus not only due to quark spins
but also to their orbital angular momentum.

In the IMF language one has to use the ``good'' components $\mu=0,3$
of the axial-vector current operator. This operator does not flip
the spin but treats differently quarks with spin up and quarks with
spin down. We have then
$M^{\tau}_{\sigma}=(\sigma_3)^{\tau}_{\sigma}$.

\section{Scalar overlap integrals and quark distributions}

From the expression (\ref{Discrete level IMF}) and if we concentrate
on the spin part the contraction of two valence wave functions $F$
with the axial vector operator gives
\begin{equation}\label{AxialOQL}
F^\dag(\sigma_3)F\propto
h^2(p)+2h(p)\frac{p_z}{|\up|}j(p)+\frac{2p_z^2-p^2}{p^2}j^2(p).
\end{equation}
Like the vector operator the axial-vector operator does not flip
quark spin. However it treats differently quarks with spin up and
quarks with spin down. A quark with total angular momentum
$J_z=+1/2$ may have orbital angular momentum $L_z=+1$ and has thus
spin $S_z=-1/2$. Only the third term in (\ref{AxialOQL}) has
components with $L_z\neq 0$ expressed by a factor $\up_\perp$. Since
the spin for those components is opposed to the total angular
momentum of the quark the sign in front of $p_z^2$ (no orbital
angular momentum in the $z$ direction) is opposed to the one in
front of $\up_\perp^2$ (non-zero orbital angular momentum in the $z$
direction). The structure of the third term in the axial sector is
thus $p_z^2-\up_\perp^2=2p_z^2-p^2$ while it was
$p_z^2+\up_\perp^2=p^2$ in the vector sector.

The axial valence quark distribution is obtained by the
multiplication of two factors with structure (\ref{VectOQL}) where
the momentum is respectively $p_1$ and $p_2$ and a third factor with
structure (\ref{AxialOQL}) and momentum $p_3$. The expansion gives
the following function $D$
\begin{eqnarray}
&D^A(p_1,p_2,p_3)=h^2(p_1)h^2(p_2)h^2(p_3)+6h^2(p_1)h^2(p_2)\left[h(p_3)\frac{p_{3z}}{|\up_3|}j(p_3)\right]+h^2(p_1)h^2(p_2)\frac{2p_{3z}^2+p_3^2}{p_3^2}j^2(p_3)&\nonumber\\
&+12h^2(p_1)\left[h(p_2)\frac{p_{2z}}{|\up_2|}j(p_2)\right]\left[h(p_3)\frac{p_{3z}}{|\up_3|}j(p_3)\right]+4h^2(p_1)\left[h(p_2)\frac{p_{2z}}{|\up_2|}j(p_2)\right]\frac{2p_{3z}^2+p_3^2}{p_3^2}j^2(p_3)&\nonumber\\
&+8\left[h(p_1)\frac{p_{1z}}{|\up_1|}j(p_1)\right]\left[h(p_2)\frac{p_{2z}}{|\up_2|}j(p_2)\right]\left[h(p_3)\frac{p_{3z}}{|\up_3|}j(p_3)\right]+h^2(p_1)j^2(p_2)\frac{4p_{3z}^2-p_3^2}{p_3^2}j^2(p_3)&\\
&+4\left[h(p_1)\frac{p_{1z}}{|\up_1|}j(p_1)\right]\left[h(p_2)\frac{p_{2z}}{|\up_2|}j(p_2)\right]\frac{2p_{3z}^2+p_3^2}{p_3^2}j^2(p_3)+2\left[h(p_1)\frac{p_{1z}}{|\up_1|}j(p_1)\right]j^2(p_2)\frac{4p_{3z}^2-p_3^2}{p_3^2}j^2(p_3)&\nonumber\\
&+j^2(p_1)j^2(p_2)\frac{2p_{3z}^2-p_3^2}{p_3^2}j^2(p_3).&\nonumber\label{Psi}
\end{eqnarray}
that is needed in the expression of the valence quark distribution
(\ref{Probability3q}). In the non-relativistic limit $j=0$ this
function $D$ is reduced to
\begin{equation}
D^A_{NR}(p_1,p_2,p_3)=D^V_{NR}(p_1,p_2,p_3)=h^2(p_1)h^2(p_2)h^2(p_3).
\end{equation}

\begin{figure}[h]\begin{center}\begin{minipage}[c]{6cm}\begin{center}\includegraphics[width=5.3cm]{Phiplot3.eps}
\end{center}\end{minipage}\hspace{1cm}\begin{minipage}[c]{6cm}\begin{center}\includegraphics[width=5.3cm]{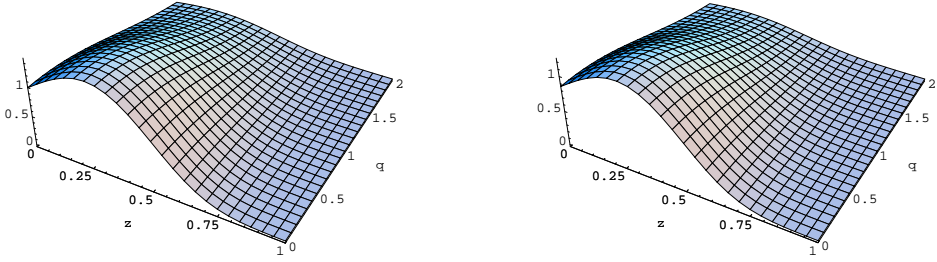}
\end{center}\end{minipage}\caption{\small{Probability distribution $\Phi^I(z,\uq_\perp)$ that three valence quarks leave the fraction $z$ of the baryon momentum and
transverse momentum $\uq_\perp$ to the quark-antiquark pair(s) in
the vector $I=V$ (left) and axial $I=A$ (right) cases plotted in
units of $M$ and normalized to $\Phi^V(0,0)=1$.
}}\label{Phiplot2}\end{center}
\end{figure}

The valence probability distribution $\Phi^I(z,\uq_\perp)$ is then
obtained by integration over the valence quark momenta, see eq.
(\ref{Probability3q}) and is depicted in Fig. \ref{Phiplot2} in
vector $I=V$ and axial $I=A$ cases.

While in the non-relativistic limit valence probability
distributions are the same, relativistic corrections (quark angular
momentum) are different in vector and axial cases. This is of course
due to the difference in structure between (\ref{AxialOQL}) and
(\ref{VectOQL}).

In the following we give the integrals appearing in each Fock sector
and the numerical values obtained for them. In the evaluation of the
scalar overlap integrals we have used the constituent quark mass
$M=345$ MeV, the Pauli-Villars mass $M_\textrm{PV}=556.8$ MeV for
the regularization of (\ref{Direct1})-(\ref{Direct4}),
(\ref{Exchange begin})-(\ref{Exchange end}) and of
(\ref{int1})-(\ref{int12}) and the baryon mass $\uM=1207$ MeV as it
follows for the ``classical'' mass in the mean field approximation
\cite{Approximation}.

\subsection{$3Q$ scalar integral}

In the $3Q$ sector there is no quark-antiquark pair and thus only
one integral is involved. It corresponds to the valence quark
distribution without momentum left to the sea $\Phi^A(0,0)$. We
remind that the normalization chosen for $h(p)$ and $j(p)$ is such
that $\Phi^V(0,0)=1$. From Figure \ref{Phiplot2} one can see that
$\Phi^A(0,0)<1$. The precise value is
\begin{equation}
\Phi^A(0,0)=0.86115.
\end{equation}
This means that in a simple $3Q$ picture all NQM axial charges have
to be multiplied by this factor. Not surprisingly this is the same
prescription as the one encountered in a standard light-cone
approach based on Melosh rotation. Valence quark motion is a
relativistic effect and is responsible for a noticeable reduction of
NQM predictions. The Melosh factor 3/4 is of course smaller than the
one we have obtained because of the function $h(p)$ absent in the
Melosh approach but necessary in a fully relativistic treatment.

The NQM is recovered in the $3Q$ non-relativistic limit only where
the three valence quarks have no angular orbital momentum and thus
$\Phi^A_{NR}(0,0)=\Phi^V_{NR}(0,0)=1$.

\subsection{$5Q$ scalar integrals}

In the $5Q$ sector there is one quark-antiquark pair. Contractions
given by the direct diagram give four different integrals
$J=\pi\pi,33,\sigma\sigma,3\sigma$. If the axial-vector operator
acts on valence quarks, the quark-antiquark pair is not affected and
thus the integrals present the vector structure for the sea. If the
axial-vector operator acts on the sea, the valence quarks are not
affected and the vector valence probability distribution has to be
used. The integrals present a new structure for the sea which
describes the transition scalar$\leftrightarrow$pseudoscalar imposed
by the ``pseudo'' feature of the axial-vector operator.

In the non-relativistic case, since $\Phi^A_{NR}=\Phi^V_{NR}$ only
one new integrals has to be computed
\begin{equation}
K^A_{\pi\pi,NR}=K^V_{\pi\pi,NR},\quad K^A_{33,NR}=K^V_{33,NR},\quad
K^A_{\sigma\sigma,NR}=K^V_{\sigma\sigma,NR},\quad
K^V_{3\sigma,NR}=0.03338
\end{equation}
while in the relativistic case there are four new integrals
\begin{equation}
K^A_{\pi\pi}=0.03003,\qquad K^A_{33}= 0.01628,\qquad
K^A_{\sigma\sigma}=0.01121,\qquad K^V_{3\sigma}=0.01626.
\end{equation}
Let us have a look to the ratios $K^A_J/K^V_J$ with
$J=\pi\pi,33,\sigma\sigma$
\begin{equation}
\frac{K^A_{\pi\pi}}{K^V_{\pi\pi}}=0.82228,\qquad\frac{K^A_{33}}{K^V_{33}}=0.82458,\qquad\frac{K^A_{\sigma\sigma}}{K^V_{\sigma\sigma}}=0.80039.
\end{equation}
The reduction is of the same order as in the $3Q$ sector. It is
however different from one structure to another due to the details
of the valence probability distributions.

Contractions given by the exchange diagram lead to thirteen
different integrals $J=1$-$13$. Since these integrals are obtained
in the non-relativistic limit only six are new $J=8$-$13$
\begin{eqnarray}
&K_8=0.00431,\qquad K_9=0.00309,\qquad K_{10}=0.00693,&\nonumber\\
&K_{11}=0.00172,\qquad K_{12}=0.00570,\qquad K_{13}=0.00230.&
\end{eqnarray}
All exchange integrals are one order of magnitude smaller than the
direct ones. This is however not sufficient to conclude that they
can be neglected. Combinatoric factors play an important role as we
have seen in the vector case.

\subsection{$7Q$ scalar integrals}

In the $7Q$ sector there are two quark-antiquark pairs. Contractions
given by the direct diagram give twelve different integrals
$J=\pi\pi\pi\pi,\pi\pi\pi\pi 2,\pi\pi 33,3333,\pi 3\pi
3,\sigma\sigma\pi\pi,\sigma\sigma 33,\sigma\sigma\sigma\sigma,\pi\pi
3\sigma,333\sigma,\pi 3\pi\sigma,\sigma\sigma 3\sigma$. Like in the
$5Q$ sector all vector structures of the sea are associated with the
axial valence probability distribution. One then gets eight
integrals. The last four structures are the new axial structures of
the sea
\begin{eqnarray}
&K^A_{\pi\pi\pi\pi}=0.00066,\qquad K^A_{\pi\pi\pi\pi
2}=0.00021,\qquad K^A_{\pi\pi
33}=0.00031,\qquad K^A_{3333}=0.00015,&\\
&K^A_{\pi 3\pi 3}=0.00013,\qquad
K^A_{\sigma\sigma\pi\pi}=0.00021,\qquad K^A_{\sigma\sigma
33}=0.00010,\qquad
K^A_{\sigma\sigma\sigma\sigma}=0.00007,&\\
&K^V_{\pi\pi 3\sigma}=0.00031,\qquad K^V_{333\sigma}=0.00014,\qquad
K^V_{\pi 3\pi\sigma}=0.00011,\qquad K^V_{\sigma\sigma
3\sigma}=0.00010.&
\end{eqnarray}

By analogy with the $5Q$ sector exchange diagram contributions are
neglected and thus have not been computed.

\section{Combinatoric Results}

Axial matrix elements are linear combinations of the axial scalar
overlap integrals. These specific combinations are obtained by
contracting the baryon rotational wave functions with the
axial-vector operator. In the following we give for each multiplet
the combinations obtained.

\subsection{Octet baryons}

In the $3Q$ sector there is no quark-antiquark pair and thus only
valence quarks contribute to the charges
\begin{equation}
\alpha_{A,q_\textrm{val}}^{(3)}=12\,\Phi^A(0,0),\qquad
\beta_{A,q_\textrm{val}}^{(3)}=-3\,\Phi^A(0,0),\qquad
\gamma_{A,q_\textrm{val}}^{(3)}=0.
\end{equation}
In the $5Q$ sector one has for the direct diagram
\begin{eqnarray}
&\alpha_{A,q_\textrm{val}}^{(5)}=\frac{6}{5}\left(29K^A_{\pi\pi}+2K^A_{33}+91K^A_{\sigma\sigma}\right),\quad
\alpha_{A,q_\textrm{s}}^{(5)}=\frac{-168}{5}\,K^V_{3\sigma},\quad
\alpha_{A,\bar q}^{(5)}=\frac{-132}{5}\,K^V_{3\sigma},&\\
&\beta_{A,q_\textrm{val}}^{(5)}=\frac{-24}{25}\left(16K^A_{\pi\pi}-11K^A_{33}+26K^A_{\sigma\sigma}\right),\quad
\beta_{A,q_\textrm{s}}^{(5)}=\frac{408}{25}\,K^V_{3\sigma},\quad
\beta_{A,\bar q}^{(5)}=\frac{228}{25}\,K^V_{3\sigma},&\\
&\gamma_{A,q_\textrm{val}}^{(5)}=\frac{-12}{25}\left(11K^A_{\pi\pi}-16K^A_{33}+K^A_{\sigma\sigma}\right),\quad
\gamma_{A,q_\textrm{s}}^{(5)}=\frac{84}{25}\,K^V_{3\sigma},\quad
\gamma_{A,\bar q}^{(5)}=\frac{84}{25}\,K^V_{3\sigma}.&
\end{eqnarray}
The $5Q$ exchange diagram gives
\begin{eqnarray}
\alpha_{A,q_\textrm{val+s}}^{(5)\textrm{exch}}&=&\frac{-2}{5}\left(89K_1+K_2+29K_3+30K_4-2K_5-152K_6-150K_7\right),\\
\alpha_{A,\bar q}^{(5)\textrm{exch}}&=&\frac{-2}{5}\left(3K_3+11K_8-5K_9+8K_{10}+16K_{11}-12K_{12}-10K_{13}\right),\\
\beta_{A,q_\textrm{val+s}}^{(5)\textrm{exch}}&=&\frac{4}{25}\left(56K_1-2K_2+32K_3+21K_4+4K_5-74K_6-114K_7\right),\\
\beta_{A,\bar q}^{(5)\textrm{exch}}&=&\frac{-2}{25}\left(3K_3-13K_8+25K_9-22K_{10}-44K_{11}-12K_{12}+50K_{13}\right),\\
\gamma_{A,q_\textrm{val+s}}^{(5)\textrm{exch}}&=&\frac{-4}{25}\left(7K_1-4K_2-11K_3-3K_4+8K_5-13K_6+27K_7\right),\\
\gamma_{A,\bar
q}^{(5)\textrm{exch}}&=&\frac{-2}{25}\left(9K_3+11K_8+25K_9-16K_{10}-32K_{11}-36K_{12}+50K_{13}\right).
\end{eqnarray}
In the $7Q$ sector the combinations are
\begin{eqnarray}
\alpha_{A,q_\textrm{val}}^{(7)}&=&\frac{48}{5}\left(33K^A_{\pi\pi\pi\pi}+30K^A_{\pi\pi\pi\pi
2}-2K^A_{\pi\pi 33}+4K^A_{\pi 3\pi 3}+134K^A_{\sigma\sigma\pi\pi}\right.\nonumber\\
&+&\left.10K^A_{\sigma\sigma 33}+211K^A_{\sigma\sigma\sigma\sigma}\right),\\
\alpha_{A,q_\textrm{s}}^{(7)}&=&\frac{-96}{5}\left(32K^V_{\pi\pi
3\sigma}-K^V_{\pi
3\pi\sigma}+65K^V_{\sigma\sigma 3\sigma}\right),\\
\alpha_{A,\bar q}^{(7)}&=&\frac{-96}{5}\left(25K^V_{\pi\pi
3\sigma}+K^V_{\pi
3\pi\sigma}+52K^V_{\sigma\sigma 3\sigma}\right),\\
\beta_{A,q_\textrm{val}}^{(7)}&=&\frac{-48}{25}\left(51K^A_{\pi\pi\pi\pi}+45K^A_{\pi\pi\pi\pi
2}+38K^A_{\pi\pi 33}-82K^A_{\pi 3\pi 3}+292K^A_{\sigma\sigma\pi\pi}\right.\nonumber\\
&-&\left.214K^A_{\sigma\sigma 33}+224K^A_{\sigma\sigma\sigma\sigma}\right),\\
\beta_{A,q_\textrm{s}}^{(7)}&=&\frac{192}{25}\left(35K^V_{\pi\pi
3\sigma}+2K^V_{\pi
3\pi\sigma}+77K^V_{\sigma\sigma 3\sigma}\right),\\
\beta_{A,\bar q}^{(7)}&=&\frac{96}{25}\left(47K^V_{\pi\pi
3\sigma}-K^V_{\pi
3\pi\sigma}+92K^V_{\sigma\sigma 3\sigma}\right),\\
\gamma_{A,q_\textrm{val}}^{(7)}&=&\frac{-48}{25}\left(13K^A_{\pi\pi\pi\pi}+10K^A_{\pi\pi\pi\pi
2}+24K^A_{\pi\pi 33}-56K^A_{\pi 3\pi 3}+106K^A_{\sigma\sigma\pi\pi}\right.\nonumber\\
&-&\left.152K^A_{\sigma\sigma 33}+7K^A_{\sigma\sigma\sigma\sigma}\right),\\
\gamma_{A,q_\textrm{s}}^{(7)}&=&\frac{96}{25}\left(25K^V_{\pi\pi
3\sigma}-8K^V_{\pi
3\pi\sigma}+37K^V_{\sigma\sigma 3\sigma}\right),\\
\gamma_{A,\bar q}^{(7)}&=&\frac{288}{25}\left(7K^V_{\pi\pi
3\sigma}-K^V_{\pi 3\pi\sigma}+12K^V_{\sigma\sigma 3\sigma}\right).
\end{eqnarray}

\subsection{Decuplet baryons}

In the $3Q$ sector there is no quark-antiquark pair and thus only
valence quarks contribute to the charges
\begin{equation}
\alpha'^{(3)}_{A,q_\textrm{val},3/2}=3\alpha'^{(3)}_{A,q_\textrm{val},1/2}=\frac{18}{5}\,\Phi^A(0,0),\qquad
\beta'^{(3)}_{A,q_\textrm{val},3/2}=3\beta'^{(3)}_{A,q_\textrm{val},1/2}=0.
\end{equation}
In the $5Q$ sector one has
\begin{eqnarray}
&\alpha'^{(5)}_{A,q_\textrm{val},3/2}=\frac{9}{20}\left(43K^A_{\pi\pi}-16K^A_{33}+67K^A_{\sigma\sigma}\right),\quad
\alpha'^{(5)}_{A,q_\textrm{s},3/2}=\frac{-99}{100}\,K^V_{3\sigma},\quad
\alpha'^{(5)}_{A,\bar q,3/2}=\frac{-36}{5}\,K^V_{3\sigma},&\\
&\alpha'^{(5)}_{A,q_\textrm{val},1/2}=\frac{3}{20}\left(23K^A_{\pi\pi}+44K^A_{33}+67K^A_{\sigma\sigma}\right),\quad
\alpha'^{(5)}_{A,q_\textrm{s},1/2}=\frac{-33}{100}\,K^V_{3\sigma},\quad
\alpha'^{(5)}_{A,\bar q,1/2}=\frac{-12}{5}\,K^V_{3\sigma},&\\
&\beta'^{(5)}_{A,q_\textrm{val},3/2}=\frac{-9}{20}\left(23K^A_{\pi\pi}-32K^A_{33}-K^A_{\sigma\sigma}\right),\quad
\beta'^{(5)}_{A,q_\textrm{s},3/2}=\frac{63}{10}\,K^V_{3\sigma},\quad
\beta'^{(5)}_{A,\bar q,3/2}=\frac{18}{5}\,K^V_{3\sigma},&\\
&\beta'^{(5)}_{A,q_\textrm{val},1/2}=\frac{-3}{20}\left(19K^A_{\pi\pi}-20K^A_{33}-K^A_{\sigma\sigma}\right),\quad
\beta'^{(5)}_{A,q_\textrm{s},1/2}=\frac{21}{10}\,K^V_{3\sigma},\quad
\beta'^{(5)}_{A,\bar q,1/2}=\frac{6}{5}\,K^V_{3\sigma}.&
\end{eqnarray}
The $7Q$ sector of the decuplet has not been computed due to its far
bigger complexity.

If the decuplet was made of three quarks only then one would have
the following relations between spin-3/2 and spin-1/2 axial
contributions
\begin{equation}\label{sphericityA}
A_{3/2}=3A_{1/2}.
\end{equation}
This picture presents the $\Delta$ as a spherical particle. Things
change in the $5Q$ sector. One notices directly that the relation is
broken by a unique structure $(K^A_{\pi\pi}-3K^A_{33})$. This
structure has exactly the same sea part as the structure found in
the vector case $(K^V_{\pi\pi}-3K^V_{33})$. Notice a difference
concerning the quark-antiquark pair contribution in the vector and
axial cases. While in the former the quadrupolar structure is
present, it is absent in the latter. Therefore the sea contribution
satisfies (\ref{sphericityA}) but not (\ref{sphericityV}). The
axial-vector operator acting on the sea allows only transitions
between scalar and pseudoscalar quark-antiquark pairs.

\subsection{Antidecuplet baryons}

In the $5Q$ sector one has for the direct diagram
\begin{eqnarray}
&\alpha''^{(5)}_{A,q_\textrm{val}}=\frac{-6}{5}\left(K^A_{\pi\pi}-2K^A_{33}-K^A_{\sigma\sigma}\right),\quad
\alpha''^{(5)}_{A,q_\textrm{s}}=\frac{12}{5}\,K^V_{3\sigma},\quad
\alpha''^{(5)}_{A,\bar q}=\frac{24}{5}\,K^V_{3\sigma},&\\
&\beta''^{(5)}_{A,q_\textrm{val}}=\frac{-6}{5}\left(K^A_{\pi\pi}-2K^A_{33}-K^A_{\sigma\sigma}\right),\quad
\beta''^{(5)}_{A,q_\textrm{s}}=\frac{12}{5}\,K^V_{3\sigma},\quad
\beta''^{(5)}_{A,\bar q}=\frac{-48}{5}\,K^V_{3\sigma}.&
\end{eqnarray}
The $5Q$ exchange diagram gives
\begin{eqnarray}
&\alpha''^{(5)\textrm{exch}}
_{A,q_\textrm{val+s}}=\frac{4}{5}\left(K_1-K_2+2K_5\right),\quad
\alpha''^{(5)\textrm{exch}}_{A,\bar q}=\frac{-4}{5}\left(K_8-K_9+K_{10}+2K_{11}-2K_{13}\right),&\\
&\beta''^{(5)\textrm{exch}}_{A,q_\textrm{val+s}}=\frac{4}{5}\left(K_1-K_2+2K_5\right),\quad
\beta''^{(5)\textrm{exch}}_{A,\bar
q}=\frac{8}{5}\left(K_8-K_9+K_{10}+2K_{11}-2K_{13}\right).&
\end{eqnarray}
In the $7Q$ sector the combinations are
\begin{eqnarray}
\alpha''^{(7)}_{A,q_\textrm{val}}&=&\frac{12}{5}\left(3K^A_{\pi\pi\pi\pi
2}-2K^A_{\pi\pi 33}+10K^A_{\pi 3\pi
3}-10K^A_{\sigma\sigma\pi\pi}+34K^A_{\sigma\sigma
33}+19K^V_{\sigma\sigma\sigma\sigma}\right),\\
\alpha''^{(7)}_{A,q_\textrm{s}}&=&\frac{24}{5}\left(4K^V_{\pi\pi
3\sigma}+K^V_{\pi 3\pi\sigma}+13K^V_{\sigma\sigma
3\sigma}\right),\\
\alpha''^{(7)}_{A,\bar q}&=&\frac{12}{5}\left(41K^V_{\pi\pi
3\sigma}-K^V_{\pi 3\pi\sigma}+80K^V_{\sigma\sigma
3\sigma}\right),\\
\beta''^{(7)}_{A,q_\textrm{val}}&=&\frac{12}{5}\left(2K^A_{\pi\pi\pi\pi}-K^A_{\pi\pi\pi\pi
2}-18K^A_{\pi\pi 33}+26K^A_{\pi 3\pi
3}-22K^A_{\sigma\sigma\pi\pi}+50K^A_{\sigma\sigma
33}+17K^V_{\sigma\sigma\sigma\sigma}\right),\\
\beta''^{(7)}_{A,q_\textrm{s}}&=&\frac{24}{5}\left(10K^V_{\pi\pi
3\sigma}+K^V_{\pi 3\pi\sigma}+19K^V_{\sigma\sigma
3\sigma}\right),\\
\beta''^{(7)}_{A,\bar q}&=&\frac{-36}{5}\left(23K^V_{\pi\pi
3\sigma}+K^V_{\pi 3\pi\sigma}+48K^V_{\sigma\sigma 3\sigma}\right).
\end{eqnarray}

\section{Numerical results and discussion}

\subsection{Octet content}

We give first the results in the non-relativistic limit. They are
collected in Table \ref{Octetresult1A}.
\begin{table}[h!]\begin{center}\caption{\small{Non-relativistic axial content of the proton compared.}}
\begin{tabular}{c|cc|cc|cc}
\hline\hline
Axial&\multicolumn{2}{c|}{$\Delta u$}&\multicolumn{2}{c|}{$\Delta d$}&\multicolumn{2}{c}{$\Delta s$}\rule{0pt}{3ex}\\
&$\bar q$&$q_s+q_{val}$&$\bar q$&$q_s+q_{val}$&$\bar
q$&$q_s+q_{val}$\rule{0pt}{3ex}\\\hline \rule{0pt}{3ex}
$3Q$&0&4/3&0&-1/3&0&0\\\rule{0pt}{3ex} $3Q+5Q$
(dir)&-0.056&1.179&0.030&-0.266&0.008&0.004\\\rule{0pt}{3ex}
$3Q+5Q$ (dir+exch)&-0.056&1.180&0.032&-0.267&0.009&0.003\rule[-2ex]{0pt}{5ex}\\
\hline\hline
\end{tabular}\label{Octetresult1A}\end{center}
\end{table}
Once more one can see than exchange contributions are small and can
be neglected in other computations. Comparing these results with the
relativistic ones from Table \ref{Octetresult2A}, it is also clear
that the coincidental similarity observed in the vector case does
not work in the axial sector. Relativistic corrections are important
to understand the proton axial charges.
\begin{table}[h!]\begin{center}\caption{\small{Our axial content of the proton compared with NQM.}}
\begin{tabular}{c|ccc|ccc|ccc}
\hline\hline
Axial&\multicolumn{3}{c|}{$\Delta u$}&\multicolumn{3}{c|}{$\Delta d$}&\multicolumn{3}{c}{$\Delta s$}\rule{0pt}{3ex}\\
&$\bar q$&$q_s$&$q_{val}$&$\bar q$&$q_s$&$q_{val}$&$\bar
q$&$q_s$&$q_{val}$\rule{0pt}{3ex}\\\hline \rule{0pt}{3ex}
NQM&0&0&4/3&0&0&-1/3&0&0&0\\\rule{0pt}{3ex}
$3Q$&0&0&1.148&0&0&-0.287&0&0&0\\\rule{0pt}{3ex}
$3Q+5Q$&-0.032&-0.042&1.086&0.017&0.028&-0.275&0.005&0.005&-0.003\\\rule{0pt}{3ex}
$3Q+5Q+7Q$&-0.046&-0.060&1.056&0.026&0.040&-0.273&0.007&0.007&-0.006\rule[-2ex]{0pt}{5ex}\\
\hline\hline
\end{tabular}\label{Octetresult2A}\end{center}
\end{table}
One can see that the sea is not $SU(3)$ symmetric $\Delta\bar
u=\Delta\bar d=\Delta s=\Delta\bar s$ as naively often assumed.
Experimental results from SMC \cite{SMC}, HERMES \cite{HERMES} and
COMPASS \cite{COMPASS} favor an asymmetric light sea scenario
$\Delta\bar u=-\Delta\bar d$. Our results show indeed that
$\Delta\bar u$ and $\Delta\bar d$ have opposite sign but the
contribution of $\Delta\bar u$ is roughly twice the contribution of
$\Delta\bar d$. Concerning the sum $\Delta\bar u+\Delta\bar d$ it is
about $2\%$ experimentally and is compatible with zero. The sum we
have obtained has the same order of magnitude but has the opposite
sign. The DNS parametrization finds $\Delta\bar u>0$ and $\Delta\bar
d<0$ while the statistical model \cite{Statistical} suggests the
opposite signs like us. For the valence contribution experiments
suggest $\Delta u_v+\Delta d_v\approx 0.40$ while we have obtained
$\approx 0.78$. This would indicate that in our approach we do not
have enough antiquarks and that our valence sector is too large. On
the top of that sea contributions to axial charges appear with a
sign opposite to the one suggested by experiments.
\begin{table}[h!]\begin{center}\caption{\small{Our flavor contributions to the proton spin and axial charges compared with NQM and experimental data.\newline}}
\begin{tabular}{c|ccc|ccc}
\hline\hline &$\Delta u$&$\Delta d$&$\Delta
s$&$g_A^{(3)}$&$g_A^{(8)}$&$g_A^{(0)}$\rule{0pt}{3ex}\\\hline
\rule{0pt}{3ex} NQM&4/3&-1/3&0&5/3&$1/\sqrt{3}$&1\\\rule{0pt}{3ex}
$3Q$&1.148&-0.287&0&1.435&0.497&0.861\\\rule{0pt}{3ex}
$3Q+5Q$&1.011&-0.230&0.006&1.241&0.444&0.787\\\rule{0pt}{3ex}
$3Q+5Q+7Q$&0.949&-0.207&0.009&1.156&0.419&0.751\\\rule{0pt}{3ex}
Exp. value&$0.83\pm 0.03$&$-0.43\pm 0.04$&$-0.10\pm 0.03$&$1.257\pm 0.003$&$0.34\pm 0.02$&$0.31\pm 0.07$\rule[-2ex]{0pt}{5ex}\\
\hline\hline
\end{tabular}\label{Octetresult3A}\end{center}
\end{table}

In table \ref{Octetresult3A} we give the flavor contributions and
proton axial charges. We can see that relativistic effects (quark
orbital angular momentum) and additional quark-antiquark pairs both
bring the proton axial charges closer to experimental values. While
$g_A^{(3)}$ and $g_A^{(8)}$ are fairly well reproduced, we still
have a too large fraction of the proton spin due to quark spins
$g_A^{(0)}$. It is known in $\chi$QSM that $g_A^{(0)}$ is sensitive
to the $m_s$. This correction due to strange quark mass reduces the
fraction of spin carried by quarks \cite{mscorr}.

Let us concentrate on the strange contribution now. We have found a
non-vanishing contribution which then naturally breaks the
Ellis-Jaffe sum rule. However compared to phenomenological
extractions \cite{Globalanalysis,Ds} it has the wrong sign and is
one order of magnitude too small. Since our results show that a
negative contribution to $\Delta s$ comes from the valence part, one
could then suggest that the valence part should be larger. This
however contradicts the previous observation that the sea part is
too small. The other way is to say that the sea contribution has the
wrong sign has suggested by $\Delta\bar u$ and $\Delta\bar d$. Note
however that the individual flavor contributions are not measured
directly but obtained as combinations of the axial charges. The
extraction of these charges relies on various assumptions,
\emph{e.g.} $g_A^{(3)}$ is based on isospin $SU(2)$ symmetry while
the extraction of $g_A^{(8)}$ from hyperon semi-leptonic decays is
based on flavor $SU(3)$ symmetry.

Let us mention a puzzling result on the experimental side. The
HERMES experiment \cite{HERMES} measured the helicity distribution
of strange quarks $\Delta s(x)$ using polarized SIDIS and found
$\Delta s(x)\approx 0$ in the range $0.03<x<0.3$ and the octet axial
charge $g^{(8)}_A$ is $0.274\pm 0.026(stat.)\pm 0.011(sys.)$ which
is substantially less than the value inferred from hyperon decay.
This seems to disagree with the analysis of the inclusive DIS data.
This disagreement could be due to a failure in one or more of the
assumptions made in the analysis of the inclusive and/or the
semi-inclusive data or it could be due to a more exotic physics
mechanism such as ``polarized condensate'' at $x=0$ not observable
in DIS \cite{Polarizedcondensate}. Further results are thus needed
for a definitive conclusion.

\begin{table}[h!]\begin{center}\caption{\small{Comparison of our octet axial decay constants with NQM predictions and experimental data \cite{PDG}.}}
\begin{tabular}{c|cccc|c}
\hline\hline &NQM&$3Q$&$3Q+5Q$&$3Q+5Q+7Q$&Exp.
Value\rule{0pt}{3ex}\\\hline \rule{0pt}{3ex}
$(g_A/g_V)_{n^0_\mathbf{ 8}\to
p^+_\mathbf{8}}$&5/3&1.435&1.241&1.156&$1.2695\pm
0.0029$\\\rule{0pt}{3ex} $(g_A/g_V)_{\Sigma^-_\mathbf{ 8}\to
\Sigma^0_\mathbf{8}}$&2/3&0.574&0.503&0.470&-\\\rule{0pt}{3ex}
$(g_A)_{\Sigma^-_\mathbf{ 8}\to
\Lambda^0_\mathbf{8}}$&$\sqrt{2/3}$&0.703&0.603&0.560&-\\\rule{0pt}{3ex}
$(g_A/g_V)_{\Sigma^0_\mathbf{ 8}\to
\Sigma^+_\mathbf{8}}$&2/3&0.574&0.503&0.470&-\\\rule{0pt}{3ex}
$(g_A)_{\Lambda^0_\mathbf{ 8}\to
\Sigma^+_\mathbf{8}}$&$\sqrt{2/3}$&0.703&0.603&0.560&-\\\rule{0pt}{3ex}
$(g_A/g_V)_{\Xi^-_\mathbf{ 8}\to
\Xi^0_\mathbf{8}}$&-1/3&-0.287&-0.236&-0.215&-\\\rule{0pt}{3ex}
$(g_A/g_V)_{\Sigma^-_\mathbf{ 8}\to
n^0_\mathbf{8}}$&-1/3&-0.287&-0.236&-0.215&$-0.340\pm
0.017$\\\rule{0pt}{3ex} $(g_A/g_V)_{\Xi^-_\mathbf{ 8}\to
\Sigma^0_\mathbf{8}}$&5/3&1.435&1.241&1.156&-\\\rule{0pt}{3ex}
$(g_A/g_V)_{\Xi^-_\mathbf{ 8}\to
\Lambda^0_\mathbf{8}}$&1/3&0.287&0.256&0.242&$0.25\pm
0.05$\\\rule{0pt}{3ex} $(g_A/g_V)_{\Sigma^0_\mathbf{ 8}\to
p^+_\mathbf{8}}$&-1/3&-0.287&-0.236&-0.215&-\\\rule{0pt}{3ex}
$(g_A/g_V)_{\Lambda^0_\mathbf{ 8}\to
p^+_\mathbf{8}}$&1&0.861&0.749&0.699&$0.718\pm
0.015$\\\rule{0pt}{3ex} $(g_A/g_V)_{\Xi^0_\mathbf{ 8}\to
\Sigma^+_\mathbf{8}}$&5/3&1.435&1.241&1.156&$1.21\pm 0.05$\rule[-2ex]{0pt}{5ex}\\
\hline\hline
\end{tabular}\label{Octetresult3}\end{center}
\end{table}

In Tables \ref{Octetresult3} and \ref{Octetresult4} one can find our
results for octet axial decay constants compared with the
experimental knowledge. There is a global fair agreement. We give
also our results in terms of the $F\&D$ parametrization. Compared
with $SU(3)$ fit\footnote{There have been several attempts to
estimate $F$ and $D$ values by taking the $SU(3)$ and $SU(2)$ flavor
breaking into account, see the review \cite{Ratcliffe}.} to
experimental data $F$ is well reproduced while $D$ is too small.

\begin{table}[h!]\begin{center}\caption{\small{Comparison of our $F\& D$ parameters with NQM predictions and $SU(3)$ fits to experimental data \cite{Yamanishi}.}}
\begin{tabular}{c|cccc|c}
\hline\hline &NQM&$3Q$&$3Q+5Q$&$3Q+5Q+7Q$&$SU(3)$
fit\rule{0pt}{3ex}\\\hline \rule{0pt}{3ex}
$F$&2/3&0.574&0.503&0.470&$0.475\pm 0.004$\\\rule{0pt}{3ex}
$D$&1&0.861&0.739&0.686&$0.793\pm 0.005$\\\rule{0pt}{3ex}
$F/D$&2/3&2/3&0.680&0.686&$0.599\pm 0.006$\\\rule{0pt}{3ex}
$3F-D$&1&0.861&0.769&0.725&$0.632\pm 0.017$\rule[-2ex]{0pt}{5ex}\\
\hline\hline
\end{tabular}\label{Octetresult4}\end{center}
\end{table}

In summary we have fairly well reproduced the octet axial decay
constants, $g_A^{(3)}$ and $g_A^{(8)}$ for the proton. The
discrepancy between our value for $g_A^{(0)}$ and experimental
extractions could be in principle explained by the breaking of
flavor $SU(3)$ symmetry. Many indices also indicate that our sea is
not large enough. Nevertheless this work supports the fact that
quark orbital angular momentum and additional pairs are essential
ingredients to understand the composition of the proton.

\subsection{Decuplet content}

In Tables \ref{Decupletresult3/2}, \ref{Decupletresult1/2},
\ref{Result3/2} and \ref{Result1/2} one can find the $\Delta^{++}$
axial content with respectively $J_z=3/2,1/2$.
\begin{table}[h!]\begin{center}\caption{\small{Our axial content of the $\Delta^{++}$ with spin projection $J_z=3/2$ compared with NQM.}}
\begin{tabular}{c|ccc|ccc|ccc}
\hline\hline
Axial&\multicolumn{3}{c|}{$\Delta u$}&\multicolumn{3}{c|}{$\Delta d$}&\multicolumn{3}{c}{$\Delta s$}\rule{0pt}{3ex}\\
$J_z=3/2$&$\bar q$&$q_s$&$q_{val}$&$\bar q$&$q_s$&$q_{val}$&$\bar
q$&$q_s$&$q_{val}$\rule{0pt}{3ex}\\\hline \rule{0pt}{3ex}
NQM&0&0&3&0&0&0&0&0&0\\\rule{0pt}{3ex}
$3Q$&0&0&2.538&0&0&0&0&0&0\\\rule{0pt}{3ex}
$3Q+5Q$&-0.061&-0.079&2.423&0.012&0.021&-0.015&0.012&0.021&-0.015\rule[-2ex]{0pt}{5ex}\\
\hline\hline
\end{tabular}\label{Decupletresult3/2}\end{center}
\end{table}
\begin{table}[h!]\begin{center}\caption{\small{Our axial content of the $\Delta^{++}$ with spin projection $J_z=1/2$ compared with NQM.}}
\begin{tabular}{c|ccc|ccc|ccc}
\hline\hline
Axial&\multicolumn{3}{c|}{$\Delta u$}&\multicolumn{3}{c|}{$\Delta d$}&\multicolumn{3}{c}{$\Delta s$}\rule{0pt}{3ex}\\
$J_z=1/2$&$\bar q$&$q_s$&$q_{val}$&$\bar q$&$q_s$&$q_{val}$&$\bar
q$&$q_s$&$q_{val}$\rule{0pt}{3ex}\\\hline \rule{0pt}{3ex}
NQM&0&0&1&0&0&0&0&0&0\\\rule{0pt}{3ex}
$3Q$&0&0&0.861&0&0&0&0&0&0\\\rule{0pt}{3ex}
$3Q+5Q$&-0.020&-0.026&0.813&0.004&0.007&-0.007&0.004&0.007&-0.007\rule[-2ex]{0pt}{5ex}\\
\hline\hline
\end{tabular}\label{Decupletresult1/2}\end{center}
\end{table}
\begin{table}[h!]\begin{center}\caption{\small{Flavor contributions to the $\Delta^{++}_{J_z=3/2}$ spin and axial charges compared with NQM.}}
\begin{tabular}{c|ccc|ccc}
\hline\hline &$\Delta u$&$\Delta d$&$\Delta
s$&$g_A^{(3)}$&$g_A^{(8)}$&$g_A^{(0)}$\rule{0pt}{3ex}\\\hline
\rule{0pt}{3ex} NQM&3&0&0&3&$\sqrt{3}$&3\\\rule{0pt}{3ex}
$3Q$&2.583&0&0&2.583&1.492&2.583\\\rule{0pt}{3ex}
$3Q+5Q$&2.283&0.018&0.018&2.265&1.307&2.319\rule[-2ex]{0pt}{5ex}\\
\hline\hline
\end{tabular}\label{Result3/2}\end{center}
\end{table}
\begin{table}[h!]\begin{center}\caption{\small{Flavor contributions to the $\Delta^{++}_{J_z=1/2}$ spin and axial charges compared with NQM.}}
\begin{tabular}{c|ccc|ccc}
\hline\hline &$\Delta u$&$\Delta d$&$\Delta
s$&$g_A^{(3)}$&$g_A^{(8)}$&$g_A^{(0)}$\rule{0pt}{3ex}\\\hline
\rule{0pt}{3ex} NQM&1&0&0&1&$1/\sqrt{3}$&1\\\rule{0pt}{3ex}
$3Q$&0.861&0&0&0.861&0.497&0.861\\\rule{0pt}{3ex}
$3Q+5Q$&0.767&0.004&0.004&0.763&0.441&0.775\rule[-2ex]{0pt}{5ex}\\
\hline\hline
\end{tabular}\label{Result1/2}\end{center}
\end{table}

To the best of our knowledge there is no experimental results
concerning the axial content of decuplet baryons. Our results can
then be considered as just theoretical predictions, at least
qualitatively. Such as in the proton, quarks spins alone do not add
up to the total decuplet baryon spin. The missing spin has to be
attributed to angular momentum of quarks and additional
quark-antiquark pairs. It is also clear that eq. (\ref{sphericityA})
are not satisfied indicating a deviation from spherical shape as
discussed in the previous section. One can however observe a
different feature compared to the octet case. While in proton the
``hidden'' flavor $q=s$ gives $\Delta\bar q=\Delta q_{s}$ in the
$\Delta^{++}$ the ``hidden'' flavors $q=d,s$ give $\Delta\bar
q\neq\Delta q_{s}$.

\subsection{Antidecuplet content}

Since the $5Q$ sector with exchange diagram in the non-relativistic
limit does not affect vector charges, it is important to check that
axial charges are not too affected. This exchange contribution is
naturally smaller than the direct one but since there is no $3Q$
sector this contribution falls directly in the dominant sector and
the conclusion drawn for ordinary baryon may be wrong for exotic
ones. The non-relativistic axial content of the $\Theta^+$
pentaquark is given in Table \ref{AntidecupletcontentA}.
\begin{table}[h!]\begin{center}\caption{\small{Non-relativistic axial content of the $\Theta^+$.}}
\begin{tabular}{c|cc|cc|cc}
\hline\hline
Axial&\multicolumn{2}{c|}{$\Delta u$}&\multicolumn{2}{c|}{$\Delta d$}&\multicolumn{2}{c}{$\Delta s$}\rule{0pt}{3ex}\\
&$\bar q$&$q_s+q_{val}$&$\bar q$&$q_s+q_{val}$&$\bar
q$&$q_s+q_{val}$\rule{0pt}{3ex}\\\hline \rule{0pt}{3ex} $5Q$
(dir)&0&0.591&0&0.591&0.735&0\\\rule{0pt}{3ex}
$5Q$ (dir+exch)&0&0.616&0&0.616&0.733&0\rule[-2ex]{0pt}{5ex}\\
\hline\hline
\end{tabular}\label{AntidecupletcontentA}\end{center}
\end{table}
Definitively one can see that exchange contributions seem fairly
negligible even when the $3Q$ sector is absent. While for $\Delta s$
the correction is less than 1\%, it is roughly 4\% for $\Delta u$
and $\Delta d$. In further computations one may reasonably forget
about such corrections and concentrate on direct diagrams only. This
allows one to spare a lot of time and energy.

The study of the $7Q$ sector has mainly been motivated by the
pentaquark. We have shown that the $5Q$ component of usual baryons
has non negligible and interesting effects on the vector and axial
quantities. In the same spirit, since there is no $3Q$ component in
pentaquarks, it would be interesting to see what happens when
considering the $7Q$ component. In Tables \ref{Antidecupletresult}
and \ref{Antidecupletresult2} one can find the $\Theta^+$ axial
content.
\begin{table}[h!]\begin{center}\caption{\small{Our axial content of the $\Theta^+$.}}
\begin{tabular}{c|ccc|ccc|ccc}
\hline\hline
Axial&\multicolumn{3}{c|}{$\Delta u$}&\multicolumn{3}{c|}{$\Delta d$}&\multicolumn{3}{c}{$\Delta s$}\rule{0pt}{3ex}\\
&$\bar q$&$q_s$&$q_{val}$&$\bar q$&$q_s$&$q_{val}$&$\bar
q$&$q_s$&$q_{val}$\rule{0pt}{3ex}\\\hline \rule{0pt}{3ex}
$5Q$&0&0.322&0.136&0&0.322&0.136&0.644&0&0\\\rule{0pt}{3ex}
$5Q+7Q$&-0.020&0.276&0.113&-0.020&0.276&0.113&0.610&0.019&-0.014\rule[-2ex]{0pt}{5ex}\\
\hline\hline
\end{tabular}\label{Antidecupletresult}\end{center}
\end{table}
\begin{table}[h!]\begin{center}\caption{\small{Flavor contributions to the $\Theta^{+}$ spin and axial charges.}}
\begin{tabular}{c|ccc|ccc}
\hline\hline &$\Delta u$&$\Delta d$&$\Delta
s$&$g_A^{(3)}$&$g_A^{(8)}$&$g_A^{(0)}$\rule{0pt}{3ex}\\\hline
\rule{0pt}{3ex}
$5Q$&0.458&0.458&0.644&0&-0.215&1.560\\\rule{0pt}{3ex}
$5Q+7Q$&0.369&0.369&0.615&0&-0.284&1.353\rule[-2ex]{0pt}{5ex}\\
\hline\hline
\end{tabular}\label{Antidecupletresult2}\end{center}
\end{table}

The first interesting thing here is that contrarily to usual baryons
the sum of all quark spins is \emph{larger} than the total baryon
spin. This means that quark spins are mainly parallel to the baryon
spin and that their angular momentum is opposite in order to
compensate and form at the end a baryon with spin $1/2$. The second
interesting thing is that this $7Q$ component does not change
qualitatively the results given by the $5Q$ sector alone. This means
that a rather good estimation of pentaquark properties can be
obtained by means of the dominant sector only.

\section{Pentaquark width}

A very interesting question about pentaquark is its width. In this
model it is predicted to be very small, a few MeV and can even be
$<1$ MeV, quite unusual for baryons. In the present approach this
can be understood by the fact that since there is no $3Q$ in the
pentaquark and that in the DYW frame only diagonal transitions in
the Fock space occur, the transition is dominated by the transition
from the pentaquark $5Q$ sector to the proton $5Q$ sector, the
latter being of course not so large. Since the pentaquark production
mechanism is not known, its width is estimated by means of the axial
decay constant $\Theta^+_\mathbf{\overline{10}}\to
K^+n^0_\mathbf{8}$. If we assume the approximate $SU(3)$ chiral
symmetry one can obtain the $\Theta\to KN$ pseudoscalar coupling
from the generalized Goldberger-Treiman relation
\begin{equation}\label{GTR}
g_{\Theta KN}=\frac{g_A(\Theta\to KN)(M_\Theta+M_N)}{2F_K}
\end{equation}
where we use $M_\Theta=1530$ MeV, $M_N=940$ MeV and
$F_K=1.2F_\pi=112$ MeV. Once this transition pseudoscalar constant
is known one can evaluate the $\Theta^+$ width from the general
expression for the $\frac{1}{2}^+$ hyperon decay \cite{Width}
\begin{equation}\label{Width}
\Gamma_\Theta=2\,\frac{g^2_{\Theta
KN}|\up|}{8\pi}\frac{(M_\Theta-M_N)^2-m_K^2}{M_\Theta^2}
\end{equation}
where
$|\up|=\sqrt{(M_\Theta^2-M_N^2-m_K^2)^2-4M_N^2m_K^2}/2M_\Theta=254$
MeV is the kaon momentum in the decay ($m_K=495$ MeV) and the factor
of 2 stands for the equal probability $K^+n^0_\mathbf{8}$ and
$K^0p^+_\mathbf{8}$ decays.

Here are the combinations arising for this axial decay constant in
the $5Q$ and $7Q$ sectors
\begin{eqnarray}
A^{(5)}(\Theta^+_\mathbf{\overline{10}}\to
K^+n^0_\mathbf{8})&=&\frac{-6}{5}\sqrt{\frac{3}{5}}\left(7K^A_{\pi\pi}-8K^A_{33}+5K^A_{\sigma\sigma}-28K^V_{3\sigma}\right),\\
A^{(5)\textrm{exch}}(\Theta^+_\mathbf{\overline{10}}\to
K^+n^0_\mathbf{8})&=&\frac{2}{5}\sqrt{\frac{3}{5}}\left(7K_1-K_2+7K_3+3K_4+2K_5-4K_6-18K_7+10K_8-10K_9\right.\nonumber\\
&&\left.+10K_{10}+20K_{11}-20K_{13}\right),\\
A^{(7)}(\Theta^+_\mathbf{\overline{10}}\to
K^+n^0_\mathbf{8})&=&\frac{-48}{5}\sqrt{\frac{3}{5}}\left(7K^A_{\pi\pi\pi\pi}+7K^A_{\pi\pi\pi\pi
2}+6K^A_{\pi\pi 33}-14K^A_{\pi 3\pi
3}+40K^A_{\sigma\sigma\pi\pi}-38K^A_{\sigma\sigma
38}\right.\nonumber\\
&&\left.+22K^A_{\sigma\sigma\sigma\sigma}-71K^V_{\pi\pi
3\sigma}+K^V_{\pi 3\pi\sigma}-140K^V_{\sigma\sigma 3\sigma}\right).
\end{eqnarray}

Let us have a look to the numerical values obtained, first in the
non-relativistic limit, see Table \ref{Thetawidth}.
\begin{table}[h!]\begin{center}\caption{\small{$\Theta^{+}$
width estimation in the non-relativistic limit.\newline}}
\begin{tabular}{c|ccc}
\hline\hline
&$g_A(\Theta\to KN)$&$g_{\Theta KN}$&$\Gamma_\Theta$ (MeV)\rule{0pt}{3ex}\\
\hline\rule{0pt}{3ex} $5Q$ (dir)&0.202&2.230&4.427\\\rule{0pt}{3ex}
$5Q$ (dir+exch)&0.203&2.242&4.472\rule[-2ex]{0pt}{5ex}\\
\hline\hline
\end{tabular}\label{Thetawidth}\end{center}
\end{table}
The width is really small compared to ordinary baryon resonances
($\approx 100$ MeV) and confirms the order of a few MeV obtained by
the other approaches to $\chi$QSM. The exchange contribution does
not change much the result even after the manipulations of
(\ref{GTR}) and (\ref{Width}).

A relativistic estimation of the $\Theta^+$ pentaquark is given in
Table \ref{Thetawidth2}.
\begin{table}[h!]\begin{center}\caption{\small{$\Theta^{+}$
width estimation.\newline}}
\begin{tabular}{c|ccc}
\hline\hline
&$g_A(\Theta\to KN)$&$g_{\Theta KN}$&$\Gamma_\Theta$ (MeV)\rule{0pt}{3ex}\\
\hline\rule{0pt}{3ex} $5Q$&0.144&1.592&2.256\\\rule{0pt}{3ex}
$5Q+7Q$&0.169&1.864&3.091\rule[-2ex]{0pt}{5ex}\\
\hline\hline
\end{tabular}\label{Thetawidth2}\end{center}
\end{table}
The first observation is that valence quark orbital motion reduces
the width by one half. This has to be related with the octet
normalizations. We have seen that relativistic corrections have
increased the fraction of the proton made of $3Q$ only. This
fraction is not accessible by the pentaquark. Consequently the axial
decay constant becomes smaller and at the end the decay width is
reduced.

The $7Q$ component does not change much the estimation with $5Q$
only. Note however, as one could have expected, that the width is
slightly increased. Indeed we have just explained that the unusually
small width of pentaquarks can be understood in the present approach
by the fact that the pentaquark cannot decay into the $3Q$ sector of
the nucleon. Since the $7Q$ component reduces the weight of the $3Q$
component in the nucleon (see Tables \ref{pourcent1} and
\ref{pourcent2}) the width is expected to increase. The view of a
narrow pentaquark resonance within the $\chi$QSM is safe and appears
naturally without any parameter fixing.

It is of course not possible today to give a definite width to the
pentaquark. This is due to all approximations used. We can just
afford estimations to give an order of magnitude and try to
understand why it has such a small width. Nevertheless one thing is
clear: \emph{if the pentaquark exists its width is at most a few
MeV.}

\chapter{Tensor charges}
\section{Introduction}

The tensor charges of a baryon are defined as forward matrix
elements of the tensor current
\begin{equation}
\langle B(p)|\bar\psi
i\sigma^{\mu\nu}\gamma_5\lambda^a\psi|B(p)\rangle=g^{(a)}_T\bar
u(p)i\sigma^{\mu\nu}\gamma_5 u(p)
\end{equation}
where $a=0,3,8$ and $\lambda^3,\lambda^8$ are Gell-Mann matrices,
$\lambda^0$ is just in this context the $3\times3$ unit matrix.
These tensor charges are related to the first moment of the
transversely polarized quark distributions
\begin{equation}\label{Nucleon tensor charges}
g^{(3)}_T=\delta u-\delta d,\qquad
g^{(8)}_T=\frac{1}{\sqrt{3}}(\delta u+\delta d-2\delta s),\qquad
g^{(0)}_T=\delta u+\delta d+\delta s
\end{equation}
where $\delta q\equiv\int_0^1\ud
z\left[q_\uparrow(z)-q_\downarrow(z)-\bar q_\uparrow(z)+\bar
q_\downarrow(z)\right]$ with $q=u,d,s$ and using the transversity
basis for a baryon travelling in the $z$ direction with its
polarization in the $x$ direction
\begin{equation}
|\uparrow\rangle=\frac{1}{\sqrt{2}}\left(|+\rangle+|-\rangle\right),\qquad
|\downarrow\rangle=\frac{1}{\sqrt{2}}\left(|+\rangle-|-\rangle\right)
\end{equation}
written in terms of the usual helicity eigenstates $|\pm\rangle$. We
split the tensor charges into valence quark, sea quark and antiquark
contributions
\begin{equation}
\delta q=\delta q_\textrm{val}+\delta q_\textrm{sea}, \qquad \delta
q_\textrm{sea}=\delta q_\textrm{s}-\delta\bar q
\end{equation}
where index s refers to the quarks in the sea pairs.

A probabilistic interpretation of tensor charges is not possible in
the usual helicity basis \cite{Ralston} $q_\pm=(1\pm\gamma_5)q/2$
since they correspond to off-diagonal transitions. The probabilistic
interpretation is only possible in the transversity basis
\cite{transversitybasis} $q_{\uparrow,\downarrow}=(q_+\pm
q_-)/\sqrt{2}$. The tensor charges just count the total number of
quarks with transverse polarization aligned \emph{minus} total
number of quarks with transverse polarization anti-aligned with
baryon polarization.

Tensor charges are of particular interest for several reasons. First
one could think that $\delta q=\Delta q$. In DIS quarks in the
nucleon appear to be free. However rotational invariance has become
non-trivial since high-energy processes select a special direction.
In the IMF these rotations involve interactions \cite{Dirac}. The
difference between axial and tensor charges has a dynamical origin.
In non-relativistic quark models the transverse spin operator
commutes with a free-quark Hamiltonian and so transversely polarized
quarks are in transverse-spin eigenstates. Then rotational
invariance implies $\delta q=\Delta q$. This can also be seen from
the tensor current $\bar \psi i\sigma^{0i}\gamma_5\psi$ which
differs from the axial-vector current $\bar\psi\gamma^i\gamma_5\psi$
by a factor $\gamma^0$. This factor is reduced to 1 in the
non-relativistic limit.

Second the corresponding quark bilinear is odd under charge
conjugation, only valence quarks contribute \cite{Jaffe} following
the standard definition. That is the reason why it is often thought
that tensor charges could give informations on the valence part only
and considered as more suited for quark models than axial charges.
Moreover since there is no valence strange quark there is \emph{a
priori} no strangeness contribution to tensor charges.

Finally helicity conservation at the quark-gluon vertex prevents
mixing between quark and gluon transversity distributions under QCD
evolution \cite{commun,mixing}. Gluon transversity distributions
only exist for targets with $J\geq 1$ because measurement of gluon
transversity distribution requires that the target change helicity
by two units of angular momentum which is not possible for spin-1/2
targets \cite{commun}. Under DGLAP evolution the angular momentum
generated by the DGLAP kernels is not shared between the quark and
gluon sectors. This has thus an effect on the evolution of the
tensor charge with $Q^2$. The sign of the anomalous dimensions at
both LO and NLO is such that tensor charges fall with increasing
$Q^2$.

Soffer \cite{Soffer} has proposed an inequality among the nucleon
twist 2 quark distributions $f_1,g_1,h_1$
\begin{equation}\label{Soff}
f_1+g_1\geq 2|h_1|
\end{equation}
In contrast to the well-known inequalities and positivity
constraints among distribution functions such as $f_1\geq |g_1|$
which are general properties of lepton-hadron scattering, derived
without reference to quarks, color or QCD, this Soffer inequality
needs a parton model to QCD to be derived \cite{Sofferineq}.
Unfortunately it turned out that it does not constrain the nucleon
tensor charge. However this inequality still has to be satisfied by
models that try to estimate quark distributions.

In the IMF language we have to use the ``good'' components $\mu=0,3$
of the tensor operator. Unlike the vector and axial-vector operators
the tensor operator flips quark helicity. We have then
$M^\tau_\sigma=(\sigma_{R,L})^\tau_\sigma$ with
$\sigma_{R,L}=(\sigma_1\pm i\sigma_2)/2$.

\section{Tensor charges on the light cone}

The tensor charge can be obtained in IMF by means of the \emph{plus}
component of the tensor operator \cite{Schmidt}
\begin{equation}
\delta q=\frac{1}{2P^+}\langle
P,\frac{1}{2}|\bar\psi_{LC}\gamma^+\gamma^R\psi_{LC}|P,-\frac{1}{2}\rangle,
\end{equation}
where $\gamma^R=\gamma^1+i\gamma^2$. Using the Melosh rotation one
can see that $\delta q_{LC}$ and $\delta q_{NR}$ are related as
follows \cite{Schmidt}
\begin{equation}
\delta q_{LC}=\langle M_T\rangle\delta q_{NR}
\end{equation}
where
\begin{equation}\label{DefMqtilde}
M_T=\frac{(m_q+z_3\uM)^2}{(m_q+z_3\uM)^2+\up_{3\perp}^2}
\end{equation}
and $\langle M_T\rangle$ is its expectation value. In the
non-relativistic limit $\up_\perp=0$ and thus $M_V=M_A=M_T=1$ as it
should be. Relativistic effects $\up_\perp\neq 0$ reduce the values
of both $M_A$ and $M_T$. It is also interesting to notice that one
has
\begin{equation}
M_V+M_A=2M_T
\end{equation}
which saturates Soffer's inequality, see eq. (\ref{Soff}). Soffer's
inequality is exact in the parton model and valid for all (explicit)
flavors likewise for antiquarks \cite{Soffer}. Since $\langle
M_A\rangle=3/4$ one obtains $\langle M_T\rangle=7/8$ and thus
\begin{equation}
\delta u=7/6,\qquad\delta d=-7/24,\qquad \delta s=0.
\end{equation}
From eqs. (\ref{DefMq}) and (\ref{DefMqtilde}) one would indeed
expect that
\begin{equation}
|\delta q|>|\Delta q|.
\end{equation}

\section{Scalar overlap integrals and quark distributions}

From the expression (\ref{Discrete level IMF}) and if we concentrate
on the spin part the contraction of two valence wave functions $F$
with the tensor operator gives
\begin{equation}\label{TensOQL}
F^\dag(\sigma^R)F\propto
h^2(p)+2h(p)\,\frac{p_z}{|\up|}\,j(p)+\frac{p_z^2}{p^2}\,j^2(p).
\end{equation}
Unlike the vector and axial-vector operators the tensor operator
flips quark spin. This quark may not have orbital angular momentum
since total angular momentum is 1/2. For this reason only $L_z=0$
components survive in (\ref{TensOQL}), \emph{i.e.} no $\up_\perp$
factor. The $3Q$ sector saturates Soffer's inequality just like the
Melosh rotation approach does, which is not surprising. However
quark-antiquark pairs are susceptible to change this.

The tensor valence quark distribution is obtained by the
multiplication of two factors with structure (\ref{VectOQL}) where
the momentum is respectively $p_1$ and $p_2$ and a third factor with
structure (\ref{TensOQL}) and momentum $p_3$. The expansion gives
the following function $D$
\begin{eqnarray}
&D^T(p_1,p_2,p_3)=h^2(p_1)h^2(p_2)h^2(p_3)+6h^2(p_1)h^2(p_2)\left[h(p_3)\frac{p_{3z}}{|\up_3|}j(p_3)\right]+h^2(p_1)h^2(p_2)\frac{p_{3z}^2+2p_3^2}{p_3^2}j^2(p_3)&\nonumber\\
&+12h^2(p_1)\left[h(p_2)\frac{p_{2z}}{|\up_2|}j(p_2)\right]\left[h(p_3)\frac{p_{3z}}{|\up_3|}j(p_3)\right]+4h^2(p_1)\left[h(p_2)\frac{p_{2z}}{|\up_2|}j(p_2)\right]\frac{p_{3z}^2+2p_3^2}{p_3^2}j^2(p_3)&\nonumber\\
&+8\left[h(p_1)\frac{p_{1z}}{|\up_1|}j(p_1)\right]\left[h(p_2)\frac{p_{2z}}{|\up_2|}j(p_2)\right]\left[h(p_3)\frac{p_{3z}}{|\up_3|}j(p_3)\right]+h^2(p_1)j^2(p_2)\frac{2p_{3z}^2+rp_3^2}{p_3^2}j^2(p_3)&\\
&+4\left[h(p_1)\frac{p_{1z}}{|\up_1|}j(p_1)\right]\left[h(p_2)\frac{p_{2z}}{|\up_2|}j(p_2)\right]\frac{p_{3z}^2+2p_3^2}{p_3^2}j^2(p_3)+2\left[h(p_1)\frac{p_{1z}}{|\up_1|}j(p_1)\right]j^2(p_2)\frac{2p_{3z}^2+p_3^2}{p_3^2}j^2(p_3)&\nonumber\\
&+j^2(p_1)j^2(p_2)\frac{p_{3z}^2}{p_3^2}j^2(p_3).&\nonumber\label{Upsilon}
\end{eqnarray}
that is needed in the expression of the valence quark distribution
(\ref{Probability3q}). In the non-relativistic limit $j=0$ this
function $D$ is reduced to
\begin{equation}
D^T_{NR}(p_1,p_2,p_3)=D^A_{NR}(p_1,p_2,p_3)=D^V_{NR}(p_1,p_2,p_3)=h^2(p_1)h^2(p_2)h^2(p_3)
\end{equation}
as expected from non-relativistic rotational invariance.

\begin{figure}[h]\begin{center}\begin{minipage}[c]{6cm}\begin{center}\includegraphics[width=5.3cm]{Phiplot5.eps}
\end{center}\end{minipage}\hspace{1cm}\begin{minipage}[c]{6cm}\begin{center}\includegraphics[width=5.3cm]{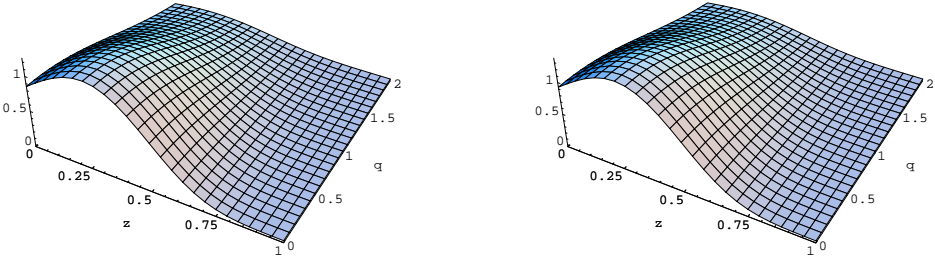}
\end{center}\end{minipage}\caption{\small{Probability distribution $\Phi^I(z,\uq_\perp)$ that three valence quarks leave the fraction $z$ of the baryon momentum and
transverse momentum $\uq_\perp$ to the quark-antiquark pair(s) in
the axial $I=A$ (left) and tensor $I=T$ (right) cases plotted in
units of $M$ and normalized to $\Phi^V(0,0)=1$.
}}\label{Phiplot3}\end{center}
\end{figure}

The valence probability distribution $\Phi^I(z,\uq_\perp)$ is then
obtained by integration over the valence quark momenta, see eq.
(\ref{Probability3q}) and is depicted in Fig. \ref{Phiplot3} in
axial $I=A$ and tensor $I=T$ cases.

While in the non-relativistic limit valence probability
distributions are the same, relativistic corrections (quark angular
momentum) are different in axial and tensor cases. This is of course
due to the difference in structure between (\ref{AxialOQL}) and
(\ref{TensOQL}).

In the following we give the integrals appearing in each Fock sector
and the numerical values obtained for them. In the evaluation of the
scalar overlap integrals we have used the constituent quark mass
$M=345$ MeV, the Pauli-Villars mass $M_\textrm{PV}=556.8$ MeV for
the regularization of (\ref{Direct1})-(\ref{Direct4}) and the baryon
mass $\uM=1207$ MeV as it follows for the ``classical'' mass in the
mean field approximation \cite{Approximation}.

\subsection{$3Q$ scalar integral}

In the $3Q$ sector there is no quark-antiquark pair and thus only
one integral is involved. It corresponds to the valence quark
distribution without momentum left to the sea $\Phi^T(0,0)$. We
remind that the normalization chosen for $h(p)$ and $j(p)$ is such
that $\Phi^V(0,0)=1$. From Figure \ref{Phiplot3} one can see that
$\Phi^A(0,0)<\Phi^T(0,0)<1$. The precise value is
\begin{equation}
\Phi^T(0,0)=0.93058.
\end{equation}
This means that in a simple $3Q$ picture all NQM tensor charges have
to be multiplied by this factor. Not surprisingly this is the same
prescription as the one encountered in a standard light-cone
approach based on Melosh rotation. Valence quark motion is a
relativistic effect and is responsible for a noticeable reduction of
NQM predictions. The Melosh reduction factor 7/8 is of course
smaller than the one we have obtained because of the function $h(p)$
absent in the Melosh approach but necessary in a fully relativistic
treatment. Since we know analytically that the pure $3Q$
contribution saturates Soffer's inequality one can notice that the
numerical accuracy of the distributions is good.

The NQM is recovered in the $3Q$ non-relativistic limit only,
\emph{i.e.} where the three valence quarks have no orbital angular
momentum and thus
$\Phi^T_{NR}(0,0)=\Phi^A_{NR}(0,0)=\Phi^V_{NR}(0,0)=1$.

\subsection{$5Q$ scalar integrals}

In the $5Q$ sector there is one quark-antiquark pair. Contractions
given by the direct diagram give three different integrals
$J=\pi\pi,33,\sigma\sigma$. If the tensor operator acts on valence
quarks, the quark-antiquark pair is not affected and thus the
integrals present the vector structure for the sea. If the tensor
operator acts on the sea, the valence quarks are not affected and
the vector valence probability distribution has to be used. However
no structure for the sea survives in this case. The integrals are
identically zero. This is due to the fact that the tensor operator
is chiral odd. Since the tensor operator flips one quark spin it
would transform a quark-antiquark pair with zero total angular
momentum into another one with one unit of angular momentum, which
is not allowed in the model. This means that only valence quarks
contribute to tensor charges, in accordance with the usual
definition of valence quarks. Note however that the usual definition
of valence quarks forbids strange quark contribution to tensor
charge while with our definition strangeness can access to the
valence sector thanks to the $5Q$ component.

Here are the numerical values obtained for the three new integrals
\begin{equation}
K^T_{\pi\pi}=0.03328,\qquad K^T_{33}=0.01802,\qquad
K^T_{\sigma\sigma}=0.01261.
\end{equation}
Let us have a look to the ratios $K^T_J/K^V_J$ with
$J=\pi\pi,33,\sigma\sigma$
\begin{equation}
\frac{K^T_{\pi\pi}}{K^V_{\pi\pi}}=0.91114,\qquad\frac{K^T_{33}}{K^V_{33}}=0.91229,\qquad\frac{K^T_{\sigma\sigma}}{K^V_{\sigma\sigma}}=0.90020.
\end{equation}
The reduction is of the same order as in the $3Q$ sector. It is
however different from one structure to another due to the details
of the valence probability distributions.

\section{Combinatoric Results}

Tensor matrix elements are linear combinations of the tensor scalar
overlap integrals. These specific combinations are obtained by
contracting the baryon rotational wave functions with the tensor
operator. In the following we give for each multiplet the
combinations obtained. We remind that for tensor charges valence
quark only contribute.

\subsection{Octet baryons}

In the $3Q$ sector we have obtained
\begin{equation}
\alpha_{T,q_\textrm{val}}^{(3)}=12\Phi^T(0,0),\qquad
\beta_{T,q_\textrm{val}}^{(3)}=-3\Phi^T(0,0),\qquad
\gamma_{T,q_\textrm{val}}^{(3)}=0.
\end{equation}
These factors are not surprising since in the non-relativistic limit
$\Phi^T_{NR}(0,0)=\Phi^A_{NR}(0,0)$ we have to recover $\Delta
q_{NR}=\delta q_{NR}$.

In the $5Q$ sector the combinations are
\begin{eqnarray}
&\alpha_{T,q_\textrm{val}}^{(5)}=\frac{6}{5}\left(30K^T_{\pi\pi}-K^T_{33}+91K^T_{\sigma\sigma}\right),&\\
&\beta_{T,q_\textrm{val}}^{(5)}=\frac{-12}{25}\left(21 K^T_{\pi\pi}+11K^T_{33}+52K^T_{\sigma\sigma}\right),&\\
&\gamma_{T,q_\textrm{val}}^{(5)}=\frac{-12}{25}\left(3
K^T_{\pi\pi}+8K^T_{33}+K^T_{\sigma\sigma}\right).&
\end{eqnarray}
The first observation is that global factors are the same as in the
axial case and that the factors in front of $K_{\sigma\sigma}$
correspond. If one looks closer one can notice that the differences
are always proportional to the structure $K_{\pi\pi}-3K_{33}$. This
means that tensor charges together with axial charges, besides
giving informations on quark spin distribution, valence quarks
number and motion, might give also informations on quadrupolar
distortion of the baryon shape.

\subsection{Decuplet baryons}

In the $3Q$ sector we have obtained
\begin{equation}
\alpha'^{(3)}_{T,q_\textrm{val},3/2}=3\alpha'^{(3)}_{T,q_\textrm{val},1/2}=\frac{18}{5}\,\Phi^T(0,0),\qquad
\beta'^{(3)}_{T,q_\textrm{val},3/2}=3\beta'^{(3)}_{T,q_\textrm{val},1/2}=0.
\end{equation}

In the $5Q$ sector the combinations are
\begin{eqnarray}
&\alpha'^{(5)}_{T,q_\textrm{val},3/2}=\frac{9}{20}\left(40K^T_{\pi\pi}-7K^T_{33}+67K^T_{\sigma\sigma}\right),&\\
&\alpha'^{(5)}_{T,q_\textrm{val},1/2}=\frac{3}{20}\left(30K^T_{\pi\pi}+23K^T_{33}+67K^T_{\sigma\sigma}\right),&\\
&\beta'^{(5)}_{T,q_\textrm{val},3/2}=\frac{-9}{20}\left(8K^T_{\pi\pi}+13K^T_{33}-K^T_{\sigma\sigma}\right),&\\
&\beta'^{(5)}_{T,q_\textrm{val},1/2}=\frac{-3}{20}\left(6K^T_{\pi\pi}+19K^T_{33}-K^T_{\sigma\sigma}\right).&
\end{eqnarray}

If the decuplet baryon was made of three quarks only then one would
have the following relations between spin-3/2 and 1/2 tensor
contributions
\begin{equation}\label{sphericity2}
T_{3/2}=3T_{1/2}.
\end{equation}
This picture presents the $\Delta$ as a spherical particle. Things
change in the $5Q$ sector. One notices directly that the relation is
broken by a unique structure $(K^T_{\pi\pi}-3K^T_{33})$. We can draw
the same conclusion as in the axial and vector cases. The pion field
is directly responsible for the deviation of decuplet baryon from
spherical shape. Moreover this structure is the only difference
between axial and tensor combinations. This supports further what we
have observed with octet baryons.

\subsection{Antidecuplet baryons}

In the $5Q$ sector one has
\begin{equation}
\alpha''^{(5)}_{T,q_\textrm{val}}=\beta''^{(5)}_{T,q_\textrm{val}}=\frac{-6}{5}\left(K^T_{33}-K^T_{\sigma\sigma}\right).
\end{equation}

In the $5Q$ sector of $\Theta^+$ pentaquark the strange flavor
appears only as an antiquark as one can see from its minimal quark
content $uudd\bar s$. That's the reason why we have found no strange
contribution. But if at least the $7Q$ sector was considered we
would have obtained a nonzero contribution due to flavor components
like $|uus(d\bar s)(d\bar s)\rangle$, $|uds(u\bar s)(d\bar
s)\rangle$ and $|dds(u\bar s)(u\bar s)\rangle$.

Even for exotic baryons the only difference between tensor and axial
combinations is proportional to the quadrupolar structure
$K_{\pi\pi}-3K_{33}$ of the pion cloud.

\section{Numerical results and discussion}
\subsection{Octet content}

We give in Table \ref{proton} the values obtained for the proton
tensor charges at the model scale $Q_0^2=0.36$ GeV$^2$.
\begin{table}[h!]\begin{center}\caption{\small{Our proton tensor charges computed at the model scale $Q_0^2=0.36$ GeV$^2$.}}
\begin{tabular}{c|ccc|ccc}
\hline\hline $p^+$&$\delta u$&$\delta d$&$\delta
s$&$g_T^{(3)}$&$g_T^{(8)}$&$g_T^{(0)}$\rule{0pt}{3ex}\\\hline
\rule{0pt}{3ex}
$3Q$&1.241&-0.310&0&1.551&0.537&0.931\\\rule{0pt}{3ex}
$3Q+5Q$&1.172&-0.315&-0.011&1.487&0.507&0.846\rule[-2ex]{0pt}{5ex}\\
\hline\hline
\end{tabular}\label{proton}\end{center}
\end{table}
Many papers are just concerned with the isovector (3) and isoscalar
(0) combinations. They assume that strangeness appears only in the
sea or even forget completely about proton strangeness. We have
shown however that the $5Q$ component introduces strangeness in the
valence sector and gives the possibility to have non-zero strange
contribution to tensor charges. This strange contribution is
\emph{negative} while it was \emph{positive} in the axial sector. In
fact only the valence contributions should be compared. In the axial
case the strange valence contribution is also \emph{negative}. The
sea was \emph{positive} and larger in magnitude.

Like all other models for the proton $\delta u$ and $\delta d$ are
not small and have a magnitude similar to $\Delta u$ and $\Delta d$.
One can also check that Soffer's inequality (\ref{Soff}) is
satisfied for explicit flavors. However the hidden flavor,
\emph{i.e.} $s$ in proton, violates the inequality.

Up to now only one experimental extraction of transversity
distributions has been achieved \cite{Exptensor}. The authors did
not give explicit values for tensor charges. They have however been
estimated to $\delta u=0.46^{+0.36}_{-0.28}$ and $\delta
d=-0.19^{+0.30}_{-0.23}$ in \cite{QuarkDiquark} at the scale
$Q^2=0.4$ GeV$^2$. These values are unexpectedly small compared to
models predictions. Further experimental results are then highly
desired to either confirm or infirm the smallness of tensor charges.
If it is confirmed then it will be very difficult to explain this
within a quark model while keeping rather large values for the axial
charges. The question is of course very intriguing and might
eventually kill a constituent quark approach.

\subsection{Decuplet content}

We give in Table \ref{delta} the values obtained for the
$\Delta^{++}$ tensor charges at the model scale $Q_0^2=0.36$
GeV$^2$.
\begin{table}[h!]\begin{center}\caption{\small{Our $\Delta^{++}$ tensor charges computed at the model scale $Q_0^2=0.36$ GeV$^2$.}}
\begin{tabular}{c|ccc|ccc}
\hline\hline $\Delta^{++}_{3/2}$&$\delta u$&$\delta d$&$\delta
s$&$g_T^{(3)}$&$g_T^{(8)}$&$g_T^{(0)}$\rule{0pt}{3ex}\\\hline
\rule{0pt}{3ex} $3Q$&2.792&0&0&2.792&1.612&2.792\\\rule{0pt}{3ex}
$3Q+5Q$&2.624&-0.046&-0.046&2.670&1.541&2.532\rule[-2ex]{0pt}{5ex}\\\hline\rule{0pt}{3ex}
$\Delta^{++}_{1/2}$&$\delta u$&$\delta d$&$\delta
s$&$g_T^{(3)}$&$g_T^{(8)}$&$g_T^{(0)}$\rule{0pt}{3ex}\\\hline
\rule{0pt}{3ex} $3Q$&0.931&0&0&0.931&0.537&0.931\\\rule{0pt}{3ex}
$3Q+5Q$&0.863&-0.016&-0.016&0.879&0.508&0.831\rule[-2ex]{0pt}{5ex}\\
\hline\hline
\end{tabular}\label{delta}\end{center}
\end{table}
Like in the proton one can see that hidden flavors, \emph{i.e.} $d$
and $s$ for $\Delta^{++}$, violate Soffer's inequality. There is no
experimental data concerning decuplet baryon tensor charges. Our
results are then just predictions.

\subsection{Antidecuplet content}

We give in Table \ref{theta} the values obtained for the $\Theta^+$
tensor charges at the model scale $Q_0^2=0.36$ GeV$^2$.
\begin{table}[h!]\begin{center}\caption{\small{Our $\Theta^+$ tensor charges computed at the model scale $Q_0^2=0.36$ GeV$^2$.}}
\begin{tabular}{c|ccc|ccc}
\hline\hline $\Theta^+$&$\delta u$&$\delta d$&$\delta
s$&$g_T^{(3)}$&$g_T^{(8)}$&$g_T^{(0)}$\rule{0pt}{3ex}\\\hline
\rule{0pt}{3ex}
$5Q$&-0.053&-0.053&0&0&-0.062&-0.107\rule[-2ex]{0pt}{5ex}\\
\hline\hline
\end{tabular}\label{theta}\end{center}
\end{table}
The vanishing value of the strange contribution is of course due to
the truncation at the $5Q$ sector. The $7Q$ component is the main
cause for the presence of strange quarks in the valence sector and
consequently in $\Theta^+$ pentaquark tensor charges.
\newpage\thispagestyle{empty}\cleardoublepage
\chapter{Magnetic and transition magnetic moments}
\section{Introduction}

The study of electromagnetic properties of the nucleon is of great
importance in understanding the structure of baryons (see reviews
\cite{Rev1}). Indeed, since electrons are point-like particles, any
observed structure in the electron-target collisions directly gives
information on the target structure. This information is encoded
within form factors which have been measured more and more
accurately throughout the last decades. Form factors measurement
revealed the role of quark orbital momentum, scale at which
perturbative QCD effects should become evident, strangeness content
of the proton and meson-cloud effects. For more than ten years the
contribution of $s$-quarks to proton electric and magnetic form
factors \cite{SAMPLEM,PVA4M,HAPPEXM,G0M} has focused interests
because this contribution is believed to come from the
quark-antiquark sea (meson cloud). There is no more doubt that
\emph{both} valence and sea-quark effects are important in the
description of electromagnetic properties of light hadrons. The only
question left now is the amplitude of this meson-cloud contribution.

We split the contribution to the moments into valence quark, sea
quark and antiquark contributions
\begin{equation}
G_{M,E,q}=G_{M,E,q_\textrm{val}}+G_{M,E,q_\textrm{sea}}, \qquad
G_{M,E,q_\textrm{sea}}=G_{M,E,q_\textrm{s}}-G_{M,E,\bar q}
\end{equation}
where index s refers to the quarks in the sea pairs.

From the NQM picture one could think that magnetic moments could be
related to the axial content of the proton because they are
proportional to the longitudinal polarization asymmetry $\Delta q$.
Actually this is not the case because of the antiquark contribution
has a different sign: axial charges are in flavor singlet
combination (quarks \emph{plus} antiquarks) $A\sim \Delta
q+\Delta\bar q$ and magnetic moments in flavor non-singlet
combination (quarks \emph{minus} antiquarks) $M\sim \Delta
q-\Delta\bar q$.

\section{Magnetic and transition magnetic moments on the light cone}

We consider in this study the interaction of an electromagnetic
field with baryons at the quark level. Constituent quarks can be
considered as quasiparticles. Their coupling with a photon is then
modulated by form factors $F_1^q$, $F_2^q$ that encode in an
effective way the other degrees of freedom, \emph{e.g.} gluons and
quark-antiquark pairs. In the present approach the gluon field has
been integrated out leaving as a by-product an effective pion mean
field that binds quarks together. The model gives a description of
the three valence quark and of the whole Dirac sea. Baryons appear
naturally as made of three quarks plus a certain number of
quark-antiquark pairs. Since this degree of freedom is explicitly
taken into account, we consider constituent quarks as point-like
particles. This means that the Dirac form factor is just the quark
charge $F_1^q=Q$ and the Pauli form factor is identically zero
$F_2^q=0$. The whole anomalous baryon magnetic moments then come
from quark orbital moment and quark-antiquark pairs.

Baryon form factors are obtained through computation of some matrix
elements of the electromagnetic current. We have used the following
vector current $J_\mu(0)=\bar qQ_q\gamma_\mu q$ where $q$ is a set
of three free Dirac spinor of definite flavor $q=u,d,s$ and $Q_q$ is
the quark charge matrix diag$=(2/3,-1/3,-1/3)$. Since we are only
concerned so far with magnetic moments, form factors are computed
only for $Q^2=-q^2=0$ (real photon) where $q=p'-p$ is the momentum
transfer, $p'$ and $p$ are respectively the momenta of outgoing and
incoming baryon.

We give below explicit definitions of form factors and recall how to
extract them when the system is described in the IMF.

\subsection{Octet form factors}

Octet baryons are spin-$\frac{1}{2}$ particles. Their interaction
with an electromagnetic field involves two form factors known as
Dirac and Pauli form factors
\begin{equation}
\langle B(p',s')|J^\mu|B(p,s)\rangle=\bar u_B(p',s')\left[\gamma^\mu
F_1(q^2)+i\frac{\sigma^{\mu\nu}q_\nu}{2M_B}F_2(q^2)\right]u_B(p,s)
\end{equation}
where $s$ and $s'$ are the component of the spin along $z$ axis. At
zero momentum transfer $F_1(0)$ and $F_2(0)$ correspond respectively
to the charge and anomalous magnetic moment of the baryon. In the
literature one also often defines another set of form factors known
as Sachs form factors \cite{Sachs} that are combinations of the
previous ones
\begin{equation}
G_E(Q^2)=F_1(Q^2)-\frac{Q^2}{4M_B^2}F_2(Q^2),\qquad
G_M(Q^2)=F_1(Q^2)+F_2(Q^2)
\end{equation}
which express the nucleon electric and magnetic form factors. In the
Breit frame and in the non-relativistic limit, their
three-dimensional Fourier transforms give electric-charge-density
distribution and magnetic-current-density distribution within the
baryon \cite{Fourier}. At zero momentum transfer $-q^2=Q^2=0$ one
obtains naturally $G_E(0)=F_1(0)=Q_B$ where $Q_B$ is the baryon
charge and $G_M(0)=G_E(0)+\kappa=F_1(0)+F_2(0)=\mu_B$ where $\mu_B$
is the baryon magnetic moment and $\kappa=F_2(0)$ is the
\emph{anomalous} baryon magnetic moment expressed in units of
$e/2M_B$.

These form factors can be extracted in the IMF from the
spin-conserving and spin-flip matrix elements of the + component of
the electromagnetic current
\begin{eqnarray}
F_1(0)&=&\langle P,\frac{1}{2}|\frac{J^+(0)}{2P^+}|P,\frac{1}{2}\rangle,\\
-q_L F_2(Q^2)&=&2M_B\langle
P,\frac{1}{2}|\frac{J^+(0)}{2P^+}|P,-\frac{1}{2}\rangle
\end{eqnarray}
with $q_L=q_1-iq_2$. It is convenient to work in the DYW frame
($q^+=0$ \cite{DYW,Brodsky}) where the photon momentum is transverse
to incident baryon momentum (chosen to be directed along the $z$
direction)
\begin{equation}
P^\mu=\left(P^+,\mathbf 0_{\perp},\frac{M_B^2}{P^+}\right),\qquad
q^\mu=\left(0,\uq_{\perp},\frac{2q\cdot P}{P^+}\right).
\end{equation}
This prevents the quark current $J^+$ from creating pairs or
annihilating the vacuum. One also has $-q_\mu q^\mu\equiv
Q^2=\uq_\perp^2$. The Pauli form factor of nucleons is therefore
computed from the overlap of light-cone wave functions differing by
one unit of orbital angular momentum $\Delta L_z=\pm 1$. The fact
that anomalous magnetic moment of the proton is not zero is an
immediate signal for the presence of non-zero angular momentum in
the proton wave function \cite{BandD}.

\subsection{Decuplet form factors}

Decuplet baryons are spin-3/2 particles. Their interaction with an
electromagnetic field involves four multipoles: a coulomb monopole
(C0), a magnetic dipole (M1), an electric quadrupole (E2) and a
magnetic octupole (M3). One has
\begin{equation}
\langle B(p',s')|J^\mu|B(p,s)\rangle=\bar
u_{B\alpha}(p',s')\mathcal{O}^{\alpha\beta\mu}u_{B\beta}(p,s)
\end{equation}
where $u_{B\beta}(p,s)$ is a Rarita-Schwinger spin vector
\cite{Rarita-Schwinger} with the subsidiary conditions $\gamma^\mu
u_\mu(p,s)=p^\mu u_\mu(p,s)=0$. The tensor has the following
Lorentz-covariant form
\begin{equation}
\mathcal{O}^{\alpha\beta\mu}=g^{\alpha\beta}\left[\gamma^\mu
F_1(q^2)+i\frac{\sigma^{\mu\nu}q_\nu}{2M_B}F_2(q^2)\right]+\frac{q^\alpha
q^\beta}{(2M_B)^2}\left[\gamma^\mu
F_3(q^2)+i\frac{\sigma^{\mu\nu}q_\nu}{2M_B}F_4(q^2)\right].
\end{equation}
This set of form factors is related to the multipole form factors as
follows when $q^2=0$
\begin{eqnarray}
Q_B&=&e\,G_{C0}(0)=e\,F_1(0)\\
\mu_B&=&\frac{e}{2M_B}G_{M1}(0)=\frac{e}{2M_B}\left[F_1(0)+F_2(0)\right]\\
\mathcal{Q}_B&=&\frac{e}{M_B^2}G_{E2}(0)=\frac{e}{M_B^2}\left[F_1(0)-\frac{1}{2}F_3(0)\right]\\
O_B&=&\frac{e}{2M_B^3}G_{M3}(0)=\frac{e}{2M_B^3}\left\{F_1(0)+F_2(0)-\frac{1}{2}\left[F_3(0)+F_4(0)\right]\right\}.
\end{eqnarray}

In this paper, we are only interested in $F_1(0)$ and $F_2(0)$ form
factors, \emph{i.e.} charge and anomalous magnetic moment
respectively. These can be extracted in the IMF from the matrix
elements of the + component of the electromagnetic current
\cite{Schlumpf}
\begin{eqnarray}
F_1(0)&=&\langle P,\frac{3}{2}|\frac{J^+(0)}{2P^+}|P,\frac{3}{2}\rangle,\\
-q_L F_2(0)&=&2\left[\sqrt{3}M_B\langle
P,\frac{3}{2}|\frac{J^+(0)}{2P^+}|P,\frac{1}{2}\rangle-q_L\langle
P,\frac{3}{2}|\frac{J^+(0)}{2P^+}|P,\frac{3}{2}\rangle\right].
\end{eqnarray}

\subsection{Octet-to-decuplet form factors}

Imposing Lorentz covariance, gauge invariance and parity
conservation, the matrix element of the vector octet-to-decuplet
transition can be parametrized in terms of three form factors only
\cite{Jones}
\begin{equation}
\langle B_{\bf 10}(p',s')|J^\mu|B_{\bf
8}(p,s)\rangle=\sqrt{\frac{2}{3}}\bar u_{B_{\bf
10}\alpha}(p',s')\mathcal{O}^{\alpha\mu}\gamma_5 u_{B_{\bf 8}}(p,s)
\end{equation}
where $\sqrt{2/3}$ is the isospin factor and
\begin{equation}
\mathcal{O}^{\alpha\mu}=(q^\alpha\gamma^\mu-g^{\alpha\mu}\qslash)G_1(q^2)+(q^\alpha
P^\mu-g^{\alpha\mu}P\cdot q)G_2(q^2)+(q^\alpha
q^\mu-g^{\alpha\mu}q^2)G_3(q^2)
\end{equation}
with $P=(p+p')/2$. Since in our case the photon is real, $G_3$ does
not contribute and only the form factors $G_1$ and $G_2$ are needed.
These form factors are related to multipole\footnote{In the
literature there are quite a few conventions for the electromagnetic
form factors of $\Delta(1232)$. We choose to use the Jones-Scadron
ones since they are free of kinematical singularities and are
dimensionless.} (Jones-Scadron \cite{Jones}) ones as follows
\begin{eqnarray}
G^*_M&=&\frac{\umo}{3}\left[(3\umd+\umo)\frac{G_1}{\umd}+(\umd-\umo)G_2\right],\\
G^*_E&=&\frac{\umo(\umd-\umo)}{3}\left[\frac{G_1}{\umd}+G_2\right]
\end{eqnarray}
where $\umd$ is the decuplet and $\umo$ the octet baryon mass. In
the literature one also defines helicity amplitudes
\begin{equation}
A_M=-e\frac{1}{\sqrt{2\omega}}\langle B_{\bf 10},M|\vec j\cdot\vec
\epsilon|B_{\bf 8},M-1\rangle,\qquad M=\frac{1}{2},\frac{3}{2}
\end{equation}
where $\omega=(\umd^2-\umo^2)/2\umd$ is the energy of the photon in
the rest frame of decuplet baryon with polarization $\vec\epsilon$.
These amplitudes can be expressed in terms of magnetic dipole $M1$
and electric quadrupole $E2$ moments
\begin{equation}
A_{3/2}=-\frac{\sqrt{3}}{2}(M1-E2), \qquad
A_{1/2}=-\frac{1}{2}(M1+3E2)
\end{equation}
which are related to multipole form factors as follows
\begin{equation}
M1=\frac{e}{2\umo}\sqrt{\frac{\umd\omega}{\umo}}G^*_M, \qquad
E2=-\frac{e}{2\umo}\sqrt{\frac{\umd\omega}{\umo}}G^*_E.
\end{equation}

The static transition magnetic moment $\mu_{B_{\bf 8}B_{\bf 10}}$ is
obtained from magnetic dipole form factor $G^*_M$ at $Q^2=0$ by
\begin{equation}
\mu_{B_{\bf 8}B_{\bf 10}}=\sqrt{\frac{\umd}{\umo}}G^*_M(0)
\end{equation}
which can be expressed in nuclear magnetons $\mu_N\equiv e/2M_N$ if
one adds the factor $M_N/\umo$ to the rhs.

The static transition quadrupole moment $\mathcal{Q}_{B_{\bf
8}B_{\bf 10}}$ is related to the electric quadrupole form factor
$G^*_E$ at $Q^2=0$ by
\begin{equation}
\mathcal{Q}_{B_{\bf 8}B_{\bf
10}}=-6\sqrt{\frac{\umd}{\umo}}\frac{1}{\umo\omega}G^*_E(0)
\end{equation}

There is a special interest in the multipole ratio which directly
indicates a deviation from spherical symmetry
\begin{equation}
R_{EM}\equiv\frac{E2}{M1}=-\frac{G^*_E}{G^*_M}=\frac{A_{1/2}-A_{3/2}/\sqrt3}{A_{1/2}+\sqrt{3}A_{3/2}}.
\end{equation}

The electromagnetic width is given by the formula
\begin{equation}
\Gamma_{B_{\bf 10}B_{\bf
8}}=\frac{\omega^2}{2\pi}\frac{\umo}{\umd}\left\{|A_{3/2}|^2+|A_{1/2}|^2\right\}=\frac{\omega^2}{2\pi}\frac{\umo}{\umd}\left\{|M1|^2+3|E2|^2\right\}=\frac{\omega^3\alpha}{2\umo^2}\left\{G_M^{*2}+3G_E^{*2}\right\}
\end{equation}
where $\alpha=e^2/4\pi=1/137$.

In the IMF, one can extract $G_1(0)$ and $G_2(0)$ from the following
matrix elements \cite{Cardarelli}
\begin{eqnarray}
I_{\frac{3}{2},\frac{1}{2}}\equiv\langle P,\frac{3}{2}|\frac{J^+(0)}{2P^+}|P,\frac{1}{2}\rangle&=&\frac{q_L}{\sqrt{3}}\left[G_1+\frac{\umd-\umo}{2}G_2\right],\\
I_{\frac{1}{2},-\frac{1}{2}}\equiv\langle
P,\frac{1}{2}|\frac{J^+(0)}{2P^+}|P,-\frac{1}{2}\rangle&=&-\frac{q_L}{3}\left[-\frac{\umo}{\umd}G_1+\frac{\umd-\umo}{2}G_2\right].
\end{eqnarray}

\section{Scalar overlap integrals and quark distributions}

From the expression (\ref{Discrete level IMF}) and if we concentrate
on the spin part the contraction of two valence wave functions $F$
gives when the baryon helicity is flipped
\begin{equation}\label{MagnOQL}
F^\dag F\propto h(p)\frac{1}{|\up|}j(p)+\frac{p_z}{p^2}j^2(p).
\end{equation}
The physical interpretation is straightforward. The magnetic
operator conserves the struck quark helicity but the total baryon
spin is flipped. The quark absorbs one unit of angular momentum and
thus jumps from a $L_z=0$ state to a $L_z\neq 0$ state. This unit of
angular momentum comes from the photon.

The magnetic valence quark distribution is obtained by the
multiplication of two factors with structure (\ref{VectOQL}) where
the momentum is respectively $p_1$ and $p_2$ and a third factor with
structure (\ref{MagnOQL}) and momentum $p_3$. The expansion gives
the following function $D$
\begin{eqnarray}
&D^M(p_1,p_2,p_3)=h^2(p_1)h^2(p_2)\left[h(p_3)\frac{1}{|\up_3|}j(p_3)\right]+h^2(p_1)h^2(p_2)\frac{p_{3z}}{p_3^2}j^2(p_3)&\nonumber\\
&+4h^2(p_1)\left[h(p_2)\frac{p_{2z}}{|\up_2|}j(p_2)\right]\left[h(p_3)\frac{1}{|\up_3|}j(p_3)\right]+2h^2(p_1)j^2(p_2)\left[h(p_3)\frac{1}{|\up_3|}j(p_3)\right]&\nonumber\\
&+4h^2(p_1)\left[h(p_2)\frac{p_{2z}}{|\up_2|}j(p_2)\right]\frac{p_{3z}}{p_3^2}j^2(p_3)+4\left[h(p_1)\frac{p_{1z}}{|\up_1|}j(p_1)\right]\left[h(p_2)\frac{p_{2z}}{|\up_2|}j(p_2)\right]\left[h(p_3)\frac{1}{|\up_3|}j(p_3)\right]&\\
&+2h^2(p_1)j^2(p_2)\frac{p_{3z}}{p_3^2}j^2(p_3)+4\left[h(p_1)\frac{p_{1z}}{|\up_1|}j(p_1)\right]j^2(p_2)\left[h(p_3)\frac{1}{|\up_3|}j(p_3)\right]&\nonumber\\
&+4\left[h(p_1)\frac{p_{1z}}{|\up_1|}j(p_1)\right]\left[h(p_2)\frac{p_{2z}}{|\up_2|}j(p_2)\right]\frac{p_{3z}}{p_3^2}j^2(p_3)+j^2(p_1)j^2(p_2)\left[h(p_3)\frac{1}{|\up_3|}j(p_3)\right]&\nonumber\\
&+4\left[h(p_1)\frac{p_{1z}}{|\up_1|}j(p_1)\right]j^2(p_2)\frac{p_{3z}}{p_3^2}j^2(p_3)+j^2(p_1)j^2(p_2)\frac{p_{3z}}{p_3^2}j^2(p_3).&\nonumber
\end{eqnarray}
that is needed in the expression of the valence quark distribution
(\ref{Probability3q}). In the non-relativistic limit $j=0$ this
function $D$ is identically zero. Magnetic moment is a purely
relativistic property on the light cone. Relativistic corrections
(quark angular momentum) are clearly essential to compute baryon
magnetic and transition magnetic moments.

\begin{figure}[h]\begin{center}\begin{minipage}[c]{6cm}\begin{center}\includegraphics[width=5.3cm]{Phiplot3.eps}
\end{center}\end{minipage}\hspace{1cm}\begin{minipage}[c]{6cm}\begin{center}\includegraphics[width=5.3cm]{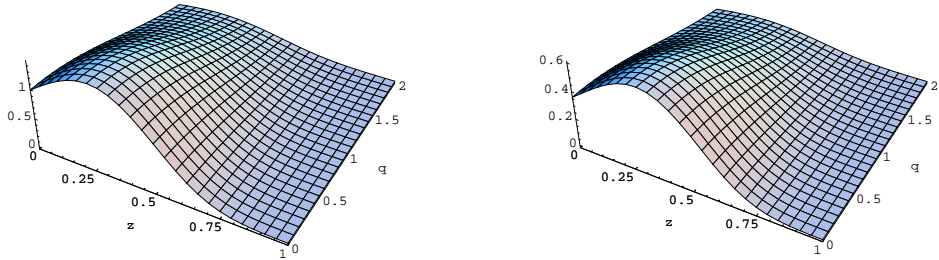}
\end{center}\end{minipage}\caption{\small{Probability distribution $\Phi^I(z,\uq_\perp)$ that three valence quarks leave the fraction $z$ of the baryon momentum and
transverse momentum $\uq_\perp$ to the quark-antiquark pair(s) in
the vector $I=V$ (left) and magnetic $I=M$ (right) cases plotted in
units of $M$ and normalized to $\Phi^V(0,0)=1$.
}}\label{Phiplot4}\end{center}
\end{figure}

The magnetic valence probability distribution $\Phi^M(z,\uq_\perp)$
is then obtained by integration over the valence quark momenta, see
eq. (\ref{Probability3q}) and is depicted in Fig. \ref{Phiplot4} in
vector $I=V$ and magnetic $I=M$ cases.

In the following we give the integrals appearing in each Fock sector
and the numerical values obtained for them. In the evaluation of the
scalar overlap integrals we have used the constituent quark mass
$M=345$ MeV, the Pauli-Villars mass $M_\textrm{PV}=556.8$ MeV for
the regularization of (\ref{Direct1})-(\ref{Direct4}) and the baryon
mass $\uM=1207$ MeV as it follows for the ``classical'' mass in the
mean field approximation \cite{Approximation}.

\subsection{$3Q$ scalar integral}

In the $3Q$ sector there is no quark-antiquark pair and thus only
one integral is involved. It corresponds to the valence quark
distribution without momentum left to the sea $\Phi^M(0,0)$. Its
precise value is
\begin{equation}
\Phi^M(0,0)=0.36210.
\end{equation}
In the non-relativistic limit $j=0$ it is zero. The extraction of
magnetic moment from spin-flip matrix elements is essentially a
relativistic procedure since it implies quarks to have orbital
angular momentum. In the usual light-cone approach this quark
orbital motion is introduced by the Melosh rotation.

\subsection{$5Q$ scalar integrals}

In the $5Q$ sector there is one quark-antiquark pair. Contractions
given by the direct diagram give four different integrals
$J=\pi\pi,33,\sigma\sigma,332$. If the magnetic operator acts on
valence quarks, the quark-antiquark pair is not affected and thus
the integrals present the vector structure for the sea. If the
magnetic operator acts on the sea, the valence quark are not
affected and the vector valence probability distribution has to be
used. The integrals present a new structure for the sea. Here the
quark-antiquark pair cannot use the unit of orbital angularmomentum
in its internal motion but can use it in its orbital motion.

Here are the numerical values obtained for the four new integrals
\begin{equation}
K^M_{\pi\pi}=0.01445,\qquad K^M_{33}= 0.00799,\qquad
K^M_{\sigma\sigma}=0.00566,\qquad K^V_{332}=0.03542.
\end{equation}
Let us have a look to the ratios $K^M_J/K^V_J$ with
$J=\pi\pi,33,\sigma\sigma$
\begin{equation}
\frac{K^M_{\pi\pi}}{K^V_{\pi\pi}}=0.39559,\qquad
\frac{K^M_{33}}{K^V_{33}}= 0.40482,\qquad
\frac{K^M_{\sigma\sigma}}{K^V_{\sigma\sigma}}=0.40420.
\end{equation}
The reduction is of the same order as in the $3Q$ sector. It is
however different from one structure to another due to the details
of the valence probability distributions. Note also that contrarily
to axial and tensor cases, the reduction factor is larger in the
$5Q$ sector than in the $3Q$ sector.

\section{Combinatoric Results}
\subsection{Octet baryons}

In the $3Q$ sector there is no quark-antiquark pair and thus only
valence quarks contribute to magnetic moments
\begin{equation}
\alpha_{M,q_\textrm{val}}^{(3)}=12\,\Phi^M(0,0)\,2\uM,\qquad
\beta_{M,q_\textrm{val}}^{(3)}=-3\,\Phi^M(0,0)\,2\uM,\qquad
\gamma_{M,q_\textrm{val}}^{(3)}=0.
\end{equation}
as it is expected from NQM. In the $5Q$ sector one has
\begin{eqnarray}
&\alpha_{M,q_\textrm{val}}^{(5)}=\frac{6}{5}\left(30K^M_{\pi\pi}-K^M_{33}+91K^M_{\sigma\sigma}\right)\,2\uM,\quad
\alpha_{M,q_\textrm{s}}^{(5)}=\frac{-96}{5}\,K^V_{332}\,2\uM,\quad
\alpha_{M,\bar q}^{(5)}=-6\,K^V_{332}\,2\uM,&\\
&\beta_{M,q_\textrm{val}}^{(5)}=\frac{-12}{25}\left(21K^M_{\pi\pi}+11K^M_{33}+52K^M_{\sigma\sigma}\right)\,2\uM,\quad
\beta_{M,q_\textrm{s}}^{(5)}=\frac{-48}{25}\,K^V_{332}\,2\uM,\quad
\beta_{M,\bar q}^{(5)}=6\,K^V_{332}\,2\uM,&\\
&\gamma_{M,q_\textrm{val}}^{(5)}=\frac{-12}{25}\left(3K^M_{\pi\pi}+8K^M_{33}+K^M_{\sigma\sigma}\right)\,2\uM,\quad
\gamma_{M,q_\textrm{s}}^{(5)}=\frac{6}{25}\,K^V_{332}\,2\uM,\quad
\gamma_{M,\bar q}^{(5)}=6\,K^V_{332}\,2\uM.&
\end{eqnarray}
Notice that the valence combinations are exactly the same as the
tensor ones. This is not a coincidence. In both cases one has to
compute spin-flip matrix elements. In the tensor case the operator
directly flips one valence quark spin while in the magnetic case the
valence quark jumps to another orbital state. At the end the result
is the same: the total baryon spin has been reversed. The
combinations are thus the same and the details concerning the
spin-flip are encoded in the integrals.

\subsection{Decuplet baryons}

In the $3Q$ sector there is no quark-antiquark pair and thus only
valence quarks contribute to the charges
\begin{equation}
\alpha'^{(3)}_{M,q_\textrm{val}}=\frac{18}{5}\,\Phi^M(0,0)\,2\uM,\qquad
\beta'^{(3)}_{M,q_\textrm{val}}=0.
\end{equation}
In the $5Q$ sector one has
\begin{eqnarray}
&\alpha'^{(5)}_{M,q_\textrm{val}}=\frac{9}{20}\left(40K^M_{\pi\pi}-7K^M_{33}+67K^M_{\sigma\sigma}\right)\,2\uM,\quad
\alpha'^{(5)}_{M,q_\textrm{s}}=\frac{171}{20}\,K^M_{332}\,2\uM,\quad
\alpha'^{(5)}_{M,\bar q}=\frac{-18}{5}\,K^M_{332}\,2\uM,&\\
&\beta'^{(5)}_{M,q_\textrm{val}}=\frac{-9}{20}\left(8K^M_{\pi\pi}+13K^M_{33}-K^M_{\sigma\sigma}\right)\,2\uM,\quad
\beta'^{(5)}_{M,q_\textrm{s}}=\frac{9}{20}\,K^M_{332}\,2\uM,\quad
\beta'^{(5)}_{M,\bar q}=\frac{63}{5}\,K^M_{332}\,2\uM.
\end{eqnarray}

\subsection{Antidecuplet baryons}

In the $5Q$ sector one has
\begin{eqnarray}
&\alpha''^{(5)}_{M,q_\textrm{val}}=\frac{-6}{5}\left(K^M_{33}-K^M_{\sigma\sigma}\right)\,2\uM,\quad
\alpha''^{(5)}_{M,q_\textrm{s}}=\frac{6}{5}\,K^M_{332}\,2\uM,\quad
\alpha''^{(5)}_{M,\bar q}=\frac{-12}{5}\,K^M_{332}\,2\uM,&\\
&\beta''^{(5)}_{M,q_\textrm{val}}=\frac{-6}{5}\left(K^M_{33}-K^M_{\sigma\sigma}\right)\,2\uM,\quad
\beta''^{(5)}_{M,q_\textrm{s}}=\frac{6}{5}\,K^M_{332}\,2\uM,\quad
\beta''^{(5)}_{M,\bar q}=\frac{24}{5}\,K^M_{332}\,2\uM.&
\end{eqnarray}

\section{Numerical results and discussion}

\subsection{Octet content}

We present in Table \ref{OctMagn1} the results we have obtained for
the proton magnetic form factors.
\begin{table}[h!]\begin{center}\caption{\small{Our magnetic content of the proton.}}
\begin{tabular}{c|ccc|ccc|ccc}
\hline\hline
Magnetic FF&\multicolumn{3}{c|}{$G_M^u$}&\multicolumn{3}{c|}{$G_M^d$}&\multicolumn{3}{c}{$G_M^s$}\rule{0pt}{3ex}\\
&$\bar q$&$q_s$&$q_{val}$&$\bar q$&$q_s$&$q_{val}$&$\bar
q$&$q_s$&$q_{val}$\rule{0pt}{3ex}\\\hline \rule{0pt}{3ex}
$3Q$&0&0&3.378&0&0&-0.845&0&0&0\\\rule{0pt}{3ex}
$3Q+5Q$&0&0.415&3.267&0.256&-0.036&-0.886&0.128&0.005&-0.033\rule[-2ex]{0pt}{5ex}\\
\hline\hline
\end{tabular}\label{OctMagn1}\end{center}
\end{table}
The $5Q$ component naturally introduces a strange contribution to
magnetic moments. In this computation it appears that most of
strange magnetic moment is due to the strange antiquark. One can see
that the $5Q$ component contributes significantly to magnetic
moments. Note that we have obtained a \emph{negative} strange
magnetic form factor $G_M^s(0)=-0.156\,\mu_N$ while SAMPLE
experiment \cite{SAMPLEM} by measuring the parity-violating
asymmetry at backward angles indicated a \emph{positive} value
$G_M^{(p)s}(0.1 \textrm{(GeV/c)}^2)=0.37\pm 0.20\pm 0.26\pm
0.07\,\mu_N$. This can probably be attributed to the fact that we
did not consider flavor $SU(3)$ symmetry breaking \footnote{This
supposition is motivated by the fact that a previous estimation in
the $\chi$QSM with the standard approach had given a positive value
\cite{Silva}.}. Another possible explanation of our bad strange
magnetic moment is the violation of Lorentz covariance due to the
impulse approximation (only one-body current and not many-body
current is considered) and the truncation of the Fock space
\cite{PartFock}. It would not be the case if one includes the
complete Fock space \cite{FullFock}. This introduces unphysical form
factors which may have a strong impact on the evaluation of the
strange magnetic moment and even change its sign \cite{Meffect}.

Octet magnetic moments are rather well known. In Table
\ref{OctMagn2} we compare our results with experimental values given
by the Particle Data Group.
\begin{table}[h!]\begin{center}\caption{\small{Our octet magnetic moments compared with experimental data \cite{PDG}.}}
\begin{tabular}{c|cccc|cccc|c}
\hline\hline
&\multicolumn{4}{c|}{$3Q$}&\multicolumn{4}{c|}{$3Q+5Q$}&\rule{0pt}{3ex}\\
$B$&$G_M^{(B)u}$&$G_M^{(B)d}$&$G_M^{(B)s}$&$\mu_B$&$G_M^{(B)u}$&$G_M^{(B)d}$&$G_M^{(B)s}$&$\mu_B$&Exp.\rule{0pt}{3ex}\\\hline
\rule{0pt}{3ex}
$p^+_\mathbf{8}$&3.378&-0.845&0&2.534&3.683&-1.178&-0.156&2.900&2.793\\\rule{0pt}{3ex}
$n^0_\mathbf{8}$&-0.845&3.378&0&-1.689&-1.178&3.683&-0.156&-1.961&-1.913\\\rule{0pt}{3ex}
$\Lambda^0_\mathbf{8}$&0&0&2.534&-0.845&-0.198&-0.198&2.744&-0.981&-$0.613\pm0.004$\\\rule{0pt}{3ex}
$\Sigma^+_\mathbf{8}$&3.378&0&-0.845&2.534&3.683&-0.156&-1.178&2.900&$2.458\pm0.010$\\\rule{0pt}{3ex}
$\Sigma^0_\mathbf{8}$&1.689&1.689&-0.845&0.845&1.763&1.763&-1.178&0.981&-\\\rule{0pt}{3ex}
$\Sigma^-_\mathbf{8}$&0&3.378&-0.845&-0.845&-0.156&3.683&-1.178&-0.939&-$1.160\pm0.025$\\\rule{0pt}{3ex}
$\Xi^0_\mathbf{8}$&-0.845&0&3.378&-1.689&-1.178&-0.156&3.683&-1.961&-$1.250\pm0.014$\\\rule{0pt}{3ex}
$\Xi^-_\mathbf{8}$&0&-0.845&3.378&-0.845&-0.156&-1.178&3.683&-0.939&-$0.651\pm0.003$\\\rule{0pt}{3ex}
$\Sigma^0_\mathbf{8}\to\Lambda^0_\mathbf{8}$&4.388&-4.388&0&1.463&5.095&-5.095&0&1.698&$1.61\pm0.08$\rule[-2ex]{0pt}{5ex}\\
\hline\hline
\end{tabular}\label{OctMagn2}\end{center}
\end{table}
At the $3Q$ level, all relations specific to $SU(6)$ NQM are
reproduced. One can notice that the $5Q$ component improves the
agreement between theoretical and experimental ratios\footnote{Since
in our approach flavor $SU(3)$ symmetry is not broken, we
concentrate our attention on particles with no explicit strange
quark, \emph{i.e.} nucleons and $\Delta$.}
\begin{equation}
\frac{\mu_{p^+_\mathbf{8}}}{\mu_{n^0_\mathbf{8}}}=-1.4786\qquad(\textrm{Exp.:
} -1.4599, \textrm{ NQM: } -1.5).
\end{equation}
One can also see that the $5Q$ component is essential in order to
reproduce fairly well proton and neutron magnetic moments. The
agreement with other particles is less good. Note that the other
particles contain explicitly strange quarks and that we have
computed only in the flavor $SU(3)$ symmetry limit where
$\mu_{p^+_\mathbf{8}}=\mu_{\Sigma^+_\mathbf{8}}$.

\subsection{Decuplet content}

We present in Table \ref{DecMagn1} the results we have obtained for
the $\Delta^{++}$ magnetic form factors.
\begin{table}[h!]\begin{center}\caption{\small{Our magnetic content of the $\Delta^{++}$.\newline}}
\begin{tabular}{c|ccc|ccc|ccc}
\hline\hline
Magnetic FF&\multicolumn{3}{c|}{$G_M^u$}&\multicolumn{3}{c|}{$G_M^d$}&\multicolumn{3}{c}{$G_M^s$}\rule{0pt}{3ex}\\
&$\bar q$&$q_s$&$q_{val}$&$\bar q$&$q_s$&$q_{val}$&$\bar
q$&$q_s$&$q_{val}$\rule{0pt}{3ex}\\\hline \rule{0pt}{3ex}
$3Q$&0&0&7.601&0&0&0&0&0&0\\\rule{0pt}{3ex}
$3Q+5Q$&0.093&1.347&7.332&0.650&0.023&-0.140&0.650&0.023&-0.140\rule[-2ex]{0pt}{5ex}\\
\hline\hline
\end{tabular}\label{DecMagn1}\end{center}
\end{table}
The $5Q$ component naturally introduces a contribution from down and
strange quarks to magnetic moments. In this computation it appears
that most of these hidden flavor magnetic moments is due to the down
and strange antiquarks.

Decuplet magnetic moments are not well known. Experiments on
$\Delta$ are notoriously difficult due to its short mean life time
of only about $6\times 10^{-34}$s. However decuplet magnetic moments
have been theoretically investigated in several models such as
quenched lattice QCD theory \cite{DecuLatt}, quark models
\cite{DecuQmod}, chiral bag model \cite{DecuChBmod}, $\chi$PT
\cite{DecuchiPT}, QCD sum rules \cite{DecuQCDsum} and $\chi$QM
\cite{DecuchiQmod}. In Table \ref{DecMagn2} we compare our results
with experimental values given by the Particle Data Group.
\begin{table}[h!]\begin{center}\caption{\small{Our decuplet magnetic moments compared with experimental data \cite{PDG}.}}
\begin{tabular}{c|cccc|cccc|c}
\hline\hline
&\multicolumn{4}{c|}{$3Q$}&\multicolumn{4}{c|}{$3Q+5Q$}&\rule{0pt}{3ex}\\
$B$&$G_M^{(B)u}$&$G_M^{(B)d}$&$G_M^{(B)s}$&$\mu_B$&$G_M^{(B)u}$&$G_M^{(B)d}$&$G_M^{(B)s}$&$\mu_B$&Exp.\rule{0pt}{3ex}\\\hline
\rule{0pt}{3ex}
$\Delta^{++}_\mathbf{10}$&7.601&0&0&5.067&8.586&-0.767&-0.767&6.236&3.7
to 7.5\\\rule{0pt}{3ex}
$\Delta^+_\mathbf{10}$&5.067&2.534&0&2.534&5.468&2.350&-0.767&3.118&$2.7^{+1.0}_{-1.3}\pm
1.5\pm 3$\\\rule{0pt}{3ex}
$\Delta^0_\mathbf{10}$&2.534&5.067&0&0&2.350&5.468&-0.767&0&-\\\rule{0pt}{3ex}
$\Delta^-_\mathbf{10}$&0&7.601&0&-2.534&-0.767&8.586&-0.767&-3.118&-\\\rule{0pt}{3ex}
$\Sigma^+_\mathbf{10}$&5.067&0&2.534&2.534&5.468&-0.767&2.350&3.118&-\\\rule{0pt}{3ex}
$\Sigma^0_\mathbf{10}$&2.534&2.534&2.534&0&2.350&2.350&2.350&0&-\\\rule{0pt}{3ex}
$\Sigma^-_\mathbf{10}$&0&5.067&2.534&-2.534&-0.767&5.468&2.350&-3.118&-\\\rule{0pt}{3ex}
$\Xi^0_\mathbf{10}$&2.534&0&0&5.067&2.350&-0.767&5.468&0&-\\\rule{0pt}{3ex}
$\Xi^-_\mathbf{10}$&0&2.534&5.067&-2.534&-0.767&2.350&5.468&-3.118&-\\\rule{0pt}{3ex}
$\Omega^-_\mathbf{10}$&0&0&7.601&-2.534&-0.767&-0.767&8.586&-3.118&-$2.02\pm0.05$\rule[-2ex]{0pt}{5ex}\\
\hline\hline
\end{tabular}\label{DecMagn2}\end{center}
\end{table}
As suggested by the approximate $SU(6)$ symmetry we have obtained
$\mu_{\Delta^+_\mathbf{10}}\approx\mu_{p^+_\mathbf{8}}$. Our
computation indicates that $\mu_{\Delta^+_\mathbf{10}}$ is a bit
larger than $\mu_{p^+_\mathbf{8}}$ while the present experimental
value suggests a smaller value. Experimental error bars are still
large and do not exclude at all
$\mu_{\Delta^+_\mathbf{10}}>\mu_{p^+_\mathbf{8}}$.

\subsection{Antidecuplet content}

We present in Table \ref{ADecMagn1} the results we have obtained for
the $\Theta^+$ magnetic form factors.
\begin{table}[h!]\begin{center}\caption{\small{Our magnetic content of the $\Theta^+$.}}
\begin{tabular}{c|ccc|ccc|ccc}
\hline\hline
Magnetic FF&\multicolumn{3}{c|}{$G_M^u$}&\multicolumn{3}{c|}{$G_M^d$}&\multicolumn{3}{c}{$G_M^s$}\rule{0pt}{3ex}\\
&$\bar q$&$q_s$&$q_{val}$&$\bar q$&$q_s$&$q_{val}$&$\bar
q$&$q_s$&$q_{val}$\rule{0pt}{3ex}\\\hline \rule{0pt}{3ex}
$5Q$&0&2.452&-0.161&0&2.245&-0.161&4.905&0&0\rule[-2ex]{0pt}{5ex}\\
\hline\hline
\end{tabular}\label{ADecMagn1}\end{center}
\end{table}
Contrarily to ordinary baryons most of pentaquark magnetic moment is
due to the sea and not valence quarks.

Antidecuplet magnetic moments are of course not known. In Table
\ref{ADecMagn2} we give our predictions for pentaquark magnetic
moments.
\begin{table}[h!]\begin{center}\caption{\small{Our antidecuplet magnetic moments.}}
\begin{tabular}{c|cccc}
\hline\hline
&\multicolumn{4}{c}{$5Q$}\rule{0pt}{3ex}\\
$B$&$G_M^{(B)u}$&$G_M^{(B)d}$&$G_M^{(B)s}$&$\mu_B$\rule{0pt}{3ex}\\\hline
\rule{0pt}{3ex}
$\Theta^+_\mathbf{\overline{10}}$&2.291&2.291&4.905&2.398\\\rule{0pt}{3ex}
$p^+_\mathbf{\overline{10}}$&2.291&3.162&4.033&2.398\\\rule{0pt}{3ex}
$n^0_\mathbf{\overline{10}}$&3.162&2.291&4.033&0\\\rule{0pt}{3ex}
$\Sigma^+_\mathbf{\overline{10}}$&2.291&4.033&3.162&2.398\\\rule{0pt}{3ex}
$\Sigma^0_\mathbf{\overline{10}}$&3.162&3.162&3.162&0\\\rule{0pt}{3ex}
$\Sigma^-_\mathbf{\overline{10}}$&4.033&2.291&3.162&-2.398\\\rule{0pt}{3ex}
$\Xi^+_\mathbf{\overline{10}}$&2.291&4.905&2.291&2.398\\\rule{0pt}{3ex}
$\Xi^0_\mathbf{\overline{10}}$&3.162&4.033&2.291&0\\\rule{0pt}{3ex}
$\Xi^-_\mathbf{\overline{10}}$&4.033&3.162&2.291&-2.398\\\rule{0pt}{3ex}
$\Xi^{--}_\mathbf{\overline{10}}$&4.905&2.291&2.291&-4.797\rule[-2ex]{0pt}{5ex}\\
\hline\hline
\end{tabular}\label{ADecMagn2}\end{center}
\end{table}
We have obtained for the positively charged pentaquarks a magnetic
moment a bit smaller than the proton one. Such a large value is
intriguing if we compare with other studies where the pentaquark
magnetic moment is either small or negative \cite{Thetamagn}. Since
there is an explicit strange antiquark, flavor $SU(3)$ symmetry
breaking may have a non negligible impact on the results.
Nevertheless it would be quite surprising that at the end the
magnetic moment becomes very small or even changes its sign. A naive
estimation of the pentaquark magnetic moment
$\mu_{\Theta^+_\mathbf{\overline{10}}}\approx\mu_{p^+_\mathbf{\overline{10}}}\approx\mu_{\Delta^+_\mathbf{10}}\approx\mu_{p^+_\mathbf{8}}$
would in fact support our result: a large positive $\Theta^+$
magnetic moment.

\section{Octet-to-decuplet transition moments}

Beside magnetic moments, octet-to-decuplet transition magnetic
moments have especially focused attention since 1979. The proton
being a spin-$1/2$ particle, no intrinsic quadrupole moment can be
directly measured because angular momentum conservation forbids a
non-zero element of a ($L=2$) quadrupole operator between spin 1/2
states. On the contrary, $\Delta$ is a spin-$3/2$ particle where
such quadrupole can be in principle measured. For collective
rotation of the deformed intrinsic state \cite{Bohrbook}, the
relation between the spectroscopic quadrupole moment $\uQcal$
measured in the laboratory frame and the intrinsic moment $\uQcal_0$
in the body-fixed intrinsic frame is given by
\begin{equation}
\uQcal=\frac{3K^2-J(J+1)}{(J+1)(2J+3)}\,\uQcal_0.
\end{equation}
where $J$ is the total angular momentum of the system in the
laboratory frame, K is the projection of $J$ onto the $z$-axis of
the body-fixed intrinsic frame and the sub-state with azimuthal
quantum number $M=J$ has been considered. The ratio between
$\uQcal_0$ and $\uQcal$ represents the averaging of the
non-spherical distribution due to the rotational motion as seen in
the laboratory frame. For spin-$1/2$ particle one has indeed
$\uQcal=0$ even for $\uQcal_0\neq 0$ \cite{Alexandrou}.

The electromagnetic transition $\gamma^*N\to\Delta$ allows one to
access to quadrupole moments of both proton and $\Delta$. Spin and
parity conservation permit only 3 multipoles to contribute to the
transition: magnetic dipole ($M1$), electric quadrupole ($E2$) and
Coulomb quadrupole ($C2$).

As we have seen, only one magnetic $\alpha_M^{8\to 10}$ and one
electric $\alpha_E^{8\to 10}$ parameters are needed to describe in
the flavor $SU(3)$ limit all magnetic dipole and electric quadrupole
form factors of the octet-to-decuplet transitions. Here are the
combinations obtained
\begin{eqnarray}
\alpha^{(3)8\to 10}_M&=&\frac{12}{\sqrt{5}}\,\Phi^M(0,0)\,2\uM,\\
\alpha^{(3)8\to 10}_E&=&0,\\
\alpha^{(5)8\to 10}_M&=&\frac{18}{5\sqrt{5}}(13K^M_{\pi\pi}+5K^M_{33}+8K^M_{332}+28K^M_{\sigma\sigma m})\,2\uM,\\
\alpha^{(5)8\to
10}_E&=&\frac{-18}{5\sqrt{5}}(K^M_{\pi\pi}-3K^M_{33})\,2\uM.
\end{eqnarray}
The $3Q$ sector, being similar to NQM, does not provide us with a
non-zero electric quadrupole. A non-zero contribution comes from the
$5Q$ sector. We have already discussed the quadrupolar deformation
of decuplet baryons in the vector, axial and tensor cases. All was
connected to a unique quadrupolar structure of sea
$K_{\pi\pi}-3K_{33}$. One can see that the quadrupole electric
transition parameter is proportional to this pion quadrupolar
structure. There is no direct contribution from the sea since no
term in $K^M_{332}$ is present. The deformation of the system is
explicitly due to the pion quadrupole moment in our approach.
Concerning the magnetic transition moment one can easily check that
in the $3Q$ sector we reproduce the $SU(6)$ prediction
$\mu_{p^+_\mathbf{8}\Delta^+_\mathbf{10}}=\frac{2\sqrt{2}}{3}\,\mu_{p^+_\mathbf{8}}$.

Let us have a look to the numerical values obtained and collected in
Table \ref{OtoDtrans}.
\begin{table}[h!]\begin{center}\caption{\small{Our octet-to-decuplet moments.}}
\begin{tabular}{r@{$\to$\,}l|cccc|cccc}
\hline\hline
\multicolumn{2}{c|}{$3Q$}&$G_M^{*u}$&$G_M^{*d}$&$G_M^{*s}$&$G_M^*$&$G_E^{*u}$&$G_E^{*d}$&$G_E^{*s}$&$G_E^*$\rule{0pt}{3ex}\\\hline\rule{0pt}{3ex}
$p^+_\mathbf{8}$&$\Delta^+_\mathbf{10}$&2.389&-2.389&0&2.389&0&0&0&0\\\rule{0pt}{3ex}
$n^0_\mathbf{8}$&$\Delta^0_\mathbf{10}$&2.389&-2.389&0&2.389&0&0&0&0\\\rule{0pt}{3ex}
$\Sigma^+_\mathbf{8}$&$\Sigma^+_\mathbf{10}$&-2.389&0&2.389&-2.389&0&0&0&0\\\rule{0pt}{3ex}
$\Lambda^0_\mathbf{8}$&$\Sigma^0_\mathbf{10}$&2.069&-2.069&0&2.069&0&0&0&0\\\rule{0pt}{3ex}
$\Sigma^0_\mathbf{8}$&$\Sigma^0_\mathbf{10}$&-1.194&-1.194&2.389&-1.194&0&0&0&0\\\rule{0pt}{3ex}
$\Sigma^-_\mathbf{8}$&$\Sigma^-_\mathbf{10}$&0&-2.389&2.389&0&0&0&0&0\\\rule{0pt}{3ex}
$\Xi^0_\mathbf{8}$&$\Xi^0_\mathbf{10}$&-2.389&0&2.389&-2.389&0&0&0&0\\\rule{0pt}{3ex}
$\Xi^-_\mathbf{8}$&$\Xi^-_\mathbf{10}$&0&-2.389&2.389&0&0&0&0&0\rule[-2ex]{0pt}{5ex}\\
\hline
\multicolumn{2}{c|}{$3Q+5Q$}&$G_M^{*u}$&$G_M^{*d}$&$G_M^{*s}$&$G_M^*$&$G_E^{*u}$&$G_E^{*d}$&$G_E^{*s}$&$G_E^*$\rule{0pt}{3ex}\\\hline\rule{0pt}{3ex}
$p^+_\mathbf{8}$&$\Delta^+_\mathbf{10}$&2.820&-2.820&0&2.820&0.026&-0.026&0&0.026\\\rule{0pt}{3ex}
$n^0_\mathbf{8}$&$\Delta^0_\mathbf{10}$&2.820&-2.820&0&2.820&0.026&-0.026&0&0.026\\\rule{0pt}{3ex}
$\Sigma^+_\mathbf{8}$&$\Sigma^+_\mathbf{10}$&-2.820&0&2.820&-2.820&-0.026&0&0.026&-0.026\\\rule{0pt}{3ex}
$\Lambda^0_\mathbf{8}$&$\Sigma^0_\mathbf{10}$&2.443&-2.443&0&2.443&0.022&-0.022&0&0.022\\\rule{0pt}{3ex}
$\Sigma^0_\mathbf{8}$&$\Sigma^0_\mathbf{10}$&-1.410&-1.410&2.820&-1.410&-0.013&-0.013&0.026&-0.013\\\rule{0pt}{3ex}
$\Sigma^-_\mathbf{8}$&$\Sigma^-_\mathbf{10}$&0&-2.820&2.820&0&0&-0.026&0.026&0\\\rule{0pt}{3ex}
$\Xi^0_\mathbf{8}$&$\Xi^0_\mathbf{10}$&-2.820&0&2.820&-2.820&-0.026&0&0.026&-0.026\\\rule{0pt}{3ex}
$\Xi^-_\mathbf{8}$&$\Xi^-_\mathbf{10}$&0&-2.820&2.820&0&0&-0.026&0.026&0\rule[-2ex]{0pt}{5ex}\\
\hline\hline
\end{tabular}\label{OtoDtrans}\end{center}
\end{table}
For the nucleon-to-$\Delta$ transition the Particle Data Group gives
values for helicity amplitudes instead of Jones and Scadron
multipole form factors $G_M^*$ and $G_E^*$. Table \ref{Comparaison}
gives the comparison with our computed observables.
\begin{table}[h!]\begin{center}\caption{\small{Comparison between theoretical and experimental transition observables \cite{PDG}.}}
\begin{tabular}{c|cccccc}
\hline\hline $p^+_\mathbf{8}\to\Delta^+_\mathbf{10}$&$A_{3/2}$
(GeV$^{-1/2}$)&$A_{1/2}$ (GeV$^{-1/2}$)&$G_M^*$ ($\mu_N$)&$G_E^*$
($\mu_N$)&$R_{EM}$&$\Gamma_{p\Delta}$ (MeV)\rule{0pt}{3ex}\\\hline
\rule{0pt}{3ex} \rule{0pt}{3ex}
$3Q$&-0.296&-0.171&2.389&0&0&0.411\\\rule{0pt}{3ex}
$3Q+5Q$&-0.232&-0.129&2.820&0.026&-0.9\%&0.573\\\rule{0pt}{3ex}
Exp.&-$0.250\pm 0.008$&-$0.135\pm 0.006$&2.798&0.046&-1.6\%&0.564\rule[-2ex]{0pt}{5ex}\\
\hline\hline
\end{tabular}\label{Comparaison}\end{center}
\end{table}
The $5Q$ sector is once more essential to reproduce experimental
data, \emph{i.e.} magnitude of $G_M^*$, correct sign for $G_E^*$ and
an electromagnetic decay width in fair agreement with experiments.
Even the ratio between $\mu_{p^+_\mathbf{8}\Delta^+_\mathbf{10}}$
and $\mu_{p^+_\mathbf{8}}$ is improved
\begin{equation}
\frac{\mu_{p^+_\mathbf{8}\Delta^+_\mathbf{10}}}{\mu_{p^+_\mathbf{8}}}=0.9727\qquad
(\textrm{Exp.: } 1.0017, \textrm{ NQM: } 0.9428)
\end{equation}
compared to the $SU(6)$ prediction. We have obtained quite good
results from an \emph{ab initio} computation and showed the
importance and the direct link between the baryon non-spherical
shape and the quadrupole structure of the pion cloud. The importance
of the pion cloud contribution is supported by lattice QCD
\cite{Lattpioncloud}, chiral perturbation theory
\cite{ChPTpioncloud} and phenomenological approaches
\cite{Modelpioncloud}. Restricted to the $3Q$ sector only, our
calculations reproduce all $SU(6)$ results: $\mu_p/\mu_n=-3/2$,
$\mu_{\Delta^+}=\mu_p$, $\mu_{N\Delta}/\mu_p=2\sqrt{2}/3$ and
$E2/M1=0$. Adding a quark-antiquark pair improves the agreement with
experimental ratios.

\section{Octet-to-antidecuplet transition moments}

Exotic members of the antidecuplet can easily be recognized because
their quantum numbers cannot be obtained from three quarks only. We
are concerned with the problem of the identification of a nucleon
resonance to a non-exotic member of this antidecuplet. It is then
interesting to study the electromagnetic transitions between octet
and antidecuplet\footnote{We remind that we will not discuss
decuplet-to-antidecuplet transition since they are forbidden by
flavor $SU(3)$ symmetry.}.

There is nowadays a lot of interest in the eta photoproduction on
nucleon. A resonance structure is seen in the photoproduction on the
neutron while it is absent on proton. Moreover if this corresponds
to a new resonance it seems to have a rather small width. The
questions to solve are first to check that it is indeed a new
resonance and second check if it could be a non-exotic partner of
$\Theta^+_\mathbf{\overline{10}}$ \cite{candidate}. Flavor $SU(3)$
symmetry and antidecuplet naturally explains the suppression on
proton but this is of course not enough to prove that it is a
non-exotic pentaquark. This interpretation is however very appealing
because of its elegant simplicity.

Like in the octet-to-decuplet case, only one parameter
$\alpha_M^{8\to\overline{10}}$ is needed to describe all
octet-to-antidecuplet magnetic dipole form factors
\begin{equation}
\alpha_M^{(5)8\to\overline{10}}=\frac{-2}{5}\sqrt{\frac{3}{5}}\left(3K_{\pi\pi
}+4K_{33}+9K_{332}+5K_{\sigma\sigma}\right)
\end{equation}

Let us have a look to the numerical values obtained collected in
Table \ref{OtoAtrans}.
\begin{table}[h!]\begin{center}\caption{\small{Our octet-to-antidecuplet transition magnetic moments.\newline}}
\begin{tabular}{r@{$\to$\,}l|cccc}
\hline\hline
\multicolumn{2}{c|}{$5Q$}&$G_M^{u}$&$G_M^{d}$&$G_M^{s}$&$G_M$\rule{0pt}{3ex}\\\hline\rule{0pt}{3ex}
$p^+_\mathbf{8}$&$p^+_\mathbf{\overline{10}}$&0&-0.45&0.45&0\\\rule{0pt}{3ex}
$n^0_\mathbf{8}$&$n^0_\mathbf{\overline{10}}$&-0.45&0&0.45&-0.45\\\rule{0pt}{3ex}
$\Sigma^+_\mathbf{8}$&$\Sigma^+_\mathbf{\overline{10}}$&0&-0.45&0.45&0\\\rule{0pt}{3ex}
$\Lambda^0_\mathbf{8}$&$\Sigma^0_\mathbf{\overline{10}}$&-0.22&-0.22&0.45&-0.22\\\rule{0pt}{3ex}
$\Sigma^0_\mathbf{8}$&$\Sigma^0_\mathbf{\overline{10}}$&0.39&-0.39&0&0.39\\\rule{0pt}{3ex}
$\Sigma^-_\mathbf{8}$&$\Sigma^-_\mathbf{\overline{10}}$&-0.45&0&0.45&-0.45\\\rule{0pt}{3ex}
$\Xi^0_\mathbf{8}$&$\Xi^0_\mathbf{\overline{10}}$&0.45&-0.45&0&0.45\\\rule{0pt}{3ex}
$\Xi^-_\mathbf{8}$&$\Xi^-_\mathbf{\overline{10}}$&0.45&-0.45&0&0.45\rule[-2ex]{0pt}{5ex}\\
\hline\hline
\end{tabular}\label{OtoAtrans}\end{center}
\end{table}

The value obtained for $|\mu_{n^0_{\overline{10}}\to n^0_8}|=0.45$
$\mu_N$ is consistent with previous expectation $(0.10-0.56)$
$\mu_N$ \cite{Ghil} but is larger than the estimate $(0.13-0.37)$
$\mu_N$ \cite{Eta}. The smallness of the numerical value (for
comparison $\mu_{\Delta\to N}\approx 3$ $\mu_N$) could be explained
in the same way as for the smallness of the $\Theta^+$ width. Since
axial and vector (and thus magnetic) currents connect only Fock
states with the same number of particles, the dominant $5Q$
component of pentaquarks are connected to the subleading $5Q$
component of octet baryons. In the non-relativistic octet baryons
are composed of only three quarks and then the transition magnetic
moments vanish.

\newpage\thispagestyle{empty}\cleardoublepage
\chapter{Conclusion and Outlook}

Throughout the present thesis we managed to study light baryon
properties by means of one of the most successful baryon models on
the market, namely Chiral Quark-Soliton Model ($\chi$QSM). This
model has given quite a good description of the nucleon and other
light baryons when studied in the usual instant form of dynamics.
$\chi$QSM is based on the Spontaneous Chiral Symmetry Breaking
(SCSB) of QCD and on a large $N_C$ logic, where $N_C$ is the number
of colors in QCD, allowing one to study baryons in a relativistic
mean field approximation. While quantum fluctuations around the mean
pion field can reasonably be neglected, rotations in spin-flavor
space are not strongly suppressed. However the usual instant time
approach of $\chi$QSM is based on an expansion in angular velocity
which is considered as being small. While this is reasonable for
ordinary baryons, \emph{i.e.} baryons made of three quarks, this
assumption is questionable for exotic baryons, \emph{i.e.} baryons
made of more than three quarks such as pentaquarks.

Recently, Diakonov, Petrov and Polyakov have formulated $\chi$QSM in
the Infinite Momentum Frame (IMF) or equivalently on the light cone.
This new approach offers many advantages. Thanks to the simple
structure of the light-cone vacuum the concept of wave function
borrowed from Quantum Mechanics is well defined. Any baryon can thus
be described by its light-cone wave function. This wave function
encodes a huge amount of information, \emph{e.g.} one can in
principle obtain Parton Distribution Functions (PDF), Form Factors
(FF) or even Generalized Parton Distributions (GPD) from an overlap
of these wave functions. A general expression for all light baryon
light-cone wave functions has been obtained by Diakonov, Petrov and
Polyakov. Moreover rotations of the mean pion field are treated
exactly by integrating over the $SU(3)$ Haar measure.
\newline\newline
\textbf{What we did}

We have used this formulation of $\chi$QSM in the IMF to study light
baryon charges in the limit of flavor $SU(3)$ symmetry. Thanks to
the light-cone wave functions and a Fock expansion in multi-quark
states we have computed the accessible baryon charges at leading
twist, namely vector, axial, tensor charges and magnetic moments.
These charges can be obtained on the light cone by means of matrix
elements of the ``good'' component, \emph{i.e.} not spoiled by
dynamics, of the corresponding quark bilinears.

The charges have been computed for all the three lightest baryon
multiplets, namely octet, decuplet and exotic antidecuplet. We have
investigated the $3Q$ and $5Q$ Fock sector for all these charges.
Furthermore concerning vector and axial charges of octet and
antidecuplet baryons, the leading part of the $7Q$ Fock sector has
also been explored. This expansion in Fock space allows one to study
the effect of additional (non-perturbative) quark-antiquark pairs in
a given baryon or with other words its pion cloud. Pions being the
lightest hadrons and being required by SCSB, they are expected to
have a non-negligible role in explaining low-energy properties of
baryons.

Contrarily to the Naive Quark Model (NQM), $\chi$QSM provides a
fully relativistic description of the baryons, \emph{i.e.} quark
orbital angular momentum is a natural part of the wave function. We
have compared the non-relativistic limit of $\chi$QSM with the exact
relativistic description to estimate the magnitude of relativistic
corrections. Quarks having a sizeable velocity inside a hadron are
expected to receive important contributions from relativistic
corrections.

All computed charges have been split into flavor contributions.
Moreover since the model makes a clear distinction between valence
quarks and sea quarks it has been possible to extract individual
contribution of valence and sea quark as well as antiquarks. The
results obtained can then be compared with our present poor
knowledge of the baryon sea and give maybe not quantitative but at
least qualitative predictions for lots of unobserved charges and
contributions, especially for the decuplet and antidecuplet baryons.

We have also considered some axial and magnetic \emph{intra}- and
\emph{inter}- multiplet transitions. This allowed us to give an
estimation of the lightest pentaquark $\Theta^+$ width.
Electromagnetic transitions also allowed us to investigate a
possible deformation of the $N\Delta$ system due to a quadrupolar
moment of the pion cloud.

Diagrams in each Fock sector can be separated into two classes. The
direct class is the leading contribution and corresponds to no
specific change in the baryon content. The exchange class is the
subleading contribution and corresponds to an exchange of roles
played by the quarks inside baryon. In Appendix B we have given
general and useful tools to find all non-equivalent diagrams in a
given Fock sector and the corresponding overall factors and signs.
\newline\newline
\textbf{What we have obtained}

$\chi$QSM is often thought as an interpolation between NQM and
Skyrme model. If we restrict ourselves to the non-relativistic $3Q$
sector all NQM predictions for vector, axial, tensor charges,
magnetic moments and transitions are recovered. Moreover octet and
decuplet $3Q$ spin-flavor wave functions are similar to the
well-known $SU(6)$ ones. Allowing quark orbital angular momentum
only change charges by a common factor, in accordance with the usual
light-cone approach based on Melosh rotation. This rotation
guarantees that the baryon has definite $J$ and $J_z$ in its rest
frame. This contribution is purely kinematical. However a covariant
light-cone wave function needs also some dynamical contribution
which is naturally present in the approach we used. The result is
that the factors we have obtained are smaller than the ones from
Melosh rotation only. This means that quark orbital angular momentum
has a smaller impact than estimated with Melosh rotation but is
still essential.

We have also computed the effect of the pion cloud. First the $5Q$
contribution has been evaluated followed by the $7Q$ component. The
normalization of baryon states allowed us to estimates the actual
fraction of octet, decuplet and antidecuplet baryons made of $3Q$,
$5Q$ and even $7Q$ (but not for the decuplet because of the far
greater complexity of its $7Q$ wave function). It turned out that
roughly 3/4 of ordinary baryons are made of $3Q$, 1/5 made of $5Q$
and 1/20 made of $7Q$. On the contrary exotic pentaquark appeared as
3/5 made of $5Q$ and 2/5 made of $7Q$. This means that exotic
baryons are more sensitive to the subleading Fock component.

The effect of exchange diagrams in the non-relativistic $5Q$ sector
has been computed. It turned out that exchange diagrams have a small
impact $\sim 1\%$ contrarily to what was naively expected before.
This can be in some sense understood physically by the fact that
exchange diagrams imply a redistribution of roles played by the
quarks. The transition implies some correlation among these quarks
and thus reduces the phase space.

The definition of valence quark we used (quark filling the discrete
level) does not coincide with the usual definition of the literature
(total number of quarks \emph{minus} total number of antiquarks,
\emph{i.e.} the net number of quarks). This comes from the fact that
quark-antiquark pairs are commonly thought as produced by the
perturbative process of gluon splitting. Experiments however suggest
that there is also a non-perturbative amount of quark-antiquark
pair. This non-perturbative amount corresponds to our pion cloud.
Consequently the emission of a pion by a valence quark can change
the composition of the valence sector. This means that even if there
is no net strange quark in the nucleon, these strange quarks may
access to the valence sector leading naturally to an asymmetry in
the strangeness distribution $s(z)-\bar s(z)\neq 0$ but
$\int\ud\,z\,(s(z)-\bar s(z))=0$ effectively suggested by the
recently observed NuTeV anomaly. Another consequence is that since
tensor charges measure only valence quark transversity, the pion
cloud can generate a non-vanishing contribution of strange quarks to
these tensor charges.

Flavor $SU(3)$ symmetry does not impose a flavor symmetric sea. As
suggested by experiments we have obtained an excess of $\bar d$ over
$\bar u$ in the proton but the difference is one order of magnitude
smaller than what is suggested by the violation of the Gottfried sum
rule. Experiments also suggest that $\Delta\bar u\simeq-\Delta\bar
d$. We have indeed obtained this difference in sign between up and
down antiquark longitudinally polarized distributions but the
absolute sign obtained is opposite to what present data favor. Let
us also note the positive sign of $\Delta s$ in the nucleon in
accord with most of models but opposite to phenomenological
extractions. Moreover the magnitude obtained is one order too small.
Let us however mention the recent HERMES results in which the
strange contribution to proton spin is obtained by means of a
different technique than usual. This experiment suggests positive
and small $\Delta s$ like our results. This is quite puzzling and so
further experimental extractions from different methods are
necessary. We have fairly well reproduced the proton isovector and
octet axial charges while the singlet one is too large. This can
probably be due to the fact that flavor $SU(3)$ symmetry is not
broken in our computations.

To estimate $\Theta^+$ pentaquark width we have computed the axial
decay constant. Thanks to a generalized Goldberger-Treiman relation
and a common hyperon decay formula the width has been evaluated to a
few MeV. While relativistic effects reduce the width to about one
half, the $7Q$ component increases it roughly by a factor 3/2. Such
a small width is very unusual compared to ordinary resonance decay
widths ($\sim 100$ MeV) and makes it hard to be seen in experiments.
The reason of this small width is obscure since the mechanism is not
known. However the present approach of the model explains it by the
fact that in the IMF $\Theta^+$ cannot decay into the $3Q$ component
of the nucleon. Quark orbital angular momentum increases the $3Q$
component while higher Fock components reduce it. This explains why
the width is reduced by relativistic effects and increased due to
additional quark-antiquark pairs. There is also another
distinguishable feature of exotic baryons. While the net
contribution of quark spin is smaller than the hosting ordinary
baryon angular momentum, in pentaquark this net contribution is
larger. Since the status about the existence of pentaquark is still
unclear this prediction will probably takes a lot of time before
being confirmed experimentally (of course if the pentaquark does
exist).

Tensor charges are very poorly known experimentally since they are
not observable in the usual Deep Inelastic Scattering experiments.
However from a theoretical point of view, they are as important as
the vector and axial charges. They have been computed in several
models and they agree on the fact that tensor charges are similar to
axial charges but a bit larger. Soffer's inequality provides us with
some bound for the tensor charges but not restrictive enough to say
something starting from vector and axial charges. One of our results
concerns this inequality. While explicit flavors do satisfy this
inequality, we have found that hidden flavors, \emph{e.g.}
strangeness in nucleon, violate it. We agree with other models on
the fact that tensor charges are similar to axial ones. The only
experimental extraction however indicates small tensor charges which
cannot be understood using models based on the successful concept of
constituent quarks. Further experimental results are highly
desirable to clear the situation.

NQM model with $SU(6)$ symmetry is quite successful in describing
octet magnetic moments. For example the predicted ratio between
proton and neutron magnetic moments is very close to the
experimental one. One can be worried about the fact that the pion
cloud, by breaking explicitly all $SU(6)$ relations would lead to a
less successful description. Our results show however that the pion
is essential in order to obtain a correct magnitude for the magnetic
moment. Moreover as expected $SU(6)$ relations are effectively
broken but the ratios obtained are even closer to experimental ones.
We emphasize also the excellent agreement of our proton and neutron
magnetic moments obtained \emph{ab initio} with the experimental
values. Hyperon magnetic moments are less good due to the breaking
of flavor $SU(3)$ symmetry. The predicted magnetic moment of
$\Delta^+$ agrees well with the present experimental extraction.
Note however that we predict a large and positive magnetic moment
for $\Theta^+$ while all the other approaches suggest small and/or
negative magnetic moment. Our result is consistent with a naive
flavor $SU(3)$ estimation
$\mu_{p_\mathbf{8}^+}\simeq\mu_{\Delta_\mathbf{10}^+}\simeq\mu_{p_\mathbf{\overline{10}}^+}\simeq\mu_{\Theta_\mathbf{\overline{10}}^+}$.
If the $\Theta^+$ magnetic moment turns out to be small and/or
negative this would imply a very large effect due to flavor $SU(3)$
symmetry breaking.

We have shown that the $N\Delta$ system is not spherically
symmetric. This distortion is explicitly due to the pion cloud and
especially to a quadrupolar structure. We have thus remarkably shown
that the pion cloud is explicitly responsible for the deformation of
the system. Furthermore such a structure appeared many times when
studying the vector, axial and tensor charges. Decuplet baryons
being spin-3/2 baryons have components $J_z=1/2,3/2$. Spherical
symmetry imposes relations between these components. They appeared
in fact broken by the same quadrupolar structure of the pion cloud.
This means that $\Delta$ is deformed. This quadrupolar structure
also appeared in the spin-1/2 baryons charges (excepted vector
ones). It appeared for example that the quadrupolar contribution of
the pion cloud in axial charges is different from the one in tensor
charges. This means that valence axial contributions are not
proportional to tensor contributions. To the best of our knowledge,
this has never been discussed in the literature.

Finally we have completed the list of exact spin-flavor baryon wave
functions introduced by Diakonov and Petrov. They have given the
expression of all spin-flavor wave functions in the $3Q$ sector and
the ones for the octet and antidecuplet baryons in the $5Q$ sector.
We have added the $5Q$ sector of decuplet and given the whole set of
$7Q$ spin-flavor wave functions along with general formulae and
identities.
\newline\newline
\textbf{What can be done}

Still a lot of work can be done within this approach. For example
since theoretical errors are not known hitherto it would be
important to test the sensitivity of our results to the input
parameters, \emph{e.g.} quark mass, nucleon mass and Pauli-Villars
mass. In this thesis we have also completely neglected the
distortion of the discrete level due to the sea. Its effects is
simply unknown. The breaking of flavor $SU(3)$ symmetry appeared
also to be important to give more reliable results, especially for
hyperons. We do not think that computing the $9Q$ component would be
worthwhile due to the already small value of the $7Q$ contribution.
It would only be needed for more accurate predictions for
pentaquarks.

Since we have at our disposal the explicit expression of all light
baryon light-cone wave functions we could in principle compute also
FF, PDF and GPD, since charges correspond only to the matrix
elements of local operators in the forward limit. One could then
estimate the transverse size of baryons, charge distributions,
\ldots One could also explore axial transitions between multiplets
and see if the quadrupolar moment of the pion cloud has other
directly observable effects.

In summary a lot of original results have been obtained as well as
interesting observations have been proposed. We have found a rather
good overall agreement with experimental data though the breaking of
flavor $SU(3)$ symmetry seems essential. The link with NQM is clear
and relativity is consistently implemented. Much work awaits to be
done in this approach and would probably lead to further interesting
observations, suggestions and predictions.

\newpage
\appendix

\renewcommand{\theequation}{A\arabic{equation}}
\setcounter{equation}{0} \pagestyle{myheadings}\markboth{}{}
\chapter{Group integrals}

We give in this appendix the complete list of octet, decuplet and
antidecuplet spin-flavor wave functions up to the $7Q$ sector. They
are group integrals over the Haar measure of the $SU(N)$ group
normalized to unity $\int\ud R=1$. Part of them are copied from the
Appendix B of \cite{DiaPet}.

\section{Method}

Here is the general method to compute integrals of several matrices
$R$, $R^\dag$. The result of an integration over the invariant
measure can only be invariant tensors which, for the $SU(N)$ group,
are built solely from the Kronecker $\delta$ and Levi-Civita
$\epsilon$ tensors. One then constructs the supposed tensor of a
given rank as the most general combination of $\delta$'s and
$\epsilon$'s satisfying the symmetry relations following from the
integral in question:
\begin{itemize}\item Since
$R^f_j$ and $R^{\dag i}_h$ are just numbers one can commute them.
Therefore the same permutation among $f$'s and $j$'s (or $h$'s and
$i$'s) does not change the value of the integral, \emph{i.e.} the
structure of the tensor.\item In the special case where there are as
many $R$ as $R^\dag$ one can exchange them which amounts to exchange
$f$ and $j$ indices with respectively $i$ and $h$.
\end{itemize}
One has however to be careful to use the same ``type'' of indices in
$\delta$'s and $\epsilon$'s, \emph{i.e.} the upper (resp. lower)
indices of $R$ with the lower (resp. upper) ones of $R^\dag$. The
indefinite coefficients in the combination are found by contracting
both sides with various $\delta$'s and $\epsilon$'s and thus by
reducing the integral to a previously derived one. We will give
below explicit examples.

\section{Basic integrals and explicit examples}

Since the method is recursive let us start with the simplest group
integrals. For any $SU(N)$ group one has
\begin{equation}\label{Basic}
\int\ud R\,R^f_j=0,\qquad \int\ud R\,R^{\dag i}_h=0,\qquad \int\ud
R\, R^f_jR^{\dag i}_h=\frac{1}{N}\,\delta^f_h\delta^i_j.
\end{equation}
The last integral is a well known result but can be derived by means
of the method explained earlier. There are two upper ($f,i$) and two
lower ($j,h$) indices. In $SU(N)$ the solution of the integral can
only be constructed from the $\delta$ and the $\epsilon$ tensor with
$N$ (upper or lower) indices. There is only one possible
structure\footnote{The $\epsilon$ tensor needs $N$ indices of the
same ``type'' and position. The only possibility left is to
introduce new indices that are summed, \emph{e.g.}
$\epsilon^{fg}\epsilon_{hg}\epsilon^{ik}\epsilon_{jk}$. This is
however not a new structure since the summation over the new indices
can be performed leading to the ``old'' structure
$\epsilon^{fg}\epsilon_{hg}\epsilon^{ik}\epsilon_{jk}=\delta^f_h\delta^i_j$.}
$\delta^f_h\delta^i_j$ leaving thus only one undetermined
coefficient $A$. The latter can be determined by contracting both
sides with, say, $\delta^j_i$. Since $R^f_jR^{\dag j}_h=\delta^f_h$
($R$ matrices belong to $SU(N)$ and are thus unitary) one has for
the lhs
\begin{equation}
\delta^j_i\times\int\ud R\,R^f_jR^{\dag i}_h=\delta^f_h
\end{equation}
and for the rhs
\begin{equation}
\delta^j_i\times A\,\delta^f_h\delta^i_j=A\,N\,\delta^f_h
\end{equation}
and one concludes that $A=1/N$.

Let us proceed with the integral of two $R$'s. Here all the upper
(lower) indices have the same ``type'' and must appear in the same
symbol. Only $\epsilon$ has many indices in the same position. In
the case $N>2$ one needs more available indices. This means that for
$SU(N)$ with $N>2$ one has
\begin{equation}
\int\ud R\,R^{f_1}_{j_1}R^{f_2}_{j_2}=0.
\end{equation}
For $N=2$, the group integral is non-vanishing since the structure
$\epsilon^{f_1f_2}\epsilon_{j_1j_2}$ is allowed. The undetermined
coefficient $A$ is obtained by contracting both sides with, say,
$\epsilon^{j_1j_2}$. Since
$\epsilon^{j_1j_2}R^{f_1}_{j_1}R^{f_2}_{j_2}=\epsilon^{f_1f_2}$ ($R$
matrices belong to $SU(2)$ and have thus $\det(R)=1$) one has for
the lhs
\begin{equation}
\epsilon^{j_1j_2}\times\int\ud
R\,R^{f_1}_{j_1}R^{f_2}_{j_2}=\epsilon^{f_1f_2}
\end{equation}
and for the rhs
\begin{equation}
\epsilon^{j_1j_2}\times
A\,\epsilon^{f_1f_2}\epsilon_{j_1j_2}=2A\,\epsilon^{f_1f_2}
\end{equation}
and thus one concludes that $A=1/2$. For $SU(2)$ one then has
\begin{equation}
\int\ud
R\,R^{f_1}_{j_1}R^{f_2}_{j_2}=\frac{1}{2}\,\epsilon^{f_1f_2}\epsilon_{j_1j_2}.
\end{equation}

The $SU(3)$ analog involves the products of three $R$'s
\begin{equation}
\int\ud
R\,R^{f_1}_{j_1}R^{f_2}_{j_2}R^{f_3}_{j_3}=\frac{1}{6}\,\epsilon^{f_1f_2f_3}\epsilon_{j_1j_2j_3}
\end{equation}
which is vanishing for $N>3$ and also for $N=2$ since all the three
upper (and lower) indices cannot be used in $\epsilon$'s. This can
be easily generalized to $SU(N)$ with the product of $N$ matrices
$R$
\begin{equation}
\int\ud R\,R^{f_1}_{j_1}R^{f_2}_{j_2}\dots
R^{f_N}_{j_N}=\frac{1}{N!}\,\epsilon^{f_1f_2\dots
f_N}\epsilon_{j_1j_2\dots j_N}.
\end{equation}
This integral is vanishing for all $SU(N')$ groups with $N'$ that is
not a divisor of $N$.

Let us now consider the product of four $R$'s in $SU(2)$. Since 2 is
a divisor of 4 the integral is non-vanishing. The general tensor
structure is a linear combination of
$\epsilon^{f_af_b}\epsilon^{f_cf_d}\epsilon_{j_wj_x}\epsilon_{j_yj_z}$
with $a,b,c,d$ and $w,x,y,z$ some permutation of 1,2,3,4. There are
\emph{a priori} 9 undetermined coefficients. The integral symmetries
reduce this number to 2. Thanks to the $SU(2)$ identity
\begin{equation}\label{Simplification}
\epsilon_{j_1j_2}\epsilon_{j_3j_4}+\epsilon_{j_1j_3}\epsilon_{j_4j_2}+\epsilon_{j_1j_4}\epsilon_{j_2j_3}=0
\end{equation}
only one undetermined coefficient is left which is obtained by
contracting both sides with, say, $\epsilon^{j_1j_2}$. The result is
thus for $SU(2)$
\begin{equation}
\int\ud
R\,R^{f_1}_{j_1}R^{f_2}_{j_2}R^{f_3}_{j_3}R^{f_4}_{j_4}=\frac{1}{6}\,\left(\epsilon^{f_1f_2}\epsilon^{f_3f_4}\epsilon_{j_1j_2}\epsilon_{j_3j_4}+\epsilon^{f_1f_3}\epsilon^{f_2f_4}\epsilon_{j_1j_3}\epsilon_{j_2j_4}+\epsilon^{f_1f_4}\epsilon^{f_2f_3}\epsilon_{j_1j_4}\epsilon_{j_2j_3}\right).
\end{equation}

The identity (\ref{Simplification}) is in fact a particular case of
a general $SU(N)$ identity. It is based on the fact that for $SU(N)$
one has $\epsilon_{j_1j_2\dots j_{N+1}}=0$ and thus
\begin{equation}\label{Formule}
\epsilon_{j_1j_2\dots j_N}X_{j_{N+1}}\pm\epsilon_{j_2j_3\dots
j_{N+1}}X_{j_1}+\epsilon_{j_3j_4\dots
j_1}X_{j_2}\pm\ldots\pm\epsilon_{j_{N+1}j_1\dots j_{N-1}}X_{j_N}=0
\end{equation}
where the $+$ (resp. $-$) sign is for $N$ even (resp. odd) and $X_j$
any tensor with at least index $j$. This identity is easy to check.
Since we work in $SU(N)$ among the $N+1$ indices at least two are
equal, say $j_k$ and $j_l$. The only surviving terms are then
$-X_{j_k}+X_{j_l}$ which give zero since $j_k=j_l$. It is very
useful and simplifies a lot the search of the general tensor
structure. Since the number of indices of both ``types'' is
identical the structure in terms of $\delta$'s and $\epsilon$'s is
also the same. The indices on $\epsilon$ can be placed in a
symmetric (\emph{e.g.}
$\epsilon^{f_1f_2}\epsilon^{f_3f_4}\epsilon_{j_1j_2}\epsilon_{j_3j_4}$)
and an asymmetric manner (\emph{e.g.}
$\epsilon^{f_1f_2}\epsilon^{f_3f_4}\epsilon_{j_1j_4}\epsilon_{j_2j_3}$).
By repeated applications of (\ref{Formule}) the asymmetric part of
the tensor can be transformed into the symmetric part reducing thus
the number of undetermined coefficients by a factor 2. In the search
of the general tensor structure one has just to consider symmetric
$\epsilon$ terms only.

We give another useful identity. In $SU(2)$ one has
$\epsilon^{f_1f_2f_3}\epsilon_{h_1h_2h_3}=0$. Using the notation
$(abc)\equiv\delta^{f_1}_{h_a}\delta^{f_2}_{h_b}\delta^{f_3}_{h_c}$
this amounts to
\begin{equation}\label{Formule2}
(123)-(132)+(231)-(213)+(312)-(321)=0.
\end{equation}
This identity is easily generalized to any $SU(N)$ group where it is
based on $\epsilon^{f_1f_2\dots f_{N+1}}\epsilon_{h_1h_2\dots
h_{N+1}}=0$.

We close this section by mentioning another group integral which is
useful to obtain further ones. For any $SU(N)$ group one has
\begin{equation}
\int\ud R\,R^{f_1}_{j_1}R^{f_2}_{j_2}R^{\dag i_1}_{h_1}R^{\dag
i_2}_{h_2}=\frac{1}{N^2-1}\left[\delta^{f_1}_{h_1}\delta^{f_2}_{h_2}\left(\delta^{i_1}_{j_1}\delta^{i_2}_{j_2}-\frac{1}{N}\,\delta^{i_2}_{j_1}\delta^{i_1}_{j_2}\right)
+\delta^{f_1}_{h_2}\delta^{f_2}_{h_1}\left(\delta^{i_2}_{j_1}\delta^{i_1}_{j_2}-\frac{1}{N}\,\delta^{i_1}_{j_1}\delta^{i_2}_{j_2}\right)\right].
\end{equation}
One can easily check that by contracting it with, say,
$\delta^{h_1}_{f_1}$ it reduces to (\ref{Basic}).

\section{Notations}

In order to simplify the formulae we introduce some notations
\begin{equation}
\left[abc\right]\equiv(123)(abc)+(231)(bca)+(312)(cab)+(213)(bac)+(132)(acb)+(321)(cba),
\end{equation}
\begin{eqnarray}
\left[abcd\right]&\equiv&(1234)(abcd)+(2341)(bcda)+(3412)(cdab)+(4123)(dabc)+(2134)(bacd)+(1342)(acdb)\nonumber\\
&+&(3421)(cdba)+(4213)(dbac)+(3214)(cbad)+(2143)(badc)+(1432)(adcb)+(4321)(dcba)\nonumber\\
&+&(4231)(dbca)+(2314)(bcad)+(3142)(cadb)+(1423)(adbc)+(1324)(acbd)+(3241)(cbda)\nonumber\\
&+&(2413)(bdac)+(4132)(dacb)+(1243)(abdc)+(2431)(bdca)+(4312)(dcab)+(3124)(cabd),\nonumber\\
\end{eqnarray}
\begin{eqnarray}
\left[abcde\right]&\equiv&(12345)(abcde)+(23451)(bcdea)+(34512)(cdeab)+(45123)(deabc)+(51234)(eabcd)\nonumber\\
&+&(21345)(bacde)+(13452)(acdeb)+(34521)(cdeba)+(45213)(debac)+(52134)(ebacd)\nonumber\\
&+&(32145)(cbade)+(21453)(badec)+(14532)(adecb)+(45321)(decba)+(53214)(ecbad)\nonumber\\
&+&(42315)(dbcae)+(23154)(bcaed)+(31542)(caedb)+(15423)(aedbc)+(54231)(edbca)\nonumber\\
&+&(52341)(ebcda)+(23415)(bcdae)+(34152)(cdaeb)+(41523)(daebc)+(15234)(aebcd)\nonumber\\
&+&(13245)(acbde)+(32451)(cbdea)+(24513)(bdeac)+(45132)(deacb)+(51324)(eacbd)\nonumber\\
&+&(14325)(adcbe)+(43251)(dcbea)+(32514)(cbead)+(25143)(beadc)+(51432)(eadcb)\nonumber\\
&+&(15342)(aecdb)+(53421)(ecdba)+(34215)(cdbae)+(42153)(dbaec)+(21534)(baecd)\nonumber\\
&+&(12435)(abdce)+(24351)(bdcea)+(43512)(dceab)+(35124)(ceabd)+(51243)(eabdc)\nonumber\\
&+&(12543)(abedc)+(25431)(bedca)+(54312)(edcab)+(43125)(dcabe)+(31254)(cabed)\nonumber\\
&+&(12354)(abced)+(23541)(bcdea)+(35412)(cedab)+(54123)(edabc)+(41235)(dabce)\nonumber\\
&+&(54321)(edcba)+(43215)(dcbae)+(32154)(cbaed)+(21543)(baedc)+(15432)(aedcb)\nonumber\\
&+&(12453)(abdec)+(24531)(bdeca)+(45312)(decab)+(53124)(ecabd)+(31245)(cabde)\nonumber\\
&+&(12534)(abecd)+(25341)(becda)+(53412)(ecdab)+(34125)(cdabe)+(41253)(dabec)\nonumber\\
&+&(23514)(bcead)+(35142)(ceadb)+(51423)(eadbc)+(14235)(adbce)+(42351)(dbcea)\nonumber\\
&+&(23145)(bcade)+(31452)(cadeb)+(14523)(adebc)+(45231)(debca)+(52314)(ebcad)\nonumber\\
&+&(34251)(cdbea)+(42513)(dbeac)+(25134)(beacd)+(51342)(eacdb)+(13425)(acdbe)\nonumber\\
&+&(21435)(badce)+(14352)(adceb)+(43521)(dceba)+(35214)(cebad)+(52143)(ebadc)\nonumber\\
&+&(21354)(baced)+(13542)(acedb)+(35421)(cedba)+(54213)(edbac)+(42135)(dbace)\nonumber\\
&+&(32541)(cbeda)+(25413)(bedac)+(54132)(edacb)+(41325)(dacbe)+(13254)(acbed)\nonumber\\
&+&(35241)(cebda)+(52413)(ebdac)+(24135)(bdace)+(41352)(daceb)+(13524)(acebd)\nonumber\\
&+&(52431)(ebdca)+(24315)(bdcae)+(43152)(dcaeb)+(31524)(caebd)+(15243)(aebdc)\nonumber\\
&+&(42531)(dbeca)+(25314)(becad)+(53142)(ecabd)+(31425)(cabde)+(14253)(abdec)\nonumber\\
&+&(32415)(cbdae)+(24153)(bdaec)+(41532)(daecb)+(15324)(aecbd)+(53241)(ecbda)\nonumber\\
\end{eqnarray}
where
\begin{eqnarray}
(abc)(def)&\equiv&\delta^{f_1}_{h_a}\delta^{f_2}_{h_b}\delta^{f_3}_{h_c}\delta^{i_d}_{j_1}\delta^{i_e}_{j_2}\delta^{i_f}_{j_3},\\
(abcd)(efgh)&\equiv&\delta^{f_1}_{h_a}\delta^{f_2}_{h_b}\delta^{f_3}_{h_c}\delta^{f_4}_{h_d}\delta^{i_e}_{j_1}\delta^{i_f}_{j_2}\delta^{i_g}_{j_3}\delta^{i_h}_{j_4},\\
(abcde)(fghij)&\equiv&\delta^{f_1}_{h_a}\delta^{f_2}_{h_b}\delta^{f_3}_{h_c}\delta^{f_4}_{h_d}\delta^{f_5}_{h_e}\delta^{i_f}_{j_1}\delta^{i_g}_{j_2}\delta^{i_h}_{j_3}\delta^{i_i}_{j_4}\delta^{i_j}_{j_5}.
\end{eqnarray}

Other structures are simplified as follows
\begin{eqnarray}
[xyz,lmn]&\equiv&[lmn]\qquad\textrm{where
}(abc)(def)\equiv\delta^{f_x}_{f_a}\delta^{f_y}_{f_b}\delta^{f_z}_{f_c}\delta^{j_d}_{j_x}\delta^{j_e}_{j_y}\delta^{j_f}_{j_z},\\\nonumber\\
\{ab\}&\equiv&\delta^{f_a}_{f_8}\delta^{f_b}_{f_{10}}\left(5\delta^{j_8}_{j_a}\delta^{j_{10}}_{j_b}-\delta^{j_8}_{j_b}\delta^{j_{10}}_{j_a}\right)
+\delta^{f_a}_{f_{10}}\delta^{f_b}_{f_8}\left(5\delta^{j_{10}}_{j_a}\delta^{j_8}_{j_b}-\delta^{j_{10}}_{j_b}\delta^{j_8}_{j_a}\right),\\\nonumber\\
\{abcde\}&\equiv&\delta^{h_1}_{f_a}\left(\epsilon^{f_bf_ch_2}\epsilon^{f_df_eh_3}+\epsilon^{f_bf_ch_3}\epsilon^{f_df_eh_2}\right)+\delta^{h_2}_{f_a}\left(\epsilon^{f_bf_ch_3}\epsilon^{f_df_eh_1}+\epsilon^{f_bf_ch_1}\epsilon^{f_df_eh_3}\right)\nonumber\\
&+&\delta^{h_3}_{f_a}\left(\epsilon^{f_bf_ch_1}\epsilon^{f_df_eh_2}+\epsilon^{f_bf_ch_2}\epsilon^{f_df_eh_1}\right),\\\nonumber\end{eqnarray}\begin{eqnarray}
\{abc,de\}&\equiv&\epsilon^{f_af_bf_c}\epsilon_{j_aj_bj_c}\left[\delta^{f_d}_{f_5}\delta^{f_e}_{f_7}\left(4\delta^{j_5}_{j_d}\delta^{j_7}_{j_e}-\delta^{j_5}_{j_e}\delta^{j_7}_{j_d}\right)
+\delta^{f_d}_{f_7}\delta^{f_e}_{f_5}\left(4\delta^{j_7}_{j_d}\delta^{j_5}_{j_e}-\delta^{j_7}_{j_e}\delta^{j_5}_{j_d}\right)
\right],\nonumber\\
\{abcdef\}&\equiv&\epsilon^{f_af_bf_c}\epsilon^{f_df_ef_f}\epsilon_{j_aj_bj_c}\epsilon_{j_dj_ej_f}+\epsilon^{f_af_bf_d}\epsilon^{f_cf_ef_f}\epsilon_{j_aj_bj_d}\epsilon_{j_cj_ej_f}+\epsilon^{f_af_bf_e}\epsilon^{f_cf_df_f}\epsilon_{j_aj_bj_e}\epsilon_{j_cj_dj_f}\nonumber\\
&+&\epsilon^{f_af_bf_f}\epsilon^{f_cf_df_e}\epsilon_{j_aj_bj_f}\epsilon_{j_cj_dj_e}+\epsilon^{f_af_cf_d}\epsilon^{f_bf_ef_f}\epsilon_{j_aj_cj_d}\epsilon_{j_bj_ej_f}+\epsilon^{f_af_cf_e}\epsilon^{f_bf_df_f}\epsilon_{j_aj_cj_e}\epsilon_{j_bj_dj_f}\nonumber\\
&+&\epsilon^{f_af_cf_f}\epsilon^{f_bf_df_e}\epsilon_{j_aj_cj_f}\epsilon_{j_bj_dj_e}+\epsilon^{f_af_df_e}\epsilon^{f_bf_cf_f}\epsilon_{j_aj_dj_e}\epsilon_{j_bj_cj_f}+\epsilon^{f_af_df_f}\epsilon^{f_bf_cf_e}\epsilon_{j_aj_dj_f}\epsilon_{j_bj_cj_e}\nonumber\\
&+&\epsilon^{f_af_ef_f}\epsilon^{f_bf_cf_d}\epsilon_{j_aj_ej_f}\epsilon_{j_bj_cj_d},\\\nonumber\\
\{abc,def\}&\equiv&\epsilon^{f_af_bf_c}\epsilon_{j_aj_bj_c}\left(7\,[def,579]-2\,([def,597]+[def,975]+[def,759])+([def,795]+[def,957])\right).\nonumber\\
\end{eqnarray}

\section{Group integrals and projections onto Fock
states}

Spin-flavor wave functions are constructed from the projection of
Fock states onto rotational wave functions. The rotational wave
functions are given in the main text along with the particle
representation, see Section 3.3. The $3Q$ state involves three
quarks that are rotated by three $R$ matrices. The $5Q$ state
involves four quarks and one antiquark that are rotated by four $R$
and one $R^\dag$ matrices. So a general $nQ$ state involves
$(n+3)/2$ quarks and $(n-3)/2$ antiquarks that are rotated by
$(n+3)/2$ $R$ and $(n-3)/2$ $R^\dag$ matrices.

\subsection{Projections of the $3Q$ state}

The first integral corresponds to the projection of the $3Q$ state
onto the octet quantum numbers for the $SU(3)$ group
\begin{eqnarray}
&&\int\ud
R\,R^{f_1}_{j_1}R^{f_2}_{j_2}R^{f_3}_{j_3}\left(R^{f_4}_{j_4}R^{\dag
j_5}_{f_5}\right)\nonumber\\
&=&\frac{1}{24}\left(\delta^{f_1}_{f_5}\delta^{j_5}_{j_1}\epsilon^{f_2f_3f_4}\epsilon_{j_2j_3j_4}+\delta^{f_2}_{f_5}\delta^{j_5}_{j_2}\epsilon^{f_1f_3f_4}\epsilon_{j_1j_3j_4}
+\delta^{f_3}_{f_5}\delta^{j_5}_{j_3}\epsilon^{f_1f_2f_4}\epsilon_{j_1j_2j_4}+\delta^{f_4}_{f_5}\delta^{j_5}_{j_4}\epsilon^{f_1f_2f_3}\epsilon_{j_1j_2j_3}\right).
\end{eqnarray}
This integral is zero for any other $SU(N)$ group.

The second integral corresponds to the projection of the $3Q$ state
onto the decuplet quantum numbers for any $SU(N)$ group
\begin{eqnarray}
\int\ud R\,R^{f_1}_{j_1}R^{f_2}_{j_2}R^{f_3}_{j_3}R_{h_1}^{\dag
i_1}R_{h_2}^{\dag i_2}R_{h_3}^{\dag
i_3}&=&\frac{1}{N(N^2-1)(N^2-4)}\,\{(N^2-2)\,[123]\nonumber\\
&-&N\,([213]+[132]+[321])+2\,([231]+[312])\}.
\end{eqnarray}
There is no problem in the case $N=2$ thanks to (\ref{Formule2})
\begin{equation}
\int\ud R\,R^{f_1}_{j_1}R^{f_2}_{j_2}R^{f_3}_{j_3}R_{h_1}^{\dag
i_1}R_{h_2}^{\dag i_2}R_{h_3}^{\dag
i_3}=\frac{1}{24}\,\{3\,[123]-([231]+[312])\}.
\end{equation}

The third integral corresponds to the projection of the antidecuplet
onto the $3Q$ state for the $SU(3)$ group
\begin{equation}
\int\ud
R\,R^{f_1}_{j_1}R^{f_2}_{j_2}R^{f_3}_{j_3}R^{f_4}_{j_4}R^{f_5}_{j_5}R^{f_6}_{j_6}=\frac{1}{72}\,\{123456\}.\label{Three
quarks antidecuplet}
\end{equation}
This integral is also non-vanishing in only two other cases $N=2$
and $N=6$. The (conjugated) rotational wave function of the
antidecuplet is
\begin{equation}
A_k^{*\{h_1h_2h_3\}}(R)=\frac{1}{3}\left(R^{h_1}_3R^{h_2}_3R^{h_3}_k+R^{h_2}_3R^{h_3}_3R^{h_1}_k+R^{h_3}_3R^{h_1}_3R^{h_2}_k\right).\label{Tensor
pentaquarks}
\end{equation}
Due to the antisymmetric structure of (\ref{Three quarks
antidecuplet}) one can see that the projection of the antidecuplet
on the $3Q$ sector is vanishing and thus that pentaquarks cannot be
made of three quarks only.

\subsection{Projections of the $5Q$ state}

The first integral corresponds to the projection of the $5Q$ state
onto the octet quantum numbers for the $SU(3)$ group
\begin{eqnarray}
\int\ud
R\,R^{f_1}_{j_1}R^{f_2}_{j_2}R^{f_3}_{j_3}\left(R^{f_4}_{j_4}R^{\dag
j_5}_{f_5}\right)\left(R^{f_6}_{j_6}R^{\dag
j_7}_{f_7}\right)&=&\frac{1}{360}[\{123,46\}+\{124,36\}+\{126,34\}+\{134,26\}+\{136,24\}\nonumber\\
&+&\{146,23\}+\{346,12\}+\{246,13\}+\{236,14\}+\{234,16\}].\nonumber\\
\end{eqnarray}
This integral is zero for any other $SU(N)$ group.

The second integral corresponds to the projection of the $5Q$ state
onto the decuplet quantum numbers for any $SU(N)$ group
\begin{eqnarray}
&&\int\ud
R\,R^{f_1}_{j_1}R^{f_2}_{j_2}R^{f_3}_{j_3}R^{f_4}_{j_4}R_{h_1}^{\dag
i_1}R_{h_2}^{\dag i_2}R_{h_3}^{\dag i_3}R_{h_4}^{\dag
i_4}=\frac{1}{N^2(N^2-1)(N^2-4)(N^2-9)}\nonumber\\
&\times&\{(N^4-8N^2+6)\,[1234]-5N\,([2341]+[4123]+[3421]+[4312]+[3142]+[2413])\nonumber\\
&+&(N^2+6)\,([3412]+[2143]+[4321])-N(N^2-4)\,([2134]+[3214]+[1432]+[1324]+[1243]+[4231])\nonumber\\
&+&(2N^2-3)\,([1342]+[4213]+[3241]+[2314]+[3124]+[4132]+[2431]+[1423])\}.
\end{eqnarray}
No problem arises either in the case $N=3$
\begin{eqnarray}
&&\int\ud
R\,R^{f_1}_{j_1}R^{f_2}_{j_2}R^{f_3}_{j_3}R^{f_4}_{j_4}R_{h_1}^{\dag
i_1}R_{h_2}^{\dag i_2}R_{h_3}^{\dag i_3}R_{h_4}^{\dag
i_4}\nonumber\\
&=&\frac{1}{2160}\,\{48\,[1234]+7\,([2341]+[4123]+[3421]+[4312]+[3142]+[2413])\nonumber\\
&-&6\,([3412]+[2143]+[4321])-11\,([2134]+[3214]+[1432]+[1324]+[1243]+[4231])\}.
\end{eqnarray}
or in the case $N=2$ thanks to (\ref{Formule2})
\begin{eqnarray}
&&\int\ud
R\,R^{f_1}_{j_1}R^{f_2}_{j_2}R^{f_3}_{j_3}R^{f_4}_{j_4}R_{h_1}^{\dag
i_1}R_{h_2}^{\dag i_2}R_{h_3}^{\dag i_3}R_{h_4}^{\dag
i_4}\nonumber\\
&=&\frac{1}{240}\,\{8\,[1234]-3\,([2341]+[4123]+[3421]+[4312]+[3142]+[2413])\nonumber\\
&+&4\,([3412]+[2143]+[4321])\}.
\end{eqnarray}

The third integral corresponds to the projection of the $5Q$ state
onto the antidecuplet quantum numbers for the $SU(3)$ group
\begin{eqnarray}
&&\int\ud
R\,R^{f_1}_{j_1}R^{f_2}_{j_2}R^{f_3}_{j_3}R^{f_4}_{j_4}R^{f_5}_{j_5}R^{f_6}_{j_6}\left(R^{f_7}_{j_7}R^{\dag
j_8}_{f_8}\right)=\frac{1}{360}\left[\delta^{f_1}_{f_8}\delta^{j_8}_{j_1}\{234567\}+\delta^{f_2}_{f_8}\delta^{j_8}_{j_2}\{134567\}+\delta^{f_3}_{f_8}\delta^{j_8}_{j_3}\{124567\}\right.\nonumber\\
&&\qquad\qquad\qquad\qquad\qquad+\delta^{f_4}_{f_8}\delta^{j_8}_{j_4}\{123567\}+\delta^{f_5}_{f_8}\delta^{j_8}_{j_5}\{123467\}+\delta^{f_6}_{f_8}\delta^{j_8}_{j_6}\{123457\}+\delta^{f_7}_{f_8}\delta^{j_8}_{j_7}\{123456\}\Big].\nonumber\\
\end{eqnarray}
This integral is also non-vanishing in only two other cases $N=2$
and $N=6$. The (conjugated) rotational wave function of the
antidecuplet (\ref{Tensor pentaquarks}) is symmetric with respect to
three flavor indices $h_1,h_2,h_3$. The projection of the $5Q$ state
is thus reduced to
\begin{equation}
\begin{split}
\int\ud
R\,&R^{f_1}_{j_1}R^{f_2}_{j_2}R^{f_3}_{j_3}\left(R^{f_4}_{j_4}R^{\dag
j_5}_{f_5}\right)A_k^{*\{h_1h_2h_3\}}(R)\\
=&\,\frac{1}{1080}\Big\{\{51234\}\left(\delta^{j_5}_k\epsilon_{j_1j_23}\epsilon_{j_3j_43}+\delta^{j_5}_3\epsilon_{j_1j_2k}\epsilon_{j_3j_43}+\delta^{j_5}_3\epsilon_{j_1j_23}\epsilon_{j_3j_4k}\right)\\
&+\{52341\}\left(\delta^{j_5}_k\epsilon_{j_2j_33}\epsilon_{j_4j_13}+\delta^{j_5}_3\epsilon_{j_2j_3k}\epsilon_{j_4j_13}+\delta^{j_5}_3\epsilon_{j_2j_33}\epsilon_{j_4j_1k}\right)\\
&+\{51324\}\left(\delta^{j_5}_k\epsilon_{j_1j_33}\epsilon_{j_2j_43}+\delta^{j_5}_3\epsilon_{j_1j_3k}\epsilon_{j_2j_43}+\delta^{j_5}_3\epsilon_{j_1j_33}\epsilon_{j_2j_4k}\right)\Big\}.
\end{split}
\end{equation}

\subsection{Projections of the $7Q$ state}

The first integral corresponds to the projection of the $7Q$ state
onto the octet quantum numbers for the $SU(3)$ group
\begin{eqnarray}
&&\int\ud
R\,R^{f_1}_{j_1}R^{f_2}_{j_2}R^{f_3}_{j_3}\left(R^{f_4}_{j_4}R^{\dag
j_5}_{f_5}\right)\left(R^{f_6}_{j_6}R^{\dag
j_7}_{f_7}\right)\left(R^{f_8}_{j_8}R^{\dag
j_9}_{f_9}\right)\nonumber\\
&=&\frac{1}{2160}\,(\{123,468\}+\{124,368\}+\{126,348\}+\{128,346\}+\{134,268\}+\{136,248\}+\{138,246\}\nonumber\\
&+&\{146,238\}+\{148,236\}+\{168,234\}+\{468,123\}+\{368,124\}+\{348,126\}+\{346,128\}\nonumber\\
&+&\{268,134\}+\{248,136\}+\{246,138\}+\{238,146\}+\{236,148\}+\{234,168\}).
\end{eqnarray}
This integral is zero for any other $SU(N)$ group.

The second integral corresponds to the projection of the $7Q$ state
onto the decuplet quantum numbers for any $SU(N)$ group
\begin{eqnarray}
&&\int\ud
R\,R^{f_1}_{j_1}R^{f_2}_{j_2}R^{f_3}_{j_3}R^{f_4}_{j_4}R^{f_5}_{j_5}R_{h_1}^{\dag
i_1}R_{h_2}^{\dag i_2}R_{h_3}^{\dag i_3}R_{h_4}^{\dag
i_4}R_{h_5}^{\dag
i_5}\nonumber\\
&=&\frac{1}{N^2(N^2-1)(N^2-4)(N^2-9)(N^2-16)}\,\{N(N^4-20N^2+78)\,[12345]-(N^4-14N^2+24)\nonumber\\
&\times&([21345]+[52341]+[12354]+[12435]+[13245]+[14325]+[32145]+[15342]+[42315]+[12543])\nonumber\\
&-&2(N^2+12)\,([34521]+[34152]+[35412]+[43512]+[24513]+[54123]+[35124]+[45132]+[45213]\nonumber\\
&&+[41523]+[21534]+[54231]+[31254]+[51432]+[53214]+[25431]+[43251]+[21453]\nonumber\\
&&+[53421]+[23154])\nonumber\end{eqnarray}\begin{eqnarray}
&+&2N(N^2-9)\,([12453]+[23145]+[42351]+[15324]+[15243]+[32415]+[24315]+[14352]+[14235]\nonumber\\
&&+[51342]+[52314]+[13425]+[25341]+[52143]+[42135]+[41325]+[13542]+[32541]\nonumber\\
&&+[12534]+[31245])\nonumber\\
&+&N(N^2-2)\,([54321]+[32154]+[15432]+[43215]+[21543]+[45312]+[42513]+[14523]+[34125]\nonumber\\
&&+[35142]+[21354]+[52431]+[13254]+[21435]+[53241])+14N\,([23451]+[31452]+[53412]\nonumber\\
&&+[23514]+[24531]+[34251]+[41253]+[51423]+[53124]+[25134]+[45231]+[51234]+[25413]\nonumber\\
&&+[43521]+[24153]+[35421]+[43152]+[41532]+[54213]+[31524]+[54132]+[35214]\nonumber\\
&&+[45123]+[34512])\nonumber\\
&-&(5N^2-24)\,([13452]+[23415]+[23541]+[24351]+[32451]+[41352]+[52413]+[13524]+[24135]\nonumber\\
&&+[35241]+[53142]+[25314]+[42531]+[14253]+[31425]+[15234]+[41235]+[51243]+[51324]\nonumber\\
&&+[52134]+[15423]+[43125]+[25143]+[45321]+[42153]+[14532]+[34215]+[31542]\nonumber\\
&&+[54312]+[32514])\}.
\end{eqnarray}
No problem arises in the case $N=4$
\begin{eqnarray}
&&\int\ud
R\,R^{f_1}_{j_1}R^{f_2}_{j_2}R^{f_3}_{j_3}R^{f_4}_{j_4}R^{f_5}_{j_5}R_{h_1}^{\dag
i_1}R_{h_2}^{\dag i_2}R_{h_3}^{\dag i_3}R_{h_4}^{\dag
i_4}R_{h_5}^{\dag
i_5}=\frac{1}{80640}\,\{179\,[12345]\nonumber\\
&-&52\,([21345]+[52341]+[12354]+[12435]+[13245]+[14325]+[32145]+[15342]+[42315]+[12543])\nonumber\\
&+&12\,([34521]+[34152]+[35412]+[43512]+[24513]+[54123]+[35124]+[45132]+[45213]+[41523]\nonumber\\
&&+[21534]+[54231]+[31254]+[51432]+[53214]+[25431]+[43251]+[21453]+[53421]+[23154])\nonumber\\
&+&19\,([12453]+[23145]+[42351]+[15324]+[15243]+[32415]+[24315]+[14352]+[14235]+[51342]\nonumber\\
&&+[52314]+[13425]+[25341]+[52143]+[42135]+[41325]+[13542]+[32541]+[12534]+[31245])\nonumber\\
&+&3\,([54321]+[32154]+[15432]+[43215]+[21543]+[45312]+[42513]+[14523]+[34125]+[35142]\nonumber\\
&&+[21354]+[52431]+[13254]+[21435]+[53241])-13\,([23451]+[31452]+[53412]+[23514]+[24531]\nonumber\\
&&+[34251]+[41253]+[51423]+[53124]+[25134]+[45231]+[51234]+[25413]+[43521]+[24153]\nonumber\\
&&+[35421]+[43152]+[41532]+[54213]+[31524]+[54132]+[35214]+[45123]+[34512])\},
\end{eqnarray}
in the case $N=3$
\begin{eqnarray}
&&\int\ud
R\,R^{f_1}_{j_1}R^{f_2}_{j_2}R^{f_3}_{j_3}R^{f_4}_{j_4}R^{f_5}_{j_5}R_{h_1}^{\dag
i_1}R_{h_2}^{\dag i_2}R_{h_3}^{\dag i_3}R_{h_4}^{\dag
i_4}R_{h_5}^{\dag
i_5}=\frac{1}{15120}\,\{151\,[12345]\nonumber\\
&-&38\,([21345]+[52341]+[12354]+[12435]+[13245]+[14325]+[32145]+[15342]+[42315]+[12543])\nonumber\\
&-&2\,([34521]+[34152]+[35412]+[43512]+[24513]+[54123]+[35124]+[45132]+[45213]+[41523]\nonumber\\
&&+[21534]+[54231]+[31254]+[51432]+[53214]+[25431]+[43251]+[21453]+[53421]+[23154])\nonumber\\
&+&10\,([12453]+[23145]+[42351]+[15324]+[15243]+[32415]+[24315]+[14352]+[14235]+[51342]\nonumber\\
&&+[52314]+[13425]+[25341]+[52143]+[42135]+[41325]+[13542]+[32541]+[12534]+[31245])\nonumber\\
&+&5\,([54321]+[32154]+[15432]+[43215]+[21543]+[45312]+[42513]+[14523]+[34125]+[35142]\nonumber\\
&&+[21354]+[52431]+[13254]+[21435]+[53241])\}
\end{eqnarray}
or in the case $N=2$ thanks to (\ref{Formule2})
\begin{eqnarray}
&&\int\ud
R\,R^{f_1}_{j_1}R^{f_2}_{j_2}R^{f_3}_{j_3}R^{f_4}_{j_4}R^{f_5}_{j_5}R_{h_1}^{\dag
i_1}R_{h_2}^{\dag i_2}R_{h_3}^{\dag i_3}R_{h_4}^{\dag
i_4}R_{h_5}^{\dag
i_5}=\frac{1}{1440}\,\{57\,[12345]\nonumber\\
&-&11\,([21345]+[52341]+[12354]+[12435]+[13245]+[14325]+[32145]+[15342]+[42315]+[12543])\nonumber\\
&+&2\,([12453]+[23145]+[42351]+[15324]+[15243]+[32415]+[24315]+[14352]+[14235]+[51342]\nonumber\\
&&+[52314]+[13425]+[25341]+[52143]+[42135]+[41325]+[13542]+[32541]+[12534]+[31245])\nonumber\\
&+&([54321]+[32154]+[15432]+[43215]+[21543]+[45312]+[42513]+[14523]+[34125]+[35142]\nonumber\\
&&+[21354]+[52431]+[13254]+[21435]+[53241])\}.
\end{eqnarray}

The third integral corresponds to the projection of the $7Q$ state
onto the antidecuplet quantum numbers for the $SU(3)$ group
\begin{eqnarray}
&&\int\ud
R\,R^{f_1}_{j_1}R^{f_2}_{j_2}R^{f_3}_{j_3}R^{f_4}_{j_4}R^{f_5}_{j_5}R^{f_6}_{j_6}\left(R^{f_7}_{j_7}R^{\dag
j_8}_{f_8}\right)\left(R^{f_9}_{j_9}R^{\dag
j_{10}}_{f_{10}}\right)=\frac{1}{8640}\nonumber\\
&\times&[\{123456\}\{79\}+\{123457\}\{69\}+\{123467\}\{59\}+\{123567\}\{49\}+\{124567\}\{39\}+\{134567\}\{29\}\nonumber\\
&+&\{234567\}\{19\}+\{123459\}\{67\}+\{123469\}\{57\}+\{123569\}\{47\}+\{124569\}\{37\}+\{134569\}\{27\}\nonumber\\
&+&\{234569\}\{17\}+\{123479\}\{56\}+\{123579\}\{46\}+\{124579\}\{36\}+\{134579\}\{26\}+\{234579\}\{16\}\nonumber\\
&+&\{123679\}\{45\}+\{124679\}\{35\}+\{134679\}\{25\}+\{234679\}\{15\}+\{125679\}\{34\}+\{135679\}\{24\}\nonumber\\
&+&\{235679\}\{14\}+\{145679\}\{23\}+\{245679\}\{13\}+\{345679\}\{12\}].
\end{eqnarray}
This integral is also non-vanishing in only two other cases $N=2$
and $N=6$. The (conjugated) rotational wave function of the
antidecuplet (\ref{Tensor pentaquarks}) is symmetric with respect to
three flavor indices $h_1,h_2,h_3$. The projection onto the $7Q$
state is thus reduced to
\begin{eqnarray}
&&\int\ud
R\,R^{f_1}_{j_1}R^{f_2}_{j_2}R^{f_3}_{j_3}\left(R^{f_4}_{j_4}R^{\dag
j_5}_{f_5}\right)\left(R_{j_6}^{\dag f_6}R_{f_7}^{\dag
j_7}\right)A_k^{*\{h_1h_2h_3\}}(R)=\frac{1}{25920}\nonumber\\
&\times&\Big\{\left[\delta^{f_1}_{f_5}\{72346\}\left(5\delta^{j_5}_{j_1}\delta^{j_7}_k-\delta^{j_7}_{j_1}\delta^{j_5}_k\right)+\delta^{f_1}_{f_7}\{52346\}\left(5\delta^{j_7}_{j_1}\delta^{j_5}_k-\delta^{j_5}_{j_1}\delta^{j_7}_k\right)\right]\nonumber\\
&+&\left[\delta^{f_1}_{f_5}\{72436\}\left(5\delta^{j_5}_{j_1}\delta^{j_7}_k-\delta^{j_7}_{j_1}\delta^{j_5}_k\right)+\delta^{f_1}_{f_7}\{52436\}\left(5\delta^{j_7}_{j_1}\delta^{j_5}_k-\delta^{j_5}_{j_1}\delta^{j_7}_k\right)\right]\nonumber\\
&+&\left[\delta^{f_1}_{f_5}\{72634\}\left(5\delta^{j_5}_{j_1}\delta^{j_7}_k-\delta^{j_7}_{j_1}\delta^{j_5}_k\right)+\delta^{f_1}_{f_7}\{52634\}\left(5\delta^{j_7}_{j_1}\delta^{j_5}_k-\delta^{j_5}_{j_1}\delta^{j_7}_k\right)\right]\nonumber\\
&+&\left[\delta^{f_2}_{f_5}\{71346\}\left(5\delta^{j_5}_{j_2}\delta^{j_7}_k-\delta^{j_7}_{j_2}\delta^{j_5}_k\right)+\delta^{f_2}_{f_7}\{51346\}\left(5\delta^{j_7}_{j_2}\delta^{j_5}_k-\delta^{j_5}_{j_2}\delta^{j_7}_k\right)\right]\nonumber\\
&+&\left[\delta^{f_2}_{f_5}\{71436\}\left(5\delta^{j_5}_{j_2}\delta^{j_7}_k-\delta^{j_7}_{j_2}\delta^{j_5}_k\right)+\delta^{f_2}_{f_7}\{51436\}\left(5\delta^{j_7}_{j_2}\delta^{j_5}_k-\delta^{j_5}_{j_2}\delta^{j_7}_k\right)\right]\nonumber\\
&+&\left[\delta^{f_2}_{f_5}\{71634\}\left(5\delta^{j_5}_{j_2}\delta^{j_7}_k-\delta^{j_7}_{j_2}\delta^{j_5}_k\right)+\delta^{f_2}_{f_7}\{51634\}\left(5\delta^{j_7}_{j_2}\delta^{j_5}_k-\delta^{j_5}_{j_2}\delta^{j_7}_k\right)\right]\nonumber\\
&+&\left[\delta^{f_3}_{f_5}\{71246\}\left(5\delta^{j_5}_{j_3}\delta^{j_7}_k-\delta^{j_7}_{j_3}\delta^{j_5}_k\right)+\delta^{f_3}_{f_7}\{51246\}\left(5\delta^{j_7}_{j_3}\delta^{j_5}_k-\delta^{j_5}_{j_3}\delta^{j_7}_k\right)\right]\nonumber\\
&+&\left[\delta^{f_3}_{f_5}\{71426\}\left(5\delta^{j_5}_{j_3}\delta^{j_7}_k-\delta^{j_7}_{j_3}\delta^{j_5}_k\right)+\delta^{f_3}_{f_7}\{51426\}\left(5\delta^{j_7}_{j_3}\delta^{j_5}_k-\delta^{j_5}_{j_3}\delta^{j_7}_k\right)\right]\nonumber\end{eqnarray}\begin{eqnarray}
&+&\left[\delta^{f_3}_{f_5}\{71624\}\left(5\delta^{j_5}_{j_3}\delta^{j_7}_k-\delta^{j_7}_{j_3}\delta^{j_5}_k\right)+\delta^{f_3}_{f_7}\{51624\}\left(5\delta^{j_7}_{j_3}\delta^{j_5}_k-\delta^{j_5}_{j_3}\delta^{j_7}_k\right)\right]\nonumber\\
&+&\left[\delta^{f_4}_{f_5}\{71236\}\left(5\delta^{j_5}_{j_4}\delta^{j_7}_k-\delta^{j_7}_{j_4}\delta^{j_5}_k\right)+\delta^{f_4}_{f_7}\{51236\}\left(5\delta^{j_7}_{j_4}\delta^{j_5}_k-\delta^{j_5}_{j_4}\delta^{j_7}_k\right)\right]\nonumber\\
&+&\left[\delta^{f_4}_{f_5}\{71326\}\left(5\delta^{j_5}_{j_4}\delta^{j_7}_k-\delta^{j_7}_{j_4}\delta^{j_5}_k\right)+\delta^{f_4}_{f_7}\{51326\}\left(5\delta^{j_7}_{j_4}\delta^{j_5}_k-\delta^{j_5}_{j_4}\delta^{j_7}_k\right)\right]\nonumber\\
&+&\left[\delta^{f_4}_{f_5}\{71623\}\left(5\delta^{j_5}_{j_4}\delta^{j_7}_k-\delta^{j_7}_{j_4}\delta^{j_5}_k\right)+\delta^{f_4}_{f_7}\{51623\}\left(5\delta^{j_7}_{j_4}\delta^{j_5}_k-\delta^{j_5}_{j_4}\delta^{j_7}_k\right)\right]\nonumber\\
&+&\left[\delta^{f_6}_{f_5}\{71234\}\left(5\delta^{j_5}_{j_6}\delta^{j_7}_k-\delta^{j_7}_{j_6}\delta^{j_5}_k\right)+\delta^{f_6}_{f_7}\{51234\}\left(5\delta^{j_7}_{j_6}\delta^{j_5}_k-\delta^{j_5}_{j_6}\delta^{j_7}_k\right)\right]\nonumber\\
&+&\left[\delta^{f_6}_{f_5}\{71324\}\left(5\delta^{j_5}_{j_6}\delta^{j_7}_k-\delta^{j_7}_{j_6}\delta^{j_5}_k\right)+\delta^{f_6}_{f_7}\{51324\}\left(5\delta^{j_7}_{j_6}\delta^{j_5}_k-\delta^{j_5}_{j_6}\delta^{j_7}_k\right)\right]\nonumber\\
&+&\left[\delta^{f_6}_{f_5}\{71423\}\left(5\delta^{j_5}_{j_6}\delta^{j_7}_k-\delta^{j_7}_{j_6}\delta^{j_5}_k\right)+\delta^{f_6}_{f_7}\{51423\}\left(5\delta^{j_7}_{j_6}\delta^{j_5}_k-\delta^{j_5}_{j_6}\delta^{j_7}_k\right)\right]\Big\}.
\end{eqnarray}

\renewcommand{\theequation}{B\arabic{equation}}
\setcounter{equation}{0}
\renewcommand{\thefigure}{B\arabic{figure}}
\setcounter{figure}{0}

\chapter{General tools for the $nQ$ Fock component}

In this appendix we will give general remarks and ``tricks'' that
help to derive easily the contributions of \emph{any} Fock
component. We will show that schematic diagrams drawn by Diakonov
and Petrov \cite{DiaPet} are a key tool that allows one to rapidly
give the sign, the spin-flavor structure, the number of equivalent
annihilation-creation operator contractions and the factor coming
from color contractions for any such diagram. We first give the
general rules and then apply them to the $7Q$ Fock component.

\begin{enumerate}
\item
First remember that dark gray rectangles of the diagrams stand for
the three valence quarks and light gray rectangles for
quark-antiquark pairs. Each line represents the color, flavor and
spin contractions
\begin{equation}
\delta^{\alpha_i}_{\alpha'_i}\delta^{f_i}_{f'_i}\delta^{\sigma_i}_{\sigma'_i}
\int\ud
z'_i\,\ud^2\up'_{i\perp}\delta(z_i-z'_i)\delta^{(2)}(\up_{i\perp}-\up'_{i\perp}).
\end{equation}
The reversed arrow stands for the antiquark.
\item
For any $nQ$ Fock component there are $(n+3)/2$ quark creation
operators and $(n-3)/2$ antiquark creation operators. The total
number of annihilation-creation operator contractions is then
\begin{equation}
\left(\frac{n+3}{2}\right)!\left(\frac{n-3}{2}\right)!
\end{equation}
This means that for the $3Q$ component there are 6
annihilation-creation operator contractions and 24 for the $5Q$
component.
\item
The number of line crossings $N$ gives the sign of the
annihilation-creation operator contractions $(-1)^N$. Indeed, any
line crossing represents an anticommutation of operators.
\item
The color structure of the valence quarks is
$\epsilon^{\alpha_1\alpha_2\alpha_3}$ and for the quark-antiquark
pair it is $\delta^{\alpha_4}_{\alpha_5}$. So if one considers
color, the antiquark line and the quark line of the same pair can be
connected and then belong to the same circuit. The color factor is
at least 3! due to the contraction of both $\epsilon$'s with
possibly a minus sign. There is another factor of 3 for any circuit
that is not connected to the valence quarks.
\item
The valence quarks are equivalent which means that different
contractions of the same valence quarks are equivalent. Indeed any
sign coming from the crossings in rule 3 is compensated by the same
sign coming from the $\epsilon$ color contraction in rule 4. That is
the reason why one needs to draw only one diagram for the $3Q$
component.
\item
The quark-antiquark pairs are equivalent which means that any
vertical exchange of the light gray rectangles (quark and antiquark
lines stay fixed to the rectangles) does not produce a new type of
diagram. This appears only from the $7Q$ component since one needs
at least two quark-antiquark pairs.
\end{enumerate}
So for the $5Q$ component there are only two types of diagrams. The
direct one has no crossing and is thus positive while the exchange
one is negative due to one crossing. There are 6 equivalent direct
annihilation-creation contractions and the color factor is $3!\cdot
3$ (there is an independent color circuit within the quark-antiquark
pair). There are 18 equivalent exchange annihilation-creation
contractions but the color factor is only 3! since the pair lines
belong to a valence circuit. This is exactly what was said in
subsection 3.6.2. Of course there are $6+18=24$
annihilation-creation operator contractions for the $5Q$ component
as stated by rule 2.
\begin{figure}[h]\begin{center}\includegraphics[width=15cm]{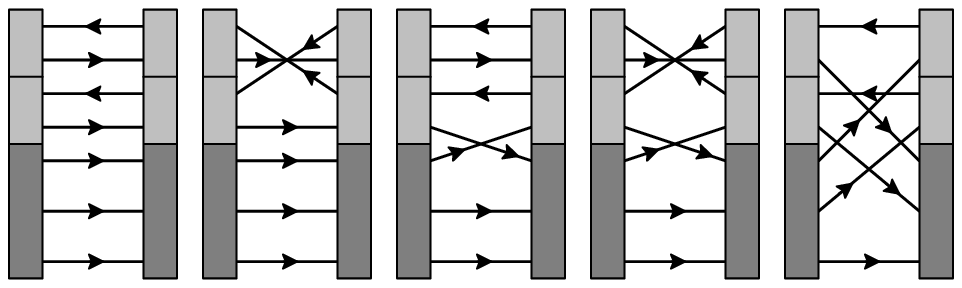}
\caption{\small{Schematic representation of the $7Q$ contributions
to the normalization.}}\label{Sevenquarks}\end{center}
\end{figure}

Let us now apply these rules to see what happens when one considers
the $7Q$ Fock component. From rules 5 and 6 we obtain that there are
only five types of diagrams, see Fig \ref{Sevenquarks}.

Let us find the signs. These prototype diagrams have been chosen
such that color contractions do not affect the sign. The first
diagram is obviously positive (no crossing).
\begin{figure}[h!]\begin{center}\begin{minipage}[c]{7cm}\begin{center}\includegraphics[width=3cm]{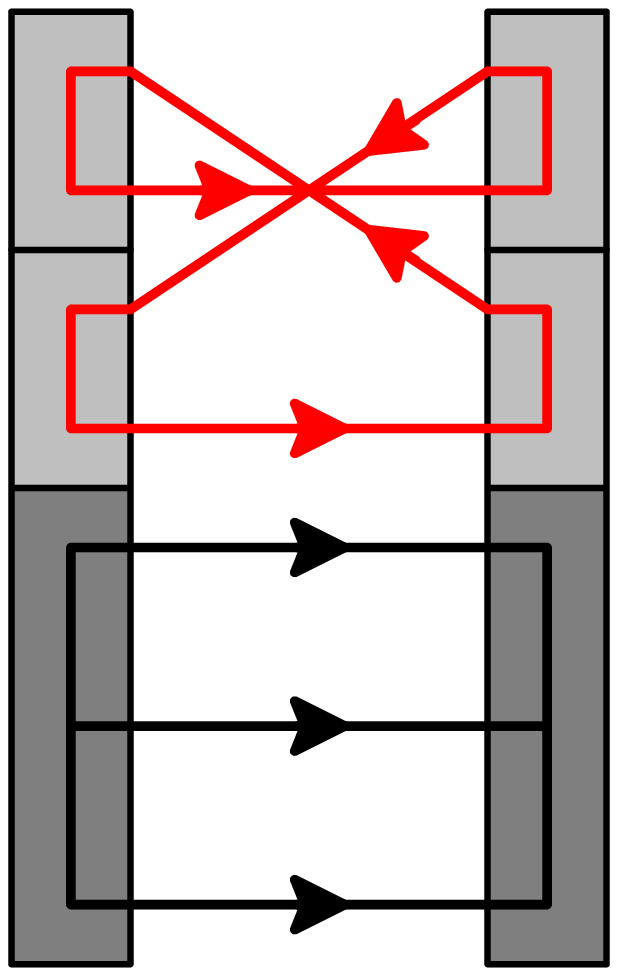}
\caption{\small{The color factor of this diagram is $3!\cdot 3$
since one has the valence circuit and an independent
circuit.\newline}}\label{Independent circuit}
\end{center}\end{minipage}\hspace{1cm}
\begin{minipage}[c]{7cm}\begin{center}\includegraphics[width=3cm]{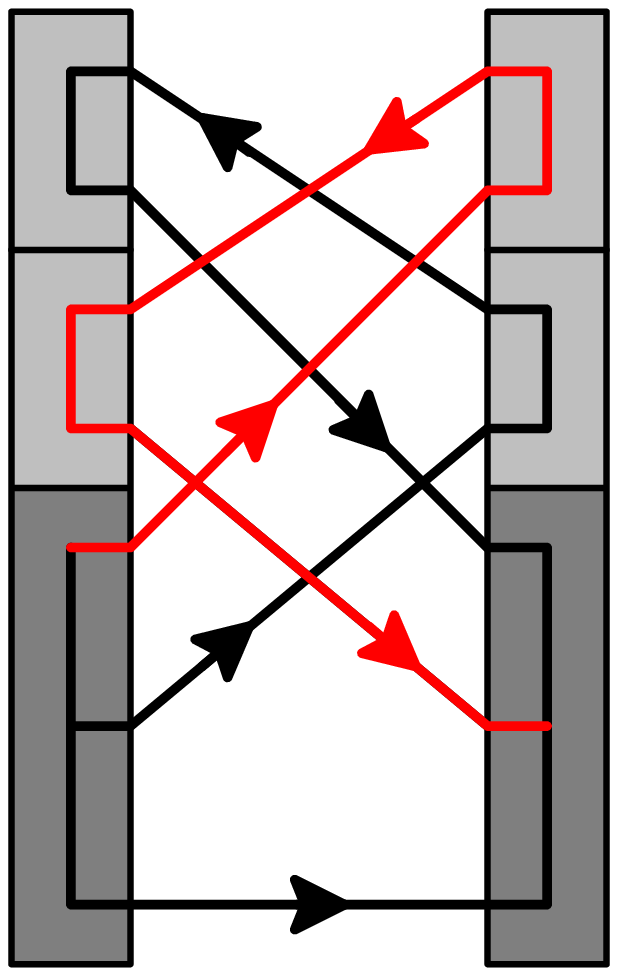}
\caption{\small{The color contractions in this diagram give a minus
factor because of interchange of two valence
quarks.\newline}}\label{Color sign}
\end{center}\end{minipage}\end{center}
\end{figure}The second one has three crossings (they are
degenerate in the drawing but it does not change anything
considering one or three crossings since the important thing is that
it is odd) and is thus negative. So is the third one with its unique
crossing. The fourth diagram has four crossings and is thus
positive. The last one has six crossings and is thus also positive.

Following rule 2 there must be $5!2!=240$ contractions. Indeed,
there are 12 of the first and second types while there are 72 of the
other ones. Thus we have $2\cdot 12+3\cdot 72=240$ contractions as
expected.

The color factor of the first diagram is $3!\cdot 3\cdot 3=54$ since
there are two independent circuits. The color factor of the second
one is only $3!\cdot 3=18$ since there is only one independent
circuit as one can see on Fig. \ref{Independent circuit}. The third
diagram has also a unique independent circuit and thus a color
factor of $3!\cdot 3=18$. For the two last diagrams there are no
more independent circuit and we have consequently a color factor of
$3!=6$.

We close this appendix by considering the diagram in Fig. \ref{Color
sign}. Since two valence quarks are exchanged, it must belong to the
fifth type of diagrams. There are seven crossings and thus a
negative sign while the fifth type of diagrams is positive. In fact,
for this particular diagram, the color contractions gives an
additional minus sign since the third quark on the left is
contracted with the second on the right
$\epsilon^{\alpha_1\alpha_2\alpha_3}\epsilon_{\alpha_1\alpha_3\alpha_2}=-6$.

\newpage\thispagestyle{empty}\cleardoublepage 

\end{document}